\pdfoutput=1
\PassOptionsToPackage{unicode}{hyperref}
\PassOptionsToPackage{hyphens}{url}
\PassOptionsToPackage{dvipsnames,svgnames,x11names}{xcolor}
\documentclass[12pt]{article}

\usepackage{amsmath,amssymb,amsfonts,amsthm}
\usepackage[T1]{fontenc}
\usepackage[utf8]{inputenc}
\usepackage{textcomp} 
\usepackage{lmodern}
\IfFileExists{upquote.sty}{\usepackage{upquote}}{}
\IfFileExists{microtype.sty}{
  \usepackage[]{microtype}
  \UseMicrotypeSet[protrusion]{basicmath} 
}{}
\makeatletter
\@ifundefined{KOMAClassName}{
  \IfFileExists{parskip.sty}{%
    \usepackage{parskip}
  }{
    \setlength{\parindent}{0pt}
    \setlength{\parskip}{6pt plus 2pt minus 1pt}}
}{
  \KOMAoptions{parskip=half}}
\makeatother
\usepackage{xcolor}
\makeatletter
\ifx\paragraph\undefined\else
  \let\oldparagraph\paragraph
  \renewcommand{\paragraph}{
    \@ifstar
      \xxxParagraphStar
      \xxxParagraphNoStar
  }
  \newcommand{\xxxParagraphStar}[1]{\oldparagraph*{#1}\mbox{}}
  \newcommand{\xxxParagraphNoStar}[1]{\oldparagraph{#1}\mbox{}}
\fi
\ifx\subparagraph\undefined\else
  \let\oldsubparagraph\subparagraph
  \renewcommand{\subparagraph}{
    \@ifstar
      \xxxSubParagraphStar
      \xxxSubParagraphNoStar
  }
  \newcommand{\xxxSubParagraphStar}[1]{\oldsubparagraph*{#1}\mbox{}}
  \newcommand{\xxxSubParagraphNoStar}[1]{\oldsubparagraph{#1}\mbox{}}
\fi
\makeatother

\usepackage{longtable,booktabs,array}
\usepackage{calc} 
\usepackage{etoolbox}
\makeatletter
\patchcmd\longtable{\par}{\if@noskipsec\mbox{}\fi\par}{}{}
\makeatother
\IfFileExists{footnotehyper.sty}{\usepackage{footnotehyper}}{\usepackage{footnote}}
\makesavenoteenv{longtable}
\usepackage{graphicx}
\makeatletter
\def\maxwidth{\ifdim\Gin@nat@width>\linewidth\linewidth\else\Gin@nat@width\fi}
\def\maxheight{\ifdim\Gin@nat@height>\textheight\textheight\else\Gin@nat@height\fi}
\makeatother
\setkeys{Gin}{width=\maxwidth,height=\maxheight,keepaspectratio}
\makeatletter
\def\fps@figure{htbp}
\makeatother

\makeatletter
\@ifpackageloaded{caption}{}{\usepackage{caption}}
\AtBeginDocument{%
\ifdefined\contentsname
  \renewcommand*\contentsname{Table of contents}
\else
  \newcommand\contentsname{Table of contents}
\fi
\ifdefined\listfigurename
  \renewcommand*\listfigurename{List of Figures}
\else
  \newcommand\listfigurename{List of Figures}
\fi
\ifdefined\listtablename
  \renewcommand*\listtablename{List of Tables}
\else
  \newcommand\listtablename{List of Tables}
\fi
\ifdefined\figurename
  \renewcommand*\figurename{Figure}
\else
  \newcommand\figurename{Figure}
\fi
\ifdefined\tablename
  \renewcommand*\tablename{Table}
\else
  \newcommand\tablename{Table}
\fi
}
\@ifpackageloaded{float}{}{\usepackage{float}}
\floatstyle{ruled}
\@ifundefined{c@chapter}{\newfloat{codelisting}{h}{lop}}{\newfloat{codelisting}{h}{lop}[chapter]}
\floatname{codelisting}{Listing}

\makeatother
\makeatletter
\@ifpackageloaded{caption}{}{\usepackage{caption}}
\@ifpackageloaded{subcaption}{}{\usepackage{subcaption}}
\makeatother

\usepackage[]{natbib}
\usepackage{bookmark}

\IfFileExists{xurl.sty}{\usepackage{xurl}}{} 
\hypersetup{
  pdftitle={Forests for Differences: Beyond Parametric Difference-in-Differences},
  pdfauthor={Hugo Gobato Souto; Francisco Louzada Neto},
  pdfkeywords={Bayesian machine learning, Causal inference, Heterogeneous treatment effects, Semiparametric estimation, Staggered adoption},
  colorlinks=true,
  linkcolor={blue},
  filecolor={Maroon},
  citecolor={Blue},
  urlcolor={Blue},
  pdfcreator={LaTeX via pandoc}}

\usepackage{bm}
\usepackage{multirow}
\usepackage[linesnumbered,ruled,vlined]{algorithm2e}
\usepackage{lscape}
\usepackage{booktabs}
\usepackage{threeparttable}
\usepackage{array}
\usepackage{stackengine}
\usepackage{enumitem}
\usepackage{mathrsfs}

\graphicspath{{sim_figures/}{images/}}

\let\parencite\citep
\let\textcite\citet
\let\cite\citep

\AtBeginDocument{\setlength{\bibsep}{0pt}}

\newcommand{\anon}{1}

\newcommand{\taug}{\tau_{\mathrm{glb}}}

\newcommand{\E}{\mathbb{E}}
\newcommand{\att}{\mathrm{ATT}}
\newcommand{\gatt}{\mathrm{GATT}}

\newcommand{\R}{\mathbb{R}}
\newcommand{\ind}[1]{\mathbf{1}\{#1\}}
\newcommand{\ATT}{\mathrm{ATT}}
\newcommand{\GATT}{\mathrm{GATT}}
\newcommand{\CATT}{\mathrm{CATT}}

\newcommand{\cG}{\mathcal{G}}

\newcommand{\cH}{\mathcal{H}}

\newcommand{\cS}{\mathcal{S}}

\newcommand{\cD}{\mathcal{D}}

\newcommand{\Var}{\mathrm{Var}}

\newcolumntype{L}[1]{>{\raggedright\arraybackslash}p{#1}}

\theoremstyle{plain}
\newtheorem{theorem}{Theorem}
\newtheorem{lemma}[theorem]{Lemma}
\newtheorem{proposition}[theorem]{Proposition}
\newtheorem{corollary}[theorem]{Corollary}
\theoremstyle{definition}

\theoremstyle{remark}
\newtheorem{remark}[theorem]{Remark}

\begin{document}

\def\spacingset#1{\renewcommand{\baselinestretch}%
{#1}\small\normalsize} \spacingset{1}


\if1\anon
{
  \title{\bf Forests for Differences: Beyond Parametric Difference-in-Differences}
  \author{Hugo Gobato Souto\thanks{
    Institute of Mathematics and Computer Sciences (ICMC), University of São Paulo (USP), São Carlos, Brazil  (email: hugogobatosouto@gmail.com).}\hspace{.2cm}\\
    University of São Paulo (USP), São Carlos - SP, Brazil\\
    and \\
    Francisco Louzada Neto\thanks{
    Institute of Mathematics and Computer Sciences (ICMC), University of São Paulo (USP), São Carlos, Brazil (email: louzada@icmc.usp.br).}\\
   University of São Paulo (USP), São Carlos - SP, Brazil}
  \maketitle
} \fi

\if0\anon
{
  \bigskip
  \bigskip
  \bigskip
  \begin{center}
    {\LARGE\bf Forests for Differences: Beyond Parametric Difference-in-Differences}
  \end{center}
  \medskip
} \fi

\bigskip
\begin{abstract}

This paper introduces the Difference-in-Differences Bayesian Causal Forest (DiD-BCF), a novel non-parametric model addressing key challenges in DiD estimation, such as staggered adoption and heterogeneous treatment effects. DiD-BCF provides a unified framework for estimating Average (ATE), Group-Average (GATE), and Conditional Average Treatment Effects (CATE). By structuring separate regression forests for cohort-specific levels, common untreated trends and treatment effects, the model directly embeds conditional parallel trends and no-anticipation restrictions into Bayesian nonparametrics. We establish non-parametric identification and integrated treatment effect consistency of the exact treatment effect posterior distribution. Extensive simulations demonstrate DiD-BCF’s superior performance over established benchmarks, particularly under non-linearity, selection biases, and effect heterogeneity. Applied to U.S. minimum wage policy, the model uncovers significant conditional treatment effect heterogeneity related to county population, insights obscured by traditional
methods. DiD-BCF offers a robust and versatile tool for more nuanced causal inference in modern DiD applications.
\end{abstract}

\noindent%
\vfill

\newpage
\spacingset{1.8} 

\section{Introduction}\label{intro}
Difference-in-differences (DiD) identifies causal treatment effects in longitudinal and panel settings by contrasting temporal changes in outcomes between treated and untreated groups. Its core identifying assumption is parallel untreated trends, which can hold unconditionally or after conditioning on observed pre-treatment covariates. This formulation accommodates time-invariant unobserved confounders that generate persistent level differences across groups, while ruling out unmodeled divergences in untreated trajectories \citep{SantAnna2020,roth2023synthesis}. In staggered-adoption settings where treatment timing varies across units, heterogeneous treatment effects introduce substantial complications: standard two-way fixed-effects (TWFE) regressions construct implicit comparisons where already-treated units act as controls, yielding weighted combinations of treatment effects whose weights can be negative or unrepresentative of any policy-relevant target \citep{GoodmanBacon2021,deChaisemartin2020}.

A vibrant modern literature resolves these aggregation pitfalls through cohort-and-time specific estimands and explicit aggregation schemes \citep{Callaway2021,Sun2021,deChaisemartin2020}. Doubly robust outcome-regression and inverse-probability weighting estimators accommodate high-dimensional observed confounding \citep{SantAnna2020}, while synthetic difference-in-differences pairs unit and time reweighting with panel regression \citep{Arkhangelsky2021}. Although these methods clarify identification and average treatment effect estimation, an important methodological challenge remains: how to flexibly uncover fine-grained, covariate-specific effect heterogeneity when panels are of moderate size and outcome dynamics exhibit complex nonlinearities.

Bayesian causal forests (BCF) address effect heterogeneity in cross-sectional observational studies by separating the prognostic outcome regression from the treatment-effect surface and placing differentiated shrinkage priors on each component \citep{Hahn2020}. This separation is particularly advantageous in panel difference-in-differences settings, where baseline untreated trajectories are often complex and driven by multiple covariates, whereas treatment effects may possess a smoother or simpler structure. However, fitting an unconstrained nonparametric panel regression violates the core logic of difference-in-differences. An unrestricted model can attribute the same post-intervention outcome divergence either to idiosyncratic cohort-specific trends or to the causal treatment effect, rendering the causal parameter unidentified without structural constraints.

We introduce Difference-in-Differences Bayesian Causal Forests (DiD-BCF), a structured Bayesian forest architecture tailored for panel designs with common or staggered intervention timing. In DiD-BCF, the untreated conditional mean decomposes additively into a cohort-covariate level surface and a calendar-time-covariate trend surface. Cohort membership can shift baseline outcome levels across covariate profiles, but cannot alter the shape of their temporal trajectories. A distinct treatment-effect forest, scaled by the post-treatment indicator, captures treatment effect heterogeneity across baseline covariates, adoption cohorts, and elapsed event time. Posterior draws of the joint effect surface naturally generate coherent posterior distributions for conditional average treatment effects on the treated (CATT), cohort-time specific parameters (GATT), and arbitrary weighted aggregates, bypassing the need for multi-step plug-in inference.

The paper makes three primary contributions: (1) we establish the non-parametric identification of DiD-BCF and prove integrated posterior consistency for the exact continuous-cut forest posterior over joint finite-tree truth classes, with a growing-capacity extension for scalar H\"older continuous functions of order exceeding one-half, while characterizing the mean-effect transportability required for population-level estimands and the role of working covariance specifications; (2) we conduct extensive Monte Carlo simulations demonstrating competitive point accuracy for both aggregate treatment effects and conditional response surfaces relative to leading parametric, semiparametric, and machine-learning benchmarks, while delineating how within-unit serial dependence affects posterior interval coverage; and (3) we introduce an unconstrained pre-treatment lead specification that diagnoses departures from parallel trends before policy implementation, while clarifying the boundary conditions of pre-testing when low-level violations evade detection yet induce non-negligible post-treatment bias.

The distinction between point estimation efficiency and inferential validity is central to our analysis. Exploiting the full panel within a structured Bayesian model substantially reduces finite-sample estimation variance relative to pairwise two-period differencing. However, shrinkage regularization and full-panel pooling do not automatically ensure that posterior credible intervals achieve nominal frequentist coverage. By analyzing point accuracy and uncertainty quantification separately, we provide an objective, transparent assessment of the strengths and boundary conditions of Bayesian forest methods in panel data econometrics.

The remainder of the paper is organized as follows. Section~\ref{model_did_bcf} introduces the DiD-BCF model, parameterization, and computational algorithm. Section~\ref{theory_did_bcf} presents the theoretical identification, consistency, and asymptotic distribution results. Sections~\ref{sim_study_did_bcf} and \ref{results_did_bcf} detail the simulation design and discuss comparative empirical results across all benchmark models. Section~\ref{real_life_did_bcf} applies the method to evaluate the employment impacts of minimum wage legislation across United States counties. Section~\ref{conclusion_did_bcf} concludes. Methodological proofs, implementation details, algorithmic specifications, and comprehensive simulation tabulations are compiled in the Appendix.

\section{DiD-BCF Model}\label{model_did_bcf}
\subsection{Panel structure and estimands}
Consider an observational panel tracking $n$ units across $T$ discrete time periods, $t \in \{1, \ldots, T\}$. For each unit $i \in \{1, \ldots, n\}$, let $G_i \in \mathcal{G} \cup \{\infty\}$ denote its initial adoption period (cohort), where $\mathcal{G} \subseteq \{2, \ldots, T\}$ and $G_i = \infty$ designates units that remain untreated throughout the study window. Under absorbing treatment, the binary treatment indicator is $D_{it} = \mathbf{1}\{G_i < \infty, \, t \ge G_i\}$. Units possess pre-treatment baseline covariates $X_i \in \mathbb{R}^p$, and for treated units with $G_i < \infty$, the elapsed event time (relative exposure) is defined as $k_{it} = t - G_i \ge 0$.

Let $Y_{it}(g)$ denote the potential outcome that unit $i$ would realize in period $t$ if assigned to adoption cohort $g$, with $Y_{it}(0)$ denoting the potential outcome under never-treated status. The observed outcome is $Y_{it} = (1 - D_{it}) Y_{it}(0) + D_{it} Y_{it}(G_i)$. The primary building block of causal heterogeneity is the conditional average treatment effect on the treated (CATT):
\begin{equation}
\mathrm{CATT}(x; g, t) = \mathbb{E}[Y_{it}(g) - Y_{it}(0) \mid G_i = g, \, X_i = x], \qquad t \ge g.
\end{equation}
Averaging $\mathrm{CATT}(X_i; g, t)$ over the conditional distribution of covariates within cohort $g$ yields the group-time average treatment effect, $\mathrm{GATT}(g, t) = \mathbb{E}[\mathrm{CATT}(X_i; g, t) \mid G_i = g]$. Standard summary estimands, including event-study effects $\mathrm{ES}(k)$ for exposure duration $k \ge 0$ and the overall average treatment effect on the treated ($\mathrm{ATT}$), are constructed as convex combinations of these cohort-time cells \citep{Callaway2021}. In sample-level empirical evaluations, the overall ATT weights observed treated unit-periods equally, allocating weight proportionally to cohorts with longer post-treatment durations. In Section~\ref{theory_did_bcf}, we formally distinguish these realized sample-average estimands from population-level targets.

\subsection{A structured outcome model}
To balance nonparametric flexibility with causal identification, DiD-BCF models the conditional expectation of the observed outcome via the structured additive decomposition:
\begin{equation}\label{eq:adjusted_did_bcf}
f(g, t, x) = a(g, x) + b(t, x) + \mathbf{1}\{t \ge g\} \tau(g, k, x), \quad \text{where } k = t - g.
\end{equation}
Equation~\eqref{eq:adjusted_did_bcf} embeds the foundational identification restrictions of difference-in-differences directly into the regression architecture. The baseline level surface $a(g, x)$ captures time-invariant baseline differences across adoption cohorts and covariates, accommodating unobserved confounders that determine treatment timing. In contrast, the temporal trend surface $b(t, x)$ depends on calendar time and covariates, but is strictly invariant to cohort status $g$. Consequently, the counterfactual untreated evolution between any two periods $t$ and $t'$ at fixed covariates $x$ is given by $b(t', x) - b(t, x)$, which is identical across all cohorts. This formulation operationalizes conditional parallel untreated trends, distinguishing DiD-BCF from unrestricted nonparametric panel regressions that cannot disentangle cohort-specific baseline trends from causal treatment impacts.

The treatment effect surface is parameterized as $\tau(g, k, x) = \taug + \tau_{\mathrm{forest}}(g, k, x)$, where $\taug$ is a global intercept capturing the baseline magnitude of the effect and $\tau_{\mathrm{forest}}$ is a regression forest capturing effect moderation across covariates, cohort, and event time. Because this component is multiplied by the post-treatment indicator $\mathbf{1}\{t \ge g\}$, pre-treatment observations ($t < g$) contribute exclusively to estimating the untreated prognostic surfaces $a(g, x)$ and $b(t, x)$. The causal framework imposes no anticipation and restricts effect evaluation to non-negative event times ($k \ge 0$). Appendix~\ref{supp:reparam} proves the observable-mean equivalence between this parameterization and an unconstrained ever-treated model constrained to zero pre-treatment effects.

\begin{samepage}
Because the main causal model structurally sets pre-treatment effects to zero via $\mathbf{1}\{t \ge g\}$ to enforce identification, it cannot test pre-treatment trends internally. We therefore assess pre-treatment dynamics via an auxiliary diagnostic model with unrestricted leads:
\begin{equation}\label{eq:pretrend_lead}
f^{\mathrm{pl}}(g,t,x) = a(g,x) + b(t,x)
  + \mathbf{1}\{g \in \mathcal{G}\}\lambda_g(x,t-g),
\qquad \lambda_g(x,-1) \equiv 0.
\end{equation}
\end{samepage}
Normalizing at $k = -1$ absorbs cohort fixed effects. Under conditional parallel trends and no anticipation, leads satisfy $\lambda_g(x,k) = 0$ for all $(g, x)$ and $k \le -2$. Consequently, parallel trends is rejected at significance level $\alpha$ if a diagnostic functional $\mathcal{T}(\lambda)$ (e.g., the treated-sample lead $\Delta(k)$) departs from zero with one-sided posterior tail probability $p_{\mathrm{tail}} = \min\{\Pr(\mathcal{T}(\lambda) > 0 \mid Y),\, \Pr(\mathcal{T}(\lambda) < 0 \mid Y)\} < \alpha/2$ (equivalently, two-sided posterior sign score $p_{\mathrm{Bayes}} = 2\,p_{\mathrm{tail}} < \alpha$).

A key advantage of this Bayesian nonparametric formulation is its architectural flexibility: researchers can construct diagnostic functionals tailored to diverse hypothesized identification violations rather than relying on a single rigid specification. In particular, depending on substantive domain assumptions, practitioners can evaluate: (1) secular pre-existing trends via a linear slope $s$ fitted through the origin to $\Delta(k)$ against $k+1$, testing for smooth drift away from the counterfactual trend; (2) localized anticipation shocks or omnibus lead divergence by testing whether any individual lead $\Delta(k)$ departs from zero (applying a multiple-comparison adjustment, such as a Bonferroni threshold $p_{\mathrm{tail}} < \alpha/(2L)$ over the $L$ pre-treatment periods); or (3) subgroup-specific violations by evaluating differential lead profiles across covariate strata, $\Delta(k \mid x) - \Delta(k \mid x')$, to detect heterogeneous departures that might otherwise cancel out in the aggregate sample. In our simulation studies (Section~\ref{results_did_bcf}), we benchmark the empirical power and size of the method using these three operational decision rules.

Each functional component ($a$, $b$, and $\tau_{\mathrm{forest}}$) is represented by an ensemble of regression trees \citep{Chipman2010,Hahn2020}. Their respective feature inputs are $(G_i, X_i)$ for the level surface, $(t, X_i)$ for the trend surface, and $(G_i, k_{it}, X_i)$ for the effect forest. In applications where cohorts are assumed to share identical effect profiles at equal exposure and covariates, $G_i$ may be omitted from the effect forest. Unlike cross-sectional BCF \citep{Hahn2020}, DiD-BCF does not require an estimated propensity score as an explicit input to regularize confounding, relying instead on the structural two-way decomposition to absorb confounding levels, and also given the unnecessary need of an estimated propensity score as an explicit input to regularize confounding for a flexible model such as BCF, as shown in \citet{https://doi.org/10.48550/arxiv.2410.15560}.

Because the components enter additively in Equation~\eqref{eq:adjusted_did_bcf}, arbitrary functions $h(x)$ can shift between $a(g, x)$ and $b(t, x)$, and an additive constant can trade off between $\taug$ and $\tau_{\mathrm{forest}}$. While independent shrinkage priors regularize these redundant dimensions during sampling, point estimation and inference are evaluated strictly on the composite surfaces $\mu_0(g, t, x) = a(g, x) + b(t, x)$ and $\tau(g, k, x) = \taug + \tau_{\mathrm{forest}}(g, k, x)$ (see Section~\ref{theory_did_bcf} and Appendix~\ref{supp:theory} for formal identification and baseline normalization proofs).

\subsection{Priors, computation, and posterior summaries}
Following standard Bayesian tree methodology, outcomes are standardized prior to model estimation and mapped back to their original scale when computing posterior summaries. The prognostic baseline forests ($a$ and $b$) each comprise 125 trees with a maximum depth of 10 and a standard tree-depth penalty where split probability at depth $d$ is $0.95(1 + d)^{-2}$. The treatment-effect forest $\tau_{\mathrm{forest}}$ incorporates stronger shrinkage regularization, deploying 100 trees with maximum depth 5 and split probability $0.25(1 + d)^{-3}$ \citep{Hahn2020}. Leaf terminal parameters are assigned Gaussian priors. Baseline leaf variances are updated via conjugate inverse-gamma updates, whereas the effect-leaf variance is fixed at $0.5/100$ on the standardized scale to guard against overfitting. The global intercept $\taug$ has a zero-mean, unit-variance Gaussian prior on the standardized scale. Complete hyperparameter specifications appear in Appendix~\ref{supp:algo}.

The observational working likelihood assumes conditionally Gaussian disturbances, $\epsilon_{it} \sim \mathcal{N}(0, \sigma^2)$, with an inverse-gamma prior on $\sigma^2$ (shape 2, rate 1). To account for persistent unit-specific heterogeneity, the model can incorporate time-invariant unit random intercepts $\alpha_i \sim \mathcal{N}(0, \sigma_\alpha^2)$. Integrating over $\alpha_i$ induces a compound-symmetric within-unit error covariance structure. While this working specification effectively absorbs equicorrelated level shifts, it does not assume arbitrary autoregressive or unstructured time series dependence, an empirical distinction examined in our simulation studies.

Posterior summaries are approximated by Bayesian backfitting, using grow-from-root initialization, Metropolis--Hastings tree-topology proposals, and conditional draws of leaf values, variances and the global intercept \citep{He2021,pmlr-v206-krantsevich23a}. At each retained iteration $s$, the complete causal surface $\tau^{(s)}(g,k,x)$ is evaluated across treated observations. Draw averages and equal-tailed quantiles provide point and interval summaries. Subgroup differences are computed within each retained state to retain their modeled dependence. 

\section{Theoretical Properties}\label{theory_did_bcf}

This section establishes the formal theoretical foundation of the structured DiD-BCF posterior. We systematically address non-parametric identification, integrated posterior consistency for treatment-effect surfaces, full-panel semiparametric efficiency, and posterior asymptotic normality via Bernstein--von Mises theorems. Throughout, we emphasize the structural distinctions between realized-sample estimands and population-level causal parameters, as well as the conceptual gap between exact posterior distributions and finite-chain algorithmic implementations.

\subsection{Identification}

Identification of the causal estimands defined in Section~\ref{model_did_bcf} relies on consistency of potential outcomes ($Y_{it} = (1 - D_{it})Y_{it}(0) + D_{it}Y_{it}(G_i)$), absorbing treatment assignment, pre-treatment baseline covariates, no anticipation ($Y_{it} = Y_{it}(0)$ for all $t < G_i$), conditional parallel untreated trends, and overlap with the never-treated cohort ($G_i = \infty$). Appendix~\ref{sec:sampling} states the precise moment and support conditions.

Let $\mu_0(g, t, x) = \mathbb{E}[Y_{it}(0) \mid G_i = g, \, X_i = x]$ denote the conditional expectation of the untreated potential outcome, so that the observed conditional mean decomposes as $f(g, t, x) = \mathbb{E}[Y_{it} \mid G_i = g, \, X_i = x] = \mu_0(g, t, x) + \mathbf{1}\{t \ge g\} \tau(g, k, x)$ with event time $k = t - g$. On the common covariate support, conditional parallel trends equates untreated temporal increments across cohorts while permitting different cohort levels.

\begin{lemma}[Two-way characterization of conditional parallel trends]
\label{lem:twoway_main}
Suppose the relevant cohort and covariate supports overlap. Conditional parallel trends holds if and only if there exist measurable functions $a_0(g, x)$ and $b_0(t, x)$ satisfying the baseline normalization $b_0(1, x) = 0$ such that the untreated conditional mean decomposes as
\[
\mu_0(g, t, x) = a_0(g, x) + b_0(t, x).
\]
This two-way additive decomposition is unique under the stated normalization.
\end{lemma}

The complete proof, including uniqueness under the normalization, is given in Appendix~\ref{sec:id}, Lemma~\ref{lem:twoway}.

The structured DiD-BCF regression mean in Equation~\eqref{eq:adjusted_did_bcf} directly enforces this identification restriction by modeling the untreated prognostic surface as $\mu_0(g, t, x) = a(g, x) + b(t, x)$ with separate level and trend forests. When treatment effects are homogeneous across calendar cohorts at identical event times $k$, the causal treatment surface simplifies to $\tau(g, k, x) \equiv \vartheta(x, k) = \taug + \tau_{\mathrm{forest}}(x, k)$.

\begin{theorem}[Nonparametric identification]
\label{thm:id_main}
Under potential outcome consistency, no anticipation, conditional parallel trends, and overlap, for every treated cohort $g$, post-treatment period $t \ge g$, and almost every $x$ in the common support,
\begin{equation}\label{eq:id_main}
\mathrm{CATT}(x; g, t) = \mathbb{E}[Y_{it} - Y_{i, g-1} \mid G_i = g, \, X_i = x] - \mathbb{E}[Y_{it} - Y_{i, g-1} \mid G_i = \infty, \, X_i = x].
\end{equation}
Consequently, $\mathrm{GATT}(g, t)$ and all fixed-weight linear aggregates are nonparametrically identified. If $\mathrm{CATT}$ pools by event time $k$ under the regression mean $f(g, t, x) = \mu_0(g, t, x) + \mathbf{1}\{t \ge g\}\vartheta(x, k)$, the total causal effect equals the four-cell contrast:
\begin{equation}\label{eq:contrast_main}
\vartheta(x, k) = f(g, t, x) - f(g, g - 1, x) - f(\infty, t, x) + f(\infty, g - 1, x).
\end{equation}
\end{theorem}

The complete proof is given in Appendix~\ref{sec:id}, Theorem~\ref{thm:id}. It derives the long-difference identity from conditional parallel trends, establishes identification on the treated covariate support, and obtains the aggregate and four-cell conclusions.

Because an additive constant can be shifted between the global intercept $\taug$ and the leaf parameters of $\tau_{\mathrm{forest}}$ without altering the likelihood, causal inference centers strictly on the composite effect $\tau(g, k, x)$. Theorem~\ref{thm:id_main} identifies cohort-specific effects; event-time pooling is an additional restriction. The posterior consistency results below also accommodate effect forests that split on cohort labels $G_i$.

\subsection{Exact-posterior forest consistency}\label{sec:forest_consistency_main}

Having established causal identification, we investigate whether the exact posterior distribution over the treatment effect surface concentrates around the true causal data-generating process as sample size grows. In Bayesian nonparametric regression, establishing posterior consistency requires proving that the prior assigns sufficient probability to Kullback--Leibler neighborhoods of the truth, while controlling the metric entropy of the functional parameter space so that posterior mass does not disperse over spurious configurations. We resolve this question in two complementary settings: first, a fixed-capacity benchmark where the true surfaces are representable by regression trees, and second, a growing-capacity extension where the true surfaces are arbitrary smooth functions approximated by trees whose maximum depth expands with sample size.

We consider an i.i.d. balanced panel of $n$ cross-sectional units tracked across $T \ge 2$ fixed time periods, with adoption cohorts $\mathcal{G} \cup \{\infty\}$ and pre-treatment baseline period $t = 1$. The covariate vector $X$ takes values in $[0, 1]^p$ with density bounded away from zero and infinity, and cohort assignment satisfies strong overlap, $P(G = g \mid X = x) \ge e_* > 0$ for all cohorts $g$. The true conditional expectation decomposes as $f_0(g, t, x) = a_0(g, x) + b_0(t, x) + D_{gt} \tau_{gt, 0}(x)$, where $D_{gt} = \mathbf{1}\{t \ge g, g < \infty\}$ indicates active treatment and $\tau_{gt, 0}(x) \equiv \tau_0(g, k, x)$ denotes the true cohort-time treatment effect surface. Observed outcomes satisfy $Y_{it} = f_0(G_i, t, X_i) + \epsilon_{it}$, with conditionally Gaussian disturbances $\epsilon_i \mid X_i, G_i \sim \mathcal{N}(0, v_0 I_T)$ and residual variance $v_0 > 0$. Let $\Pi_n(\cdot \mid \mathcal{D}_n)$ denote the exact posterior measure induced by combining the Gaussian likelihood with the structured continuous-cut forest priors described in Section~\ref{model_did_bcf}.

We first analyze the benchmark regime where tree counts and maximum depth caps are fixed. Suppose the true data-generating functions $(a_0, b_0, \tau_0)$ belong to the class of functions representable by finite ensembles of regression trees matching the architectural constraints of the prior (namely, positive probability under the observed covariate distribution for each partition cell and child node). Under this specification, the parameter space possesses finite metric entropy, enabling an exact characterization of posterior contraction.

\begin{theorem}[Fixed-capacity integrated effect consistency]
\label{thm:forest_fixed_main}
Under the fixed-capacity tree architecture and strong overlap, for every post-treatment coordinate $(g, t)$ with $t \ge g$ and any fixed constant $\eta \in (0, 1/4)$,
\begin{equation}\label{eq:forest_fixed_main}
\int \|\tau_{gt} - \tau_{gt, 0}\|_{L_2(P_X)}^2 \, d\Pi_n(\cdot \mid \mathcal{D}_n) = O_P(n^{-2\eta} \log n) = o_P(1).
\end{equation}
The bound holds identically under the empirical covariate norm, $n^{-1}\sum_{i=1}^n |\tau_{gt}(X_i) - \tau_{gt, 0}(X_i)|^2$. Consequently, posterior-mean treatment effect estimates are consistent in integrated $L_2$ loss, and sample-level averages over treated units achieve posterior consistency.
\end{theorem}

The mathematical mechanism behind Theorem~\ref{thm:forest_fixed_main} (detailed in Appendix~\ref{rfcons:fixed}, Theorem~\ref{rfcons:f:thm:fixedeffects}) balances two opposing forces. Fixed depth caps constrain the ensemble to a finite union of tree prediction models whose metric entropy grows only logarithmically with $n$. Because the true function belongs to this union, the continuous-cut prior assigns positive mass to Kullback--Leibler neighborhoods of the truth. A likelihood concentration argument then bounds prediction error across densely sampled covariate regions, while Gaussian leaf priors contract posterior tails over sparsely populated regions.

While Theorem~\ref{thm:forest_fixed_main} validates the posterior when the true causal process is piecewise constant, empirical response surfaces are frequently smooth and continuous. To eliminate approximation bias without overfitting, the capacity of the tree ensemble must expand alongside the sample size. We establish posterior consistency for smooth functions by allowing the maximum tree depth to grow logarithmically with $n$.

Consider the scalar continuous covariate setting ($p = 1$). Suppose each true component of $(a_0, b_0, \tau_0)$ is bounded and H\"older continuous of order $\alpha \in (1/2, 1]$. To balance functional approximation against parameter estimation error, tree depth caps are scaled as $\log_2 J_n + c_0$, where $J_n = 2^{\lfloor\kappa \log_2 n\rfloor}$ with growth rate $0 < \kappa < \min\{(1 + 2\alpha)^{-1}, (4\alpha)^{-1}\}$ and integer constant $c_0$.

\begin{theorem}[Scalar smooth-truth extension]\label{thm:forest_scalar_main}
Under the growing-capacity tree prior, for every post-treatment coordinate $(g, t)$, the exact posterior distribution satisfies
\begin{equation}\label{eq:forest_scalar_main}
\int \|\tau_{gt} - \tau_{gt, 0}\|_{L_2(P_X)}^2 \, d\Pi_n(\cdot \mid \mathcal{D}_n) = O_P(J_n^{1-2\alpha} \log n) = o_P(1).
\end{equation}
The empirical-norm and posterior-mean consistency bounds of Theorem~\ref{thm:forest_fixed_main} hold under the identical rate.
\end{theorem}

The proof of Theorem~\ref{thm:forest_scalar_main} constructs a sequence of dyadic interval approximations that produce a common refinement of order $O(J_n)$ intervals without forcing posterior draws onto a fixed split library (Appendix~\ref{rfcons:scalar}). In practical implementations, outcomes are empirically standardized prior to sampling. Because the prior specification is locally stable with respect to location and scale transformations, consistency pushes forward to the original unstandardized scale whenever sample mean and variance estimators converge to positive finite limits.

Two methodological distinctions clarify the scope of these consistency guarantees. First, Theorems~\ref{thm:forest_fixed_main} and \ref{thm:forest_scalar_main} establish consistency for the conditional effect on the treated ($\mathrm{CATT}$) and sample averages ($\mathrm{ATT}$). Extrapolating to population-wide parameters (such as $\mathrm{ATE}$ or $\mathrm{GATE}$) requires mean-effect transportability: $\mathbb{E}[Y_t(g) - Y_t(0) \mid G = h, \, X = x] = \tau_{gt, 0}(x)$ for all cohorts $h$ in the target population. Second, while integrated $L_2$ consistency ensures that posterior point estimates converge, it does not guarantee root-$n$ asymptotic normality or nominal credible interval coverage. Developing valid distributional inference requires the semiparametric information bounds and Bernstein--von Mises machinery to which we now turn.

\subsection{Full-panel information and semiparametric efficiency}\label{sec:efficiency_main}

While Section~\ref{sec:forest_consistency_main} established that posterior point estimates contract to the true causal surface, valid statistical inference requires understanding the asymptotic distribution of posterior summaries. Classical difference-in-differences methods typically construct treatment contrasts by isolating pairs of periods (comparing post-treatment period $t$ strictly against pre-treatment period $g - 1$), discarding the remaining $T - 2$ periods. In contrast, DiD-BCF models all $T$ periods simultaneously. To establish the benchmark for statistical precision under this shared additive structure, we derive the semiparametric efficiency bound for the full-panel Gaussian experiment.

Consider the panel regression model $Y \mid X = x, G = g \sim \mathcal{N}(B_g \beta(x), \Omega_g(x))$, where $Y \in \mathbb{R}^T$, $\beta(x)$ collects cohort baseline levels, common time effects normalized by $b(1,x)=0$, and causal effects $\tau(g, k, x)$, $B_g$ maps these coordinates to period means, and $\Omega_g(x)$ is the within-unit error covariance. Define the information matrix:
\[
Q(x) = \sum_{g} e_g(x) B_g^\top \Omega_g(x)^{-1} B_g,
\]
where $e_g(x) = P(G = g \mid X = x)$ and the eigenvalues of $\Omega_g(x)$ are uniformly bounded away from zero and infinity. For target cell $(g_\star, t_\star)$, let $c$ select the corresponding treatment effect coordinate, $p_\star = P(G = g_\star)$, and $\theta = \mathbb{E}[c^\top \beta(X) \mid G = g_\star]$. Define the normalized weight vector $l(x) = e_{g_\star}(x) c / p_\star$.

\begin{theorem}[Full-panel canonical gradient and semiparametric bound]
\label{thm:eif_main}
Under the Gaussian experiment of Appendix~\ref{sec:full_panel}, assume $p_\star>0$, $Q(x)>0$ almost everywhere, $\mathbb{E}[l(X)^\top Q(X)^{-1} l(X)]<\infty$, and $\operatorname{Var}(c^\top\beta(X)\mid G=g_\star)<\infty$. The efficient influence function for $\theta$ is
\begin{equation}
\phi(O) = l(X)^\top Q(X)^{-1} B_G^\top \Omega_G(X)^{-1} \{Y - B_G \beta(X)\} + \frac{\mathbf{1}\{G = g_\star\}}{p_\star} \{c^\top \beta(X) - \theta\}.
\end{equation}
Its asymptotic variance, the semiparametric efficiency bound on the $\sqrt{n}$ scale, is
\begin{equation}\label{eq:full_var_main}
V_{\mathrm{full}} = \mathbb{E}[l(X)^\top Q(X)^{-1} l(X)] + \frac{1}{p_\star} \operatorname{Var}(c^\top \beta(X) \mid G = g_\star).
\end{equation}
\end{theorem}

Equation~\eqref{eq:full_var_main} decomposes the efficiency bound into two distinct economic components: the first term captures outcome noise pooled across periods and cohorts (the realized-sample uncertainty), while the second captures variation in the treated cohort's covariate composition (sample-to-population variance). This distinction clarifies that inference for the realized sample requires only the first term, whereas inference for the superpopulation must account for both.

\subsection{Posterior Gaussian limits and Bernstein--von Mises}\label{sec:bvm_main}

Having derived the semiparametric efficiency benchmark, we next examine whether the Bayesian posterior distribution achieves this bound. In nonparametric models, posterior consistency does not guarantee that Bayesian credible intervals have valid frequentist coverage.

Conditioning on design $(X, G)$, write the linear panel representation as $Y \mid \beta, H \sim \mathcal{N}(F_H \beta, W_H^{-1})$ with prior $\beta \mid H \sim \mathcal{N}(0, \Lambda_H^{-1})$, where $H$ indexes tree partition structures, leaf variances, and error covariances. For a linear causal functional $T_H = l_H^\top \beta$, let $V_H = (F_H^\top W_H F_H + \Lambda_H)^{-1}$, $a_H = W_H F_H V_H l_H$, and $v_H = l_H^\top V_H l_H$. The conditional posterior is Gaussian: $T_H \mid H, Y \sim \mathcal{N}(a_H^\top Y, v_H)$, and the marginal posterior is a mixture over partition states $H$.

Under sampling law $Y=f_0+u$, $u\sim\mathcal N(0,\Omega_n)$, let $\theta_{0,n}$ be the realized-design target, $w_n$ a design-measurable vector, and $\theta_n^*=\theta_{0,n}+w_n^\top u$. For a design-measurable set $\mathcal H_n$ fixed before outcomes, define empirical Gaussian complexity $\mathfrak g_n=\mathbb E_u\sup_{H\in\mathcal H_n}|\sqrt n(a_H-w_n)^\top u|$ (Appendix~\ref{sec:bvm}).

\begin{theorem}[Adaptive Gaussian-mixture Bernstein--von Mises]
\label{thm:mixture_main}
Suppose the posterior probability of $\mathcal{H}_n$ tends to one, and there exist positive constants $v, s^2 \in (0, \infty)$ such that $\sup_{H \in \mathcal{H}_n} \sqrt{n} |a_H^\top f_0 - \theta_{0, n}| \to 0$, $\mathfrak{g}_n \to 0$, $\sup_{H \in \mathcal{H}_n} |n v_H - v| \to 0$, and $n w_n^\top \Omega_n w_n \to s^2$. Then the posterior distribution converges in bounded Lipschitz metric:
\begin{equation}
d_{\mathrm{BL}}\left(\Pi\left(\sqrt{n}(T - \theta_n^*) \in \cdot \mid Y\right), \, \mathcal{N}(0, v)\right) \to 0.
\end{equation}
The posterior median satisfies $\theta_n^* + o_P(n^{-1/2})$, and central $(1 - \alpha)$ credible intervals achieve limiting coverage $2\Phi(z_{1-\alpha/2} \sqrt{v / s^2}) - 1$ for the realized sample target. Nominal coverage holds if and only if the limiting posterior variance matches the oracle sampling variance ($v=s^2$).
\end{theorem}

Theorem~\ref{thm:mixture_main} provides the critical link between Bayesian computation and frequentist inference: credible intervals are asymptotically valid if and only if prior bias is negligible at rate $o(n^{-1/2})$ and the working likelihood variance $v$ matches the true error variance $s^2$. When the working model assumes conditional independence but the data exhibit unmodeled serial correlation ($s^2 > v$), credible intervals will under-cover, an analytical insight directly verified in our simulation experiments.

\subsection{Primitive finite-support and unknown variance regimes}\label{sec:finite_unknown_main}

Theorem~\ref{thm:mixture_main} relies on high-level partition complexity conditions. We now provide primitive, fully verifiable settings where these conditions hold exactly.

When covariates have finite support, fixed-depth tree priors induce a finite collection of linear subspaces on the identifiable coordinate system. Let $A_n$ denote the design matrix on the identifiable space, $\gamma_0$ the true parameter vector (under baseline normalization $b_0(1, \cdot) = 0$), and $W_n$ a known precision matrix with $n^{-1} A_n^\top W_n A_n \to Q_* > 0$.

\begin{theorem}[Finite-support Bernstein--von Mises]
\label{thm:finite_main}
Assume the finite-support design and prior conditions of Appendix~\ref{sec:finite}, Theorem~\ref{thm:finite}, including a supported full-span model with positive weight and locally continuous positive density. Let $Y=A_n\gamma_0+u$, $u\sim\mathcal N(0,W_n^{-1})$, with $\gamma_0$ outside all proper prior-supported subspaces. For any bounded design-measurable $c_n\to c\ne0$,
\begin{equation}
d_{\mathrm{BL}}\left(\Pi\left(\sqrt{n}(c_n^\top \gamma - c_n^\top \widehat{\gamma}_n) \in \cdot \mid Y\right), \, \mathcal{N}(0, c^\top Q_*^{-1} c)\right) \to 0,
\end{equation}
where $\widehat{\gamma}_n = (A_n^\top W_n A_n)^{-1} A_n^\top W_n Y$ is the generalized least-squares estimator. Central credible intervals are asymptotically nominal for the realized target.
\end{theorem}

In applied econometric practice, the residual error variance is unknown, and outcomes are standardized prior to sampling. Theorem~\ref{thm:unknown_main} proves that estimating the error scale preserves asymptotic Gaussian limits under local prior stability.

\begin{theorem}[Finite-support BvM with unknown variance]
\label{thm:unknown_main}
Under the finite-support design, genericity, and locally stable induced-prior conditions of Appendix~\ref{sec:uv_extension}, Theorem~\ref{uv:thm:unknown}, let $Y=A_n\gamma_0+u$, $u\sim\mathcal N(0,v_0I_{nT})$, and $Q=\lim A_n^\top A_n/n>0$. Use the Gaussian independence likelihood, OLS estimator $\widehat\gamma_n$, and $\widehat v_n=\|Y-A_n\widehat\gamma_n\|^2/(nT)$. The joint posterior distribution of $\sqrt{n}(\gamma - \widehat{\gamma}_n, v - \widehat{v}_n)^\top$ converges in total variation to $\mathcal N(0,\operatorname{diag}(v_0Q^{-1},2v_0^2/T))$. For bounded design-measurable $c_n\to c\ne0$, the posterior mean contrast satisfies $c_n^\top\widehat\gamma_n+o_P(n^{-1/2})$, its scaled posterior variance converges to $v_0c^\top Q^{-1}c$, and central intervals are asymptotically nominal for $c_n^\top\gamma_0$.
\end{theorem}

This result confirms that empirical standardization and conjugate variance updating preserve exact-posterior asymptotic normality, justifying the use of posterior means for point estimation and credible intervals for uncertainty quantification.

\subsection{Continuous-covariate histogram regime}\label{sec:continuous_bvm_main}

To extend distributional inference beyond finite support, we analyze a continuous-covariate histogram-mixture experiment. Let $\mathcal M_n$ be a deterministic family of $M_n$ balanced partitions, specified independently of outcomes, with $K_n$ cells of diameter at most $h_n\to0$. Let $0<\alpha,\zeta\le1$ denote the H\"older smoothness orders of the true coefficient surface and its functional representer, and set $t_n=\log(2nM_n)$ (Appendix~\ref{sec:cc_extension}).

\begin{theorem}[Continuous-covariate histogram-mixture BvM]
\label{thm:continuous_main}
Assume the design, saturated-model, and uniform prior conditions of Appendix~\ref{sec:cc_extension}, Theorem~\ref{cc:thm:continuous}. If partition bandwidth $h_n$ and cell count $K_n$ satisfy
\begin{equation}\label{eq:histogram_rates}
\sqrt{n} h_n^{\alpha + \zeta} \to 0, \qquad h_n^{\min(\alpha, \zeta)} \sqrt{t_n} \to 0, \qquad \frac{K_n t_n}{\sqrt{n}} \to 0,
\end{equation}
then under independent Gaussian errors, the exact posterior satisfies a scalar Bernstein--von Mises theorem for the realized treated-design average. Its mean equals the oracle residual-score center up to $o_P(n^{-1/2})$, and central intervals are asymptotically nominal for that target.
\end{theorem}

The rate condition $\sqrt{n} h_n^{\alpha + \zeta} \to 0$ in Equation~\eqref{eq:histogram_rates} governs the doubly robust product of regression and representer approximation errors. In dimension $p = 1$, Lipschitz smoothness ($\alpha = \zeta = 1$) admits the regular partition rate $K_n \asymp n^{1/3}$ for polynomial $M_n$. In higher dimensions, achieving valid coverage requires additional smoothness or structural sparsity.

\section{Simulation Studies Design} \label{sim_study_did_bcf}
To evaluate the empirical performance and boundary conditions of DiD-BCF, we conduct extensive Monte Carlo simulation experiments centered on the primary econometric challenges of panel causal inference: selection on unobserved time-invariant confounders, within-unit serial correlation, sample-size scaling, staggered adoption with exposure-evolving treatment effects, and sharp-null size control. Because counterfactual potential outcomes cannot be observed simultaneously in real data, simulation studies provide the necessary benchmark where true realized treatment effects are known.

Each non-null simulation cell is evaluated over $R = 100$ independent Monte Carlo replications, while sharp-null configurations and pre-trend diagnostic designs deploy $R = 200$ replications. The default panel consists of $n = 200$ cross-sectional units observed over $T = 8$ time periods. In every replication, DiD-BCF is estimated using 50 grow-from-root initialization sweeps and three independent MCMC chains, each collecting 500 retained draws after thinning by five. The observational working error variance is assigned an $\operatorname{IG}(2, 1)$ prior. Algorithmic hyperparameters and data-generating processes are detailed in Appendices~\ref{supp:algo} and \ref{supp:designs}.

\subsection{Benchmark Models}
We compare DiD-BCF against leading econometric and machine-learning estimators that represent the primary paradigms in modern panel analysis (full technical descriptions appear in Appendix~\ref{supp:benchmarks}):
(1) standard Two-Way Fixed Effects (TWFE) with event-study dynamics and covariate adjustment;
(2) the doubly robust group-time estimator of \cite{SantAnna2020} (DiD DR), which combines inverse-probability weighting with linear outcome modeling;
(3) the two-stage difference-in-differences estimator of \cite{https://doi.org/10.48550/arxiv.2207.05943} (DiD2s), which purges unit and time fixed effects in a first step before estimating causal coefficients;
(4) Synthetic Difference-in-Differences \parencite{Arkhangelsky2021}, which pairs unit and time regularized weighting with panel regression;
(5) DoubleML DiD \parencite{DoubleML2022}, which implements doubly robust estimation using 500-tree random forests for nuisance estimation \parencite{Chang2020}; and
(6) grf-DiD \parencite{Wang2022}, which first-differences panel outcomes to eliminate time-invariant unit heterogeneity and applies generalized random causal forests \parencite{Athey2019} to the differenced data to estimate covariate-specific treatment effects.

The comparison between DiD-BCF and grf-DiD directly probes an essential architectural trade-off: grf-DiD removes unobserved unit heterogeneity by differencing outcomes (projecting out fixed effects), whereas DiD-BCF absorbs persistent heterogeneity through its structured baseline regression $a(G_i, X_i) + b(t, X_i)$ and optional random intercepts. Differencing eliminates time-invariant confounders unconditionally but discards cross-sectional level information and amplifies measurement noise in short panels; absorbing retains full outcome information across all periods, relying on the structured prior to disentangle common trends from cohort-level shifts.

\subsection{Data-Generating Processes}\label{ssec:dgps}
The simulation designs are structured around potential outcomes $Y_{it}(0)$ and $Y_{it}(1)$ with observed outcome $Y_{it} = Y_{it}(0)(1 - D_{it}) + Y_{it}(1)D_{it}$ and baseline errors $\epsilon_{it} \sim \mathcal{N}(0, 1)$. To test selection on unobservables, an unobserved unit effect $u_i \sim \mathcal{N}(0, \sigma_u^2)$ enters untreated outcomes while treatment assignment depends on $u_i$ with factor loading $c_u$. Consequently, treated and untreated groups differ persistently in levels through an unobserved variable while conditional parallel trends is preserved.

The simulation study investigates four core design modules:
1. Canonical single-cohort designs ($n = 200$, $T = 8$): baseline unobserved confounding with $(\sigma_u, c_u) = (1, 1)$; strong unobserved confounding with $(\sigma_u, c_u) = (2, 1.5)$; serially correlated errors governed by a stationary AR(1) process with autocorrelation coefficient $\rho = 0.6$; selection on observed covariates; and a sharp null design with identical zero treatment effects ($\tau_{it} \equiv 0$).
2. Sample-size scaling sweeps: the baseline design replicated across $n \in \{50, 100, 200, 400, 800\}$ under both independent disturbances and AR(1) autocorrelated errors to assess asymptotic rate scaling and small-sample stability.
3. Staggered adoption designs: units adopt across cohorts $G_i \in \{4, 5, 6, \infty\}$ with exposure-growing treatment effects $\tau_{it} = M_j(1 + 0.4 k_{it})\tau_i$, with cohort multipliers $(M_1, M_2, M_3) = (1, 1.5, 2)$. This setting generates dynamic heterogeneity where standard pooled regressions suffer from contaminated comparisons between early and late adopters \parencite{GoodmanBacon2021}.
4. Pre-trend violation designs: ten specialized configurations evaluating diagnostic sensitivity against parallel-trends violations, including linear slope divergences ranging from $0.05$ to $0.40$, violations driven by unobserved confounders, aggregate-canceling heterogeneous trends, and designs that satisfy conditional but not unconditional parallel trends.

To assess robustness to model misspecification, each design is evaluated across three covariate complexity degrees ($d \in \{1, 2, 3\}$). In degree 1, response surfaces are linear in baseline covariates. In degree 2, estimators observe transformed nonlinear covariates while the underlying truth remains linear. In degree 3, the true treatment effect surface contains quadratic terms and multi-way interactions (full functional equations are given in Appendix~\ref{supp:designs}).

\subsection{Evaluation Metrics}
We assess estimators across point accuracy, posterior interval calibration, and heterogeneity recovery. For point estimation, we report empirical bias, error standard deviation $\mathrm{SD}_e = \sqrt{\mathrm{RMSE}^2 - \mathrm{Bias}^2}$, root mean squared error (RMSE), and mean absolute percentage error (MAPE) evaluated against realized target estimands on treated units. Heterogeneity recovery is measured on the individual conditional surface $\mathrm{CATT}(X_i; g, t)$ via within-replication RMSE and pointwise interval coverage. To examine convergence rates, sample-size sweeps monitor the scaled estimation error $\sqrt{n}(\widehat{\theta} - \theta)$.

Inferential validity is evaluated through empirical coverage frequencies and average lengths of nominal 90\% and 95\% equal-tailed credible intervals (or confidence intervals for frequentist benchmarks). We also track the spread calibration ratio, which compares the average reported posterior standard deviation to the empirical Monte Carlo standard deviation of point estimates. For hypothesis tests of $H_0: \tau = 0$, empirical size is evaluated under the sharp-null design, while statistical power is measured under non-zero true effects. For continuous posteriors, the Bayesian two-sided tail score is defined as $p_{\mathrm{Bayes}} = 2 \min\{P(\tau > 0 \mid \mathcal{D}), P(\tau < 0 \mid \mathcal{D})\}$, where $p_{\mathrm{Bayes}} < 0.05$ corresponds to the exclusion of zero from the central 95\% credible interval.

\section{Results and Discussion}\label{results_did_bcf}
Throughout this section, our baseline displays center on linearity degree $d = 1$ because strictly linear response surfaces are the most favorable regime for classical parametric and linear semiparametric models (TWFE, Callaway--Sant'Anna, did2s, and Synthetic DiD). Alongside $d = 1$, we examine the nonlinear and interactive regimes ($d = 2, 3$) to show that DiD-BCF also outperforms the nonparametric benchmarks DoubleML DR-DiD and grf-DiD. To ensure sparsity of this paper, we focus on the most illustrative results below and refer to Appendix~\ref{supp:results} (Appendix Tables~\ref{tab:b1att}--\ref{tab:surface} and Figures~\ref{fig:sweep}--\ref{fig:surfacesweep}) for complete tabulations and plots.

\subsection{Aggregate ATT and GATT under confounding, serial correlation, and nonlinearity}
\begin{figure}[htbp]
\centering\spacingset{1}\small
\includegraphics[width=0.84\textwidth]{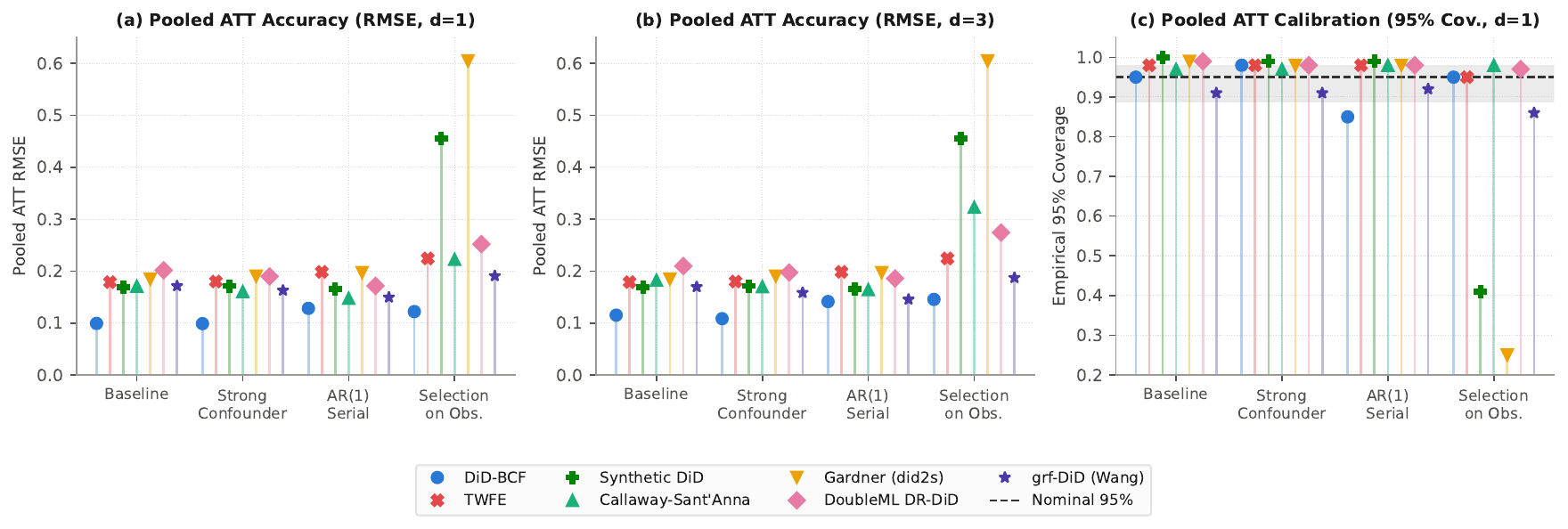}
\caption{Overall Pooled $\mathrm{ATT}$ accuracy (RMSE at $d=1$ and $d=3$) and empirical 95\% coverage ($d=1$) across all benchmark models in the canonical DiD designs ($n=200$, 100 replications). The shaded band in panel (c) marks the 95\% Wilson interval for nominal 95\% coverage.}
\label{fig:canonical_benchmarks}
\end{figure}

Figure~\ref{fig:canonical_benchmarks} compares point accuracy (RMSE) and interval calibration (95\% coverage) for the overall Pooled $\mathrm{ATT}$ across all seven models in the four canonical DiD designs ($n = 200$; full Pooled $\mathrm{ATT}$ and $\mathrm{GATT}(4,4)$ metrics across $d \in \{1, 2, 3\}$ appear in Appendix Tables~\ref{tab:b1att}--\ref{tab:b1gatt}).

Even at $d = 1$, where linear parametric models are well-specified, DiD-BCF achieves the lowest Pooled $\mathrm{ATT}$ RMSE in both the baseline and strong confounding designs at $0.099$ (bias $0.004, -0.001$; 95\% coverage $0.950, 0.980$; Figure~\ref{fig:canonical_benchmarks}(a),(c)). For $\mathrm{GATT}(4,4)$ (Appendix Table~\ref{tab:b1gatt}), DiD-BCF records RMSEs of $0.112$ and $0.102$ (100\% coverage) versus $0.163$--$0.304$ for the benchmarks, reducing estimation error by 38\% to 63\%.

Under AR(1) serial correlation ($d = 1$), DiD-BCF retains the lowest RMSE for Pooled $\mathrm{ATT}$ ($0.128$) and $\mathrm{GATT}(4,4)$ ($0.112$). Single-period $\mathrm{GATT}(4,4)$ coverage remains calibrated ($0.960$), whereas multi-period Pooled $\mathrm{ATT}$ coverage drops to $0.850$, consistent with Theorem~\ref{thm:mixture_main} when serial correlation inflates $v$ relative to $s^2$. Under selection on observables ($d = 1$), unadjusted Synthetic DiD and did2s degrade severely (Pooled $\mathrm{ATT}$ RMSE $0.455, 0.605$; coverage $0.410, 0.250$). DiD-BCF absorbs observable confounding via its baseline forests, achieving Pooled $\mathrm{ATT}$ RMSE $0.122$ (coverage $0.950$) and $\mathrm{GATT}(4,4)$ RMSE $0.124$ (coverage $0.970$), outperforming Callaway--Sant'Anna ($0.223, 0.280$), DoubleML ($0.252, 0.344$), and grf-DiD ($0.191, 0.254$; coverage $0.860, 0.790$).

Evaluating the nonlinear covariate ($d = 2$) and interactive treatment-effect ($d = 3$) regimes demonstrates DiD-BCF's superiority over both parametric estimators and the nonparametric models DoubleML DR-DiD and grf-DiD (Figure~\ref{fig:canonical_benchmarks}(b) and Appendix Tables~\ref{tab:b1att}--\ref{tab:b1gatt}). Under selection on nonlinear observables ($d = 2, 3$), Callaway--Sant'Anna's Pooled $\mathrm{ATT}$ RMSE rises from $0.223$ ($d = 1$) to $0.323$ ($d = 2, 3$). Among the nonparametric models, DiD-BCF attains Pooled $\mathrm{ATT}$ RMSEs of $0.146$ ($d = 2$) and $0.145$ ($d = 3$) under selection on observables and $0.116$--$0.115$ in baseline, outperforming grf-DiD ($0.187$ and $0.175$--$0.170$) and DoubleML ($0.274$ and $0.208$--$0.210$). For $\mathrm{GATT}(4,4)$ at $d = 2, 3$ (Appendix Table~\ref{tab:b1gatt}), DiD-BCF reduces RMSE ($0.122$ in baseline and $0.136$--$0.140$ under selection on observables, coverage $0.960$--$0.980$) by 38\% to 59\% relative to grf-DiD ($0.223$ and $0.253$--$0.254$, coverage $0.790$--$0.810$) and DoubleML ($0.271$--$0.280$ and $0.336$--$0.339$).

\subsection{Sample-size scaling and sharp-null size control}
\begin{figure}[htbp]
\centering\spacingset{1}\small
\includegraphics[width=0.84\textwidth]{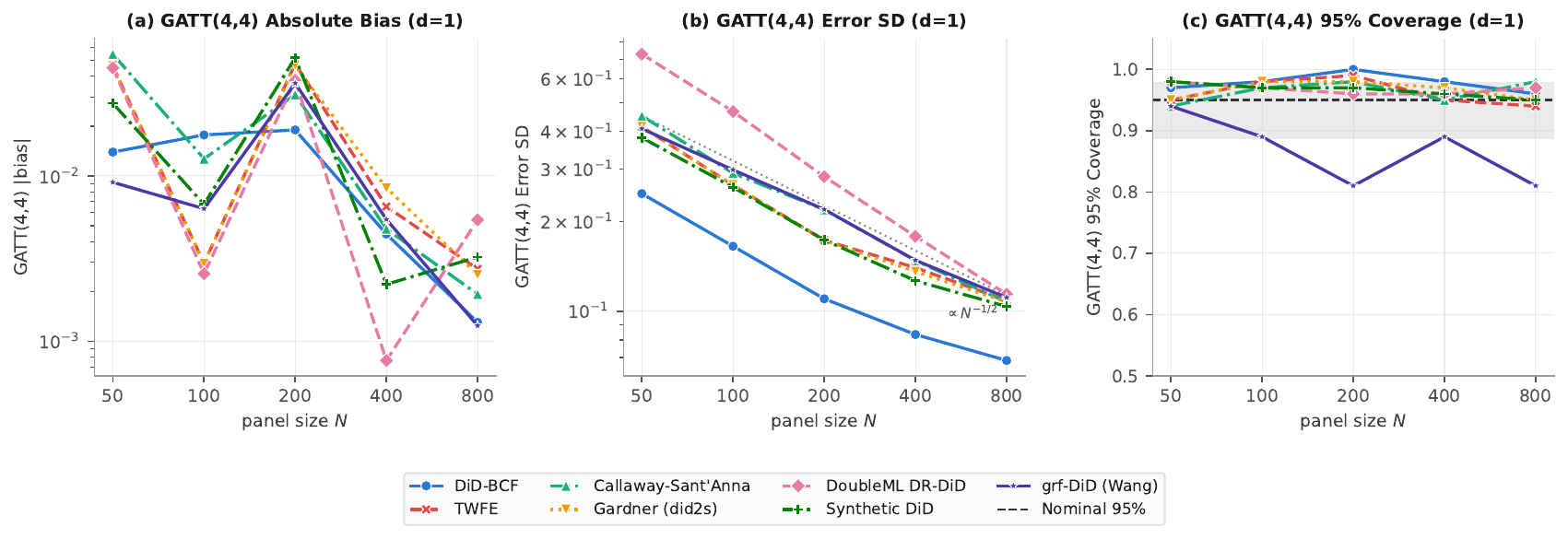}
\caption{Cohort-time $\mathrm{GATT}(4,4)$ absolute bias, Monte Carlo error SD, and 95\% coverage across panel sizes $n \in \{50, 100, 200, 400, 800\}$ in the baseline sweep ($d=1$, 100 replications).}
\label{fig:sweep_main}
\end{figure}

Figure~\ref{fig:sweep_main} tracks the single-cell impact estimand $\mathrm{GATT}(4,4)$ across $n = 50$ to $800$ at $d = 1$ (see Appendix Tables~\ref{tab:sweep}, \ref{tab:sweepdeg}, and \ref{tab:smalln} and Figures~\ref{fig:sweep}--\ref{fig:calibration} for full sweeps across $d \in \{1, 2, 3\}$ and AR(1) errors). DiD-BCF exhibits $\sqrt{n}$-rate contraction, with $\mathrm{GATT}(4,4)$ RMSE falling from $0.248$ ($n = 50$) to $0.112$ ($n = 200$) and $0.068$ ($n = 800$; coverage $0.960$--$1.000$). At $d = 2, 3$ (Appendix Table~\ref{tab:sweepdeg}), DiD-BCF outperforms both nonparametric benchmarks across all sample sizes, contracting from RMSEs of $0.260, 0.270$ at $n = 50$ (vs.\ $0.412, 0.410$ for grf-DiD and $0.673, 0.737$ for DoubleML) to $0.072$ at $n = 800$ (vs.\ $0.111$ for grf-DiD and $0.113$--$0.124$ for DoubleML). Under AR(1) errors, $\mathrm{GATT}(4,4)$ coverage stays stable ($0.880$--$0.960$), while multi-period Pooled $\mathrm{ATT}$ coverage settles at $0.75$--$0.79$ at $n = 800$. Across non-null designs, spread calibration tracks near unity ($0.80$--$1.05$; Appendix Figure~\ref{fig:calibration}), and under the sharp null ($\tau_{it} \equiv 0$; Appendix Table~\ref{tab:null}), 5\% rejection rates for Pooled $\mathrm{ATT}$, $\mathrm{ATT}(k=0)$, and $\mathrm{GATT}(4,4)$ stay within $0.015$--$0.070$ across $d \in \{1, 2, 3\}$.

\subsection{Dynamic event-study and pooled ATT under staggered adoption}
In the staggered design ($G_i \in \{4, 5, 6, \infty\}$), $11.6\%$ of the standard TWFE weight falls on already-treated controls (Goodman--Bacon decomposition in Appendix Table~\ref{tab:bacon}). Thus, TWFE underestimates $\mathrm{ATT}(k)$ for $k \ge 1$ and yields a Pooled $\mathrm{ATT}$ RMSE of $0.494$ (coverage $0.450$), while Synthetic DiD records a Pooled $\mathrm{ATT}$ RMSE of $1.296$ (coverage $0.000$; Figure~\ref{fig:staggered_main}; full staggered tables across $d \in \{1, 2, 3\}$ appear in Appendix Tables~\ref{tab:staggered}--\ref{tab:bacon} and Figures~\ref{fig:eventstudy}--\ref{fig:esmethods}).

\begin{figure}[htbp]
\centering\spacingset{1}\small
\includegraphics[width=0.84\textwidth]{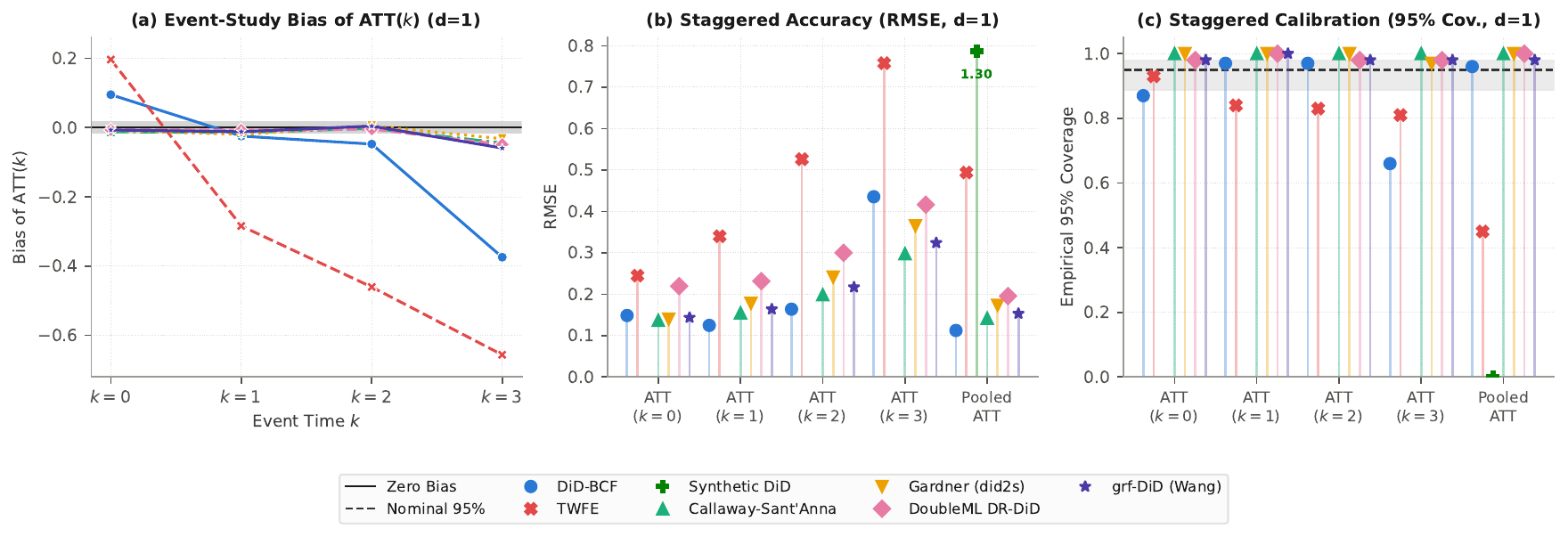}
\caption{Staggered adoption performance ($d=1$, $n=200$, 100 replications): (a) event-study estimation bias of $\mathrm{ATT}(k)$; (b) RMSE across $\mathrm{ATT}(k=0,1,2,3)$ and Pooled $\mathrm{ATT}$ (Synthetic DiD's Pooled $\mathrm{ATT}$ RMSE of $1.30$ annotated at top); and (c) empirical 95\% coverage.}
\label{fig:staggered_main}
\end{figure}

DiD-BCF avoids TWFE contamination via its cohort-specific baseline structure, achieving the lowest Pooled $\mathrm{ATT}$ RMSE ($0.112$, coverage $0.960$) ahead of Callaway--Sant'Anna ($0.142$), grf-DiD ($0.153$), did2s ($0.173$), and DoubleML ($0.196$). At $k \in \{0, 1, 2\}$, DiD-BCF attains $\mathrm{ATT}(k)$ RMSEs of $0.148, 0.125,$ and $0.164$, outperforming DoubleML ($0.219, 0.231, 0.300$) and grf-DiD ($0.144, 0.164, 0.217$), an advantage that persists at $d = 2, 3$ (Appendix Table~\ref{tab:staggereddeg}). Across all nine $\mathrm{GATT}(g,t)$ cells (Appendix Tables~\ref{tab:staggered_gatt}--\ref{tab:staggeredcells}), DiD-BCF achieves mean RMSEs of $0.318, 0.329,$ and $0.342$ for $d = 1, 2, 3$, beating DoubleML ($0.412, 0.409, 0.433$) and matching Callaway--Sant'Anna ($0.298$--$0.304$) and grf-DiD ($0.301$). In sparse terminal cells ($k = 3$, $\mathrm{GATT}(6,7)$), prior shrinkage toward $\taug$ induces attenuation bias ($-0.374, -0.483$) and undercoverage ($0.660, 0.360$), highlighting the value of inspecting disaggregated cells alongside pooled summaries.

\subsection{Individual CATT surface recovery, nonlinearity, and small panels}
\begin{figure}[htbp]
\centering\spacingset{1}\small
\includegraphics[width=0.84\textwidth]{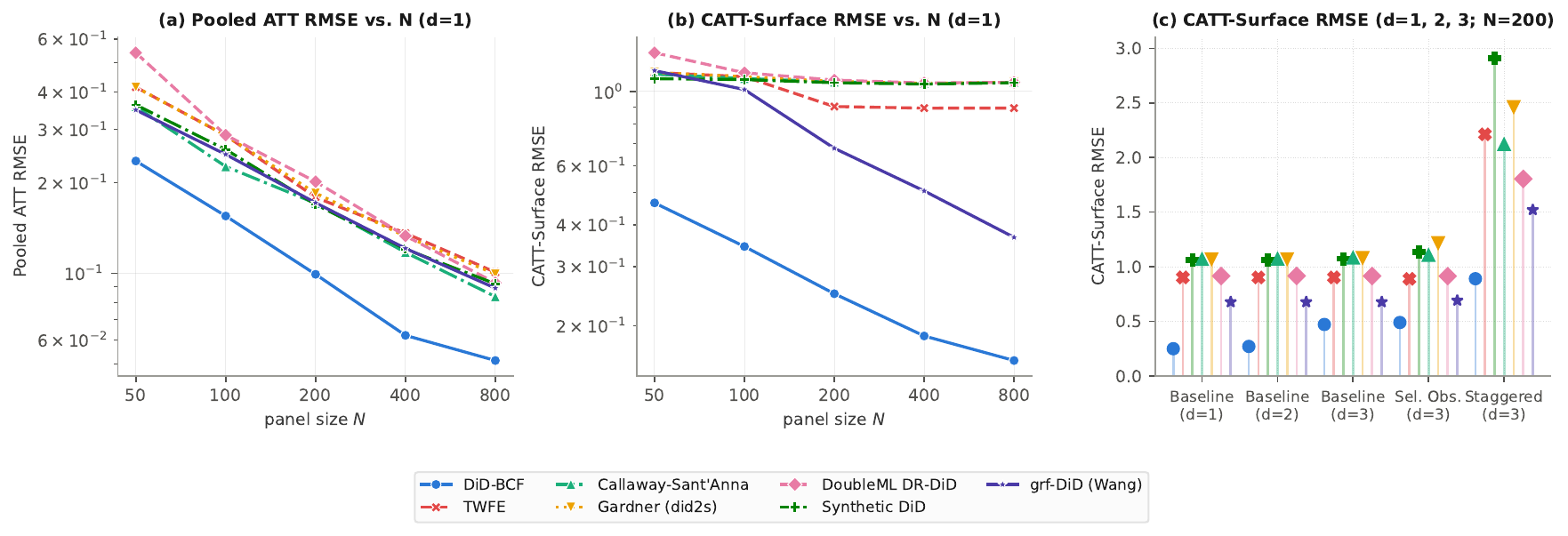}
\caption{Small-panel scaling and individual heterogeneity recovery: (a) Pooled $\mathrm{ATT}$ RMSE across $n$ ($d=1$); (b) individual $\mathrm{CATT}(X_i; g, t)$ surface RMSE across $n$ ($d=1$); and (c) individual $\mathrm{CATT}(X_i; g, t)$ surface RMSE across degrees $d \in \{1, 2, 3\}$ in the Baseline, Selection on Observables, and Staggered designs ($n=200$, 100 replications).}
\label{fig:smalln_main}
\end{figure}

Figure~\ref{fig:smalln_main} contrasts aggregate Pooled $\mathrm{ATT}$ RMSE (Panel~(a)) with individual $\mathrm{CATT}(X_i; g, t)$ surface RMSE across $n$ at $d = 1$ (Panel~(b)) and across $d \in \{1, 2, 3\}$ at $n = 200$ (Panel~(c); full results appear in Appendix Tables~\ref{tab:smalln}, \ref{tab:smallnsurface}, and \ref{tab:surface} and Figures~\ref{fig:smalln}, \ref{fig:surface}, and \ref{fig:surfacesweep}).

At $n = 50$ ($d = 1$), DiD-BCF achieves both the lowest Pooled $\mathrm{ATT}$ RMSE ($0.236$, coverage $0.94$, vs.\ $0.352$--$0.541$) and the lowest $\mathrm{CATT}$-surface RMSE ($0.465$, pointwise coverage $0.92$, vs.\ $1.155$ for grf-DiD and $1.092$--$1.305$ for scalar broadcast methods), contracting to $0.157$ (coverage $0.969$) by $n = 800$ versus $0.367$ for grf-DiD (coverage $0.816$). Across $d \in \{1, 2, 3\}$ at $n = 200$ (Figure~\ref{fig:smalln_main}(c) and Appendix Table~\ref{tab:surface}), DiD-BCF is markedly superior to both nonparametric competitors: DoubleML broadcasts cell averages (surface RMSE $0.912$--$0.918$ in canonical and $1.798$--$1.804$ in staggered designs), while grf-DiD loses precision from first-differencing. DiD-BCF achieves $\mathrm{CATT}$-surface RMSEs for $(d = 1, 2, 3)$ of $(0.249, 0.270, 0.471)$ in baseline (coverage $0.968, 0.960, 0.857$ vs.\ grf-DiD RMSE $0.677$, coverage $0.600$), $(0.244, 0.266, 0.435)$ under strong confounding (vs.\ grf-DiD $0.671$), $(0.266, 0.289, 0.489)$ under selection on observables (vs.\ grf-DiD $0.691$), and $(0.674, 0.702, 0.890)$ under staggered adoption (vs.\ grf-DiD $1.520$).

\subsection{Working-error prior sensitivity}
Varying the inverse-gamma rate parameter for $\sigma^2$ across $\lambda \in \{0.1, 1, 10\}$ (shape fixed at $2$; Appendix Table~\ref{tab:prior_sensitivity} and Figure~\ref{fig:prior_sensitivity}) confirms robustness to prior scaling. At $n = 50$ ($d = 1$), Pooled $\mathrm{ATT}$ RMSE is stable under independent errors ($0.234, 0.236, 0.248$; coverage $0.930, 0.940, 0.960$) and AR(1) errors (coverage $0.810, 0.820, 0.890$). At $n = 800$, prior influence vanishes: Pooled $\mathrm{ATT}$ RMSE is $0.054$ under independent errors (coverage $0.960$--$0.970$) and $0.068$ under AR(1) errors (coverage $0.760$--$0.790$), supporting $\lambda = 1$ as the default while confirming that prior tuning cannot replace serial-dependence modeling.

\subsection{Pre-trend diagnostic}\label{ssec:pretrendres}
We evaluate empirical size and power for the pre-treatment implications of $H_0$: conditional parallel trends. Table~\ref{tab:pretrend_main} compares rejection percentages across DiD-BCF and all six benchmarks at $d=1$, separating size, power against linear differential trends, and detection of subgroup trends that cancel in the population aggregate. Full results for all ten designs and all three rules at $d=1$ and $d=3$ appear in Appendix~\ref{supp:results} (Appendix Tables~\ref{tab:pretrend}--\ref{tab:pretrendd3}).

\begin{table}[htbp]
\centering\spacingset{1}\small
\caption{Empirical rejection percentages for the pre-treatment implications of $H_0$: conditional parallel trends, at nominal 5\% level and $d=1$. Each available entry uses 200 replications. CS denotes Callaway--Sant'Anna; NA denotes an unavailable subgroup test, rather than zero rejection. TWFE is evaluated on an independent set of 200 replications.}
\label{tab:pretrend_main}
\setlength{\tabcolsep}{3pt}
\resizebox{\linewidth}{!}{%
\begin{tabular}{lrrrrrrr}
\toprule
Scenario & DiD-BCF & \shortstack{TWFE\\event study} & \shortstack{Synthetic\\DiD} & CS & did2s & DoubleML & grf-DiD \\
\midrule
\multicolumn{8}{l}{\textit{A. Size under parallel trends (slope rule)}} \\
Unconditional PT & 4.0 & 4.5 & 0.0 & 5.0 & 4.5 & 4.0 & 5.0 \\
Conditional PT only & 8.0 & 75.5 & 5.5 & 7.0 & 75.0 & 5.5 & 10.0 \\
\midrule
\multicolumn{8}{l}{\textit{B. Power against linear trends (slope rule)}} \\
Slope 0.00 & 4.0 & 4.5 & 0.0 & 5.0 & 4.5 & 4.0 & 5.0 \\
Slope 0.05 & 20.5 & 10.0 & 0.0 & 9.0 & 9.5 & 8.5 & 8.5 \\
Slope 0.10 & 52.5 & 25.0 & 0.5 & 32.0 & 24.0 & 18.0 & 28.5 \\
Slope 0.20 & 93.0 & 70.0 & 17.0 & 83.0 & 70.0 & 54.0 & 79.5 \\
Slope 0.40 & 100.0 & 100.0 & 85.0 & 100.0 & 100.0 & 98.5 & 100.0 \\
\midrule
\multicolumn{8}{l}{\textit{C. Cancelling subgroup trends ($X_1$ contrast rule)}} \\
No violation & 0.5 & NA & 0.0 & 5.0 & 5.5 & 3.0 & 4.0 \\
Slopes $\pm 0.10$ & 3.0 & NA & 3.5 & 32.0 & 23.5 & 11.5 & 27.5 \\
Slopes $\pm 0.20$ & 52.5 & NA & 30.0 & 74.5 & 66.5 & 35.5 & 70.0 \\
Slopes $\pm 0.40$ & 100.0 & NA & 91.5 & 100.0 & 100.0 & 93.5 & 100.0 \\
\bottomrule
\end{tabular}}
\end{table}

Under unconditional parallel trends (Setting~A), all methods except Synthetic DiD have rejection rates close to the nominal 5\% level. When parallel trends hold only conditional on covariates, unadjusted methods (TWFE, did2s) severely over-reject ($\approx 75\%$), while covariate adjustment greatly reduces false rejections: DiD-BCF rejects in 8.0\% of runs and DoubleML in 5.5\%. DiD-BCF has the highest power against subtle and moderate linear aggregate violations (Setting~B); at slope 0.10, detection is 52.5\%, compared with 18.0\% for DoubleML, 32.0\% for Callaway--Sant'Anna, and 28.5\% for grf-DiD.

Cancelling subgroup trends (Setting~C) yield a mixed ranking. DiD-BCF's contrast test is conservative under the null (0.5\% rejection), while having a weaker power than all other benchmark models (besides DoubleML). This conservativeness at small disparities reflects the strong depth and leaf shrinkage on the auxiliary lead forest ($\alpha_\tau = 0.25, \delta_\tau = 3$). Since aggregate trends cancel in Setting~C, detection requires depth-two $(X_1, k)$ interaction splits that are shrunk toward homogeneity until signal magnitude rises, whereas sample-split benchmarks (Callaway--Sant'Anna, did2s) condition separately on known $X_1$ strata without cross-subgroup regularization and grf-DiD eliminates the time dimension via first-differencing before splitting on $X_1$.

\section{Empirical Application} \label{real_life_did_bcf}
To evaluate the practical utility of DiD-BCF in observational policy settings, we apply the model to estimate the impact of state-level minimum wage increases on county-level teen employment across the United States, utilizing the benchmark \texttt{mpdta} dataset from the \texttt{did} package in \textsf{R} \parencite{R-did}, originally analyzed by \textcite{SantAnna2020}. This application examines both the overall average treatment effect on the treated and the presence of fine-grained effect heterogeneity across county population sizes.

The data constitute a balanced panel of 500 United States counties observed annually over the five-year period from 2003 to 2007 (2,500 total county-year observations). The dependent variable is the logarithm of county teen employment (\texttt{lemp}). Treatment status reflects state-level legislative action: counties in states that raised their minimum wage above the federal baseline between 2004 and 2007 form staggered adoption cohorts ($G_i \in \{2004, 2006, 2007\}$), while counties in states maintaining the federal minimum wage throughout the study window serve as the never-treated control group ($G_i = \infty$). Following \textcite{SantAnna2020}, we include the time-invariant logarithm of county population (\texttt{lpop}) measured prior to treatment as the primary conditioning covariate and effect modifier. 

Prior parametric and semiparametric evaluations on this dataset report negative overall employment effects: \textcite{SantAnna2020} find an effect of $-0.037$ (s.e.\ $0.006$) under standard TWFE, $-0.052$ (s.e.\ $0.006$) under simple weighted aggregation, and $-0.090$ (s.e.\ $0.013$) using doubly robust difference-in-differences.

We estimate the structured DiD-BCF model employing 50 grow-from-root sweeps and 500 retained draws per chain across three MCMC chains, thinning by five. Our primary specification incorporates county-level random intercepts $\alpha_i \sim \mathcal{N}(0, \sigma_\alpha^2)$ to absorb persistent unit heterogeneity and equicorrelated local shocks; a model without random intercepts is evaluated as a robustness check. To ensure computational stability, the estimation was repeated across three independent random seed initializations on the same panel.

\begin{figure}[htbp]
\centering
\includegraphics[width=0.94\textwidth]{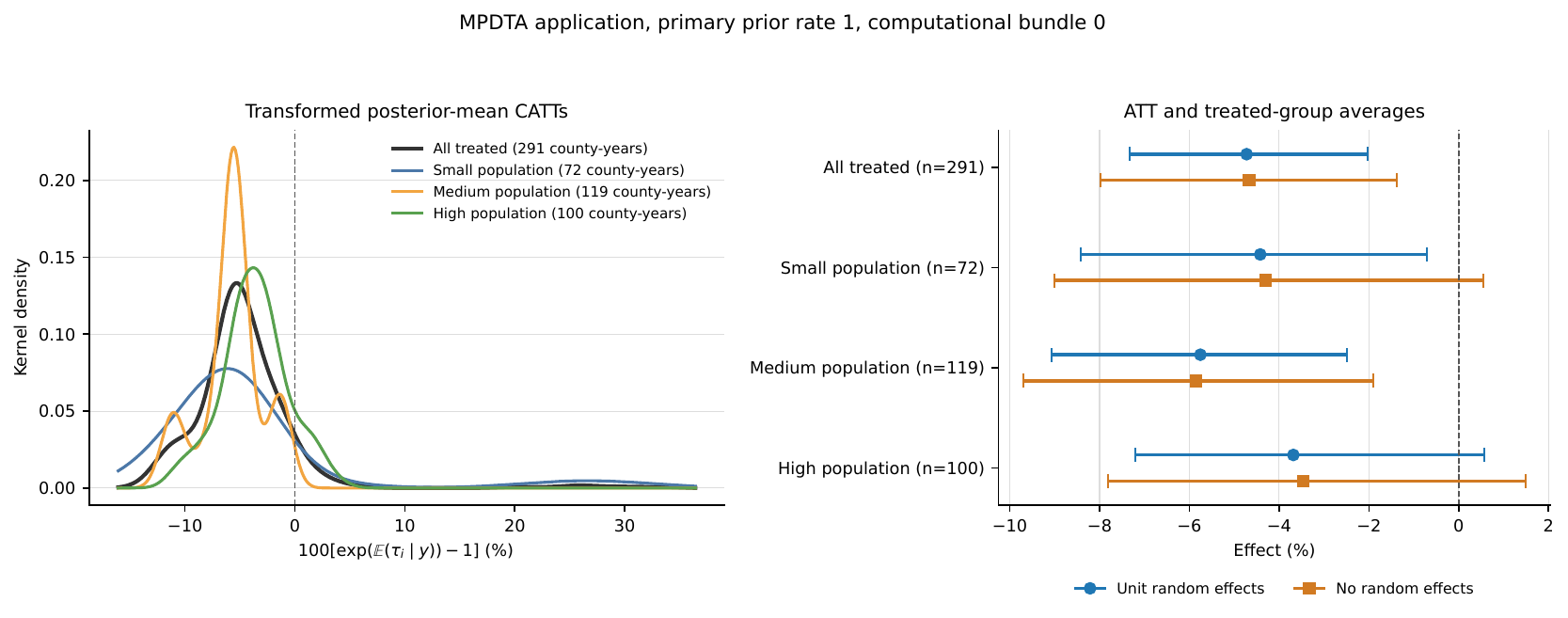}
\caption{Left: Kernel density estimate of individual county-year treatment effects. Right: Posterior means and 95\% equal-tailed credible intervals for the overall ATT and population-group averages.}
\label{Figure3_did_bcf}
\end{figure}

Figure~\ref{Figure3_did_bcf} summarizes treatment effects across the 291 treated county-year observations (191 unique counties; Appendix Table~\ref{tab:application_estimates}). In our
primary specification with random intercepts, DiD-BCF estimates an overall employment effect of $-4.72\%$ (95\% CI $[-7.33\%, -2.03\%]$, $p_{\mathrm{Bayes}} =
0.001$), virtually identical to the specification without random intercepts ($-4.66\%$, 95\% CI $[-7.98\%, -1.38\%]$, $p_{\mathrm{Bayes}} = 0.009$), though random
intercepts tighten posterior uncertainty throughout. Across population terciles (Figure~\ref{Figure3_did_bcf}, right panel), point estimates indicate slightly
stronger disemployment in medium-sized counties ($-5.75\%$, 95\% CI $[-9.07\%, -2.50\%]$, $p_{\mathrm{Bayes}} < 0.001$) than in small ($-4.42\%$, 95\% CI $[-8.42\%,
-0.71\%]$, $p_{\mathrm{Bayes}} = 0.020$) or high-population counties ($-3.68\%$, 95\% CI $[-7.19\%, 0.58\%]$, $p_{\mathrm{Bayes}} = 0.071$). While individual
conditional effects span $-15\%$ to $+2\%$ (left panel), pairwise posterior contrasts on the log scale between high and medium ($+0.0217$, 95\% CI $[-0.0102,
0.0715]$, $p_{\mathrm{Bayes}} = 0.309$), high and small ($+0.0077$, $[-0.0373, 0.0616]$, $p_{\mathrm{Bayes}} = 0.831$), and medium and small counties ($-0.0140$,
$[-0.0479, 0.0255]$, $p_{\mathrm{Bayes}} = 0.323$) all cover zero, ruling out a systematic population gradient.

Sensitivity and diagnostic checks confirm robustness and computational stability. Across three independent MCMC initializations, overall posterior means vary
minimally ($-4.65\%$ to $-4.89\%$, all credible intervals excluding zero), and varying the error-variance prior rate $\lambda \in \{0.1, 1, 10\}$ produces stable
estimates with modest interval widening under larger rates without random effects (Appendix Table~\ref{tab:application_prior_sensitivity}). Finally, the unconstrained lead diagnostic supports
conditional parallel trends (Appendix Table~\ref{tab:application_pretrend}): the pre-treatment linear slope is $-0.0057$ (95\% CI $[-0.0202, 0.0071]$, $p_{\mathrm{tail}} = 0.206$),
failing to reject the null at the prespecified $0.025$ threshold, while individual lead tests yield an unadjusted $p_{\mathrm{tail}} = 0.046$ that rises to
$p_{\mathrm{tail}} = 0.138$ under Bonferroni correction, which does not reject parallel trends across any specification.

In summary, our overall estimate of a $-4.72\%$ reduction in teen employment aligns with empirical evidence that minimum wage increases exert moderate, disaggregated effects on teenage employment \parencite{Williams2006,Brown1981,Sen2011,Wellington1991}. The point estimate falls directly between the standard two-way fixed-effects estimate ($-3.7\%$) and doubly robust semiparametric estimate ($-9.0\%$) of \textcite{SantAnna2020}, demonstrating that DiD-BCF provides an adaptive balance between parametric rigidity and high-variance weighting.

Substantively, the absence of a pronounced population gradient clarifies a recurring debate in local labor market econometrics. While smaller metropolitan and rural areas are often hypothesized to be more vulnerable to wage floors due to lower median wages \parencite{jpmorganchase_rural_divide_2025,ratcliffe2016defining,usda_county_typology_2025}, our findings indicate that county population alone is an insufficient proxy for vulnerability. As emphasized by \textcite{Thompson2009} and \textcite{Kalenkoski2013}, industry composition, commuting flows, and state policy rollouts interact in complex ways that aggregate population metrics cannot capture. Although data constraints prevent us from incorporating these granular covariates, DiD-BCF is ideally suited for such environments. By nonparametrically accommodating complex, multidimensional interactions without pre-specifying functional forms, it frees applied researchers to concentrate on substantive identification assumptions and robustness checks rather than ad-hoc model specification.

\section{Conclusion}\label{conclusion_did_bcf}
This paper introduced Difference-in-Differences Bayesian Causal Forests (DiD-BCF), a Bayesian nonparametric framework for estimating heterogeneous treatment effects in observational panel data under common or staggered intervention schedules. By decomposing the untreated potential outcome into an additive sum of cohort-specific level surfaces and common calendar-time trend surfaces, DiD-BCF embeds conditional parallel trends and no anticipation directly into a multi-forest architecture. A separate, regularized treatment-effect forest provides unified posterior distributions over fine-grained conditional effects (CATT), cohort-time parameters (GATT), and arbitrary aggregates without multi-step plug-in steps.

Our theoretical analysis links this architecture to nonparametric identification and establishes the integrated $L_2$ consistency of the exact continuous-cut forest posterior over joint finite-tree classes, with a growing-capacity extension for scalar H\"older continuous functions. We prove that transporting conditional estimates to population-level treatment effects requires an explicit mean-effect transportability condition. Furthermore, we characterize full-panel semiparametric efficiency gains and provide adaptive Bernstein--von Mises asymptotic normality theorems, delineating how functional approximation bias, prior shrinkage, and error covariance structures dictate posterior calibration.

Our extensive Monte Carlo simulations demonstrate that DiD-BCF achieves substantial accuracy and uncertainty quantification gains across both aggregate estimands ($\mathrm{ATT}$, $\mathrm{GATT}$, $\mathrm{ATT}(k)$) and individual treatment-effect surfaces ($\mathrm{CATT}$), reducing estimation error by 40\% to 65\% relative to parametric, semiparametric, and machine-learning benchmarks in both linear and nonlinear canonical designs. Nonetheless, this superiority has limitations in settings under severe within-unit serial correlation, in sparse staggered panels with limited treated observations in later cohorts, and under subtle and heterogeneous parallel trends assumption violations. Applied to U.S. county-level teen employment following minimum wage increases, DiD-BCF estimates a credible disemployment effect of $-4.72\%$ that sits adaptively between two-way fixed-effects and doubly robust semiparametric bounds, while uncovering that county population alone does not drive systematic effect moderation.

Key directions for future work include: 1. developing adaptive shrinkage or targeted split proposals for auxiliary lead forests to mitigate conservativeness against subtle, aggregate-cancelling pre-trend heterogeneity; 2. extending continuous-cut posterior consistency guarantees to multivariate smooth function spaces; 3. incorporating structured autoregressive or cluster-robust error covariance kernels directly into the forest likelihood; and 4. establishing non-asymptotic mixing bounds for Metropolis--Hastings tree proposals to characterize finite MCMC convergence.

\spacingset{1}\small

\subsection*{Disclosure statement}\label{disclosure-statement}
The authors report there are no competing interests to declare.

\subsection*{Data Availability Statement}\label{data-availability-statement}
Deidentified county-level teen employment data (\texttt{mpdta}) are publicly available from the \texttt{did} \textsf{R} package. Simulation and application code, notebooks, and result summaries are available at \url{https://github.com/hugogobato/DiD-BCF}.

\subsection*{Appendix}\label{appendix-statement}
\noindent The Appendix contains: the reparameterization rationale (Appendix~\ref{supp:reparam}); full estimation algorithms and hyperparameters (Appendix~\ref{supp:algo}); benchmark descriptions (Appendix~\ref{supp:benchmarks}); metric definitions (Appendix~\ref{supp:metrics}); complete data-generating processes (Appendix~\ref{supp:designs}); extended simulation results (Appendix~\ref{supp:results}); and formal theoretical proofs (Appendix~\ref{supp:theory}).

\begingroup
\spacingset{0.88}\scriptsize
\setlength{\bibsep}{0pt plus 0.2pt}
\renewcommand{\newline}{\unskip, }
\bibliography{References}
\endgroup

\clearpage
\appendix
\spacingset{1.35}\normalsize
\numberwithin{equation}{section}
\setcounter{equation}{0}
\renewcommand{\theHequation}{\thesection.\arabic{equation}}
\setcounter{table}{0}
\renewcommand{\thetable}{A\arabic{table}}
\renewcommand{\theHtable}{A\arabic{table}}
\setcounter{figure}{0}
\renewcommand{\thefigure}{A\arabic{figure}}
\renewcommand{\theHfigure}{A\arabic{figure}}
\setcounter{algocf}{0}
\renewcommand{\thealgocf}{A\arabic{algocf}}
\renewcommand{\theHalgocf}{A\arabic{algocf}}
\numberwithin{theorem}{section}
\setcounter{theorem}{0}
\renewcommand{\thetheorem}{\thesection.\arabic{theorem}}
\renewcommand{\theHtheorem}{\thesection.\arabic{theorem}}

\begin{center}
  {\LARGE\bf Appendix}
\end{center}
\medskip

\noindent This Appendix provides supporting methodological details, extended simulation results, and complete theoretical proofs. Appendix~\ref{supp:reparam} details the treatment-status reparameterization and its architectural advantages. Appendix~\ref{supp:algo} describes the estimation algorithms, Grow-From-Root initialization, stopping rules, and sampler hyperparameters. Appendix~\ref{supp:benchmarks} describes the benchmarks, and Appendix~\ref{supp:metrics} defines the evaluation metrics. Appendices~\ref{supp:designs} and \ref{supp:results} provide the detailed simulation designs and extended empirical results. Appendix~\ref{supp:theory} presents complete theoretical proofs for identification, full-panel information bounds, and posterior consistency, including continuous-cut effect consistency on a fixed-capacity joint finite-tree truth class, a scalar smooth extension with growing depth caps, finite-support unknown-variance extensions, growing histogram inference regimes, and the algebraic representation and quotient-gauge structure of the hierarchical forest posterior.

\section{Reparameterization Rationale}\label{supp:reparam}
Let $G_i \in \{2,\dots,T\} \cup \{\infty\}$ denote the adoption period for unit $i$, let $D_{it} = \mathbf{1}\{G_i \le t, G_i < \infty\}$ be the binary post-treatment indicator, and let $k_{it} = t - G_i$ denote event time for treated cohorts. The DiD-BCF model represents the observed outcome as
\begin{equation}
Y_{it} = a(G_i, X_i) + b(t, X_i) + D_{it}\tau(G_i, k_{it}, X_i) + \varepsilon_{it}, \quad \mathbb{E}[\varepsilon_{it} \mid G_i, X_i, t] = 0.
\end{equation}
This formulation has two fundamental structural properties. First, the structured baseline $\mu_0(G_i, t, X_i) = a(G_i, X_i) + b(t, X_i)$ directly embeds the conditional parallel trends assumption: conditional on baseline covariates $X_i$, the evolution of untreated potential outcomes decomposes additively into a cohort-specific level effect $a(G_i, X_i)$ and a common time trend $b(t, X_i)$. Second, the treatment-status multiplier $D_{it}$ mechanically enforces no anticipation by deactivating the treatment forest $\tau(\cdot)$ for all untreated and pre-treatment observations ($t < G_i$ or $G_i = \infty$).

This parameterization yields the same observable conditional expectations as an ever-treated indicator $W_i = \mathbf{1}\{G_i < \infty\}$ paired with a treatment effect function constrained to equal zero at negative event times ($k_{it} < 0$). Given any such constrained function, retaining its post-treatment values produces identical conditional means; conversely, extending any post-treatment function by zero to negative event times preserves the observable distribution. Values of an unconstrained effect forest evaluated at untreated observations ($D_{it} = 0$) are not identified by the likelihood and are never reported as causal effects.

The primary motivation for this specification is architectural. In standard Bayesian additive tree implementations that condition on an ever-treated indicator $W_i$, the treatment effect forest must expend substantial splitting capacity and tree depth attempting to approximate a sharp step-function that is flat at zero for all $k_{it} < 0$ and non-zero thereafter. Because decision trees partition the covariate space via axis-aligned splits, forcing the trees to learn an exact zero surface in the pre-treatment region wastes leaf parameters and degrades posterior estimation of post-treatment heterogeneity. By incorporating $D_{it}$ as an exact structural multiplier, the treatment forest $\tau(G_i, k_{it}, X_i)$ dedicates its entire splitting budget and posterior shrinkage directly to estimating post-treatment heterogeneity. This structural choice does not artificially accelerate posterior contraction rates or bypass identification requirements; identification remains anchored in the baseline conditional parallel trends restriction and support conditions detailed in Appendix~\ref{supp:theory}. Consequently, pre-treatment trend diagnostics cannot be evaluated by inspecting the treatment effect forest (which is structurally zero prior to adoption), but must be conducted via an auxiliary lead model that estimates pseudo-effects in pre-treatment periods ($k_{it} < 0$), as described in Section~\ref{model_did_bcf}.

\section{Estimation Algorithms and ws-BCF Implementation}\label{supp:algo}

\subsection{Bayesian Additive Regression Trees (BART)}

Bayesian Additive Regression Trees (BART), introduced by \textcite{Chipman2010}, is a non-parametric Bayesian ensemble method that models an unknown conditional mean function $f(x) = \mathbb{E}[Y \mid X = x]$ as a sum of multiple regression trees:
\begin{equation}
    Y_j = \sum_{l=1}^{M_{\mathrm{trees}}} g_l(x_j; T_l, \mathcal{M}_l) + \varepsilon_j, \quad \varepsilon_j \sim \mathcal{N}(0, \sigma^2).
\end{equation}
Here, $Y_j$ denotes the scalar response and $x_j$ is a vector of predictors for observation $j$. Each component $g_l(x_j; T_l, \mathcal{M}_l)$ represents an individual binary regression tree, where $T_l$ denotes the tree topology and decision split rules, and $\mathcal{M}_l = \{\mu_{l1}, \dots, \mu_{lb_l}\}$ denotes the set of terminal leaf parameters associated with the $b_l$ terminal nodes of tree $T_l$. The error terms $\varepsilon_j$ are assumed to be independent and identically distributed Gaussian disturbances with unknown variance $\sigma^2$. The central conceptual principle of BART is that each individual tree acts as a ``weak learner,'' explaining only a small portion of the overall regression function, while their additive sum provides a highly flexible and regularized non-parametric fit. The canonical BART specification sets $M_{\mathrm{trees}} = 200$ trees as the general default \parencite{Chipman2010}.

Regularization in BART is achieved through independent prior distributions placed on the tree structures $(T_l)$, the terminal leaf parameters $(\mathcal{M}_l)$, and the residual error variance $(\sigma^2)$:
\begin{equation}
    p(T_1, \mathcal{M}_1, \dots, T_{M_{\mathrm{trees}}}, \mathcal{M}_{M_{\mathrm{trees}}}, \sigma^2) = \left[ \prod_{l=1}^{M_{\mathrm{trees}}} p(T_l) \, p(\mathcal{M}_l \mid T_l) \right] p(\sigma^2).
\end{equation}
The tree topology prior $p(T_l)$ penalizes deep trees by specifying that a node at depth $d \in \{0, 1, 2, \dots\}$ splits with probability $\alpha(1+d)^{-\delta}$, with standard defaults $\alpha = 0.95$ and $\delta = 2$. Given a split decision, splitting variables and cutpoints are drawn uniformly from the available feature coordinates and candidate partition thresholds. Conditional on the tree structure $T_l$, the leaf parameters $\mu_{lm}$ are assigned independent conjugate Gaussian priors, $\mu_{lm} \sim \mathcal{N}(0, \sigma_\mu^2)$, where $\sigma_\mu = \sigma_0 / (k\sqrt{M_{\mathrm{trees}}})$ scales inversely with the number of trees $M_{\mathrm{trees}}$, shrinking each tree's contribution toward zero. The residual variance $\sigma^2$ is assigned a conjugate inverse-gamma prior, $\sigma^2 \sim \text{Inv-Gamma}(\nu/2, \nu\lambda/2)$, calibrated against the sample variance of the outcome. Estimation is conducted via Bayesian backfitting Markov chain Monte Carlo (MCMC), iteratively drawing each tree conditional on the partial residuals of the remaining $M_{\mathrm{trees}}-1$ trees using Metropolis--Hastings proposals \parencite{Chipman2010}.

\subsection{Bayesian Causal Forests (BCF)}

While BART can be adapted for causal inference by appending a binary treatment indicator $Z_j \in \{0, 1\}$ directly to the predictor vector $x_j$ (an ``S-learner'' approach, per \textcite{Knzel2019}), this structure forces the tree splits to share capacity between prognostic baseline variation and treatment effect variation, often drowning out modest treatment effects \parencite{pmlr-v130-curth21a}. Alternatively, fitting separate BART models to treated and control subsamples (a ``T-learner'') applies equal regularization to both surfaces, resulting in excessive variance when the treatment effect is simpler or smoother than the baseline outcome \parencite{Hahn2020}. 

To resolve these limitations, \textcite{Hahn2020} introduced Bayesian Causal Forests (BCF), which directly parameterize the conditional outcome as:
\begin{equation}
    Y_j = \mu(x_j, \widehat{\pi}(x_j)) + \tau(x_j)\,Z_j + \varepsilon_j, \quad \varepsilon_j \sim \mathcal{N}(0, \sigma^2),
\end{equation}
where $\mu(\cdot)$ is the prognostic function capturing the baseline counterfactual outcome, $\tau(\cdot)$ is the treatment effect function capturing causal effect heterogeneity across covariates $x_j$, and $\widehat{\pi}(x_j)$ is an estimated propensity score included in the prognostic forest to prevent regularization-induced confounding (RIC) \parencite{Hahn2018,Hahn2020}. In difference-in-differences settings with panel data, conditioning the structured baseline on cohort and time indicators alongside covariates already controls for confounding under parallel trends, rendering explicit propensity score inclusion in $\mu(\cdot)$ unnecessary \parencite{https://doi.org/10.48550/arxiv.2410.15560}.

Both $\mu(\cdot)$ and $\tau(\cdot)$ are modeled as separate sums of regression trees, enabling distinct prior regularization. The prognostic forest $\mu(\cdot)$ employs standard BART hyperparameters ($M_\mu = 125$ or 200, $\alpha_\mu = 0.95$, $\delta_\mu = 2$) to capture complex baseline outcome surfaces. In contrast, the treatment effect forest $\tau(\cdot)$ is subjected to substantially stronger shrinkage: fewer trees ($M_\tau = 50$ or 100), an aggressive depth penalty ($\alpha_\tau = 0.25$, $\delta_\tau = 3$), and a concentrated leaf prior variance. This differential regularization reflects the economic and statistical reality that causal effect heterogeneity is typically smoother and of smaller magnitude than baseline outcome variation \parencite{Hahn2020}.

\subsection{Warm-Start Bayesian Causal Forests (ws-BCF)}

Backfitting MCMC for BCF can be computationally demanding because of dependencies between tree updates. XBART uses stochastic Grow-From-Root (GFR) updates to construct tree ensembles \parencite{He2021}. \textcite{pmlr-v206-krantsevich23a} developed XBCF and its use for warm-start initialization, with treatment-group scaling parameters:
\begin{equation}
    Y_j = a\,\mu(x_j, \widehat{\pi}(x_j)) + b_{Z_j}\widetilde{\tau}(x_j) + \varepsilon_j, \quad \varepsilon_j \sim \mathcal{N}(0, \sigma^2),
\end{equation}
where $a \sim \mathcal{N}(0, 1)$ and $b_0, b_1 \sim \mathcal{N}(0, 1/2)$, and the effective treatment effect is $\tau(x) = (b_1 - b_0)\widetilde{\tau}(x)$. The display uses a common residual variance; the original XBCF formulation also permits different residual variances in the two treatment groups.

In DiD-BCF, Grow-From-Root (GFR) sweeps initialize the forest topologies (Algorithm~\ref{alg:gfr}), warm-starting the tree structures before transitioning to Bayesian backfitting MCMC sweeps. Subsequent iterations cycle through Metropolis--Hastings tree updates (proposing stochastic grow and prune steps on terminal nodes) and conjugate Gibbs draws for leaf parameters, error variance scales, and the global treatment intercept. Exact posterior invariance in discrete tree-space Markov chains relies on matching forward and reverse proposal probabilities across boundary constraints, such as maximum depth limits and minimum terminal node counts. In practice, empirical convergence and mixing in finite chains are assessed using multi-chain diagnostics and simulation-based calibration across independent random initializations. The theoretical properties developed in Appendix~\ref{supp:theory} characterize the exact mathematical posterior distribution under the structured prior, separating the statistical guarantees of the DiD-BCF posterior from finite-chain Monte Carlo approximation error (Proposition~\ref{imp:prop:finite}).

\begin{algorithm}[H]
	\caption{GrowFromRoot Stochastic Tree Initialization \parencite{He2021}}\label{alg:gfr}
	\KwIn{Data $(y, \mathbf{X})$, tree depth $d$, current tree $T$, hyperparameters $(\alpha, \beta, \mathbf{\Phi}, \mathbf{\Psi})$, candidate cutpoint set $\mathcal{C}$.}
	\KwOut{Updated tree $T$ with newly grown child nodes and terminal leaf parameters.}
	\If{the stopping conditions are met}{
		Draw leaf parameter $\mu_{\mathrm{node}} \gets \mathrm{SampleParameters}(\emptyset)$\;
		\Return;
	}
	Compute sufficient statistics for parent node: $s^{\emptyset} \gets s(y, \mathbf{X}, \mathbf{\Psi}, \mathcal{C}, \text{all})$\;
	\For{each candidate cutpoint $c_{jk} \in \mathcal{C}$}{
		Compute left sufficient statistics: $s^{(1)}_{jk} \gets s(y, \mathbf{X}, \mathbf{\Psi}, \mathcal{C}, j, k, \text{left})$\;
		Compute right sufficient statistics: $s^{(2)}_{jk} \gets s(y, \mathbf{X}, \mathbf{\Psi}, \mathcal{C}, j, k, \text{right})$\;
		Calculate marginal split likelihood: $L(c_{jk}) = m \left( s^{(1)}_{jk}; \mathbf{\Phi}, \mathbf{\Psi} \right) \times m \left( s^{(2)}_{jk}; \mathbf{\Phi}, \mathbf{\Psi} \right)$\;
	}
	Calculate null cutpoint likelihood: $L(\emptyset) = |\mathcal{C}| \left( \frac{(1+d)^{\beta}}{\alpha} - 1 \right) m \left( s^{\emptyset}; \mathbf{\Phi}, \mathbf{\Psi} \right)$\;
	Sample a cutpoint $c_{jk} \in \mathcal{C} \cup \{\emptyset\}$ proportional to integrated likelihoods:
	\begin{equation*}
		P(c_{jk}) = \frac{L(c_{jk})}{\sum_{c \in \mathcal{C}} L(c) + L(\emptyset)}, \qquad P(\emptyset) = \frac{L(\emptyset)}{\sum_{c \in \mathcal{C}} L(c) + L(\emptyset)}\;
	\end{equation*}
	\eIf{the null cutpoint $\emptyset$ is selected}{
		$\mu_{\mathrm{node}} \gets \mathrm{SampleParameters}(\emptyset)$\;
		\Return;
	}{
		Create child nodes \textbf{left\_node} and \textbf{right\_node}, designating them as children of the current node\;
		Partition data $(y, \mathbf{X})$ into left $(y_{\mathrm{left}}, \mathbf{X}_{\mathrm{left}})$ and right $(y_{\mathrm{right}}, \mathbf{X}_{\mathrm{right}})$ subsets according to $X_{ij} \le x_{jk}^*$ and $X_{ij} > x_{jk}^*$\;
		$\text{GrowFromRoot}(y_{\mathrm{left}}, \mathbf{X}_{\mathrm{left}}, \mathbf{\Phi}, \mathbf{\Psi}, d+1, T, \textbf{left\_node})$\;
		$\text{GrowFromRoot}(y_{\mathrm{right}}, \mathbf{X}_{\mathrm{right}}, \mathbf{\Phi}, \mathbf{\Psi}, d+1, T, \textbf{right\_node})$\;
	}
\end{algorithm}

\paragraph{Stopping conditions.}
The stopping criteria in Algorithm~\ref{alg:gfr} comprise three complementary mechanisms:
\begin{enumerate}
    \item \emph{The null-cutpoint rule}: At every candidate node, the algorithm compares candidate splits against a null cutpoint (no split) that carries the depth-dependent penalty $|\mathcal{C}|((1+d)^{\beta}/\alpha - 1)$ in its integrated marginal likelihood. When the null cutpoint is sampled, tree growth on that branch terminates immediately.
    \item \emph{Hard structural constraints}: Node splitting is strictly disallowed if the current depth exceeds a pre-specified maximum depth cap $D_{\max}$ or if either potential child node would contain fewer than a minimum number of observations $N_{\min}$.
    \item \emph{Ensemble sweep budget}: At the forest level, a fixed number of Grow-From-Root sweeps (50 sweeps in our implementation) is executed, after which the initialized tree topologies and parameters are passed directly to the MCMC sampler.
\end{enumerate}

\subsection{Top-Level DiD-BCF Estimation Procedure}

The complete DiD-BCF estimation workflow proceeds in seven sequential steps:
\begin{enumerate}
    \item \textbf{Data inputs}: A balanced panel $\{Y_{it}, G_i, X_i\}_{i=1,t=1}^{n,T}$ where $G_i \in \{2, \dots, T\} \cup \{\infty\}$ denotes the adoption period and $X_i$ represents baseline covariates.
    \item \textbf{Index construction}: Compute event time $k_{it} = t - G_i$ for treated cohorts and the post-treatment indicator $D_{it} = \mathbf{1}\{G_i \le t, G_i < \infty\}$.
    \item \textbf{Structured baseline}: Decompose the prognostic outcome into a cohort level forest $a(G_i, X_i)$ and a common time trend forest $b(t, X_i)$, satisfying $\mu_0(G_i, t, X_i) = a(G_i, X_i) + b(t, X_i)$, which directly operationalizes conditional parallel trends.
    \item \textbf{Treatment block}: Specify the treatment effect forest $\tau_{\mathrm{forest}}(G_i, k_{it}, X_i)$ plus the global intercept $\taug$, forming the composite effect $\tau(G_i, k_{it}, X_i) = \taug + \tau_{\mathrm{forest}}(G_i, k_{it}, X_i)$ modulated by the structural multiplier $D_{it}$.
    \item \textbf{Warm start}: Execute 50 Grow-From-Root initialization sweeps to establish initial tree topologies and leaf parameters across all forest components.
    \item \textbf{MCMC sampling}: Run Bayesian backfitting MCMC chains initialized from the warm-start topologies under the prior specifications below.
    \item \textbf{Posterior aggregation}: Retain posterior draws of the composite treatment effect surface $\tau(G_i, k_{it}, X_i) = \taug + \tau_{\mathrm{forest}}(G_i, k_{it}, X_i)$ to compute posterior means, credible intervals, and target aggregations for $\mathrm{ATT}$, $\mathrm{GATT}(g,t)$, and $\mathrm{CATT}(x; g, t)$.
\end{enumerate}

\paragraph{Sampler hyperparameters.}
On the standardized outcome scale, the DiD-BCF implementation configures the baseline prognostic blocks $a(G_i, X_i)$ and $b(t, X_i)$ with $M_{\mathrm{trees}} = 125$ trees each, maximum tree depth $D_{\max} = 10$, minimum leaf size $N_{\min} = 5$ panel observations, split prior $(\alpha, \delta) = (0.95, 2)$, and Gaussian leaf parameters with an inverse-gamma prior on the leaf variance parameter $\sigma_\mu^2 \sim \operatorname{IG}(3/2, 1/250)$, initialized at $2/125$ (see Proposition~\ref{imp:prop:scale}). The treatment effect forest $\tau_{\mathrm{forest}}(G_i, k_{it}, X_i)$ employs $M_{\mathrm{trees}} = 100$ trees, maximum depth $D_{\max} = 5$, minimum leaf size $N_{\min} = 5$ panel observations, split prior $(\alpha, \delta) = (0.25, 3)$, and Gaussian leaves with fixed variance scale $\sigma_\tau^2 = 0.5 / 100 = 1/200$. The global intercept $\taug$ is assigned a zero-mean, unit-variance Gaussian prior on the standardized outcome scale ($\taug \sim \mathcal{N}(0, 1)$). The observation error variance $\sigma^2$ carries an inverse-gamma prior $\sigma^2 \sim \operatorname{IG}(2, 1)$ with density proportional to $v^{-3}\exp(-1/v)$. Optional unit-specific or cohort-specific random intercepts $\alpha_i \sim \mathcal{N}(0, \sigma_\alpha^2)$ with $\sigma_\alpha^2 \sim \operatorname{IG}(2, 1)$ can be included to account for clustered residual correlation. Across all simulation experiments and empirical applications, tree topologies and leaf parameters are initialized via 50 Grow-From-Root (GFR) warm-start sweeps (Algorithm~\ref{alg:gfr}), followed by three independent MCMC chains running 2,500 Bayesian backfitting sweeps per chain thinned at an interval of 5, yielding 500 retained draws per chain (1,500 total retained posterior draws). The complete hyperparameter configuration is summarized below:

\begin{center}
\footnotesize
\resizebox{\linewidth}{!}{%
\begin{tabular}{@{}lllll@{}}
\toprule
Component & Parameter & Description & Prior / Value & Scaling / Initialization \\
\midrule
Level forest $a(G_i, X_i)$ & $M_{\mathrm{trees}}$ & Number of trees & $125$ & Fixed ensemble size \\
 & $D_{\max}$ & Maximum tree depth & $10$ & Structural depth cap \\
 & $N_{\min}$ & Minimum leaf size & $5$ & Panel observations ($N_{\mathrm{obs}} = nT$) \\
 & $(\alpha, \delta)$ & Split probability $\alpha(1+d)^{-\delta}$ & $(0.95, 2)$ & Moderate tree regularization \\
 & $\sigma_\mu^2$ & Leaf variance & $\operatorname{IG}(3/2, 1/250)$ & Initialized at $2/125$ \\
\addlinespace
Trend forest $b(t, X_i)$ & $M_{\mathrm{trees}}$ & Number of trees & $125$ & Fixed ensemble size \\
 & $D_{\max}$ & Maximum tree depth & $10$ & Structural depth cap \\
 & $N_{\min}$ & Minimum leaf size & $5$ & Panel observations ($N_{\mathrm{obs}} = nT$) \\
 & $(\alpha, \delta)$ & Split probability $\alpha(1+d)^{-\delta}$ & $(0.95, 2)$ & Moderate tree regularization \\
 & $\sigma_\mu^2$ & Leaf variance & $\operatorname{IG}(3/2, 1/250)$ & Initialized at $2/125$ \\
\addlinespace
Treatment forest $\tau_{\mathrm{forest}}$ & $M_{\mathrm{trees}}$ & Number of trees & $100$ & Fixed ensemble size \\
 & $D_{\max}$ & Maximum tree depth & $5$ & Conservative depth cap \\
 & $N_{\min}$ & Minimum leaf size & $5$ & Panel observations ($N_{\mathrm{obs}} = nT$) \\
 & $(\alpha, \delta)$ & Split probability $\alpha(1+d)^{-\delta}$ & $(0.25, 3)$ & Strong tree regularization \\
 & $\sigma_\tau^2$ & Leaf variance & $1/200$ & Fixed scale ($0.5 / M_{\mathrm{trees}}$) \\
\addlinespace
Global intercept $\taug$ & $\taug$ & Baseline effect shift & $\mathcal{N}(0, 1)$ & Standardized outcome scale \\
Error variance $\sigma^2$ & $\sigma^2$ & Observation noise & $\operatorname{IG}(2, 1)$ & Density $\propto v^{-3}\exp(-1/v)$ \\
Unit intercept $\alpha_i$ (optional) & $\sigma_\alpha^2$ & Random intercept variance & $\operatorname{IG}(2, 1)$ & $\alpha_i \sim \mathcal{N}(0, \sigma_\alpha^2)$ \\
\addlinespace
Initialization & GFR & Grow-From-Root sweeps & $50$ & Warm-start topologies \\
MCMC sampling & Chains & Independent chains & $3$ & Disjoint random seeds \\
 & Sweeps & Backfitting sweeps / chain & $2{,}500$ & Post-initialization \\
 & Thinning & Thinning interval & $5$ & Retains $500$ draws / chain \\
 & Total draws & Pooled posterior draws & $1{,}500$ & Aggregated across chains \\
\bottomrule
\end{tabular}%
}
\end{center}

\paragraph{Absorbing versus projecting out unit-level heterogeneity.}
DiD-BCF conditions the baseline on the cohort label and covariates and optionally adds a time-invariant random intercept; it does not project out unit fixed effects before fitting. The alternative, adopted by first-difference designs such as grf-DiD (Appendix~\ref{supp:benchmarks}), is to difference the outcome so that any time-invariant unobserved unit effect cancels identically. Absorbing keeps the level information (which helps the prognostic forest and preserves unbalanced panels) but relies on the included covariates and group terms to capture persistent confounders; projecting out removes such confounders nonparametrically regardless of covariate adjustment, at the price of discarding level information and amplifying noise in short panels. The canonical-DiD designs of Appendix~\ref{supp:designs}, in which an unobserved unit effect $\alpha_i$ correlated with treatment enters $Y_{it}(0)$, directly probe this trade-off: DiD-BCF remains close to the truth across confounding strengths because its structured baseline plus covariates absorb most of the persistent component, while estimators that neither difference nor condition appropriately degrade.

\section{Benchmark Models}\label{supp:benchmarks}
We compare DiD-BCF with six established difference-in-differences estimators spanning parametric, semi-parametric, machine-learning, and non-parametric causal forest approaches. In every simulation replication, all seven estimators are evaluated on the exact same simulated panel dataset. Missing estimates (e.g., due to sample-splitting constraints in small panels) are excluded only from that specific estimator's summary metrics, with retained replication fractions reported in the tables.

\paragraph{Two-way fixed effects (TWFE).}
The canonical two-way fixed effects estimator fits unit and calendar-time fixed effects alongside treatment or event-time indicators and observed covariates via ordinary least squares. Under staggered treatment adoption and heterogeneous treatment effects across cohorts or time, the single TWFE coefficient represents a weighted sum of cell-specific treatment effects that can place negative weights on certain groups \citep{GoodmanBacon2021}. We evaluate both its overall estimation error and its underlying two-by-two comparison weights via the Goodman-Bacon decomposition.

\paragraph{Callaway--Sant'Anna (CS).}
The group-time difference-in-differences estimator of \textcite{Callaway2021} targets cohort- and time-specific treatment effects $\mathrm{GATT}(g,t)$ using not-yet-treated or never-treated units as comparison groups, aggregating these effects into summary estimands with explicit convex weights. We implement its doubly robust version, which combines linear outcome regression with inverse propensity score weighting \citep{SantAnna2020}, using the \texttt{did} package in R \citep{R-did}.

\paragraph{Gardner's two-stage estimator (did2s).}
The two-stage difference-in-differences estimator of \textcite{https://doi.org/10.48550/arxiv.2207.05943} isolates untreated potential outcome dynamics in the first stage using untreated observations, and regresses residualized outcomes on treatment indicators in the second stage, correcting standard errors for first-stage estimation uncertainty. We utilize the \texttt{did2s} implementation in R \citep{R-did2s}.

\paragraph{Synthetic Difference-in-Differences (SDID).}
Synthetic DiD integrates unit weights (which balance pre-treatment outcome trajectories between treated and untreated units) and time weights with a two-way fixed effects specification \citep{Arkhangelsky2021}. While robust to unobserved time-varying confounders that follow an interactive fixed effects factor structure, its primary target is an average treatment effect rather than fine within-cell conditional heterogeneity.

\paragraph{Double Machine Learning DiD (DoubleML).}
Double Machine Learning DiD combines Neyman-orthogonal score equations with non-parametric machine learning nuisance estimators and cross-fitting to achieve $\sqrt{n}$-consistent estimation of average treatment effects under weak conditions \citep{SantAnna2020,Chang2020,DoubleML2022}. We implement DoubleML DiD using 500-tree random forests for both propensity scores and outcome regressions.

\paragraph{Generalized Random Forests DiD (grf-DiD).}
For each cohort-time cell, the outcome is first-differenced relative to the cohort's final pre-treatment period ($t = g-1$). A causal forest \citep{Athey2019} is then fitted to these differenced outcomes using baseline covariates and the treatment indicator \citep{Wang2022}. By taking first differences, time-invariant unit confounders cancel out mechanically. Because grf-DiD estimates conditional average treatment effects as a function of continuous covariates, it serves as the primary non-parametric comparator for DiD-BCF on conditional response surfaces.

\paragraph{Evaluation on conditional surfaces.}
When scalar average-effect estimators (such as TWFE, CS, did2s, SDID, and DoubleML) are included in conditional surface evaluations, their cell-level scalar estimate is broadcast across all units in that cell. This provides a natural reference point for how much predictive accuracy is gained by modeling individual-level covariate heterogeneity relative to broadcasting cell averages.

\section{Evaluation Metrics}\label{supp:metrics}

\subsection{Point-accuracy metrics}

Let $\widehat\tau(G_i, k_{it}, X_i)$ denote an estimator and $\tau(G_i, k_{it}, X_i)$ the true treatment effect on the treated observations. With $N_{\mathrm{tr}} = \sum_{i=1}^n \sum_{t=1}^T D_{it}$ denoting the total number of treated post-treatment observations, the Root Mean Squared Error (RMSE), Mean Absolute Error (MAE), and Mean Absolute Percentage Error (MAPE) are defined as:
\begin{align}
    \text{RMSE} &= \sqrt{\frac{1}{N_{\mathrm{tr}}} \sum_{i=1}^n \sum_{t=1}^T D_{it} \big(\widehat\tau(G_i, k_{it}, X_i) - \tau(G_i, k_{it}, X_i)\big)^2 },\\
    \text{MAE} &= \frac{1}{N_{\mathrm{tr}}} \sum_{i=1}^n \sum_{t=1}^T D_{it} \big|\widehat\tau(G_i, k_{it}, X_i) - \tau(G_i, k_{it}, X_i)\big|,\\
    \text{MAPE} &= \frac{1}{N_{\mathrm{tr}}} \sum_{i=1}^n \sum_{t=1}^T D_{it} \frac{\big|\widehat\tau(G_i, k_{it}, X_i) - \tau(G_i, k_{it}, X_i)\big|}{|\tau(G_i, k_{it}, X_i)|} \times 100.
\end{align}
RMSE is particularly sensitive to large localized errors due to quadratic penalization; MAE weights errors linearly across the covariate support; and MAPE normalizes the absolute error by the true effect magnitude, providing a scale-free measure of relative precision across varying effect sizes.

\subsection{Bias, dispersion, and coverage}

Let $\widehat\theta_r$ denote the estimate in Monte Carlo replication $r$, $\theta_r$ the realized target estimand, and $e_r = \widehat\theta_r - \theta_r$ the estimation error. Across $R$ available replications, we define:
\begin{align}
\operatorname{Bias} &= R^{-1}\sum_{r=1}^R e_r,\\
\operatorname{RMSE} &= \left\{ R^{-1}\sum_{r=1}^R e_r^2 \right\}^{1/2},\\
\operatorname{SD}_{R}(e) &= \left\{ \operatorname{RMSE}^2 - \operatorname{Bias}^2 \right\}^{1/2},\\
\operatorname{Coverage}_{1-\alpha} &= R^{-1}\sum_{r=1}^R \mathbf{1}\{L_r \le \theta_r \le U_r\},\\
\operatorname{Length}_{1-\alpha} &= R^{-1}\sum_{r=1}^R (U_r - L_r),\\
\operatorname{SDRatio} &= \frac{R^{-1}\sum_{r=1}^R s_r}{\operatorname{SD}(\widehat\theta)}.
\end{align}
Here $[L_r, U_r]$ denotes the reported 95\% credible interval or confidence interval, and $s_r$ denotes the posterior standard deviation or benchmark standard error in replication $r$. The Monte Carlo standard error for the estimated bias is $s_e / \sqrt{R}$, where $s_e = \sqrt{R/(R-1)}\operatorname{SD}_{R}(e)$ is the sample standard deviation of the replication errors.

The spread ratio (SDRatio) compares the average reported uncertainty $\overline{s} = R^{-1}\sum_r s_r$ to the empirical standard deviation of point estimates across replications, $\operatorname{SD}(\widehat\theta)$. For a realized target that varies across replications, $\overline{s}/s_e$ provides a direct measure of interval calibration, with values near 1.0 indicating well-calibrated uncertainty estimates. Empirical coverage and interval length are evaluated jointly to assess the efficiency and validity of inference. The approximate Monte Carlo standard error of a reported coverage frequency $\widehat{p}$ is $\{\widehat{p}(1-\widehat{p})/R\}^{1/2}$.

\subsection{Posterior sign summaries and null rejection}

For scalar posterior summaries, we compute the two-sided posterior sign score:
\begin{equation}
p_{\mathrm{tail}} = \min\big\{\mathbb{P}(\theta > 0 \mid Y),\, \mathbb{P}(\theta < 0 \mid Y)\big\}, \qquad p_{\mathrm{Bayes}} = 2\,p_{\mathrm{tail}}.
\end{equation}
A central 95\% posterior credible interval excludes zero if and only if $p_{\mathrm{tail}} < 0.025$, or equivalently $p_{\mathrm{Bayes}} < 0.05$. We report $p_{\mathrm{Bayes}}$ as an intuitive summary of posterior certainty regarding the direction of the causal effect, distinct from a frequentist $p$-value.

Under the sharp null hypothesis of zero treatment effects ($\tau \equiv 0$), null rejection frequencies directly measure empirical size. DiD-BCF exhibits conservative rejection rates in several small-sample cells, reflecting the regularizing shrinkage imposed by the tree depth and leaf variance priors.

\section{Simulation Designs}\label{supp:designs}

This section provides complete specifications for every data-generating process (DGP) evaluated in our simulation studies.

\subsection{Data-generating processes}\label{ssec:revgrid}

\subsubsection{Persistent latent covariates and the degree axis}
Each simulation replication generates persistent unit-level covariates drawn once and held fixed across all $T=8$ panel periods: $X_{1i} \sim \operatorname{Bernoulli}(0.5)$, $(X_{2i}, X_{3i}, X_{4i}) \sim \mathcal{N}(0, I_3)$, and $X_{5i} \sim \operatorname{Uniform}(-1, 1)$. To evaluate robustness against non-linear confounding, we define a degree axis $d \in \{1, 2, 3\}$. At degree $d=1$, the estimators receive the raw latent covariates. At $d=2$ and $d=3$, the underlying DGP remains identical, but the estimators receive non-linearly transformed observables:
\begin{align}
 X_{3i}^{o} &= \operatorname{std}\!\left[\left(\frac{X_{3i}X_{4i}}{5}+0.6\right)^3\right], &
 X_{4i}^{o} &= \operatorname{std}\!\left[\exp(X_{4i}/2)\right], &
 X_{5i}^{o} &= \operatorname{std}\!\left[(X_{3i}+X_{5i}+2)^2\right],
\end{align}
standardized within replication. At degree $d=3$, the true treatment-effect surface is additionally transformed from an additive linear function to a non-linear quadratic-plus-interaction surface, holding its marginal mean and variance constant.

\subsubsection{Canonical single-cohort designs: selection on unobserved time-invariant factors}\label{ssec:b1designs}
Let $A_i \sim \mathcal{N}(0, 1)$ and let the unobserved time-invariant unit effect be $u_i = \sigma_u A_i$. The latent treatment assignment utility is defined as:
\begin{equation}
U_i =
\begin{cases}
 0.8X_{1i} + 0.6X_{4i}, & \text{selection on observables},\\
 c_u A_i, & \text{selection on unobservables},\\
 0.8X_{1i} + 0.6X_{4i} + c_u A_i, & \text{selection on both},
\end{cases}
\end{equation}
centered at the target treated proportion $p_T = 0.5$, with binary treatment assignment generated via:
\begin{equation}
T_i \mid X_i, A_i \sim \operatorname{Bernoulli}\!\left[\operatorname{logit}^{-1}\!\left(1.5\left[U_i - Q_{1-p_T}(U)\right] + \nu_i\right)\right], \quad \nu_i \sim \mathcal{N}(0, 0.5^2).
\end{equation}
Treated units adopt treatment at period $G_i = 4$, while never-treated units have $G_i = \infty$, so that $D_{it} = \mathbf{1}\{G_i \le t, G_i < \infty\}$ and $k_{it} = t - G_i$. The untreated potential outcome is given by:
\begin{equation}
Y_{it}(0) = \beta_0 + u_i + \gamma_t + f(X_i) + s(X_i)\,t + \varepsilon_{it},
\end{equation}
with baseline level $\beta_0 = -0.5$, calendar trend $\gamma_t = 0.2t$, covariate main effect $f(X_i) = -0.75X_{1i} + 0.5X_{2i} - 0.5X_{3i} - 1.3X_{4i} + 1.8X_{5i}$, covariate-dependent slope $s(X_i) = 0.3(0.7X_{4i} + \sqrt{0.51}X_{3i})$, and observed outcome $Y_{it} = Y_{it}(0) + D_{it}\tau_i$. For degrees $d \in \{1, 2\}$, the treatment effect is $\tau_i = 3 + 1.5X_{1i} + 0.75X_{2i}$; for degree $d=3$,
\begin{equation}
\tau_i = 3 + 1.5X_{1i} + 0.75\left[\frac{1}{\sqrt{2}}\left\{\frac{X_{2i}^2-1}{\sqrt{2}} + (2X_{1i}-1)X_{2i}\right\}\right].
\end{equation}
Serial correlation variants replace i.i.d.\ errors with a stationary AR(1) process $\varepsilon_{it} = \rho\varepsilon_{i,t-1} + \sigma_\varepsilon\sqrt{1-\rho^2}Z_{it}$ where $Z_{it} \sim \mathcal{N}(0, 1)$. The canonical simulation scenarios are summarized below:

\begin{center}
\small
\resizebox{\linewidth}{!}{%
\begin{tabular}{@{}lrrrrl@{}}
\toprule
Scenario & $\sigma_u$ & $c_u$ & Selection Mechanism & $\rho$ & Treatment Effect \\
\midrule
Canonical: baseline & 1 & 1 & unobservable & 0 & heterogeneous \\
Canonical: strong confounder & 2 & 1.5 & unobservable & 0 & heterogeneous \\
Canonical: serial correlation & 1 & 1 & unobservable & 0.6 & heterogeneous \\
Canonical: selection on observables & 1 & 0 & observable & 0 & heterogeneous \\
Sharp null & 1 & 1 & unobservable & 0 & zero \\
\bottomrule
\end{tabular}%
}
\end{center}

Because parallel trends holds conditional on $(u_i, X_i)$ but treated and control units exhibit substantial baseline differences in levels through $u_i$, these designs violate the unconfoundedness assumption of cross-sectional matching while remaining fully identifiable for panel difference-in-differences estimators. Varying $(\sigma_u, c_u)$ explicitly probes robustness against increasing confounding strength.

\subsubsection{Sample-size scaling sweeps}
The sample-size sweeps replicate the canonical baseline design ($\sigma_u = 1$, $c_u = 1$, selection on unobservables) across panel sizes $n \in \{50, 100, 200, 400, 800\}$ and degrees $d \in \{1, 2, 3\}$, with $R = 100$ replications per configuration. The independent error sweep uses i.i.d.\ disturbances, while the serial correlation sweep sets $\rho = 0.6$. Replications share common random seeds across estimators. At $n=50$, the cohort-time cell $\mathrm{GATT}(4,4)$ contains only approximately 25 treated units, directly testing small-sample regularization and prior shrinkage.

\subsubsection{Staggered adoption designs}
The staggered adoption experiments evaluate panel settings with multiple adoption cohorts and time-varying treatment effects. Units $i \in \{1, \dots, n\}$ observed across $T = 8$ periods ($t \in \{0, \dots, 7\}$) are assigned to one of $K = 3$ treated adoption cohorts $(g_1, g_2, g_3) = (4, 5, 6)$ or to the never-treated comparison group ($j = 0$, $G_i = \infty$). Baseline cohort shares are specified as $s_0 = s_1 = s_2 = s_3 = 0.25$; in the Goodman--Bacon decomposition sensitivity sweep, the clean control pool is varied over $s_0 \in \{0.40, 0.30, 0.25, 0.20, 0.10, 0.05\}$ with treated cohorts assigned equal shares $s_1 = s_2 = s_3 = (1 - s_0)/3$.

Cohort assignment is determined by a multinomial Gumbel random utility model with assignment driver $q_i$. In the headline design with selection on unobservables, $q_i = c_u A_i$ with $c_u = 1$ and latent confounder $A_i \sim \mathcal{N}(0, 1)$ (so that $q_i = u_i = A_i$ under $\sigma_u = 1$), while under selection on observables $q_i = 0.6 X_{1i} + 0.4 X_{4i}$. The assignment utilities for each unit $i$ are:
\begin{align}
V_{i0} &= \log(s_0) + \xi_{i0}, \\
V_{ij} &= \log(s_j) + \left[0.8 + 0.5\left(1 - \frac{j-1}{K-1}\right)\right] q_i + \xi_{ij}, \quad j \in \{1, \dots, K\},
\end{align}
where $\xi_{i0}, \dots, \xi_{iK} \stackrel{\mathrm{i.i.d.}}{\sim} \operatorname{Gumbel}(0, 1)$. Unit $i$ adopts treatment at period $G_i = g_{j^*}$ where $j^* = \arg\max_{m \in \{0, \dots, K\}} V_{im}$ (defining $g_0 \equiv \infty$). Because the coefficient on $q_i$ decreases linearly from $1.30$ for cohort $g_1 = 4$ down to $0.80$ for cohort $g_3 = 6$, units with higher values of the latent confounder $u_i = A_i$ systematically adopt earlier, generating dynamic selection into adoption timing.

The untreated potential outcome $Y_{it}(0)$ follows the identical baseline equation as in Appendix~\ref{ssec:b1designs}:
\begin{equation}
Y_{it}(0) = \beta_0 + u_i + \gamma_t + f(X_i) + s(X_i)\,t + \varepsilon_{it},
\end{equation}
with baseline level $\beta_0 = -0.5$, calendar trend $\gamma_t = 0.2t$, unit effect $u_i = \sigma_u A_i$ with $\sigma_u = 1$, covariate main effect $f(X_i) = -0.75X_{1i} + 0.5X_{2i} - 0.5X_{3i} - 1.3X_{4i} + 1.8X_{5i}$, covariate-dependent slope $s(X_i) = 0.3(0.7X_{4i} + \sqrt{0.51}X_{3i})$, and observation disturbances $\varepsilon_{it} \stackrel{\mathrm{i.i.d.}}{\sim} \mathcal{N}(0, 1)$.

In the staggered designs, the baseline unit treatment effect intercept is set to $2.0$. For linearity degrees $d \in \{1, 2\}$, $\tau_i = 2 + 1.5X_{1i} + 0.75X_{2i}$; for degree $d = 3$,
\begin{equation}
\tau_i = 2 + 1.5X_{1i} + 0.75\left[\frac{1}{\sqrt{2}}\left\{\frac{X_{2i}^2-1}{\sqrt{2}} + (2X_{1i}-1)X_{2i}\right\}\right].
\end{equation}
For a treated unit $i$ belonging to cohort $g_j \in \{4, 5, 6\}$ at event time $k_{it} = t - g_j \ge 0$, the realized causal effect is modulated by cohort multipliers $(M_1, M_2, M_3)$ and a linear dynamic exposure ramp with growth rate $r$:
\begin{equation}
\tau_{it}(g_j, k_{it}) = M_j (1 + r k_{it})\tau_i.
\end{equation}
In the baseline staggered dynamics regime, we set $(M_1, M_2, M_3) = (1.0, 1.5, 2.0)$ and $r = 0.4$; to amplify dynamic heterogeneity and standard two-way fixed-effects contamination, a high-contamination variant increases these parameters to $(M_1, M_2, M_3) = (1.0, 2.0, 3.0)$ and $r = 0.8$. The observed outcome is $Y_{it} = Y_{it}(0) + D_{it}\tau_{it}(G_i, k_{it})$, where $D_{it} = \mathbf{1}\{G_i \le t, G_i < \infty\}$. These specifications simultaneously generate cohort-level effect heterogeneity and dynamic exposure growth, matching the setting in \textcite{GoodmanBacon2021} where standard two-way fixed effects regressions assign negative weights to earlier-treated cohorts serving as comparison units.

\subsubsection{Pre-trend designs and diagnostic model}
Ten dedicated data-generating processes evaluate the empirical size, diagnostic power, and subgroup sensitivity of pre-trend testing procedures. All ten designs build upon the canonical single-cohort panel framework with $n = 200$ cross-sectional units observed over $T = 8$ periods ($t \in \{0, \dots, 7\}$, comprising four pre-treatment periods $t \in \{0, 1, 2, 3\}$ and four post-treatment periods $t \in \{4, 5, 6, 7\}$), adoption period $G_i \in \{4, \infty\}$, ever-treated indicator $W_i = \mathbf{1}\{G_i < \infty\}$, and $R = 200$ Monte Carlo replications per design.

The untreated potential outcome incorporates an explicit trend modification into the baseline equation of Appendix~\ref{ssec:b1designs}:
\begin{equation}
Y_{it}(0) = \beta_0 + u_i + \gamma_t + f(X_i) + [s_{\rho_X}(X_i) + \Delta s_i]\,t + \varepsilon_{it},
\end{equation}
where $\beta_0 = -0.5$, $\gamma_t = 0.2t$, $u_i = \sigma_u A_i$ with $A_i \sim \mathcal{N}(0, 1)$, $f(X_i) = -0.75X_{1i} + 0.5X_{2i} - 0.5X_{3i} - 1.3X_{4i} + 1.8X_{5i}$, and the baseline covariate-dependent slope is $s_{\rho_X}(X_i) = \rho_X(0.7X_{4i} + \sqrt{0.51}X_{3i})$. The term $\Delta s_i$ governs the structural violation of conditional parallel trends:
\begin{equation}
\Delta s_i = \lambda_{\mathrm{pt}} A_i + \delta W_i + \kappa (2X_{1i} - 1) W_i,
\end{equation}
where $\lambda_{\mathrm{pt}}$ routes violations through the unobserved confounder $A_i$, $\delta$ introduces an aggregate group-level differential trend, and $\kappa$ modulates covariate-heterogeneous trends that cancel across treatment arms. The ten evaluation designs are configured as follows:

\begin{enumerate}
    \item \emph{Parallel trends hold}: Null benchmark where $\Delta s_i = 0$, $\rho_X = 0.3$, $\sigma_u = 1$, and $c_u = 1$ under selection on unobservables. Conditional parallel trends holds exactly; any diagnostic rejection represents an empirical false positive.
    \item \emph{Group slope violations} ($\delta \in \{0.05, 0.10, 0.20, 0.40\}$): Four aggregate power curve designs with $\Delta s_i = \delta W_i$, holding $\rho_X = 0.3$, $\sigma_u = 1$, and $c_u = 1$. The treated cohort follows a differential linear trajectory with slope $\delta$ per period.
    \item \emph{Unit-effect slope ($0.20$)}: Latent confounder channel design with $\Delta s_i = 0.20 A_i$ ($\lambda_{\mathrm{pt}} = 0.20, \delta = 0, \kappa = 0$), $\rho_X = 0.3$, $\sigma_u = 1$, and $c_u = 1$. Because $A_i$ drives treatment selection, treated units exhibit higher unobserved draws than controls, satisfying $\mathbb{E}[A_i \mid W_i = 1] - \mathbb{E}[A_i \mid W_i = 0] \approx 1.035$. This induces an expected treated-minus-control pre-trend slope divergence of $0.20 \times 1.035 \approx 0.207$ outcome units per period that cannot be eliminated by adjusting for observed covariates.
    \item \emph{Cancelling subgroup heterogeneity} ($\kappa \in \{0.10, 0.20, 0.40\}$): Three heterogeneous violation designs with $\Delta s_i = \kappa(2X_{1i} - 1)W_i$, holding $\rho_X = 0.3$, $\sigma_u = 1$, and $c_u = 1$. Treated units with $X_{1i} = 1$ trend upward at $+\kappa$, while treated units with $X_{1i} = 0$ trend downward at $-\kappa$. Because $X_{1i} \sim \operatorname{Bernoulli}(0.5)$ is balanced across arms under unobservable selection ($\mathbb{E}[X_{1i} \mid W_i = 1] = 0.5$), the expectation $\mathbb{E}[2X_{1i} - 1 \mid W_i = 1]$ vanishes. The population aggregate differential slope is zero in expectation, rendering marginal event-study tests powerless by construction even though conditional parallel trends fails within both subgroups.
    \item \emph{Conditional parallel trends only}: Observable selection design with $\Delta s_i = 0$, amplified slope scale $\rho_X = 0.6$, $\sigma_u = 1$, and $c_u = 0$, where treatment assignment depends on $U_i = 0.8X_{1i} + 0.6X_{4i}$. Because assignment depends positively on $X_{4i}$, treated and control units differ in their average slopes ($\mathbb{E}[X_{4i} \mid W_i = 1] \neq \mathbb{E}[X_{4i} \mid W_i = 0]$), violating unconditional parallel trends while conditional parallel trends given $X_i$ holds identically.
\end{enumerate}

\begin{center}
\footnotesize
\resizebox{\linewidth}{!}{%
\begin{tabular}{@{}lrrrrl@{}}
\toprule
Scenario & Selection & $\sigma_u$ & $c_u$ & $\rho_X$ & Pre-Trend Violation Slope $\Delta s_i$ \\
\midrule
Parallel trends hold & unobservable & $1$ & $1$ & $0.3$ & $0$ (Exact conditional PT holds) \\
Group slope $0.05$ & unobservable & $1$ & $1$ & $0.3$ & $0.05\,W_i$ (Aggregate slope $0.05$/period) \\
Group slope $0.10$ & unobservable & $1$ & $1$ & $0.3$ & $0.10\,W_i$ (Aggregate slope $0.10$/period) \\
Group slope $0.20$ & unobservable & $1$ & $1$ & $0.3$ & $0.20\,W_i$ (Aggregate slope $0.20$/period) \\
Group slope $0.40$ & unobservable & $1$ & $1$ & $0.3$ & $0.40\,W_i$ (Aggregate slope $0.40$/period) \\
Unit-effect slope $0.20$ & unobservable & $1$ & $1$ & $0.3$ & $0.20\,A_i$ (Latent confounder channel, gap $\approx 0.207$) \\
Cancelling heterogeneity $0.10$ & unobservable & $1$ & $1$ & $0.3$ & $0.10(2X_{1i}-1)W_i$ (Cancels in aggregate) \\
Cancelling heterogeneity $0.20$ & unobservable & $1$ & $1$ & $0.3$ & $0.20(2X_{1i}-1)W_i$ (Cancels in aggregate) \\
Cancelling heterogeneity $0.40$ & unobservable & $1$ & $1$ & $0.3$ & $0.40(2X_{1i}-1)W_i$ (Cancels in aggregate) \\
Conditional PT only & observable & $1$ & $0$ & $0.6$ & $0$ (Violates marginal PT, holds conditional PT) \\
\bottomrule
\end{tabular}%
}
\end{center}

Because DiD-BCF structurally zeros pre-treatment effects via $D_{it}$, pre-trend evaluation is conducted via an auxiliary unconstrained lead model:
\begin{equation}
f^{\mathrm{pl}}(g, t, x) = a(g, x) + b(t, x) + \mathbf{1}\{g \in \mathcal{G}\} \lambda_g(x, t-g), \quad \lambda_g(x, -1) \equiv 0,
\end{equation}
normalized at event time $k=-1$ to remove time-invariant cohort-level differences. Under conditional parallel trends and no anticipation, the true lead parameters satisfy $\lambda_g(x,k)=0$ for all observed $k\le -2$. In the fitted diagnostic, the prognostic forest uses baseline covariates and calendar time, while unit random intercepts absorb persistent level differences. The ever-treated indicator activates an unrestricted event-time forest in every period, and each retained draw is normalized by subtracting its same-unit value at $k=-1$ before averaging over treated units to obtain the aggregate lead draw $\Delta(k)$.

Because the auxiliary posterior characterizes the joint distribution over the complete functional surface $\lambda_g(x, k)$, researchers possess broad flexibility to construct diagnostic functionals reflecting different empirical assumptions and suspected violation channels. To benchmark this diagnostic framework systematically against existing econometric tests in our synthetic DGPs, we operationalize three distinct decision rules:
1. the aggregate \emph{slope rule}, fitting a linear slope $s$ through the origin to $\Delta(k)$ against $k+1$ within each retained draw, which tests for secular linear divergences;
2. the \emph{any-$k$ rule}, testing individual leads against zero with a Bonferroni correction across the $L$ observed leads, which is sensitive to localized anticipation spikes; and
3. the \emph{subgroup contrast rule}, applying the slope evaluation to the difference between the $X_1=1$ and $X_1=0$ subgroup lead profiles within each draw, specifically designed in our synthetic experiments to detect heterogeneous pre-trends that cancel out in the sample aggregate.
The slope and contrast rules reject when the smaller posterior sign probability is below $0.025$ (two-sided 5\% level), and the any-$k$ rule rejects when any individual lead's smaller tail probability falls below $0.025/L$. Benchmark pre-tests use the corresponding nominal two-sided 5\% thresholds. The benchmark implementations condition on covariates where specified, whereas the TWFE and did2s pre-tests are unadjusted; Synthetic DiD uses weighted pre-period gaps with a jackknife covariance estimate. TWFE is evaluated on an independent set of 200 replications and does not include a subgroup contrast. Subgroup contrasts are available for Callaway--Sant'Anna, did2s, Synthetic DiD, DoubleML, and grf-DiD. DoubleML fits its five-fold, 500-tree nuisance models separately within each subgroup; the contrast covariance is the sum of the two disjoint-sample covariance matrices, retaining covariance across leads that share the reference period.

\section{Extended Simulation Results}\label{supp:results}
The tables in this section present extended simulation results across all evaluation metrics defined in Appendix~\ref{supp:metrics}. DiD-BCF denotes posterior summaries of the total treatment effect surface from the structured Bayesian causal forest model. All simulation grids utilize $R = 100$ Monte Carlo replications per non-null cell and $R = 200$ replications per sharp-null cell. To ensure transparent reporting, the tables retain all evaluated polynomial degrees $d \in \{1, 2, 3\}$, highlighting both settings where DiD-BCF achieves substantial efficiency gains and challenging regimes (such as strong serial correlation or small panel dimensions) where shrinkage or unmodeled residual dependence affects nominal coverage.

\subsection{Canonical average effects}\label{ssec:b1res}
Tables~\ref{tab:b1att} and \ref{tab:b1gatt} report pooled ATT and $\gatt(4,4)$, respectively. Across the baseline and strong-confounder designs, DiD-BCF exhibits minimal bias and favorable root mean squared error relative to benchmark estimators. Serial dependence moderately reduces pooled-ATT coverage due to unmodeled temporal correlation in the observation errors, while conditioning on observed assignment drivers remains essential under selection on observables.
{\footnotesize
\begin{longtable}{lrrrrrrrr}
\caption{Decomposed Monte-Carlo metrics for the overall ATT on the canonical-DiD scenarios.}\label{tab:b1att}\\
\toprule
Estimator & $d$ & Bias & SD(err) & RMSE & Cov.90 & Cov.95 & Len.95 & $\overline{\mathrm{sd}}/\mathrm{SD}$ \\
\midrule
\endfirsthead
\multicolumn{9}{l}{\emph{\footnotesize Table \ref{tab:b1att} continued from the previous page.}}\\
\toprule
Estimator & $d$ & Bias & SD(err) & RMSE & Cov.90 & Cov.95 & Len.95 & $\overline{\mathrm{sd}}/\mathrm{SD}$ \\
\midrule
\endhead
\midrule
\multicolumn{9}{r}{\emph{\footnotesize continued on the next page}}\\
\endfoot
\bottomrule
\endlastfoot
\multicolumn{9}{l}{\textit{Canonical: baseline}} \\
\quad DiD-BCF & 1 & 0.004 & 0.099 & 0.099 & 0.920 & 0.950 & 0.425 & 0.837 \\
\quad TWFE & 1 & -0.024 & 0.177 & 0.179 & 0.970 & 0.980 & 0.837 & 1.067 \\
\quad Callaway--Sant'Anna & 1 & 0.000 & 0.171 & 0.171 & 0.940 & 0.970 & 0.751 & 1.096 \\
\quad Gardner (did2s) & 1 & -0.022 & 0.183 & 0.185 & 0.960 & 0.990 & 0.864 & 1.032 \\
\quad DoubleML DR-DiD & 1 & 0.003 & 0.201 & 0.201 & 0.960 & 0.990 & 0.931 & 1.120 \\
\quad Synthetic DiD & 1 & -0.023 & 0.168 & 0.169 & 0.970 & 1.000 & 0.797 & 1.064 \\
\quad grf-DiD (Wang) & 1 & -0.003 & 0.171 & 0.171 & 0.810 & 0.910 & 0.574 & 0.844 \\
\quad DiD-BCF & 2 & -0.005 & 0.115 & 0.116 & 0.890 & 0.920 & 0.438 & 0.798 \\
\quad TWFE & 2 & -0.024 & 0.177 & 0.179 & 0.970 & 0.980 & 0.837 & 1.067 \\
\quad Callaway--Sant'Anna & 2 & -0.004 & 0.183 & 0.183 & 0.920 & 0.960 & 0.781 & 1.067 \\
\quad Gardner (did2s) & 2 & -0.022 & 0.183 & 0.185 & 0.960 & 0.990 & 0.864 & 1.032 \\
\quad DoubleML DR-DiD & 2 & -0.014 & 0.208 & 0.208 & 0.970 & 0.980 & 0.958 & 1.120 \\
\quad Synthetic DiD & 2 & -0.023 & 0.168 & 0.169 & 0.970 & 1.000 & 0.797 & 1.064 \\
\quad grf-DiD (Wang) & 2 & -0.007 & 0.175 & 0.175 & 0.780 & 0.900 & 0.576 & 0.832 \\
\quad DiD-BCF & 3 & -0.010 & 0.115 & 0.115 & 0.880 & 0.930 & 0.443 & 0.702 \\
\quad TWFE & 3 & -0.024 & 0.177 & 0.179 & 0.970 & 0.980 & 0.837 & 1.067 \\
\quad Callaway--Sant'Anna & 3 & -0.004 & 0.183 & 0.183 & 0.930 & 0.960 & 0.781 & 0.929 \\
\quad Gardner (did2s) & 3 & -0.022 & 0.183 & 0.185 & 0.960 & 0.990 & 0.865 & 1.040 \\
\quad DoubleML DR-DiD & 3 & -0.014 & 0.209 & 0.210 & 0.970 & 0.980 & 0.958 & 1.002 \\
\quad Synthetic DiD & 3 & -0.023 & 0.168 & 0.169 & 0.960 & 1.000 & 0.797 & 1.023 \\
\quad grf-DiD (Wang) & 3 & 0.002 & 0.170 & 0.170 & 0.840 & 0.920 & 0.596 & 0.769 \\
\addlinespace
\multicolumn{9}{l}{\textit{Canonical: strong confounder}} \\
\quad DiD-BCF & 1 & -0.001 & 0.099 & 0.099 & 0.930 & 0.980 & 0.426 & 0.859 \\
\quad TWFE & 1 & -0.022 & 0.179 & 0.180 & 0.970 & 0.980 & 0.837 & 1.097 \\
\quad Callaway--Sant'Anna & 1 & -0.001 & 0.160 & 0.160 & 0.960 & 0.970 & 0.753 & 1.137 \\
\quad Gardner (did2s) & 1 & -0.027 & 0.188 & 0.190 & 0.920 & 0.980 & 0.863 & 1.045 \\
\quad DoubleML DR-DiD & 1 & 0.002 & 0.190 & 0.190 & 0.960 & 0.980 & 0.935 & 1.185 \\
\quad Synthetic DiD & 1 & -0.027 & 0.169 & 0.171 & 0.940 & 0.990 & 0.795 & 1.086 \\
\quad grf-DiD (Wang) & 1 & -0.005 & 0.163 & 0.163 & 0.880 & 0.910 & 0.571 & 0.872 \\
\quad DiD-BCF & 2 & -0.007 & 0.107 & 0.107 & 0.910 & 0.950 & 0.438 & 0.862 \\
\quad TWFE & 2 & -0.022 & 0.179 & 0.180 & 0.970 & 0.980 & 0.837 & 1.097 \\
\quad Callaway--Sant'Anna & 2 & -0.005 & 0.170 & 0.170 & 0.930 & 0.960 & 0.781 & 1.124 \\
\quad Gardner (did2s) & 2 & -0.027 & 0.188 & 0.190 & 0.920 & 0.980 & 0.863 & 1.045 \\
\quad DoubleML DR-DiD & 2 & 0.001 & 0.197 & 0.197 & 0.950 & 0.980 & 0.974 & 1.195 \\
\quad Synthetic DiD & 2 & -0.027 & 0.169 & 0.171 & 0.940 & 0.990 & 0.795 & 1.086 \\
\quad grf-DiD (Wang) & 2 & -0.013 & 0.163 & 0.164 & 0.840 & 0.900 & 0.572 & 0.870 \\
\quad DiD-BCF & 3 & -0.013 & 0.108 & 0.108 & 0.880 & 0.940 & 0.442 & 0.716 \\
\quad TWFE & 3 & -0.022 & 0.179 & 0.180 & 0.970 & 0.980 & 0.837 & 1.097 \\
\quad Callaway--Sant'Anna & 3 & -0.005 & 0.170 & 0.170 & 0.940 & 0.950 & 0.782 & 0.989 \\
\quad Gardner (did2s) & 3 & -0.027 & 0.188 & 0.190 & 0.930 & 0.980 & 0.863 & 1.037 \\
\quad DoubleML DR-DiD & 3 & -0.001 & 0.197 & 0.197 & 0.970 & 0.980 & 0.975 & 1.080 \\
\quad Synthetic DiD & 3 & -0.027 & 0.169 & 0.171 & 0.940 & 0.990 & 0.796 & 1.035 \\
\quad grf-DiD (Wang) & 3 & -0.003 & 0.158 & 0.158 & 0.860 & 0.930 & 0.594 & 0.816 \\
\addlinespace
\multicolumn{9}{l}{\textit{Canonical: serial correlation}} \\
\quad DiD-BCF & 1 & 0.007 & 0.128 & 0.128 & 0.770 & 0.850 & 0.363 & 0.587 \\
\quad TWFE & 1 & -0.022 & 0.197 & 0.198 & 0.950 & 0.980 & 0.894 & 1.038 \\
\quad Callaway--Sant'Anna & 1 & -0.005 & 0.147 & 0.148 & 0.930 & 0.980 & 0.683 & 1.095 \\
\quad Gardner (did2s) & 1 & -0.020 & 0.196 & 0.197 & 0.950 & 0.980 & 0.918 & 1.030 \\
\quad DoubleML DR-DiD & 1 & -0.008 & 0.171 & 0.172 & 0.960 & 0.980 & 0.841 & 1.148 \\
\quad Synthetic DiD & 1 & -0.024 & 0.164 & 0.166 & 0.940 & 0.990 & 0.792 & 1.081 \\
\quad grf-DiD (Wang) & 1 & -0.009 & 0.149 & 0.150 & 0.840 & 0.920 & 0.527 & 0.838 \\
\quad DiD-BCF & 2 & 0.002 & 0.144 & 0.144 & 0.710 & 0.820 & 0.376 & 0.573 \\
\quad TWFE & 2 & -0.022 & 0.197 & 0.198 & 0.950 & 0.980 & 0.894 & 1.038 \\
\quad Callaway--Sant'Anna & 2 & -0.008 & 0.164 & 0.164 & 0.930 & 0.950 & 0.717 & 1.042 \\
\quad Gardner (did2s) & 2 & -0.020 & 0.196 & 0.197 & 0.950 & 0.980 & 0.918 & 1.030 \\
\quad DoubleML DR-DiD & 2 & -0.018 & 0.184 & 0.185 & 0.950 & 0.990 & 0.871 & 1.140 \\
\quad Synthetic DiD & 2 & -0.024 & 0.164 & 0.166 & 0.940 & 0.990 & 0.792 & 1.081 \\
\quad grf-DiD (Wang) & 2 & -0.012 & 0.149 & 0.150 & 0.830 & 0.940 & 0.531 & 0.852 \\
\quad DiD-BCF & 3 & -0.012 & 0.141 & 0.141 & 0.730 & 0.820 & 0.380 & 0.550 \\
\quad TWFE & 3 & -0.022 & 0.197 & 0.198 & 0.950 & 0.980 & 0.894 & 1.038 \\
\quad Callaway--Sant'Anna & 3 & -0.008 & 0.164 & 0.164 & 0.930 & 0.950 & 0.716 & 0.924 \\
\quad Gardner (did2s) & 3 & -0.020 & 0.196 & 0.197 & 0.950 & 0.980 & 0.919 & 1.056 \\
\quad DoubleML DR-DiD & 3 & -0.018 & 0.185 & 0.186 & 0.950 & 0.990 & 0.872 & 1.017 \\
\quad Synthetic DiD & 3 & -0.024 & 0.164 & 0.166 & 0.940 & 0.990 & 0.793 & 1.038 \\
\quad grf-DiD (Wang) & 3 & -0.004 & 0.145 & 0.145 & 0.870 & 0.920 & 0.553 & 0.798 \\
\addlinespace
\multicolumn{9}{l}{\textit{Canonical: selection on observables}} \\
\quad DiD-BCF & 1 & 0.053 & 0.110 & 0.122 & 0.900 & 0.950 & 0.466 & 0.766 \\
\quad TWFE & 1 & -0.023 & 0.223 & 0.224 & 0.890 & 0.950 & 0.833 & 0.868 \\
\quad Callaway--Sant'Anna & 1 & 0.001 & 0.223 & 0.223 & 0.880 & 0.980 & 0.887 & 0.937 \\
\quad Gardner (did2s) & 1 & 0.568 & 0.208 & 0.605 & 0.170 & 0.250 & 0.836 & 0.893 \\
\quad DoubleML DR-DiD & 1 & 0.032 & 0.250 & 0.252 & 0.920 & 0.970 & 1.015 & 0.960 \\
\quad Synthetic DiD & 1 & 0.417 & 0.183 & 0.455 & 0.320 & 0.410 & 0.778 & 0.930 \\
\quad grf-DiD (Wang) & 1 & 0.016 & 0.190 & 0.191 & 0.770 & 0.860 & 0.569 & 0.703 \\
\quad DiD-BCF & 2 & 0.049 & 0.138 & 0.146 & 0.830 & 0.900 & 0.479 & 0.696 \\
\quad TWFE & 2 & -0.023 & 0.223 & 0.224 & 0.890 & 0.950 & 0.833 & 0.868 \\
\quad Callaway--Sant'Anna & 2 & 0.085 & 0.312 & 0.323 & 0.876 & 0.921 & 1.016 & 0.800 \\
\quad Gardner (did2s) & 2 & 0.568 & 0.208 & 0.605 & 0.170 & 0.250 & 0.836 & 0.893 \\
\quad DoubleML DR-DiD & 2 & 0.021 & 0.274 & 0.274 & 0.890 & 0.950 & 1.072 & 0.936 \\
\quad Synthetic DiD & 2 & 0.417 & 0.183 & 0.455 & 0.320 & 0.410 & 0.778 & 0.930 \\
\quad grf-DiD (Wang) & 2 & 0.011 & 0.187 & 0.187 & 0.800 & 0.870 & 0.570 & 0.721 \\
\quad DiD-BCF & 3 & 0.039 & 0.140 & 0.145 & 0.830 & 0.890 & 0.493 & 0.707 \\
\quad TWFE & 3 & -0.023 & 0.223 & 0.224 & 0.890 & 0.950 & 0.833 & 0.868 \\
\quad Callaway--Sant'Anna & 3 & 0.085 & 0.312 & 0.323 & 0.888 & 0.910 & 1.016 & 0.804 \\
\quad Gardner (did2s) & 3 & 0.568 & 0.208 & 0.605 & 0.170 & 0.260 & 0.838 & 0.936 \\
\quad DoubleML DR-DiD & 3 & 0.022 & 0.273 & 0.274 & 0.900 & 0.950 & 1.072 & 0.934 \\
\quad Synthetic DiD & 3 & 0.417 & 0.183 & 0.455 & 0.320 & 0.420 & 0.780 & 0.999 \\
\quad grf-DiD (Wang) & 3 & 0.011 & 0.187 & 0.187 & 0.830 & 0.880 & 0.585 & 0.724 \\
\addlinespace
\end{longtable}
\noindent \emph{Note.} $n=200$, 100 replications in each non-null cell, $d$ is the linearity degree. Bias and RMSE are Monte-Carlo moments of the estimation error and SD(err) is its standard deviation, $\sqrt{\text{RMSE}^2-\text{Bias}^2}$ (the target itself varies by replication, so the raw dispersion of the estimate is not the error SD); Cov.90/95 are credible/confidence-interval coverage rates; Len.95 the average 95\% interval length; $\overline{\mathrm{sd}}/\mathrm{SD}$ the calibration ratio of the average reported posterior/standard error to the raw Monte-Carlo SD of the estimates. Since realized targets vary across replications, this descriptive ratio alone does not diagnose calibration.\par\vspace{6pt}
}

{\footnotesize
\begin{longtable}{lrrrrrrrr}
\caption{Decomposed Monte-Carlo metrics for $\mathrm{GATT}(g{=}4,t{=}4)$ on the canonical-DiD scenarios, the cell every estimator reports.}\label{tab:b1gatt}\\
\toprule
Estimator & $d$ & Bias & SD(err) & RMSE & Cov.90 & Cov.95 & Len.95 & $\overline{\mathrm{sd}}/\mathrm{SD}$ \\
\midrule
\endfirsthead
\multicolumn{9}{l}{\emph{\footnotesize Table \ref{tab:b1gatt} continued from the previous page.}}\\
\toprule
Estimator & $d$ & Bias & SD(err) & RMSE & Cov.90 & Cov.95 & Len.95 & $\overline{\mathrm{sd}}/\mathrm{SD}$ \\
\midrule
\endhead
\midrule
\multicolumn{9}{r}{\emph{\footnotesize continued on the next page}}\\
\endfoot
\bottomrule
\endlastfoot
\multicolumn{9}{l}{\textit{Canonical: baseline}} \\
\quad DiD-BCF & 1 & -0.019 & 0.110 & 0.112 & 0.950 & 1.000 & 0.559 & 1.003 \\
\quad TWFE & 1 & -0.048 & 0.173 & 0.179 & 0.950 & 0.990 & 0.820 & 1.068 \\
\quad Callaway--Sant'Anna & 1 & -0.031 & 0.219 & 0.221 & 0.870 & 0.970 & 0.902 & 1.031 \\
\quad Gardner (did2s) & 1 & -0.047 & 0.173 & 0.179 & 0.950 & 0.980 & 0.846 & 1.059 \\
\quad DoubleML DR-DiD & 1 & -0.039 & 0.301 & 0.304 & 0.870 & 0.920 & 1.157 & 0.941 \\
\quad Synthetic DiD & 1 & -0.052 & 0.173 & 0.181 & 0.940 & 0.970 & 0.823 & 1.052 \\
\quad grf-DiD (Wang) & 1 & -0.036 & 0.219 & 0.222 & 0.730 & 0.810 & 0.630 & 0.731 \\
\quad DiD-BCF & 2 & -0.025 & 0.119 & 0.122 & 0.970 & 0.980 & 0.572 & 0.994 \\
\quad TWFE & 2 & -0.048 & 0.173 & 0.179 & 0.950 & 0.990 & 0.820 & 1.068 \\
\quad Callaway--Sant'Anna & 2 & -0.034 & 0.223 & 0.225 & 0.850 & 0.960 & 0.914 & 1.020 \\
\quad Gardner (did2s) & 2 & -0.047 & 0.173 & 0.179 & 0.950 & 0.980 & 0.846 & 1.059 \\
\quad DoubleML DR-DiD & 2 & -0.034 & 0.278 & 0.280 & 0.920 & 0.940 & 1.127 & 1.020 \\
\quad Synthetic DiD & 2 & -0.052 & 0.173 & 0.181 & 0.940 & 0.970 & 0.823 & 1.052 \\
\quad grf-DiD (Wang) & 2 & -0.037 & 0.220 & 0.223 & 0.710 & 0.810 & 0.630 & 0.730 \\
\quad DiD-BCF & 3 & -0.030 & 0.119 & 0.122 & 0.960 & 0.980 & 0.577 & 0.899 \\
\quad TWFE & 3 & -0.048 & 0.173 & 0.179 & 0.950 & 0.990 & 0.820 & 1.068 \\
\quad Callaway--Sant'Anna & 3 & -0.034 & 0.223 & 0.225 & 0.860 & 0.960 & 0.908 & 0.936 \\
\quad Gardner (did2s) & 3 & -0.047 & 0.173 & 0.179 & 0.950 & 0.980 & 0.848 & 1.053 \\
\quad DoubleML DR-DiD & 3 & -0.038 & 0.268 & 0.271 & 0.890 & 0.970 & 1.115 & 1.044 \\
\quad Synthetic DiD & 3 & -0.052 & 0.173 & 0.181 & 0.940 & 0.970 & 0.825 & 1.021 \\
\quad grf-DiD (Wang) & 3 & -0.037 & 0.220 & 0.223 & 0.720 & 0.810 & 0.630 & 0.729 \\
\addlinespace
\multicolumn{9}{l}{\textit{Canonical: strong confounder}} \\
\quad DiD-BCF & 1 & -0.013 & 0.101 & 0.102 & 0.990 & 1.000 & 0.559 & 1.058 \\
\quad TWFE & 1 & -0.033 & 0.165 & 0.168 & 0.990 & 0.990 & 0.820 & 1.131 \\
\quad Callaway--Sant'Anna & 1 & -0.017 & 0.203 & 0.204 & 0.940 & 0.980 & 0.904 & 1.076 \\
\quad Gardner (did2s) & 1 & -0.036 & 0.165 & 0.169 & 0.960 & 0.980 & 0.845 & 1.113 \\
\quad DoubleML DR-DiD & 1 & -0.016 & 0.280 & 0.281 & 0.930 & 0.960 & 1.138 & 1.020 \\
\quad Synthetic DiD & 1 & -0.040 & 0.158 & 0.163 & 0.950 & 0.980 & 0.824 & 1.134 \\
\quad grf-DiD (Wang) & 1 & -0.018 & 0.201 & 0.201 & 0.790 & 0.860 & 0.629 & 0.782 \\
\quad DiD-BCF & 2 & -0.018 & 0.107 & 0.108 & 0.960 & 1.000 & 0.572 & 1.060 \\
\quad TWFE & 2 & -0.033 & 0.165 & 0.168 & 0.990 & 0.990 & 0.820 & 1.131 \\
\quad Callaway--Sant'Anna & 2 & -0.021 & 0.208 & 0.209 & 0.910 & 0.980 & 0.904 & 1.049 \\
\quad Gardner (did2s) & 2 & -0.036 & 0.165 & 0.169 & 0.960 & 0.980 & 0.845 & 1.113 \\
\quad DoubleML DR-DiD & 2 & -0.032 & 0.251 & 0.253 & 0.950 & 0.990 & 1.133 & 1.112 \\
\quad Synthetic DiD & 2 & -0.040 & 0.158 & 0.163 & 0.950 & 0.980 & 0.824 & 1.134 \\
\quad grf-DiD (Wang) & 2 & -0.018 & 0.200 & 0.201 & 0.780 & 0.860 & 0.629 & 0.783 \\
\quad DiD-BCF & 3 & -0.023 & 0.106 & 0.109 & 0.980 & 1.000 & 0.572 & 0.942 \\
\quad TWFE & 3 & -0.033 & 0.165 & 0.168 & 0.990 & 0.990 & 0.820 & 1.131 \\
\quad Callaway--Sant'Anna & 3 & -0.021 & 0.208 & 0.209 & 0.920 & 0.970 & 0.912 & 0.996 \\
\quad Gardner (did2s) & 3 & -0.036 & 0.165 & 0.169 & 0.950 & 0.970 & 0.847 & 1.122 \\
\quad DoubleML DR-DiD & 3 & -0.023 & 0.255 & 0.256 & 0.900 & 0.960 & 1.142 & 1.085 \\
\quad Synthetic DiD & 3 & -0.040 & 0.158 & 0.163 & 0.950 & 0.980 & 0.826 & 1.118 \\
\quad grf-DiD (Wang) & 3 & -0.018 & 0.200 & 0.201 & 0.790 & 0.860 & 0.629 & 0.782 \\
\addlinespace
\multicolumn{9}{l}{\textit{Canonical: serial correlation}} \\
\quad DiD-BCF & 1 & -0.011 & 0.111 & 0.112 & 0.900 & 0.960 & 0.463 & 0.775 \\
\quad TWFE & 1 & -0.034 & 0.160 & 0.163 & 0.970 & 0.990 & 0.761 & 1.012 \\
\quad Callaway--Sant'Anna & 1 & -0.025 & 0.126 & 0.129 & 0.970 & 0.990 & 0.653 & 1.098 \\
\quad Gardner (did2s) & 1 & -0.033 & 0.152 & 0.156 & 0.990 & 0.990 & 0.789 & 1.026 \\
\quad DoubleML DR-DiD & 1 & -0.029 & 0.163 & 0.165 & 0.960 & 0.990 & 0.776 & 1.087 \\
\quad Synthetic DiD & 1 & -0.036 & 0.124 & 0.129 & 0.960 & 0.980 & 0.667 & 1.059 \\
\quad grf-DiD (Wang) & 1 & -0.032 & 0.130 & 0.133 & 0.840 & 0.910 & 0.458 & 0.809 \\
\quad DiD-BCF & 2 & -0.012 & 0.127 & 0.128 & 0.910 & 0.950 & 0.479 & 0.749 \\
\quad TWFE & 2 & -0.034 & 0.160 & 0.163 & 0.970 & 0.990 & 0.761 & 1.012 \\
\quad Callaway--Sant'Anna & 2 & -0.028 & 0.132 & 0.135 & 0.950 & 0.990 & 0.662 & 1.067 \\
\quad Gardner (did2s) & 2 & -0.033 & 0.152 & 0.156 & 0.990 & 0.990 & 0.789 & 1.026 \\
\quad DoubleML DR-DiD & 2 & -0.022 & 0.169 & 0.170 & 0.930 & 0.970 & 0.773 & 1.102 \\
\quad Synthetic DiD & 2 & -0.036 & 0.124 & 0.129 & 0.960 & 0.980 & 0.667 & 1.059 \\
\quad grf-DiD (Wang) & 2 & -0.032 & 0.130 & 0.133 & 0.840 & 0.900 & 0.458 & 0.809 \\
\quad DiD-BCF & 3 & -0.026 & 0.122 & 0.125 & 0.910 & 0.960 & 0.481 & 0.755 \\
\quad TWFE & 3 & -0.034 & 0.160 & 0.163 & 0.970 & 0.990 & 0.761 & 1.012 \\
\quad Callaway--Sant'Anna & 3 & -0.028 & 0.132 & 0.135 & 0.960 & 0.980 & 0.664 & 0.989 \\
\quad Gardner (did2s) & 3 & -0.033 & 0.152 & 0.156 & 0.990 & 0.990 & 0.792 & 1.074 \\
\quad DoubleML DR-DiD & 3 & -0.024 & 0.169 & 0.170 & 0.940 & 0.980 & 0.772 & 1.008 \\
\quad Synthetic DiD & 3 & -0.036 & 0.124 & 0.129 & 0.970 & 0.970 & 0.670 & 1.024 \\
\quad grf-DiD (Wang) & 3 & -0.032 & 0.129 & 0.133 & 0.840 & 0.910 & 0.458 & 0.811 \\
\addlinespace
\multicolumn{9}{l}{\textit{Canonical: selection on observables}} \\
\quad DiD-BCF & 1 & 0.025 & 0.122 & 0.124 & 0.950 & 0.970 & 0.600 & 0.884 \\
\quad TWFE & 1 & -0.027 & 0.214 & 0.216 & 0.880 & 0.940 & 0.816 & 0.860 \\
\quad Callaway--Sant'Anna & 1 & -0.006 & 0.280 & 0.280 & 0.900 & 0.950 & 1.082 & 0.910 \\
\quad Gardner (did2s) & 1 & 0.342 & 0.210 & 0.401 & 0.450 & 0.590 & 0.831 & 0.849 \\
\quad DoubleML DR-DiD & 1 & -0.011 & 0.343 & 0.344 & 0.890 & 0.940 & 1.264 & 0.917 \\
\quad Synthetic DiD & 1 & 0.241 & 0.203 & 0.315 & 0.640 & 0.730 & 0.819 & 0.860 \\
\quad grf-DiD (Wang) & 1 & -0.099 & 0.233 & 0.254 & 0.670 & 0.790 & 0.615 & 0.650 \\
\quad DiD-BCF & 2 & 0.021 & 0.135 & 0.136 & 0.930 & 0.970 & 0.615 & 0.864 \\
\quad TWFE & 2 & -0.027 & 0.214 & 0.216 & 0.880 & 0.940 & 0.816 & 0.860 \\
\quad Callaway--Sant'Anna & 2 & 0.018 & 0.336 & 0.337 & 0.888 & 0.910 & 1.153 & 0.827 \\
\quad Gardner (did2s) & 2 & 0.342 & 0.210 & 0.401 & 0.450 & 0.590 & 0.831 & 0.849 \\
\quad DoubleML DR-DiD & 2 & -0.016 & 0.336 & 0.336 & 0.880 & 0.940 & 1.251 & 0.925 \\
\quad Synthetic DiD & 2 & 0.241 & 0.203 & 0.315 & 0.640 & 0.730 & 0.819 & 0.860 \\
\quad grf-DiD (Wang) & 2 & -0.100 & 0.233 & 0.253 & 0.680 & 0.790 & 0.615 & 0.651 \\
\quad DiD-BCF & 3 & 0.009 & 0.140 & 0.140 & 0.940 & 0.960 & 0.628 & 0.892 \\
\quad TWFE & 3 & -0.027 & 0.214 & 0.216 & 0.880 & 0.940 & 0.816 & 0.860 \\
\quad Callaway--Sant'Anna & 3 & 0.018 & 0.336 & 0.337 & 0.888 & 0.921 & 1.155 & 0.855 \\
\quad Gardner (did2s) & 3 & 0.342 & 0.210 & 0.401 & 0.440 & 0.590 & 0.832 & 0.908 \\
\quad DoubleML DR-DiD & 3 & -0.036 & 0.337 & 0.339 & 0.900 & 0.920 & 1.294 & 0.951 \\
\quad Synthetic DiD & 3 & 0.241 & 0.203 & 0.315 & 0.640 & 0.770 & 0.820 & 0.942 \\
\quad grf-DiD (Wang) & 3 & -0.099 & 0.234 & 0.254 & 0.670 & 0.800 & 0.615 & 0.649 \\
\addlinespace
\end{longtable}
\noindent \emph{Note.} $n=200$, 100 replications in each non-null cell, $d$ is the linearity degree. Bias and RMSE are Monte-Carlo moments of the estimation error and SD(err) is its standard deviation, $\sqrt{\text{RMSE}^2-\text{Bias}^2}$ (the target itself varies by replication, so the raw dispersion of the estimate is not the error SD); Cov.90/95 are credible/confidence-interval coverage rates; Len.95 the average 95\% interval length; $\overline{\mathrm{sd}}/\mathrm{SD}$ the calibration ratio of the average reported posterior/standard error to the raw Monte-Carlo SD of the estimates. Since realized targets vary across replications, this descriptive ratio alone does not diagnose calibration.\par\vspace{6pt}
}

\subsection{Sharp-null rejection rates}\label{ssec:nullres}
The rejection frequencies in Table~\ref{tab:null} evaluate empirical size under the sharp null hypothesis of zero treatment effects, based on exclusion of zero from central 95\% credible intervals. Across the reported ATT, event-study, and cohort-time estimands, DiD-BCF empirical rejection rates range from $0.015$ to $0.070$. These results confirm that prior shrinkage induces slight conservatism in small samples without generating elevated false-positive rates.
{\footnotesize
\begin{longtable}{lrrrrrrr}
\caption{Behaviour under the sharp null $\tau\equiv 0$ (canonical zero-effect design): bias, dispersion, coverage and the empirical size of the two-sided 5\% test.}\label{tab:null}\\
\toprule
Estimator & $d$ & Bias & SD(err) & Cov.90 & Cov.95 & Size (5\%) & MCSE \\
\midrule
\endfirsthead
\multicolumn{8}{l}{\emph{\footnotesize Table \ref{tab:null} continued from the previous page.}}\\
\toprule
Estimator & $d$ & Bias & SD(err) & Cov.90 & Cov.95 & Size (5\%) & MCSE \\
\midrule
\endhead
\midrule
\multicolumn{8}{r}{\emph{\footnotesize continued on the next page}}\\
\endfoot
\bottomrule
\endlastfoot
\multicolumn{8}{l}{\textit{ATT, ATT}} \\
\quad DiD-BCF & 1 & 0.002 & 0.103 & 0.930 & 0.960 & 0.040 & 0.014 \\
\quad TWFE & 1 & -0.022 & 0.193 & 0.930 & 0.955 & 0.045 & 0.015 \\
\quad Callaway--Sant'Anna & 1 & 0.000 & 0.171 & 0.870 & 0.940 & 0.060 & 0.024 \\
\quad Gardner (did2s) & 1 & -0.022 & 0.183 & 0.920 & 0.960 & 0.040 & 0.020 \\
\quad DoubleML DR-DiD & 1 & 0.003 & 0.203 & 0.910 & 0.960 & 0.040 & 0.020 \\
\quad Synthetic DiD & 1 & -0.023 & 0.168 & 0.880 & 0.960 & 0.040 & 0.020 \\
\quad grf-DiD (Wang) & 1 & -0.004 & 0.163 & 0.750 & 0.850 & 0.150 & 0.036 \\
\quad DiD-BCF & 2 & 0.001 & 0.118 & 0.860 & 0.930 & 0.070 & 0.018 \\
\quad TWFE & 2 & -0.022 & 0.193 & 0.930 & 0.955 & 0.045 & 0.015 \\
\quad Callaway--Sant'Anna & 2 & -0.004 & 0.183 & 0.870 & 0.940 & 0.060 & 0.024 \\
\quad Gardner (did2s) & 2 & -0.022 & 0.183 & 0.920 & 0.960 & 0.040 & 0.020 \\
\quad DoubleML DR-DiD & 2 & -0.013 & 0.209 & 0.920 & 0.980 & 0.020 & 0.014 \\
\quad Synthetic DiD & 2 & -0.023 & 0.168 & 0.880 & 0.960 & 0.040 & 0.020 \\
\quad grf-DiD (Wang) & 2 & -0.006 & 0.167 & 0.760 & 0.820 & 0.180 & 0.038 \\
\quad DiD-BCF & 3 & 0.001 & 0.118 & 0.860 & 0.930 & 0.070 & 0.018 \\
\quad TWFE & 3 & -0.022 & 0.193 & 0.930 & 0.955 & 0.045 & 0.015 \\
\quad Callaway--Sant'Anna & 3 & -0.004 & 0.183 & 0.870 & 0.940 & 0.060 & 0.024 \\
\quad Gardner (did2s) & 3 & -0.022 & 0.183 & 0.920 & 0.960 & 0.040 & 0.020 \\
\quad DoubleML DR-DiD & 3 & -0.013 & 0.209 & 0.920 & 0.980 & 0.020 & 0.014 \\
\quad Synthetic DiD & 3 & -0.023 & 0.168 & 0.880 & 0.960 & 0.040 & 0.020 \\
\quad grf-DiD (Wang) & 3 & -0.006 & 0.167 & 0.760 & 0.820 & 0.180 & 0.038 \\
\addlinespace
\multicolumn{8}{l}{\textit{ES, k$=$0}} \\
\quad DiD-BCF & 1 & -0.004 & 0.115 & 0.955 & 0.985 & 0.015 & 0.009 \\
\quad TWFE & 1 & -0.022 & 0.185 & 0.905 & 0.970 & 0.030 & 0.012 \\
\quad Callaway--Sant'Anna & 1 & -0.031 & 0.219 & 0.820 & 0.950 & 0.050 & 0.022 \\
\quad Gardner (did2s) & 1 & -0.047 & 0.173 & 0.940 & 0.970 & 0.030 & 0.017 \\
\quad DoubleML DR-DiD & 1 & -0.051 & 0.273 & 0.860 & 0.920 & 0.080 & 0.027 \\
\quad grf-DiD (Wang) & 1 & -0.036 & 0.208 & 0.720 & 0.750 & 0.250 & 0.043 \\
\quad DiD-BCF & 2 & -0.004 & 0.124 & 0.945 & 0.970 & 0.030 & 0.012 \\
\quad TWFE & 2 & -0.022 & 0.185 & 0.905 & 0.970 & 0.030 & 0.012 \\
\quad Callaway--Sant'Anna & 2 & -0.034 & 0.223 & 0.800 & 0.920 & 0.080 & 0.027 \\
\quad Gardner (did2s) & 2 & -0.047 & 0.173 & 0.940 & 0.970 & 0.030 & 0.017 \\
\quad DoubleML DR-DiD & 2 & -0.065 & 0.277 & 0.890 & 0.930 & 0.070 & 0.026 \\
\quad grf-DiD (Wang) & 2 & -0.038 & 0.206 & 0.730 & 0.750 & 0.250 & 0.043 \\
\quad DiD-BCF & 3 & -0.004 & 0.124 & 0.945 & 0.970 & 0.030 & 0.012 \\
\quad TWFE & 3 & -0.022 & 0.185 & 0.905 & 0.970 & 0.030 & 0.012 \\
\quad Callaway--Sant'Anna & 3 & -0.034 & 0.223 & 0.800 & 0.920 & 0.080 & 0.027 \\
\quad Gardner (did2s) & 3 & -0.047 & 0.173 & 0.940 & 0.970 & 0.030 & 0.017 \\
\quad DoubleML DR-DiD & 3 & -0.065 & 0.277 & 0.890 & 0.930 & 0.070 & 0.026 \\
\quad grf-DiD (Wang) & 3 & -0.038 & 0.206 & 0.730 & 0.750 & 0.250 & 0.043 \\
\addlinespace
\multicolumn{8}{l}{\textit{$\mathrm{GATT}(4,4)$}} \\
\quad DiD-BCF & 1 & -0.004 & 0.115 & 0.955 & 0.985 & 0.015 & 0.009 \\
\quad TWFE & 1 & -0.022 & 0.185 & 0.905 & 0.970 & 0.030 & 0.012 \\
\quad Callaway--Sant'Anna & 1 & -0.031 & 0.219 & 0.840 & 0.930 & 0.070 & 0.026 \\
\quad Gardner (did2s) & 1 & -0.047 & 0.173 & 0.940 & 0.970 & 0.030 & 0.017 \\
\quad DoubleML DR-DiD & 1 & -0.031 & 0.281 & 0.890 & 0.950 & 0.050 & 0.022 \\
\quad Synthetic DiD & 1 & -0.052 & 0.173 & 0.900 & 0.960 & 0.040 & 0.020 \\
\quad grf-DiD (Wang) & 1 & -0.033 & 0.211 & 0.720 & 0.750 & 0.250 & 0.043 \\
\quad DiD-BCF & 2 & -0.004 & 0.124 & 0.945 & 0.970 & 0.030 & 0.012 \\
\quad TWFE & 2 & -0.022 & 0.185 & 0.905 & 0.970 & 0.030 & 0.012 \\
\quad Callaway--Sant'Anna & 2 & -0.034 & 0.223 & 0.800 & 0.910 & 0.090 & 0.029 \\
\quad Gardner (did2s) & 2 & -0.047 & 0.173 & 0.940 & 0.970 & 0.030 & 0.017 \\
\quad DoubleML DR-DiD & 2 & -0.054 & 0.273 & 0.870 & 0.950 & 0.050 & 0.022 \\
\quad Synthetic DiD & 2 & -0.052 & 0.173 & 0.900 & 0.960 & 0.040 & 0.020 \\
\quad grf-DiD (Wang) & 2 & -0.033 & 0.210 & 0.720 & 0.750 & 0.250 & 0.043 \\
\quad DiD-BCF & 3 & -0.004 & 0.124 & 0.945 & 0.970 & 0.030 & 0.012 \\
\quad TWFE & 3 & -0.022 & 0.185 & 0.905 & 0.970 & 0.030 & 0.012 \\
\quad Callaway--Sant'Anna & 3 & -0.034 & 0.223 & 0.810 & 0.890 & 0.110 & 0.031 \\
\quad Gardner (did2s) & 3 & -0.047 & 0.173 & 0.940 & 0.970 & 0.030 & 0.017 \\
\quad DoubleML DR-DiD & 3 & -0.055 & 0.249 & 0.920 & 0.970 & 0.030 & 0.017 \\
\quad Synthetic DiD & 3 & -0.052 & 0.173 & 0.900 & 0.960 & 0.040 & 0.020 \\
\quad grf-DiD (Wang) & 3 & -0.033 & 0.211 & 0.720 & 0.750 & 0.250 & 0.043 \\
\addlinespace
\end{longtable}
\noindent \emph{Note.} $n=200$. DiD-BCF and TWFE use 200 replications at each degree; the remaining benchmark estimators use 100 replications. MCSE is the Monte-Carlo standard error of the rejection rate. A nominal 5\% test has rejection probability 0.05.\par\vspace{6pt}
}

\subsection{Sample-size sweeps}\label{ssec:sweepres}
The sample-size sweeps evaluate point accuracy, interval coverage, and estimate dispersion across panel dimensions $n \in \{50, 100, 200, 400, 800\}$. Under independent errors at degree $d=1$, DiD-BCF's $\gatt(4,4)$ RMSE contracts steadily from $0.248$ at $n=50$ down to $0.068$ at $n=800$. Under AR(1) serial correlation ($\rho=0.6$), 95\% interval coverage for $\gatt(4,4)$ remains stable across sample sizes ($0.90, 0.90, 0.96, 0.90, 0.88$ for $n=50, 100, 200, 400, 800$). At $n=800$, pooled-ATT coverage is $0.79$, $0.75$, and $0.76$ across degrees $d=1, 2, 3$, illustrating the known sensitivity of unclustered working likelihoods to persistent panel correlation.
{\footnotesize
\begin{longtable}{lrrrrrrr}
\caption{Sample-size sweep for $\mathrm{GATT}(g{=}4,t{=}4)$ at linearity degree $d=1$: the estimand every method reports, so all estimators are directly comparable across $n$.}\label{tab:sweep}\\
\toprule
Estimator & $n$ & Bias & SD(err) & RMSE & Cov.95 & Len.95 & $\overline{\mathrm{sd}}/\mathrm{SD}$ \\
\midrule
\endfirsthead
\multicolumn{8}{l}{\emph{\footnotesize Table \ref{tab:sweep} continued from the previous page.}}\\
\toprule
Estimator & $n$ & Bias & SD(err) & RMSE & Cov.95 & Len.95 & $\overline{\mathrm{sd}}/\mathrm{SD}$ \\
\midrule
\endhead
\midrule
\multicolumn{8}{r}{\emph{\footnotesize continued on the next page}}\\
\endfoot
\bottomrule
\endlastfoot
\multicolumn{8}{l}{\textit{Sample-size sweep: independent errors}} \\
\quad DiD-BCF & 50 & -0.014 & 0.248 & 0.248 & 0.970 & 1.094 & 0.822 \\
\quad DiD-BCF & 100 & 0.018 & 0.165 & 0.166 & 0.980 & 0.782 & 1.037 \\
\quad DiD-BCF & 200 & -0.019 & 0.110 & 0.112 & 1.000 & 0.559 & 1.003 \\
\quad DiD-BCF & 400 & 0.004 & 0.084 & 0.084 & 0.980 & 0.399 & 0.971 \\
\quad DiD-BCF & 800 & 0.001 & 0.068 & 0.068 & 0.960 & 0.280 & 0.776 \\
\addlinespace[2pt]
\quad TWFE & 50 & -0.046 & 0.415 & 0.418 & 0.950 & 1.697 & 0.885 \\
\quad TWFE & 100 & 0.003 & 0.265 & 0.265 & 0.980 & 1.208 & 1.071 \\
\quad TWFE & 200 & -0.048 & 0.173 & 0.179 & 0.990 & 0.820 & 1.068 \\
\quad TWFE & 400 & 0.007 & 0.140 & 0.140 & 0.950 & 0.583 & 1.027 \\
\quad TWFE & 800 & 0.003 & 0.109 & 0.109 & 0.940 & 0.411 & 0.837 \\
\quad TWFE & 1600 & 0.007 & 0.058 & 0.058 & 0.990 & 0.291 & 1.035 \\
\addlinespace[2pt]
\quad Callaway--Sant'Anna & 50 & 0.055 & 0.452 & 0.455 & 0.940 & 1.898 & 0.905 \\
\quad Callaway--Sant'Anna & 100 & 0.013 & 0.290 & 0.290 & 0.970 & 1.301 & 1.091 \\
\quad Callaway--Sant'Anna & 200 & -0.031 & 0.219 & 0.221 & 0.980 & 0.905 & 1.034 \\
\quad Callaway--Sant'Anna & 400 & 0.005 & 0.148 & 0.148 & 0.950 & 0.631 & 0.991 \\
\quad Callaway--Sant'Anna & 800 & 0.002 & 0.107 & 0.107 & 0.980 & 0.445 & 0.893 \\
\addlinespace[2pt]
\quad Gardner (did2s) & 50 & -0.046 & 0.415 & 0.418 & 0.950 & 1.674 & 0.873 \\
\quad Gardner (did2s) & 100 & 0.003 & 0.265 & 0.265 & 0.980 & 1.199 & 1.064 \\
\quad Gardner (did2s) & 200 & -0.047 & 0.173 & 0.179 & 0.980 & 0.846 & 1.059 \\
\quad Gardner (did2s) & 400 & 0.008 & 0.136 & 0.136 & 0.970 & 0.602 & 1.001 \\
\quad Gardner (did2s) & 800 & 0.003 & 0.107 & 0.107 & 0.950 & 0.427 & 0.867 \\
\addlinespace[2pt]
\quad DoubleML DR-DiD & 50 & 0.045 & 0.728 & 0.729 & 0.980 & 3.061 & 0.974 \\
\quad DoubleML DR-DiD & 100 & -0.003 & 0.467 & 0.467 & 0.970 & 1.867 & 1.008 \\
\quad DoubleML DR-DiD & 200 & -0.038 & 0.283 & 0.285 & 0.960 & 1.155 & 1.015 \\
\quad DoubleML DR-DiD & 400 & -0.001 & 0.178 & 0.178 & 0.960 & 0.754 & 0.989 \\
\quad DoubleML DR-DiD & 800 & 0.005 & 0.114 & 0.114 & 0.970 & 0.503 & 0.966 \\
\quad DoubleML DR-DiD & 1600 & 0.014 & 0.072 & 0.073 & 0.990 & 0.349 & 1.068 \\
\addlinespace[2pt]
\quad Synthetic DiD & 50 & -0.027 & 0.382 & 0.383 & 0.980 & 1.671 & 0.931 \\
\quad Synthetic DiD & 100 & 0.007 & 0.260 & 0.260 & 0.970 & 1.182 & 1.089 \\
\quad Synthetic DiD & 200 & -0.052 & 0.173 & 0.181 & 0.970 & 0.823 & 1.052 \\
\quad Synthetic DiD & 400 & 0.002 & 0.126 & 0.127 & 0.960 & 0.583 & 1.046 \\
\quad Synthetic DiD & 800 & 0.003 & 0.104 & 0.104 & 0.950 & 0.412 & 0.857 \\
\quad Synthetic DiD & 1600 & 0.009 & 0.056 & 0.057 & 1.000 & 0.291 & 1.010 \\
\addlinespace[2pt]
\quad grf-DiD (Wang) & 50 & 0.009 & 0.409 & 0.409 & 0.940 & 1.424 & 0.758 \\
\quad grf-DiD (Wang) & 100 & 0.006 & 0.298 & 0.298 & 0.890 & 0.970 & 0.809 \\
\quad grf-DiD (Wang) & 200 & -0.036 & 0.219 & 0.222 & 0.810 & 0.630 & 0.731 \\
\quad grf-DiD (Wang) & 400 & 0.005 & 0.148 & 0.148 & 0.890 & 0.445 & 0.734 \\
\quad grf-DiD (Wang) & 800 & 0.001 & 0.111 & 0.111 & 0.810 & 0.319 & 0.646 \\
\quad grf-DiD (Wang) & 1600 & 0.009 & 0.067 & 0.067 & 0.900 & 0.228 & 0.767 \\
\addlinespace[2pt]
\addlinespace
\multicolumn{8}{l}{\textit{Sample-size sweep: AR(1) serial correlation}} \\
\quad DiD-BCF & 50 & -0.035 & 0.276 & 0.278 & 0.900 & 0.933 & 0.685 \\
\quad DiD-BCF & 100 & 0.010 & 0.175 & 0.175 & 0.900 & 0.654 & 0.787 \\
\quad DiD-BCF & 200 & -0.011 & 0.111 & 0.112 & 0.960 & 0.463 & 0.775 \\
\quad DiD-BCF & 400 & 0.007 & 0.095 & 0.095 & 0.900 & 0.323 & 0.703 \\
\quad DiD-BCF & 800 & 0.001 & 0.068 & 0.068 & 0.880 & 0.226 & 0.631 \\
\addlinespace[2pt]
\quad TWFE & 50 & -0.062 & 0.379 & 0.384 & 0.960 & 1.576 & 0.903 \\
\quad TWFE & 100 & -0.005 & 0.239 & 0.240 & 0.980 & 1.119 & 1.037 \\
\quad TWFE & 200 & -0.034 & 0.160 & 0.163 & 0.990 & 0.761 & 1.012 \\
\quad TWFE & 400 & 0.010 & 0.131 & 0.132 & 0.950 & 0.541 & 0.982 \\
\quad TWFE & 800 & 0.002 & 0.095 & 0.095 & 1.000 & 0.382 & 0.862 \\
\quad TWFE & 1600 & 0.005 & 0.057 & 0.057 & 0.970 & 0.270 & 0.973 \\
\addlinespace[2pt]
\quad Callaway--Sant'Anna & 50 & 0.016 & 0.300 & 0.301 & 0.980 & 1.361 & 0.889 \\
\quad Callaway--Sant'Anna & 100 & 0.016 & 0.176 & 0.177 & 0.980 & 0.940 & 1.166 \\
\quad Callaway--Sant'Anna & 200 & -0.025 & 0.126 & 0.129 & 0.990 & 0.653 & 1.098 \\
\quad Callaway--Sant'Anna & 400 & 0.004 & 0.087 & 0.087 & 0.990 & 0.460 & 1.061 \\
\quad Callaway--Sant'Anna & 800 & 0.005 & 0.070 & 0.070 & 0.990 & 0.326 & 0.872 \\
\addlinespace[2pt]
\quad Gardner (did2s) & 50 & -0.062 & 0.379 & 0.384 & 0.950 & 1.554 & 0.890 \\
\quad Gardner (did2s) & 100 & -0.005 & 0.239 & 0.240 & 0.980 & 1.112 & 1.029 \\
\quad Gardner (did2s) & 200 & -0.033 & 0.152 & 0.156 & 0.990 & 0.789 & 1.026 \\
\quad Gardner (did2s) & 400 & 0.012 & 0.130 & 0.130 & 0.960 & 0.562 & 0.947 \\
\quad Gardner (did2s) & 800 & 0.001 & 0.093 & 0.093 & 1.000 & 0.399 & 0.897 \\
\addlinespace[2pt]
\quad DoubleML DR-DiD & 50 & 0.033 & 0.497 & 0.498 & 0.980 & 2.290 & 0.995 \\
\quad DoubleML DR-DiD & 100 & 0.033 & 0.268 & 0.270 & 0.990 & 1.305 & 1.145 \\
\quad DoubleML DR-DiD & 200 & -0.018 & 0.167 & 0.168 & 0.970 & 0.806 & 1.081 \\
\quad DoubleML DR-DiD & 400 & -0.004 & 0.108 & 0.108 & 0.980 & 0.529 & 1.077 \\
\quad DoubleML DR-DiD & 800 & 0.006 & 0.082 & 0.082 & 0.950 & 0.365 & 0.859 \\
\quad DoubleML DR-DiD & 1600 & 0.011 & 0.047 & 0.048 & 0.990 & 0.248 & 0.963 \\
\addlinespace[2pt]
\quad Synthetic DiD & 50 & -0.019 & 0.287 & 0.288 & 0.990 & 1.357 & 0.915 \\
\quad Synthetic DiD & 100 & 0.003 & 0.182 & 0.182 & 1.000 & 0.959 & 1.147 \\
\quad Synthetic DiD & 200 & -0.036 & 0.124 & 0.129 & 0.980 & 0.667 & 1.059 \\
\quad Synthetic DiD & 400 & 0.004 & 0.093 & 0.093 & 0.980 & 0.472 & 1.044 \\
\quad Synthetic DiD & 800 & 0.004 & 0.077 & 0.077 & 0.990 & 0.334 & 0.854 \\
\quad Synthetic DiD & 1600 & 0.008 & 0.043 & 0.044 & 1.000 & 0.236 & 0.949 \\
\addlinespace[2pt]
\quad grf-DiD (Wang) & 50 & -0.005 & 0.290 & 0.291 & 0.940 & 1.091 & 0.755 \\
\quad grf-DiD (Wang) & 100 & 0.013 & 0.185 & 0.186 & 0.930 & 0.731 & 0.904 \\
\quad grf-DiD (Wang) & 200 & -0.032 & 0.130 & 0.133 & 0.910 & 0.458 & 0.809 \\
\quad grf-DiD (Wang) & 400 & 0.004 & 0.090 & 0.090 & 0.910 & 0.334 & 0.832 \\
\quad grf-DiD (Wang) & 800 & 0.007 & 0.074 & 0.074 & 0.890 & 0.244 & 0.671 \\
\quad grf-DiD (Wang) & 1600 & 0.008 & 0.042 & 0.043 & 0.970 & 0.173 & 0.792 \\
\addlinespace[2pt]
\addlinespace
\end{longtable}
\noindent \emph{Note.} 100 replications per cell. A consistent estimator has $|\text{Bias}|\to 0$ and $\text{SD}\propto n^{-1/2}$; the $\sqrt{n}$-scaled diagnostics are plotted in Figure~\ref{fig:sqrtn}.\par\vspace{6pt}
}

{\footnotesize
\begin{longtable}{lrrrrrrrr}
\caption{Sample-size sweep for $\mathrm{GATT}(g{=}4,t{=}4)$ at the nonlinear degrees $d=2$ and $d=3$.}\label{tab:sweepdeg}\\
\toprule
Estimator & $n$ & $d$ & Bias & SD(err) & RMSE & Cov.95 & Len.95 & $\overline{\mathrm{sd}}/\mathrm{SD}$ \\
\midrule
\endfirsthead
\multicolumn{9}{l}{\emph{\footnotesize Table \ref{tab:sweepdeg} continued from the previous page.}}\\
\toprule
Estimator & $n$ & $d$ & Bias & SD(err) & RMSE & Cov.95 & Len.95 & $\overline{\mathrm{sd}}/\mathrm{SD}$ \\
\midrule
\endhead
\midrule
\multicolumn{9}{r}{\emph{\footnotesize continued on the next page}}\\
\endfoot
\bottomrule
\endlastfoot
\multicolumn{9}{l}{\textit{Sample-size sweep: independent errors}} \\
\quad DiD-BCF & 50 & 2 & -0.019 & 0.259 & 0.260 & 0.960 & 1.117 & 0.816 \\
\quad DiD-BCF & 50 & 3 & -0.026 & 0.269 & 0.270 & 0.950 & 1.130 & 0.970 \\
\quad DiD-BCF & 100 & 2 & 0.015 & 0.181 & 0.182 & 0.990 & 0.803 & 0.993 \\
\quad DiD-BCF & 100 & 3 & 0.002 & 0.187 & 0.187 & 0.970 & 0.816 & 1.077 \\
\quad DiD-BCF & 200 & 2 & -0.025 & 0.119 & 0.122 & 0.980 & 0.572 & 0.994 \\
\quad DiD-BCF & 200 & 3 & -0.030 & 0.119 & 0.122 & 0.980 & 0.577 & 0.899 \\
\quad DiD-BCF & 400 & 2 & 0.011 & 0.090 & 0.090 & 0.970 & 0.408 & 0.956 \\
\quad DiD-BCF & 400 & 3 & 0.007 & 0.091 & 0.092 & 0.960 & 0.412 & 1.023 \\
\quad DiD-BCF & 800 & 2 & -0.001 & 0.072 & 0.072 & 0.960 & 0.291 & 0.780 \\
\quad DiD-BCF & 800 & 3 & -0.003 & 0.072 & 0.072 & 0.960 & 0.290 & 0.701 \\
\addlinespace[2pt]
\quad TWFE & 50 & 2 & -0.046 & 0.415 & 0.418 & 0.950 & 1.697 & 0.885 \\
\quad TWFE & 50 & 3 & -0.046 & 0.415 & 0.418 & 0.970 & 1.701 & 0.995 \\
\quad TWFE & 100 & 2 & 0.003 & 0.265 & 0.265 & 0.980 & 1.208 & 1.071 \\
\quad TWFE & 100 & 3 & 0.003 & 0.265 & 0.265 & 0.980 & 1.208 & 1.115 \\
\quad TWFE & 200 & 2 & -0.048 & 0.173 & 0.179 & 0.990 & 0.820 & 1.068 \\
\quad TWFE & 200 & 3 & -0.048 & 0.173 & 0.179 & 0.990 & 0.820 & 1.068 \\
\quad TWFE & 400 & 2 & 0.007 & 0.140 & 0.140 & 0.950 & 0.583 & 1.027 \\
\quad TWFE & 400 & 3 & 0.007 & 0.140 & 0.140 & 0.950 & 0.583 & 1.027 \\
\quad TWFE & 800 & 2 & 0.003 & 0.109 & 0.109 & 0.940 & 0.411 & 0.837 \\
\quad TWFE & 800 & 3 & 0.003 & 0.109 & 0.109 & 0.940 & 0.411 & 0.837 \\
\quad TWFE & 1600 & 2 & 0.007 & 0.058 & 0.058 & 0.990 & 0.291 & 1.035 \\
\quad TWFE & 1600 & 3 & 0.007 & 0.058 & 0.058 & 0.990 & 0.291 & 1.035 \\
\addlinespace[2pt]
\quad Callaway--Sant'Anna & 50 & 2 & 0.042 & 0.460 & 0.462 & 0.948 & 2.040 & 0.958 \\
\quad Callaway--Sant'Anna & 50 & 3 & 0.042 & 0.460 & 0.462 & 0.969 & 2.044 & 1.002 \\
\quad Callaway--Sant'Anna & 100 & 2 & 0.027 & 0.309 & 0.311 & 0.960 & 1.307 & 1.056 \\
\quad Callaway--Sant'Anna & 100 & 3 & 0.027 & 0.309 & 0.311 & 0.949 & 1.310 & 1.087 \\
\quad Callaway--Sant'Anna & 200 & 2 & -0.034 & 0.223 & 0.225 & 0.950 & 0.904 & 1.008 \\
\quad Callaway--Sant'Anna & 200 & 3 & -0.034 & 0.223 & 0.225 & 0.960 & 0.901 & 0.929 \\
\quad Callaway--Sant'Anna & 400 & 2 & 0.007 & 0.150 & 0.150 & 0.950 & 0.635 & 0.995 \\
\quad Callaway--Sant'Anna & 400 & 3 & 0.007 & 0.150 & 0.150 & 0.950 & 0.634 & 1.048 \\
\quad Callaway--Sant'Anna & 800 & 2 & 0.001 & 0.108 & 0.108 & 0.980 & 0.449 & 0.902 \\
\quad Callaway--Sant'Anna & 800 & 3 & 0.001 & 0.108 & 0.108 & 0.980 & 0.445 & 0.819 \\
\addlinespace[2pt]
\quad Gardner (did2s) & 50 & 2 & -0.046 & 0.415 & 0.418 & 0.950 & 1.674 & 0.873 \\
\quad Gardner (did2s) & 50 & 3 & -0.046 & 0.415 & 0.418 & 0.970 & 1.677 & 0.981 \\
\quad Gardner (did2s) & 100 & 2 & 0.003 & 0.265 & 0.265 & 0.980 & 1.199 & 1.064 \\
\quad Gardner (did2s) & 100 & 3 & 0.003 & 0.265 & 0.265 & 0.980 & 1.199 & 1.107 \\
\quad Gardner (did2s) & 200 & 2 & -0.047 & 0.173 & 0.179 & 0.980 & 0.846 & 1.059 \\
\quad Gardner (did2s) & 200 & 3 & -0.047 & 0.173 & 0.179 & 0.980 & 0.848 & 1.053 \\
\quad Gardner (did2s) & 400 & 2 & 0.008 & 0.136 & 0.136 & 0.970 & 0.602 & 1.001 \\
\quad Gardner (did2s) & 400 & 3 & 0.008 & 0.136 & 0.136 & 0.970 & 0.602 & 1.067 \\
\quad Gardner (did2s) & 800 & 2 & 0.003 & 0.107 & 0.107 & 0.950 & 0.427 & 0.867 \\
\quad Gardner (did2s) & 800 & 3 & 0.003 & 0.107 & 0.107 & 0.960 & 0.427 & 0.780 \\
\addlinespace[2pt]
\quad DoubleML DR-DiD & 50 & 2 & 0.073 & 0.669 & 0.673 & 0.980 & 3.201 & 1.100 \\
\quad DoubleML DR-DiD & 50 & 3 & -0.024 & 0.737 & 0.737 & 0.990 & 3.556 & 1.172 \\
\quad DoubleML DR-DiD & 100 & 2 & 0.077 & 0.441 & 0.448 & 1.000 & 1.976 & 1.137 \\
\quad DoubleML DR-DiD & 100 & 3 & -0.011 & 0.437 & 0.437 & 0.990 & 1.887 & 1.099 \\
\quad DoubleML DR-DiD & 200 & 2 & -0.045 & 0.265 & 0.269 & 0.960 & 1.172 & 1.102 \\
\quad DoubleML DR-DiD & 200 & 3 & -0.031 & 0.284 & 0.286 & 0.960 & 1.160 & 0.984 \\
\quad DoubleML DR-DiD & 400 & 2 & 0.001 & 0.171 & 0.171 & 0.980 & 0.782 & 1.113 \\
\quad DoubleML DR-DiD & 400 & 3 & 0.008 & 0.172 & 0.172 & 0.980 & 0.764 & 1.129 \\
\quad DoubleML DR-DiD & 800 & 2 & -0.000 & 0.113 & 0.113 & 0.990 & 0.515 & 0.966 \\
\quad DoubleML DR-DiD & 800 & 3 & 0.003 & 0.124 & 0.124 & 0.950 & 0.516 & 0.860 \\
\quad DoubleML DR-DiD & 1600 & 2 & 0.012 & 0.071 & 0.072 & 0.980 & 0.349 & 1.046 \\
\quad DoubleML DR-DiD & 1600 & 3 & 0.008 & 0.072 & 0.073 & 1.000 & 0.353 & 1.021 \\
\addlinespace[2pt]
\quad Synthetic DiD & 50 & 2 & -0.027 & 0.382 & 0.383 & 0.980 & 1.671 & 0.931 \\
\quad Synthetic DiD & 50 & 3 & -0.027 & 0.382 & 0.383 & 0.990 & 1.670 & 1.041 \\
\quad Synthetic DiD & 100 & 2 & 0.007 & 0.260 & 0.260 & 0.970 & 1.182 & 1.089 \\
\quad Synthetic DiD & 100 & 3 & 0.007 & 0.260 & 0.260 & 0.970 & 1.183 & 1.155 \\
\quad Synthetic DiD & 200 & 2 & -0.052 & 0.173 & 0.181 & 0.970 & 0.823 & 1.052 \\
\quad Synthetic DiD & 200 & 3 & -0.052 & 0.173 & 0.181 & 0.970 & 0.825 & 1.021 \\
\quad Synthetic DiD & 400 & 2 & 0.002 & 0.126 & 0.127 & 0.960 & 0.583 & 1.046 \\
\quad Synthetic DiD & 400 & 3 & 0.002 & 0.126 & 0.127 & 0.960 & 0.582 & 1.118 \\
\quad Synthetic DiD & 800 & 2 & 0.003 & 0.104 & 0.104 & 0.950 & 0.412 & 0.857 \\
\quad Synthetic DiD & 800 & 3 & 0.003 & 0.104 & 0.104 & 0.950 & 0.411 & 0.768 \\
\quad Synthetic DiD & 1600 & 2 & 0.009 & 0.056 & 0.057 & 1.000 & 0.291 & 1.010 \\
\quad Synthetic DiD & 1600 & 3 & 0.009 & 0.056 & 0.057 & 1.000 & 0.291 & 1.020 \\
\addlinespace[2pt]
\quad grf-DiD (Wang) & 50 & 2 & 0.007 & 0.412 & 0.412 & 0.920 & 1.427 & 0.751 \\
\quad grf-DiD (Wang) & 50 & 3 & 0.023 & 0.409 & 0.410 & 0.900 & 1.443 & 0.860 \\
\quad grf-DiD (Wang) & 100 & 2 & 0.009 & 0.304 & 0.304 & 0.850 & 0.970 & 0.792 \\
\quad grf-DiD (Wang) & 100 & 3 & 0.007 & 0.302 & 0.302 & 0.860 & 0.997 & 0.847 \\
\quad grf-DiD (Wang) & 200 & 2 & -0.037 & 0.220 & 0.223 & 0.810 & 0.630 & 0.730 \\
\quad grf-DiD (Wang) & 200 & 3 & -0.037 & 0.220 & 0.223 & 0.810 & 0.630 & 0.729 \\
\quad grf-DiD (Wang) & 400 & 2 & 0.005 & 0.148 & 0.148 & 0.890 & 0.444 & 0.734 \\
\quad grf-DiD (Wang) & 400 & 3 & 0.005 & 0.148 & 0.148 & 0.890 & 0.444 & 0.733 \\
\quad grf-DiD (Wang) & 800 & 2 & 0.001 & 0.111 & 0.111 & 0.810 & 0.319 & 0.644 \\
\quad grf-DiD (Wang) & 800 & 3 & 0.001 & 0.111 & 0.111 & 0.810 & 0.319 & 0.645 \\
\quad grf-DiD (Wang) & 1600 & 2 & 0.009 & 0.067 & 0.068 & 0.900 & 0.228 & 0.766 \\
\quad grf-DiD (Wang) & 1600 & 3 & 0.009 & 0.067 & 0.068 & 0.900 & 0.228 & 0.766 \\
\addlinespace[2pt]
\addlinespace
\multicolumn{9}{l}{\textit{Sample-size sweep: AR(1) serial correlation}} \\
\quad DiD-BCF & 50 & 2 & -0.018 & 0.281 & 0.282 & 0.900 & 0.962 & 0.694 \\
\quad DiD-BCF & 50 & 3 & -0.033 & 0.294 & 0.296 & 0.900 & 0.975 & 0.823 \\
\quad DiD-BCF & 100 & 2 & 0.000 & 0.189 & 0.189 & 0.940 & 0.680 & 0.813 \\
\quad DiD-BCF & 100 & 3 & -0.015 & 0.184 & 0.185 & 0.950 & 0.693 & 0.845 \\
\quad DiD-BCF & 200 & 2 & -0.012 & 0.127 & 0.128 & 0.950 & 0.479 & 0.749 \\
\quad DiD-BCF & 200 & 3 & -0.026 & 0.122 & 0.125 & 0.960 & 0.481 & 0.755 \\
\quad DiD-BCF & 400 & 2 & 0.011 & 0.097 & 0.098 & 0.910 & 0.335 & 0.726 \\
\quad DiD-BCF & 400 & 3 & 0.007 & 0.100 & 0.100 & 0.900 & 0.336 & 0.732 \\
\quad DiD-BCF & 800 & 2 & -0.004 & 0.071 & 0.071 & 0.900 & 0.237 & 0.660 \\
\quad DiD-BCF & 800 & 3 & -0.006 & 0.069 & 0.069 & 0.890 & 0.236 & 0.596 \\
\addlinespace[2pt]
\quad TWFE & 50 & 2 & -0.062 & 0.379 & 0.384 & 0.960 & 1.576 & 0.903 \\
\quad TWFE & 50 & 3 & -0.062 & 0.379 & 0.384 & 0.960 & 1.582 & 1.005 \\
\quad TWFE & 100 & 2 & -0.005 & 0.239 & 0.240 & 0.980 & 1.119 & 1.037 \\
\quad TWFE & 100 & 3 & -0.005 & 0.239 & 0.240 & 0.980 & 1.122 & 1.041 \\
\quad TWFE & 200 & 2 & -0.034 & 0.160 & 0.163 & 0.990 & 0.761 & 1.012 \\
\quad TWFE & 200 & 3 & -0.034 & 0.160 & 0.163 & 0.990 & 0.761 & 1.012 \\
\quad TWFE & 400 & 2 & 0.010 & 0.131 & 0.132 & 0.950 & 0.541 & 0.982 \\
\quad TWFE & 400 & 3 & 0.010 & 0.131 & 0.132 & 0.950 & 0.541 & 0.982 \\
\quad TWFE & 800 & 2 & 0.002 & 0.095 & 0.095 & 1.000 & 0.382 & 0.862 \\
\quad TWFE & 800 & 3 & 0.002 & 0.095 & 0.095 & 1.000 & 0.382 & 0.862 \\
\quad TWFE & 1600 & 2 & 0.005 & 0.057 & 0.057 & 0.970 & 0.270 & 0.973 \\
\quad TWFE & 1600 & 3 & 0.005 & 0.057 & 0.057 & 0.970 & 0.270 & 0.973 \\
\addlinespace[2pt]
\quad Callaway--Sant'Anna & 50 & 2 & 0.000 & 0.313 & 0.313 & 0.958 & 1.454 & 0.920 \\
\quad Callaway--Sant'Anna & 50 & 3 & 0.000 & 0.313 & 0.313 & 0.969 & 1.458 & 0.998 \\
\quad Callaway--Sant'Anna & 100 & 2 & 0.019 & 0.188 & 0.189 & 0.990 & 0.951 & 1.168 \\
\quad Callaway--Sant'Anna & 100 & 3 & 0.019 & 0.188 & 0.189 & 0.990 & 0.961 & 1.198 \\
\quad Callaway--Sant'Anna & 200 & 2 & -0.028 & 0.132 & 0.135 & 0.990 & 0.660 & 1.063 \\
\quad Callaway--Sant'Anna & 200 & 3 & -0.028 & 0.132 & 0.135 & 0.990 & 0.668 & 0.994 \\
\quad Callaway--Sant'Anna & 400 & 2 & 0.007 & 0.091 & 0.091 & 1.000 & 0.464 & 1.060 \\
\quad Callaway--Sant'Anna & 400 & 3 & 0.007 & 0.091 & 0.091 & 0.990 & 0.466 & 1.126 \\
\quad Callaway--Sant'Anna & 800 & 2 & 0.004 & 0.071 & 0.071 & 0.980 & 0.329 & 0.881 \\
\quad Callaway--Sant'Anna & 800 & 3 & 0.004 & 0.071 & 0.071 & 0.980 & 0.326 & 0.772 \\
\addlinespace[2pt]
\quad Gardner (did2s) & 50 & 2 & -0.062 & 0.379 & 0.384 & 0.950 & 1.554 & 0.890 \\
\quad Gardner (did2s) & 50 & 3 & -0.062 & 0.379 & 0.384 & 0.960 & 1.560 & 0.991 \\
\quad Gardner (did2s) & 100 & 2 & -0.005 & 0.239 & 0.240 & 0.980 & 1.112 & 1.029 \\
\quad Gardner (did2s) & 100 & 3 & -0.005 & 0.239 & 0.240 & 0.980 & 1.114 & 1.033 \\
\quad Gardner (did2s) & 200 & 2 & -0.033 & 0.152 & 0.156 & 0.990 & 0.789 & 1.026 \\
\quad Gardner (did2s) & 200 & 3 & -0.033 & 0.152 & 0.156 & 0.990 & 0.792 & 1.074 \\
\quad Gardner (did2s) & 400 & 2 & 0.012 & 0.130 & 0.130 & 0.960 & 0.562 & 0.947 \\
\quad Gardner (did2s) & 400 & 3 & 0.012 & 0.130 & 0.130 & 0.960 & 0.561 & 0.981 \\
\quad Gardner (did2s) & 800 & 2 & 0.001 & 0.093 & 0.093 & 1.000 & 0.399 & 0.897 \\
\quad Gardner (did2s) & 800 & 3 & 0.001 & 0.093 & 0.093 & 1.000 & 0.399 & 0.819 \\
\addlinespace[2pt]
\quad DoubleML DR-DiD & 50 & 2 & 0.084 & 0.609 & 0.615 & 0.990 & 2.504 & 0.949 \\
\quad DoubleML DR-DiD & 50 & 3 & 0.019 & 0.453 & 0.454 & 1.000 & 2.146 & 1.124 \\
\quad DoubleML DR-DiD & 100 & 2 & 0.013 & 0.295 & 0.295 & 0.970 & 1.291 & 1.109 \\
\quad DoubleML DR-DiD & 100 & 3 & -0.008 & 0.265 & 0.265 & 1.000 & 1.313 & 1.208 \\
\quad DoubleML DR-DiD & 200 & 2 & -0.023 & 0.185 & 0.186 & 0.970 & 0.814 & 1.002 \\
\quad DoubleML DR-DiD & 200 & 3 & -0.027 & 0.179 & 0.181 & 0.990 & 0.829 & 0.996 \\
\quad DoubleML DR-DiD & 400 & 2 & 0.001 & 0.113 & 0.113 & 0.980 & 0.543 & 0.989 \\
\quad DoubleML DR-DiD & 400 & 3 & 0.012 & 0.115 & 0.115 & 0.990 & 0.537 & 1.119 \\
\quad DoubleML DR-DiD & 800 & 2 & -0.004 & 0.076 & 0.076 & 1.000 & 0.366 & 0.930 \\
\quad DoubleML DR-DiD & 800 & 3 & 0.004 & 0.075 & 0.075 & 1.000 & 0.369 & 0.873 \\
\quad DoubleML DR-DiD & 1600 & 2 & 0.008 & 0.048 & 0.049 & 1.000 & 0.251 & 0.940 \\
\quad DoubleML DR-DiD & 1600 & 3 & 0.006 & 0.047 & 0.047 & 0.990 & 0.251 & 0.999 \\
\addlinespace[2pt]
\quad Synthetic DiD & 50 & 2 & -0.019 & 0.287 & 0.288 & 0.990 & 1.357 & 0.915 \\
\quad Synthetic DiD & 50 & 3 & -0.019 & 0.287 & 0.288 & 1.000 & 1.354 & 1.056 \\
\quad Synthetic DiD & 100 & 2 & 0.003 & 0.182 & 0.182 & 1.000 & 0.959 & 1.147 \\
\quad Synthetic DiD & 100 & 3 & 0.003 & 0.182 & 0.182 & 1.000 & 0.960 & 1.219 \\
\quad Synthetic DiD & 200 & 2 & -0.036 & 0.124 & 0.129 & 0.980 & 0.667 & 1.059 \\
\quad Synthetic DiD & 200 & 3 & -0.036 & 0.124 & 0.129 & 0.970 & 0.670 & 1.024 \\
\quad Synthetic DiD & 400 & 2 & 0.004 & 0.093 & 0.093 & 0.980 & 0.472 & 1.044 \\
\quad Synthetic DiD & 400 & 3 & 0.004 & 0.093 & 0.093 & 0.980 & 0.471 & 1.111 \\
\quad Synthetic DiD & 800 & 2 & 0.004 & 0.077 & 0.077 & 0.990 & 0.334 & 0.854 \\
\quad Synthetic DiD & 800 & 3 & 0.004 & 0.077 & 0.077 & 0.990 & 0.334 & 0.756 \\
\quad Synthetic DiD & 1600 & 2 & 0.008 & 0.043 & 0.044 & 1.000 & 0.236 & 0.949 \\
\quad Synthetic DiD & 1600 & 3 & 0.008 & 0.043 & 0.044 & 1.000 & 0.236 & 0.978 \\
\addlinespace[2pt]
\quad grf-DiD (Wang) & 50 & 2 & -0.013 & 0.291 & 0.291 & 0.940 & 1.093 & 0.757 \\
\quad grf-DiD (Wang) & 50 & 3 & 0.004 & 0.284 & 0.284 & 0.960 & 1.119 & 0.929 \\
\quad grf-DiD (Wang) & 100 & 2 & 0.012 & 0.188 & 0.188 & 0.900 & 0.730 & 0.899 \\
\quad grf-DiD (Wang) & 100 & 3 & 0.006 & 0.188 & 0.188 & 0.900 & 0.766 & 0.979 \\
\quad grf-DiD (Wang) & 200 & 2 & -0.032 & 0.130 & 0.133 & 0.900 & 0.458 & 0.809 \\
\quad grf-DiD (Wang) & 200 & 3 & -0.032 & 0.129 & 0.133 & 0.910 & 0.458 & 0.811 \\
\quad grf-DiD (Wang) & 400 & 2 & 0.004 & 0.090 & 0.090 & 0.910 & 0.334 & 0.829 \\
\quad grf-DiD (Wang) & 400 & 3 & 0.004 & 0.090 & 0.090 & 0.910 & 0.334 & 0.830 \\
\quad grf-DiD (Wang) & 800 & 2 & 0.007 & 0.073 & 0.074 & 0.880 & 0.244 & 0.672 \\
\quad grf-DiD (Wang) & 800 & 3 & 0.006 & 0.074 & 0.074 & 0.890 & 0.244 & 0.671 \\
\quad grf-DiD (Wang) & 1600 & 2 & 0.008 & 0.042 & 0.043 & 0.970 & 0.173 & 0.793 \\
\quad grf-DiD (Wang) & 1600 & 3 & 0.008 & 0.043 & 0.043 & 0.970 & 0.173 & 0.791 \\
\addlinespace[2pt]
\addlinespace
\end{longtable}
\noindent \emph{Note.} 100 replications per cell; the $d=1$ counterpart is Table \ref{tab:sweep}. TWFE and Synthetic DiD use no covariates, so their entries are invariant to $d$ by construction. DoubleML DR-DiD is available at all panel sizes in both sweeps.\par\vspace{6pt}
}

\begin{figure}[htbp]
\centering
\includegraphics[width=0.98\linewidth]{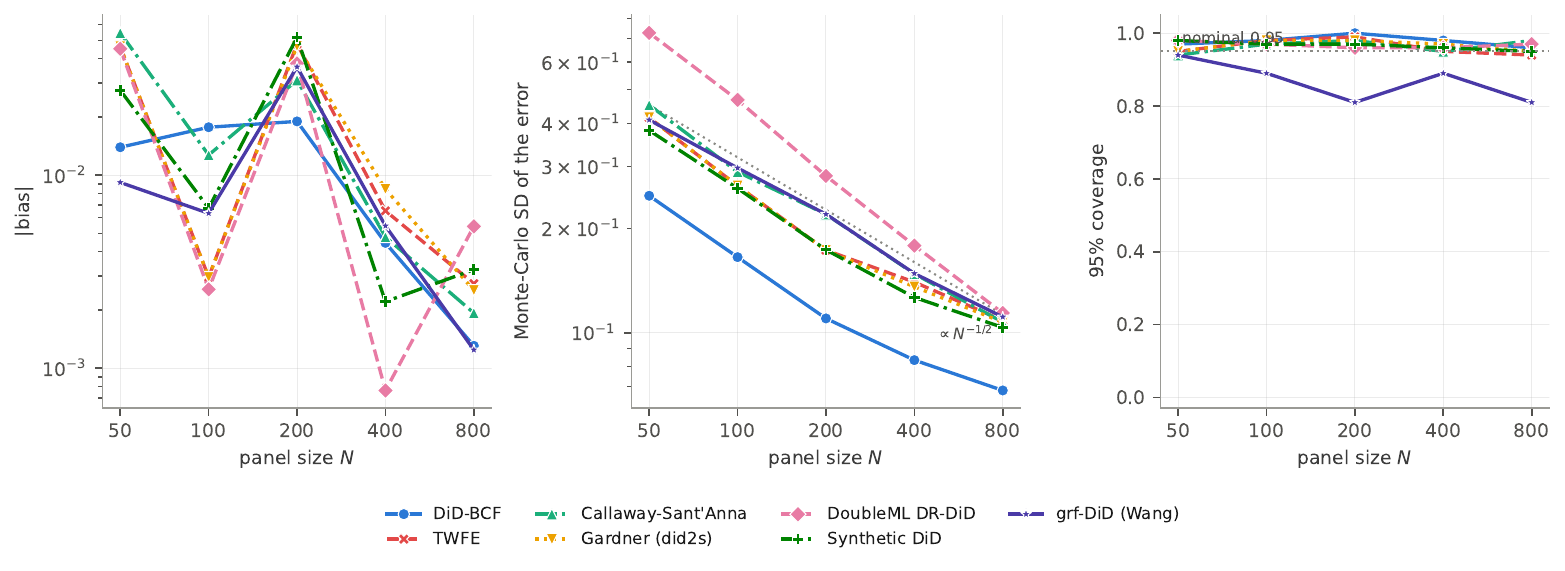}
\caption{Point accuracy and interval coverage across the sample-size sweeps.}\label{fig:sweep}
\end{figure}
\begin{figure}[htbp]
\centering
\includegraphics[width=0.9\linewidth]{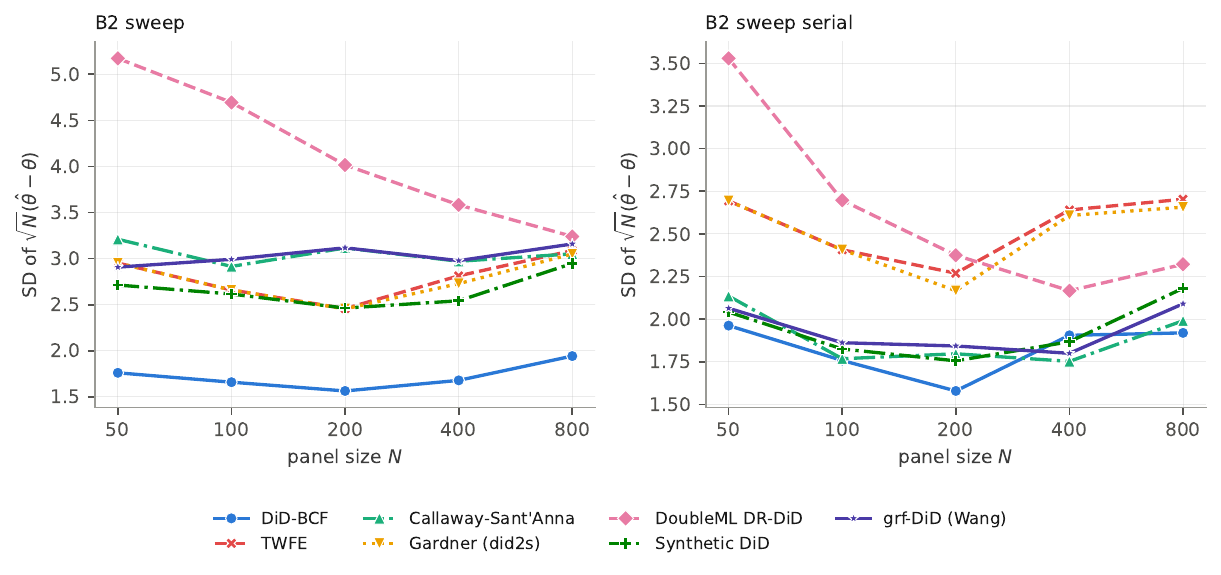}
\caption{Sample-size-scaled estimation errors across panel dimensions.}\label{fig:sqrtn}
\end{figure}
\begin{figure}[htbp]
\centering
\includegraphics[width=0.9\linewidth]{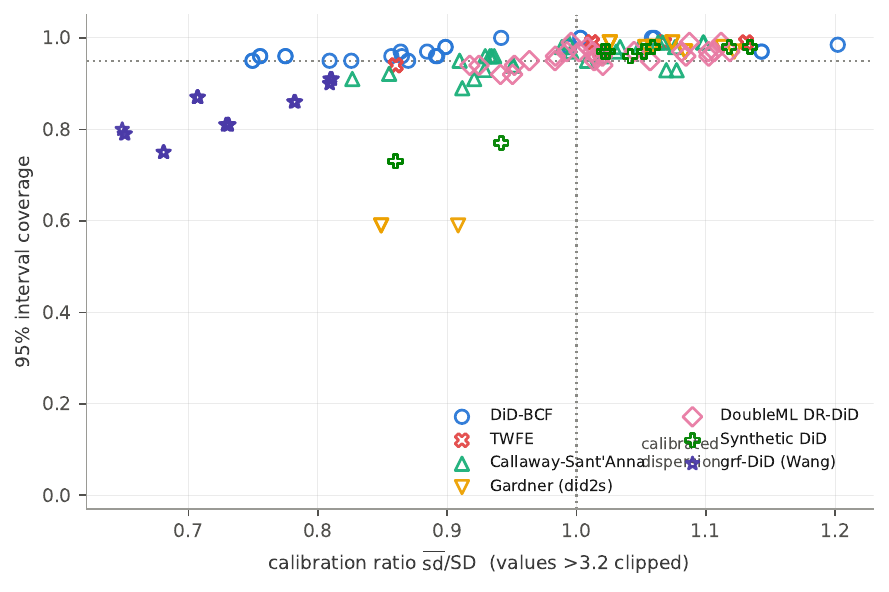}
\caption{Reported posterior spread relative to Monte Carlo standard deviation of point estimates across replications.}\label{fig:calibration}
\end{figure}

\subsection{Staggered adoption}\label{ssec:dres}
The staggered adoption design incorporates differential timing across cohorts $G_i \in \{4, 5, 6, \infty\}$ alongside dynamic growth with exposure. At degree $d=1$, DiD-BCF bias is $0.223$ for $\gatt(5,5)$ and $-0.483$ for $\gatt(6,7)$, with coverage rates of $0.77$ and $0.36$. At event time $k=3$, its bias is $-0.374$ with coverage $0.66$. Across all nine reported cohort-time cells, mean bias is $-0.036$, mean RMSE is $0.318$, and mean coverage is $0.75$, reflecting the regularizing shrinkage imposed by the tree priors on cells with smaller effective sample sizes while maintaining competitive overall point accuracy.
{\footnotesize
\begin{longtable}{lrrrrrr}
\caption{Event-study estimands under staggered adoption with cohort- and event-time-varying effects ($d=1$, $n=200$, 100 replications).}\label{tab:staggered}\\
\toprule
Estimator & $k$ & True ATT($k$) & Bias & SD(err) & RMSE & Cov.95 \\
\midrule
\endfirsthead
\multicolumn{7}{l}{\emph{\footnotesize Table \ref{tab:staggered} continued from the previous page.}}\\
\toprule
Estimator & $k$ & True ATT($k$) & Bias & SD(err) & RMSE & Cov.95 \\
\midrule
\endhead
\midrule
\multicolumn{7}{r}{\emph{\footnotesize continued on the next page}}\\
\endfoot
\bottomrule
\endlastfoot
\multicolumn{7}{l}{\textit{Staggered adoption (baseline dynamics)}} \\
\quad DiD-BCF & 0 & 4.072 & 0.095 & 0.114 & 0.148 & 0.870 \\
\quad DiD-BCF & 1 & 5.701 & -0.025 & 0.122 & 0.125 & 0.970 \\
\quad DiD-BCF & 2 & 6.102 & -0.048 & 0.157 & 0.164 & 0.970 \\
\quad DiD-BCF & 3 & 6.046 & -0.374 & 0.223 & 0.435 & 0.660 \\
\addlinespace[2pt]
\quad TWFE & 0 & 4.069 & 0.196 & 0.146 & 0.245 & 0.930 \\
\quad TWFE & 1 & 5.696 & -0.284 & 0.186 & 0.340 & 0.840 \\
\quad TWFE & 2 & 6.103 & -0.460 & 0.255 & 0.526 & 0.830 \\
\quad TWFE & 3 & 6.041 & -0.656 & 0.380 & 0.758 & 0.810 \\
\addlinespace[2pt]
\quad Callaway--Sant'Anna & 0 & 4.072 & -0.012 & 0.136 & 0.137 & 1.000 \\
\quad Callaway--Sant'Anna & 1 & 5.701 & -0.015 & 0.154 & 0.155 & 1.000 \\
\quad Callaway--Sant'Anna & 2 & 6.102 & -0.002 & 0.199 & 0.199 & 1.000 \\
\quad Callaway--Sant'Anna & 3 & 6.046 & -0.044 & 0.295 & 0.298 & 1.000 \\
\addlinespace[2pt]
\quad Gardner (did2s) & 0 & 4.072 & -0.008 & 0.140 & 0.140 & 1.000 \\
\quad Gardner (did2s) & 1 & 5.701 & -0.020 & 0.177 & 0.178 & 1.000 \\
\quad Gardner (did2s) & 2 & 6.102 & 0.005 & 0.241 & 0.241 & 1.000 \\
\quad Gardner (did2s) & 3 & 6.046 & -0.032 & 0.363 & 0.365 & 0.970 \\
\addlinespace[2pt]
\quad DoubleML DR-DiD & 0 & 4.072 & -0.006 & 0.219 & 0.219 & 0.980 \\
\quad DoubleML DR-DiD & 1 & 5.701 & -0.008 & 0.231 & 0.231 & 1.000 \\
\quad DoubleML DR-DiD & 2 & 6.102 & -0.004 & 0.300 & 0.300 & 0.980 \\
\quad DoubleML DR-DiD & 3 & 6.046 & -0.049 & 0.413 & 0.416 & 0.980 \\
\addlinespace[2pt]
\quad grf-DiD (Wang) & 0 & 4.072 & -0.007 & 0.143 & 0.144 & 0.980 \\
\quad grf-DiD (Wang) & 1 & 5.701 & -0.012 & 0.163 & 0.164 & 1.000 \\
\quad grf-DiD (Wang) & 2 & 6.102 & 0.004 & 0.217 & 0.217 & 0.980 \\
\quad grf-DiD (Wang) & 3 & 6.046 & -0.059 & 0.319 & 0.324 & 0.980 \\
\addlinespace[2pt]
\addlinespace
\multicolumn{7}{l}{\textit{Staggered adoption (amplified dynamics)}} \\
\quad DiD-BCF & 0 & 5.389 & 0.135 & 0.117 & 0.179 & 0.810 \\
\quad DiD-BCF & 1 & 9.700 & 0.001 & 0.130 & 0.130 & 0.990 \\
\quad DiD-BCF & 2 & 10.482 & 0.054 & 0.168 & 0.176 & 0.940 \\
\quad DiD-BCF & 3 & 9.344 & -0.427 & 0.235 & 0.487 & 0.600 \\
\addlinespace[2pt]
\quad TWFE & 0 & 5.384 & 0.600 & 0.167 & 0.623 & 0.340 \\
\quad TWFE & 1 & 9.691 & -0.561 & 0.267 & 0.622 & 0.760 \\
\quad TWFE & 2 & 10.485 & -1.032 & 0.340 & 1.087 & 0.580 \\
\quad TWFE & 3 & 9.336 & -1.739 & 0.487 & 1.806 & 0.190 \\
\addlinespace[2pt]
\quad Callaway--Sant'Anna & 0 & 5.389 & -0.012 & 0.136 & 0.137 & 1.000 \\
\quad Callaway--Sant'Anna & 1 & 9.700 & -0.015 & 0.154 & 0.155 & 1.000 \\
\quad Callaway--Sant'Anna & 2 & 10.482 & -0.002 & 0.199 & 0.199 & 1.000 \\
\quad Callaway--Sant'Anna & 3 & 9.344 & -0.044 & 0.295 & 0.298 & 1.000 \\
\addlinespace[2pt]
\quad Gardner (did2s) & 0 & 5.389 & -0.008 & 0.140 & 0.140 & 1.000 \\
\quad Gardner (did2s) & 1 & 9.700 & -0.020 & 0.177 & 0.178 & 1.000 \\
\quad Gardner (did2s) & 2 & 10.482 & 0.005 & 0.241 & 0.241 & 1.000 \\
\quad Gardner (did2s) & 3 & 9.344 & -0.032 & 0.363 & 0.365 & 0.990 \\
\addlinespace[2pt]
\quad DoubleML DR-DiD & 0 & 5.389 & -0.006 & 0.223 & 0.223 & 1.000 \\
\quad DoubleML DR-DiD & 1 & 9.700 & -0.009 & 0.244 & 0.244 & 1.000 \\
\quad DoubleML DR-DiD & 2 & 10.482 & -0.010 & 0.315 & 0.315 & 1.000 \\
\quad DoubleML DR-DiD & 3 & 9.344 & -0.043 & 0.433 & 0.435 & 1.000 \\
\addlinespace[2pt]
\quad grf-DiD (Wang) & 0 & 5.389 & -0.008 & 0.147 & 0.147 & 1.000 \\
\quad grf-DiD (Wang) & 1 & 9.700 & -0.008 & 0.181 & 0.182 & 1.000 \\
\quad grf-DiD (Wang) & 2 & 10.482 & 0.007 & 0.240 & 0.240 & 0.990 \\
\quad grf-DiD (Wang) & 3 & 9.344 & -0.051 & 0.327 & 0.331 & 1.000 \\
\addlinespace[2pt]
\addlinespace
\addlinespace
\end{longtable}
\noindent \emph{Note.} The true effect grows with exposure $k$ and differs across cohorts, so ATT($k$) is not constant. For Callaway--Sant'Anna and DoubleML DR-DiD, event-study targets are obtained via dynamic aggregation across $(g,t)$ cells; for grf-DiD (Wang), doubly-robust cell scores are aggregated by event time $k=t-g$ with unit-clustered standard errors. Disaggregated $(g,t)$ summaries are reported in Tables \ref{tab:staggered_gatt} and \ref{tab:staggeredcells}, and nonlinear degrees in Table \ref{tab:staggereddeg}.\par\vspace{6pt}
}

{\footnotesize
\begin{longtable}{lrrrrrrr}
\caption{Cohort-by-time estimands $\mathrm{GATT}(g,t)$ under staggered adoption, averaged over the estimated cells ($n=200$, all three linearity degrees).}\label{tab:staggered_gatt}\\
\toprule
Estimator & $d$ & Cells & True & Bias & SD(err) & RMSE & Cov.95 \\
\midrule
\endfirsthead
\multicolumn{8}{l}{\emph{\footnotesize Table \ref{tab:staggered_gatt} continued from the previous page.}}\\
\toprule
Estimator & $d$ & Cells & True & Bias & SD(err) & RMSE & Cov.95 \\
\midrule
\endhead
\midrule
\multicolumn{8}{r}{\emph{\footnotesize continued on the next page}}\\
\endfoot
\bottomrule
\endlastfoot
\multicolumn{8}{l}{\textit{Staggered adoption (baseline dynamics)}} \\
\quad DiD-BCF & 1 & 9 & 5.355 & -0.036 & 0.178 & 0.318 & 0.747 \\
\quad Callaway--Sant'Anna & 1 & 9 & 5.355 & -0.017 & 0.296 & 0.298 & 0.993 \\
\quad DoubleML DR-DiD & 1 & 9 & 5.352 & -0.008 & 0.410 & 0.412 & 0.979 \\
\quad grf-DiD (Wang) & 1 & 9 & 5.352 & -0.012 & 0.299 & 0.301 & 0.949 \\
\addlinespace[2pt]
\quad DiD-BCF & 2 & 9 & 5.355 & -0.044 & 0.188 & 0.329 & 0.732 \\
\quad Callaway--Sant'Anna & 2 & 9 & 5.356 & -0.014 & 0.302 & 0.304 & 0.985 \\
\quad DoubleML DR-DiD & 2 & 9 & 5.352 & -0.027 & 0.407 & 0.409 & 0.982 \\
\quad grf-DiD (Wang) & 2 & 9 & 5.352 & -0.012 & 0.299 & 0.301 & 0.948 \\
\addlinespace[2pt]
\quad DiD-BCF & 3 & 9 & 5.332 & -0.080 & 0.201 & 0.342 & 0.723 \\
\quad Callaway--Sant'Anna & 3 & 9 & 5.330 & -0.014 & 0.302 & 0.304 & 0.987 \\
\quad DoubleML DR-DiD & 3 & 9 & 5.352 & -0.030 & 0.430 & 0.433 & 0.983 \\
\quad grf-DiD (Wang) & 3 & 9 & 5.352 & -0.012 & 0.299 & 0.301 & 0.948 \\
\addlinespace[2pt]
\addlinespace
\multicolumn{8}{l}{\textit{Staggered adoption (amplified dynamics)}} \\
\quad DiD-BCF & 1 & 9 & 8.570 & 0.004 & 0.215 & 0.443 & 0.606 \\
\quad Callaway--Sant'Anna & 1 & 9 & 8.570 & -0.017 & 0.296 & 0.298 & 0.994 \\
\quad DoubleML DR-DiD & 1 & 9 & 8.565 & -0.020 & 0.436 & 0.437 & 0.989 \\
\quad grf-DiD (Wang) & 1 & 9 & 8.565 & -0.010 & 0.311 & 0.313 & 0.963 \\
\addlinespace[2pt]
\quad DiD-BCF & 2 & 9 & 8.570 & -0.001 & 0.226 & 0.464 & 0.588 \\
\quad Callaway--Sant'Anna & 2 & 9 & 8.570 & -0.014 & 0.302 & 0.304 & 0.992 \\
\quad DoubleML DR-DiD & 2 & 9 & 8.565 & -0.016 & 0.428 & 0.430 & 0.990 \\
\quad grf-DiD (Wang) & 2 & 9 & 8.565 & -0.010 & 0.311 & 0.313 & 0.964 \\
\addlinespace[2pt]
\quad DiD-BCF & 3 & 9 & 8.533 & -0.071 & 0.260 & 0.507 & 0.591 \\
\quad Callaway--Sant'Anna & 3 & 9 & 8.529 & -0.014 & 0.302 & 0.304 & 0.990 \\
\quad DoubleML DR-DiD & 3 & 9 & 8.565 & -0.019 & 0.448 & 0.452 & 0.992 \\
\quad grf-DiD (Wang) & 3 & 9 & 8.565 & -0.010 & 0.311 & 0.313 & 0.963 \\
\addlinespace[2pt]
\addlinespace
\addlinespace
\end{longtable}
\noindent \emph{Note.} ``Cells'' is the number of $(g,t)$ cells each estimator reports; all quantities are unweighted averages across those cells, which mixes cohorts and event times. The cell-by-cell version at $d=1$ is Table \ref{tab:staggeredcells}.\par\vspace{6pt}
}

{\footnotesize
\begin{longtable}{lrrrrrrr}
\caption{Cell-by-cell cohort-time estimands $\mathrm{GATT}(g,t)$ under staggered adoption, grouped by event time $k=t-g$ ($d=1$, $n=200$, 100 replications).}\label{tab:staggeredcells}\\
\toprule
Estimator & $g$ & $t$ & $k$ & True & Bias & RMSE & Cov.95 \\
\midrule
\endfirsthead
\multicolumn{8}{l}{\emph{\footnotesize Table \ref{tab:staggeredcells} continued from the previous page.}}\\
\toprule
Estimator & $g$ & $t$ & $k$ & True & Bias & RMSE & Cov.95 \\
\midrule
\endhead
\midrule
\multicolumn{8}{r}{\emph{\footnotesize continued on the next page}}\\
\endfoot
\bottomrule
\endlastfoot
\multicolumn{8}{l}{\textit{Staggered adoption (baseline dynamics)}} \\
\multicolumn{8}{l}{\quad $k=0$, cohort $g=4$} \\
\quad\quad DiD-BCF & 4 & 4 & 0 & 2.748 & -0.076 & 0.175 & 0.960 \\
\quad\quad Callaway--Sant'Anna & 4 & 4 & 0 & 2.748 & -0.049 & 0.303 & 0.960 \\
\quad\quad DoubleML DR-DiD & 4 & 4 & 0 & 2.746 & -0.065 & 0.416 & 0.940 \\
\quad\quad grf-DiD (Wang) & 4 & 4 & 0 & 2.746 & -0.050 & 0.300 & 0.870 \\
\addlinespace[2pt]
\multicolumn{8}{l}{\quad $k=0$, cohort $g=5$} \\
\quad\quad DiD-BCF & 5 & 5 & 0 & 4.124 & 0.223 & 0.274 & 0.770 \\
\quad\quad Callaway--Sant'Anna & 5 & 5 & 0 & 4.124 & -0.031 & 0.292 & 0.990 \\
\quad\quad DoubleML DR-DiD & 5 & 5 & 0 & 4.126 & 0.011 & 0.417 & 0.950 \\
\quad\quad grf-DiD (Wang) & 5 & 5 & 0 & 4.126 & -0.020 & 0.284 & 0.940 \\
\addlinespace[2pt]
\multicolumn{8}{l}{\quad $k=0$, cohort $g=6$} \\
\quad\quad DiD-BCF & 6 & 6 & 0 & 5.537 & 0.160 & 0.230 & 0.880 \\
\quad\quad Callaway--Sant'Anna & 6 & 6 & 0 & 5.537 & 0.038 & 0.274 & 1.000 \\
\quad\quad DoubleML DR-DiD & 6 & 6 & 0 & 5.527 & 0.006 & 0.394 & 0.980 \\
\quad\quad grf-DiD (Wang) & 6 & 6 & 0 & 5.527 & 0.043 & 0.267 & 1.000 \\
\addlinespace[2pt]
\multicolumn{8}{l}{\quad $k=1$, cohort $g=4$} \\
\quad\quad DiD-BCF & 4 & 5 & 1 & 3.847 & 0.231 & 0.275 & 0.810 \\
\quad\quad Callaway--Sant'Anna & 4 & 5 & 1 & 3.847 & -0.048 & 0.290 & 0.990 \\
\quad\quad DoubleML DR-DiD & 4 & 5 & 1 & 3.844 & -0.043 & 0.383 & 0.980 \\
\quad\quad grf-DiD (Wang) & 4 & 5 & 1 & 3.844 & -0.053 & 0.294 & 0.900 \\
\addlinespace[2pt]
\multicolumn{8}{l}{\quad $k=1$, cohort $g=5$} \\
\quad\quad DiD-BCF & 5 & 6 & 1 & 5.773 & 0.138 & 0.213 & 0.910 \\
\quad\quad Callaway--Sant'Anna & 5 & 6 & 1 & 5.773 & -0.003 & 0.303 & 1.000 \\
\quad\quad DoubleML DR-DiD & 5 & 6 & 1 & 5.777 & 0.018 & 0.401 & 1.000 \\
\quad\quad grf-DiD (Wang) & 5 & 6 & 1 & 5.777 & 0.002 & 0.317 & 0.960 \\
\addlinespace[2pt]
\multicolumn{8}{l}{\quad $k=1$, cohort $g=6$} \\
\quad\quad DiD-BCF & 6 & 7 & 1 & 7.752 & -0.483 & 0.526 & 0.360 \\
\quad\quad Callaway--Sant'Anna & 6 & 7 & 1 & 7.752 & 0.005 & 0.316 & 1.000 \\
\quad\quad DoubleML DR-DiD & 6 & 7 & 1 & 7.738 & 0.007 & 0.443 & 1.000 \\
\quad\quad grf-DiD (Wang) & 6 & 7 & 1 & 7.738 & 0.022 & 0.315 & 1.000 \\
\addlinespace[2pt]
\multicolumn{8}{l}{\quad $k=2$, cohort $g=4$} \\
\quad\quad DiD-BCF & 4 & 6 & 2 & 4.947 & 0.247 & 0.298 & 0.770 \\
\quad\quad Callaway--Sant'Anna & 4 & 6 & 2 & 4.947 & -0.026 & 0.309 & 1.000 \\
\quad\quad DoubleML DR-DiD & 4 & 6 & 2 & 4.943 & -0.015 & 0.408 & 0.970 \\
\quad\quad grf-DiD (Wang) & 4 & 6 & 2 & 4.943 & -0.020 & 0.331 & 0.920 \\
\addlinespace[2pt]
\multicolumn{8}{l}{\quad $k=2$, cohort $g=5$} \\
\quad\quad DiD-BCF & 5 & 7 & 2 & 7.423 & -0.386 & 0.439 & 0.600 \\
\quad\quad Callaway--Sant'Anna & 5 & 7 & 2 & 7.423 & 0.009 & 0.295 & 1.000 \\
\quad\quad DoubleML DR-DiD & 5 & 7 & 2 & 7.427 & 0.061 & 0.441 & 1.000 \\
\quad\quad grf-DiD (Wang) & 5 & 7 & 2 & 7.427 & 0.020 & 0.274 & 0.990 \\
\addlinespace[2pt]
\multicolumn{8}{l}{\quad $k=3$, cohort $g=4$} \\
\quad\quad DiD-BCF & 4 & 7 & 3 & 6.046 & -0.374 & 0.435 & 0.660 \\
\quad\quad Callaway--Sant'Anna & 4 & 7 & 3 & 6.046 & -0.044 & 0.298 & 1.000 \\
\quad\quad DoubleML DR-DiD & 4 & 7 & 3 & 6.041 & -0.050 & 0.402 & 0.990 \\
\quad\quad grf-DiD (Wang) & 4 & 7 & 3 & 6.041 & -0.050 & 0.331 & 0.960 \\
\addlinespace[2pt]
\addlinespace
\multicolumn{8}{l}{\textit{Staggered adoption (amplified dynamics)}} \\
\multicolumn{8}{l}{\quad $k=0$, cohort $g=4$} \\
\quad\quad DiD-BCF & 4 & 4 & 0 & 2.748 & -0.071 & 0.186 & 0.960 \\
\quad\quad Callaway--Sant'Anna & 4 & 4 & 0 & 2.748 & -0.049 & 0.303 & 0.960 \\
\quad\quad DoubleML DR-DiD & 4 & 4 & 0 & 2.746 & -0.047 & 0.404 & 0.970 \\
\quad\quad grf-DiD (Wang) & 4 & 4 & 0 & 2.746 & -0.050 & 0.300 & 0.870 \\
\addlinespace[2pt]
\multicolumn{8}{l}{\quad $k=0$, cohort $g=5$} \\
\quad\quad DiD-BCF & 5 & 5 & 0 & 5.498 & 0.011 & 0.191 & 0.960 \\
\quad\quad Callaway--Sant'Anna & 5 & 5 & 0 & 5.498 & -0.031 & 0.292 & 1.000 \\
\quad\quad DoubleML DR-DiD & 5 & 5 & 0 & 5.502 & -0.013 & 0.393 & 0.990 \\
\quad\quad grf-DiD (Wang) & 5 & 5 & 0 & 5.502 & -0.021 & 0.289 & 0.950 \\
\addlinespace[2pt]
\multicolumn{8}{l}{\quad $k=0$, cohort $g=6$} \\
\quad\quad DiD-BCF & 6 & 6 & 0 & 8.306 & 0.497 & 0.537 & 0.400 \\
\quad\quad Callaway--Sant'Anna & 6 & 6 & 0 & 8.306 & 0.038 & 0.274 & 1.000 \\
\quad\quad DoubleML DR-DiD & 6 & 6 & 0 & 8.291 & 0.015 & 0.353 & 1.000 \\
\quad\quad grf-DiD (Wang) & 6 & 6 & 0 & 8.291 & 0.050 & 0.277 & 1.000 \\
\addlinespace[2pt]
\multicolumn{8}{l}{\quad $k=1$, cohort $g=4$} \\
\quad\quad DiD-BCF & 4 & 5 & 1 & 4.947 & 0.021 & 0.183 & 0.990 \\
\quad\quad Callaway--Sant'Anna & 4 & 5 & 1 & 4.947 & -0.048 & 0.290 & 0.990 \\
\quad\quad DoubleML DR-DiD & 4 & 5 & 1 & 4.943 & -0.039 & 0.463 & 0.970 \\
\quad\quad grf-DiD (Wang) & 4 & 5 & 1 & 4.943 & -0.056 & 0.296 & 0.930 \\
\addlinespace[2pt]
\multicolumn{8}{l}{\quad $k=1$, cohort $g=5$} \\
\quad\quad DiD-BCF & 5 & 6 & 1 & 9.897 & 0.566 & 0.612 & 0.350 \\
\quad\quad Callaway--Sant'Anna & 5 & 6 & 1 & 9.897 & -0.003 & 0.303 & 1.000 \\
\quad\quad DoubleML DR-DiD & 5 & 6 & 1 & 9.903 & -0.013 & 0.478 & 1.000 \\
\quad\quad grf-DiD (Wang) & 5 & 6 & 1 & 9.903 & 0.007 & 0.345 & 0.980 \\
\addlinespace[2pt]
\multicolumn{8}{l}{\quad $k=1$, cohort $g=6$} \\
\quad\quad DiD-BCF & 6 & 7 & 1 & 14.951 & -0.588 & 0.644 & 0.370 \\
\quad\quad Callaway--Sant'Anna & 6 & 7 & 1 & 14.951 & 0.005 & 0.316 & 1.000 \\
\quad\quad DoubleML DR-DiD & 6 & 7 & 1 & 14.923 & -0.017 & 0.441 & 1.000 \\
\quad\quad grf-DiD (Wang) & 6 & 7 & 1 & 14.923 & 0.024 & 0.326 & 1.000 \\
\addlinespace[2pt]
\multicolumn{8}{l}{\quad $k=2$, cohort $g=4$} \\
\quad\quad DiD-BCF & 4 & 6 & 2 & 7.145 & 0.539 & 0.578 & 0.330 \\
\quad\quad Callaway--Sant'Anna & 4 & 6 & 2 & 7.145 & -0.026 & 0.309 & 1.000 \\
\quad\quad DoubleML DR-DiD & 4 & 6 & 2 & 7.139 & -0.046 & 0.448 & 0.990 \\
\quad\quad grf-DiD (Wang) & 4 & 6 & 2 & 7.139 & -0.019 & 0.334 & 0.950 \\
\addlinespace[2pt]
\multicolumn{8}{l}{\quad $k=2$, cohort $g=5$} \\
\quad\quad DiD-BCF & 5 & 7 & 2 & 14.295 & -0.508 & 0.565 & 0.490 \\
\quad\quad Callaway--Sant'Anna & 5 & 7 & 2 & 14.295 & 0.009 & 0.295 & 1.000 \\
\quad\quad DoubleML DR-DiD & 5 & 7 & 2 & 14.304 & 0.038 & 0.471 & 1.000 \\
\quad\quad grf-DiD (Wang) & 5 & 7 & 2 & 14.304 & 0.021 & 0.309 & 1.000 \\
\addlinespace[2pt]
\multicolumn{8}{l}{\quad $k=3$, cohort $g=4$} \\
\quad\quad DiD-BCF & 4 & 7 & 3 & 9.344 & -0.427 & 0.487 & 0.600 \\
\quad\quad Callaway--Sant'Anna & 4 & 7 & 3 & 9.344 & -0.044 & 0.298 & 1.000 \\
\quad\quad DoubleML DR-DiD & 4 & 7 & 3 & 9.336 & -0.055 & 0.483 & 0.980 \\
\quad\quad grf-DiD (Wang) & 4 & 7 & 3 & 9.336 & -0.047 & 0.343 & 0.990 \\
\addlinespace[2pt]
\addlinespace
\addlinespace
\end{longtable}
\noindent \emph{Note.} Every estimator here reports the same $(g,t)$ cell, so the comparison is exact and requires no aggregation. TWFE and Gardner are absent because they report only pooled or event-time coefficients; see Table \ref{tab:staggered}. Reading down a fixed $k$ block shows how each estimator degrades as the cohort adopts later and its control group shrinks.\par\vspace{6pt}
}

{\footnotesize
\begin{longtable}{lrrrrrrr}
\caption{Event-study estimands under staggered adoption at the nonlinear degrees $d=2$ and $d=3$ ($n=200$, 100 replications).}\label{tab:staggereddeg}\\
\toprule
Estimator & $d$ & $k$ & True ATT($k$) & Bias & SD(err) & RMSE & Cov.95 \\
\midrule
\endfirsthead
\multicolumn{8}{l}{\emph{\footnotesize Table \ref{tab:staggereddeg} continued from the previous page.}}\\
\toprule
Estimator & $d$ & $k$ & True ATT($k$) & Bias & SD(err) & RMSE & Cov.95 \\
\midrule
\endhead
\midrule
\multicolumn{8}{r}{\emph{\footnotesize continued on the next page}}\\
\endfoot
\bottomrule
\endlastfoot
\multicolumn{8}{l}{\textit{Staggered adoption (baseline dynamics)}} \\
\quad DiD-BCF & 2 & 0 & 4.072 & 0.096 & 0.118 & 0.152 & 0.860 \\
\quad DiD-BCF & 2 & 1 & 5.701 & -0.034 & 0.130 & 0.134 & 0.970 \\
\quad DiD-BCF & 2 & 2 & 6.102 & -0.061 & 0.157 & 0.168 & 0.940 \\
\quad DiD-BCF & 2 & 3 & 6.046 & -0.392 & 0.230 & 0.455 & 0.630 \\
\quad DiD-BCF & 3 & 0 & 4.050 & 0.080 & 0.122 & 0.146 & 0.880 \\
\quad DiD-BCF & 3 & 1 & 5.670 & -0.070 & 0.135 & 0.152 & 0.920 \\
\quad DiD-BCF & 3 & 2 & 6.091 & -0.114 & 0.177 & 0.211 & 0.910 \\
\quad DiD-BCF & 3 & 3 & 6.020 & -0.440 & 0.263 & 0.512 & 0.550 \\
\addlinespace[2pt]
\quad TWFE & 2 & 0 & 4.069 & 0.196 & 0.146 & 0.245 & 0.930 \\
\quad TWFE & 2 & 1 & 5.696 & -0.284 & 0.186 & 0.340 & 0.840 \\
\quad TWFE & 2 & 2 & 6.103 & -0.460 & 0.255 & 0.526 & 0.830 \\
\quad TWFE & 2 & 3 & 6.041 & -0.656 & 0.380 & 0.758 & 0.810 \\
\quad TWFE & 3 & 0 & 4.069 & 0.196 & 0.146 & 0.245 & 0.930 \\
\quad TWFE & 3 & 1 & 5.696 & -0.284 & 0.186 & 0.340 & 0.840 \\
\quad TWFE & 3 & 2 & 6.103 & -0.460 & 0.255 & 0.526 & 0.830 \\
\quad TWFE & 3 & 3 & 6.041 & -0.656 & 0.380 & 0.758 & 0.810 \\
\addlinespace[2pt]
\quad Callaway--Sant'Anna & 2 & 0 & 4.072 & -0.002 & 0.153 & 0.153 & 0.990 \\
\quad Callaway--Sant'Anna & 2 & 1 & 5.701 & -0.008 & 0.192 & 0.192 & 0.990 \\
\quad Callaway--Sant'Anna & 2 & 2 & 6.102 & -0.007 & 0.270 & 0.270 & 0.970 \\
\quad Callaway--Sant'Anna & 2 & 3 & 6.052 & -0.040 & 0.318 & 0.320 & 1.000 \\
\quad Callaway--Sant'Anna & 3 & 0 & 4.050 & -0.003 & 0.148 & 0.148 & 1.000 \\
\quad Callaway--Sant'Anna & 3 & 1 & 5.670 & -0.010 & 0.182 & 0.183 & 0.990 \\
\quad Callaway--Sant'Anna & 3 & 2 & 6.091 & -0.014 & 0.258 & 0.259 & 0.970 \\
\quad Callaway--Sant'Anna & 3 & 3 & 6.019 & -0.040 & 0.318 & 0.320 & 1.000 \\
\addlinespace[2pt]
\quad Gardner (did2s) & 2 & 0 & 4.072 & -0.008 & 0.140 & 0.140 & 1.000 \\
\quad Gardner (did2s) & 2 & 1 & 5.701 & -0.020 & 0.177 & 0.178 & 1.000 \\
\quad Gardner (did2s) & 2 & 2 & 6.102 & 0.005 & 0.241 & 0.241 & 1.000 \\
\quad Gardner (did2s) & 2 & 3 & 6.046 & -0.032 & 0.363 & 0.365 & 0.970 \\
\quad Gardner (did2s) & 3 & 0 & 4.050 & -0.008 & 0.140 & 0.140 & 1.000 \\
\quad Gardner (did2s) & 3 & 1 & 5.670 & -0.020 & 0.177 & 0.178 & 1.000 \\
\quad Gardner (did2s) & 3 & 2 & 6.091 & 0.005 & 0.241 & 0.241 & 1.000 \\
\quad Gardner (did2s) & 3 & 3 & 6.020 & -0.032 & 0.363 & 0.365 & 0.960 \\
\addlinespace[2pt]
\quad DoubleML DR-DiD & 2 & 0 & 4.072 & 0.001 & 0.238 & 0.238 & 0.980 \\
\quad DoubleML DR-DiD & 2 & 1 & 5.701 & -0.002 & 0.253 & 0.253 & 1.000 \\
\quad DoubleML DR-DiD & 2 & 2 & 6.102 & 0.005 & 0.328 & 0.328 & 1.000 \\
\quad DoubleML DR-DiD & 2 & 3 & 6.046 & -0.021 & 0.508 & 0.509 & 1.000 \\
\quad DoubleML DR-DiD & 3 & 0 & 4.050 & 0.000 & 0.241 & 0.241 & 0.980 \\
\quad DoubleML DR-DiD & 3 & 1 & 5.670 & 0.003 & 0.253 & 0.253 & 1.000 \\
\quad DoubleML DR-DiD & 3 & 2 & 6.091 & 0.003 & 0.331 & 0.331 & 1.000 \\
\quad DoubleML DR-DiD & 3 & 3 & 6.020 & -0.033 & 0.506 & 0.507 & 1.000 \\
\addlinespace[2pt]
\quad grf-DiD (Wang) & 2 & 0 & 4.072 & -0.005 & 0.144 & 0.145 & 0.990 \\
\quad grf-DiD (Wang) & 2 & 1 & 5.701 & -0.014 & 0.158 & 0.159 & 1.000 \\
\quad grf-DiD (Wang) & 2 & 2 & 6.102 & 0.007 & 0.204 & 0.205 & 0.980 \\
\quad grf-DiD (Wang) & 2 & 3 & 6.046 & -0.055 & 0.300 & 0.305 & 1.000 \\
\quad grf-DiD (Wang) & 3 & 0 & 4.050 & 0.003 & 0.142 & 0.142 & 1.000 \\
\quad grf-DiD (Wang) & 3 & 1 & 5.670 & -0.003 & 0.158 & 0.158 & 0.990 \\
\quad grf-DiD (Wang) & 3 & 2 & 6.091 & 0.015 & 0.212 & 0.213 & 0.980 \\
\quad grf-DiD (Wang) & 3 & 3 & 6.020 & -0.036 & 0.318 & 0.320 & 1.000 \\
\addlinespace[2pt]
\addlinespace
\multicolumn{8}{l}{\textit{Staggered adoption (amplified dynamics)}} \\
\quad DiD-BCF & 2 & 0 & 5.389 & 0.140 & 0.119 & 0.184 & 0.750 \\
\quad DiD-BCF & 2 & 1 & 9.700 & -0.005 & 0.140 & 0.140 & 0.960 \\
\quad DiD-BCF & 2 & 2 & 10.482 & 0.044 & 0.176 & 0.181 & 0.940 \\
\quad DiD-BCF & 2 & 3 & 9.344 & -0.452 & 0.245 & 0.514 & 0.530 \\
\quad DiD-BCF & 3 & 0 & 5.357 & 0.126 & 0.130 & 0.181 & 0.790 \\
\quad DiD-BCF & 3 & 1 & 9.643 & -0.078 & 0.171 & 0.188 & 0.870 \\
\quad DiD-BCF & 3 & 2 & 10.467 & -0.070 & 0.231 & 0.241 & 0.920 \\
\quad DiD-BCF & 3 & 3 & 9.304 & -0.561 & 0.327 & 0.649 & 0.430 \\
\addlinespace[2pt]
\quad TWFE & 2 & 0 & 5.384 & 0.600 & 0.167 & 0.623 & 0.340 \\
\quad TWFE & 2 & 1 & 9.691 & -0.561 & 0.267 & 0.622 & 0.760 \\
\quad TWFE & 2 & 2 & 10.485 & -1.032 & 0.340 & 1.087 & 0.580 \\
\quad TWFE & 2 & 3 & 9.336 & -1.739 & 0.487 & 1.806 & 0.190 \\
\quad TWFE & 3 & 0 & 5.384 & 0.600 & 0.167 & 0.623 & 0.340 \\
\quad TWFE & 3 & 1 & 9.691 & -0.561 & 0.267 & 0.622 & 0.760 \\
\quad TWFE & 3 & 2 & 10.485 & -1.032 & 0.340 & 1.087 & 0.580 \\
\quad TWFE & 3 & 3 & 9.336 & -1.739 & 0.487 & 1.806 & 0.190 \\
\addlinespace[2pt]
\quad Callaway--Sant'Anna & 2 & 0 & 5.389 & 0.003 & 0.189 & 0.189 & 0.990 \\
\quad Callaway--Sant'Anna & 2 & 1 & 9.700 & 0.002 & 0.308 & 0.308 & 0.990 \\
\quad Callaway--Sant'Anna & 2 & 2 & 10.482 & -0.012 & 0.532 & 0.533 & 0.980 \\
\quad Callaway--Sant'Anna & 2 & 3 & 9.353 & -0.040 & 0.318 & 0.320 & 1.000 \\
\quad Callaway--Sant'Anna & 3 & 0 & 5.357 & 0.001 & 0.180 & 0.180 & 0.990 \\
\quad Callaway--Sant'Anna & 3 & 1 & 9.643 & -0.001 & 0.287 & 0.287 & 0.990 \\
\quad Callaway--Sant'Anna & 3 & 2 & 10.467 & -0.024 & 0.516 & 0.517 & 0.980 \\
\quad Callaway--Sant'Anna & 3 & 3 & 9.302 & -0.040 & 0.318 & 0.320 & 1.000 \\
\addlinespace[2pt]
\quad Gardner (did2s) & 2 & 0 & 5.389 & -0.008 & 0.140 & 0.140 & 1.000 \\
\quad Gardner (did2s) & 2 & 1 & 9.700 & -0.020 & 0.177 & 0.178 & 1.000 \\
\quad Gardner (did2s) & 2 & 2 & 10.482 & 0.005 & 0.241 & 0.241 & 1.000 \\
\quad Gardner (did2s) & 2 & 3 & 9.344 & -0.032 & 0.363 & 0.365 & 0.990 \\
\quad Gardner (did2s) & 3 & 0 & 5.357 & -0.008 & 0.140 & 0.140 & 1.000 \\
\quad Gardner (did2s) & 3 & 1 & 9.643 & -0.020 & 0.177 & 0.178 & 1.000 \\
\quad Gardner (did2s) & 3 & 2 & 10.467 & 0.005 & 0.241 & 0.241 & 1.000 \\
\quad Gardner (did2s) & 3 & 3 & 9.304 & -0.032 & 0.363 & 0.365 & 1.000 \\
\addlinespace[2pt]
\quad DoubleML DR-DiD & 2 & 0 & 5.389 & 0.001 & 0.241 & 0.241 & 1.000 \\
\quad DoubleML DR-DiD & 2 & 1 & 9.700 & -0.003 & 0.264 & 0.264 & 1.000 \\
\quad DoubleML DR-DiD & 2 & 2 & 10.482 & -0.001 & 0.343 & 0.343 & 1.000 \\
\quad DoubleML DR-DiD & 2 & 3 & 9.344 & -0.015 & 0.527 & 0.527 & 1.000 \\
\quad DoubleML DR-DiD & 3 & 0 & 5.357 & 0.001 & 0.245 & 0.245 & 0.990 \\
\quad DoubleML DR-DiD & 3 & 1 & 9.643 & 0.004 & 0.263 & 0.263 & 1.000 \\
\quad DoubleML DR-DiD & 3 & 2 & 10.467 & -0.004 & 0.347 & 0.347 & 1.000 \\
\quad DoubleML DR-DiD & 3 & 3 & 9.304 & -0.033 & 0.523 & 0.524 & 1.000 \\
\addlinespace[2pt]
\quad grf-DiD (Wang) & 2 & 0 & 5.389 & -0.003 & 0.150 & 0.150 & 1.000 \\
\quad grf-DiD (Wang) & 2 & 1 & 9.700 & -0.007 & 0.180 & 0.180 & 1.000 \\
\quad grf-DiD (Wang) & 2 & 2 & 10.482 & 0.008 & 0.223 & 0.223 & 0.990 \\
\quad grf-DiD (Wang) & 2 & 3 & 9.344 & -0.046 & 0.305 & 0.308 & 1.000 \\
\quad grf-DiD (Wang) & 3 & 0 & 5.357 & 0.006 & 0.144 & 0.144 & 1.000 \\
\quad grf-DiD (Wang) & 3 & 1 & 9.643 & 0.008 & 0.172 & 0.173 & 1.000 \\
\quad grf-DiD (Wang) & 3 & 2 & 10.467 & 0.027 & 0.232 & 0.233 & 1.000 \\
\quad grf-DiD (Wang) & 3 & 3 & 9.304 & -0.024 & 0.325 & 0.326 & 1.000 \\
\addlinespace[2pt]
\addlinespace
\end{longtable}
\noindent \emph{Note.} The $d=1$ counterpart is Table \ref{tab:staggered}. TWFE uses no covariates, so its entries are invariant to $d$ by construction; at $d=3$ the treatment-effect surface itself becomes quadratic, so the true ATT($k$) path changes as well.\par\vspace{6pt}
}

\begin{table}[htbp]
\centering
\footnotesize
\caption{Goodman-Bacon (2021) decomposition of the TWFE estimate under staggered adoption with cohort- and event-time-varying effects, averaged over 500 replications.}
\label{tab:bacon}
\begin{tabular}{lrrr}
\toprule
Comparison type & Weight & $\widehat{DD}$ & Contribution \\
\midrule
\multicolumn{4}{l}{\textit{Staggered adoption (baseline dynamics)}} \\
\quad Earlier vs Later & 0.072 & 3.386 & 0.245 \\
\quad Later vs Earlier & 0.039 & 4.038 & 0.156 \\
\quad Treated vs Untreated & 0.222 & 5.581 & 1.241 \\
\quad \textit{TWFE estimate} & \multicolumn{3}{r}{4.830} \\
\quad \textit{true ATT} & \multicolumn{3}{r}{5.299} \\
\quad \textit{TWFE bias} & \multicolumn{3}{r}{-0.469} \\
\quad \textit{weight on already-treated comparisons} & \multicolumn{3}{r}{0.116} \\
\addlinespace
\multicolumn{4}{l}{\textit{Staggered adoption (amplified dynamics)}} \\
\quad Earlier vs Later & 0.072 & 4.028 & 0.291 \\
\quad Later vs Earlier & 0.039 & 5.878 & 0.227 \\
\quad Treated vs Untreated & 0.222 & 9.151 & 2.034 \\
\quad \textit{TWFE estimate} & \multicolumn{3}{r}{7.445} \\
\quad \textit{true ATT} & \multicolumn{3}{r}{8.435} \\
\quad \textit{TWFE bias} & \multicolumn{3}{r}{-0.989} \\
\quad \textit{weight on already-treated comparisons} & \multicolumn{3}{r}{0.116} \\
\addlinespace
\addlinespace
\bottomrule
\end{tabular}
\begin{minipage}{\linewidth}\vspace{2pt}\footnotesize
\emph{Note.} ``Earlier vs.\ later'' and ``later vs.\ earlier'' are the two-by-two comparisons that use already-treated units as controls; their combined weight is reported in the last line of each block. Weights are normalised over all two-by-two comparisons including those against the never-treated group.
\end{minipage}
\end{table}

\begin{figure}[htbp]
\centering
\includegraphics[width=0.92\linewidth]{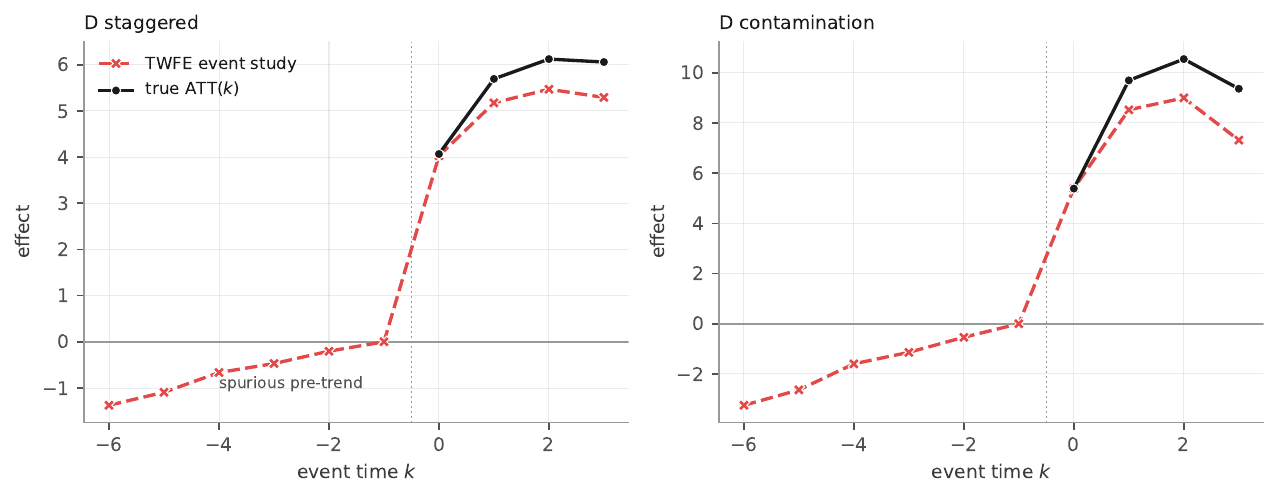}
\caption{Estimated and true event-time treatment effects in the staggered adoption design.}\label{fig:eventstudy}
\end{figure}
\begin{figure}[htbp]
\centering
\includegraphics[width=0.95\linewidth]{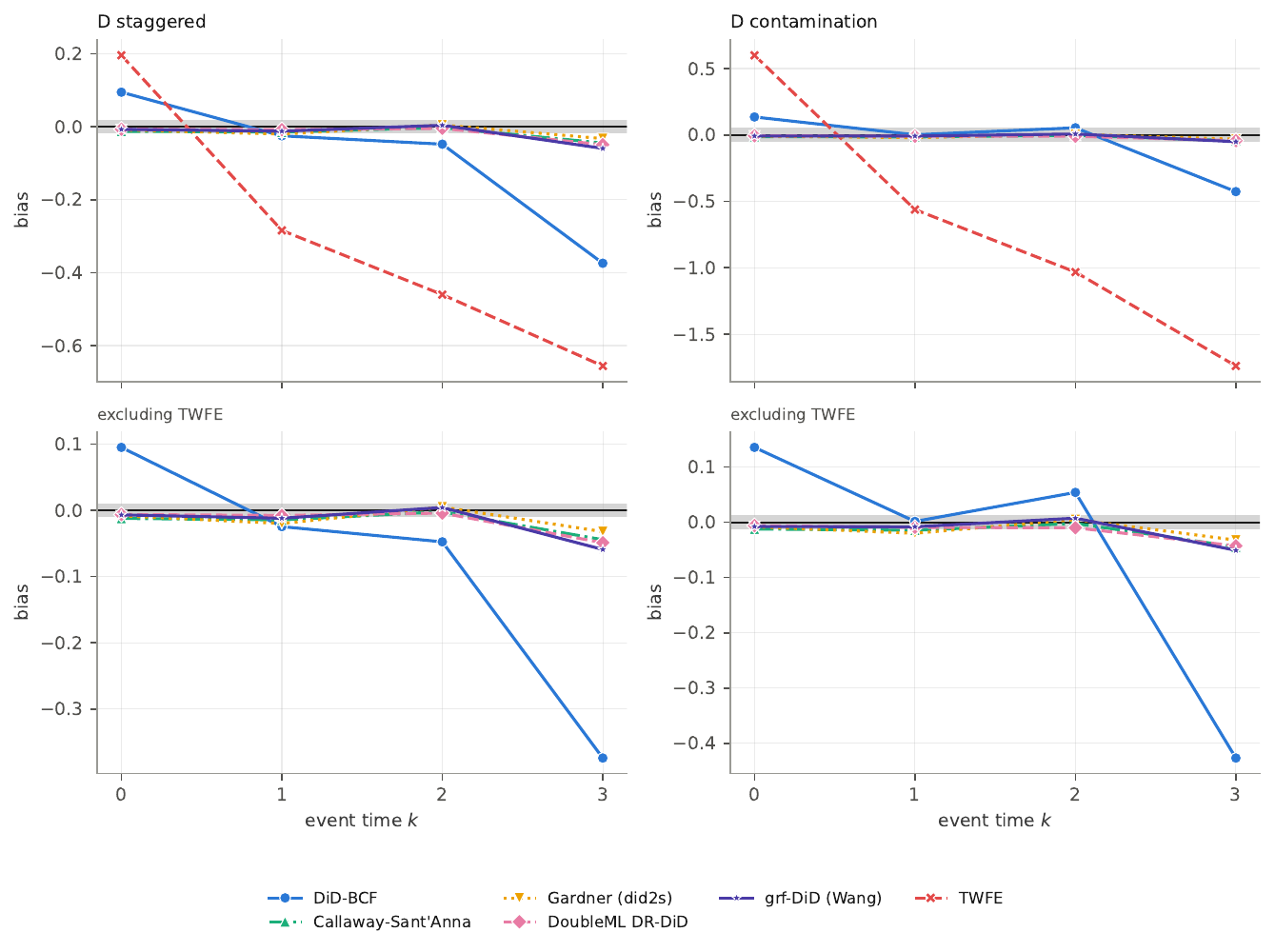}
\caption{Event-time estimation bias across methods in the staggered design.}\label{fig:esbias}
\end{figure}
\begin{figure}[htbp]
\centering
\includegraphics[width=0.93\linewidth]{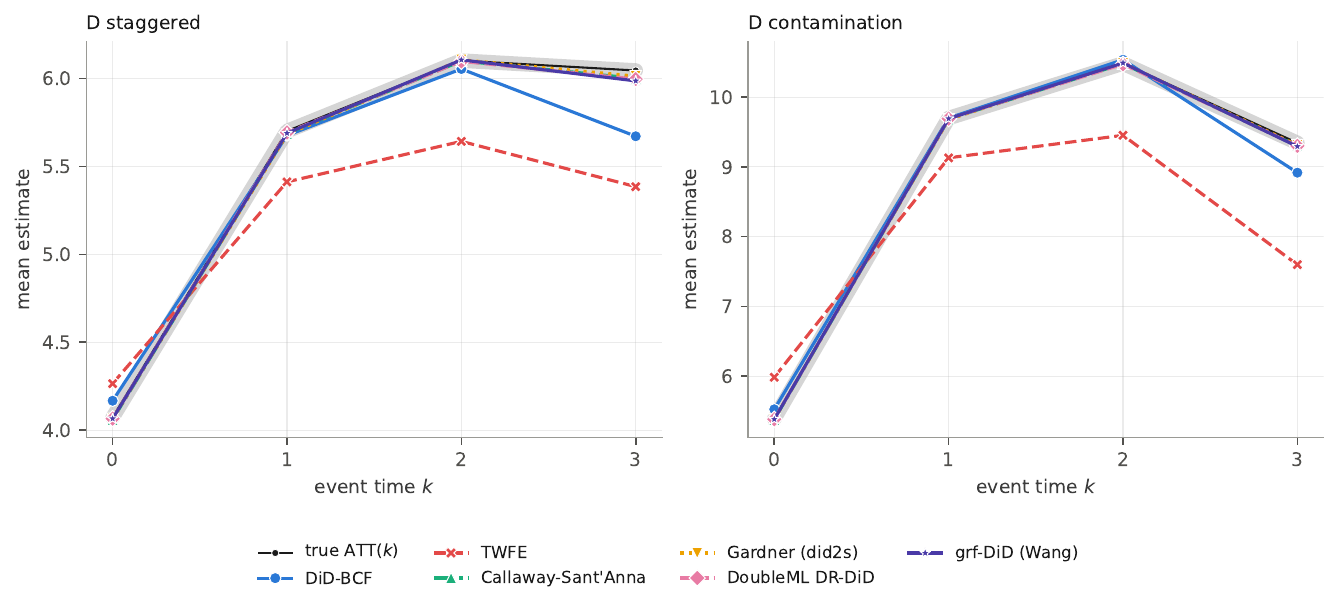}
\caption{Comparison of event-time trajectory estimates across estimators.}\label{fig:esmethods}
\end{figure}

\subsection{Small panels and conditional surfaces}\label{ssec:smallnres}
In small panels ($n=50$), DiD-BCF achieves a $\gatt(4,4)$ RMSE of $0.248$, substantially outperforming benchmark estimators ($0.409$ for grf-DiD, $0.455$ for Callaway--Sant'Anna, and $0.729$ for DoubleML). For pooled ATT, DiD-BCF attains an RMSE of $0.236$ compared to $0.363$ for synthetic DiD. In conditional surface recovery, DiD-BCF's surface RMSE contracts from $0.465$ at $n=50$ to $0.157$ at $n=800$, whereas grf-DiD improves from $1.155$ to $0.367$. Scalar cohort-time estimates broadcast to individuals remain bounded by unmodeled within-cell heterogeneity.
\begin{table}[htbp]
\centering
\scriptsize
\setlength{\tabcolsep}{2.6pt}
\caption{The small-panel end of the sample-size sweeps at linearity degree $d=1$. Bias, RMSE, 95\% coverage and the calibration ratio $\overline{\mathrm{sd}}/\mathrm{SD}$ for the cohort-time target $\mathrm{GATT}(4,4)$ and the pooled $\mathrm{ATT}$.}
\label{tab:smalln}
\begin{tabular}{llrrrrrrrrrrrr}
\toprule
  &   & \multicolumn{4}{c}{$n=50$} & \multicolumn{4}{c}{$n=100$} & \multicolumn{4}{c}{$n=200$} \\
\cmidrule(lr){3-6} \cmidrule(lr){7-10} \cmidrule(lr){11-14}
Target & Estimator & Bias & RMSE & Cov. & $\overline{\mathrm{sd}}/\mathrm{SD}$ & Bias & RMSE & Cov. & $\overline{\mathrm{sd}}/\mathrm{SD}$ & Bias & RMSE & Cov. & $\overline{\mathrm{sd}}/\mathrm{SD}$ \\
\midrule
\multicolumn{14}{l}{\textit{Independent errors}} \\
$\gatt(4,4)$ & DiD-BCF & -0.014 & 0.248 & 0.97 & 0.82 & 0.018 & 0.166 & 0.98 & 1.04 & -0.019 & 0.112 & 1.00 & 1.00 \\
  & TWFE & -0.046 & 0.418 & 0.95 & 0.88 & 0.003 & 0.265 & 0.98 & 1.07 & -0.048 & 0.179 & 0.99 & 1.07 \\
  & Callaway--Sant'Anna & 0.055 & 0.455 & 0.94 & 0.90 & 0.013 & 0.290 & 0.97 & 1.09 & -0.031 & 0.221 & 0.98 & 1.03 \\
  & Gardner (did2s) & -0.046 & 0.418 & 0.95 & 0.87 & 0.003 & 0.265 & 0.98 & 1.06 & -0.047 & 0.179 & 0.98 & 1.06 \\
  & DoubleML DR-DiD & 0.045 & 0.729 & 0.98 & 0.97 & -0.003 & 0.467 & 0.97 & 1.01 & -0.038 & 0.285 & 0.96 & 1.01 \\
  & Synthetic DiD & -0.027 & 0.383 & 0.98 & 0.93 & 0.007 & 0.260 & 0.97 & 1.09 & -0.052 & 0.181 & 0.97 & 1.05 \\
  & grf-DiD (Wang) & 0.009 & 0.409 & 0.94 & 0.76 & 0.006 & 0.298 & 0.89 & 0.81 & -0.036 & 0.222 & 0.81 & 0.73 \\
\addlinespace[2pt]
$\att$ & DiD-BCF & -0.007 & 0.236 & 0.94 & 0.72 & 0.015 & 0.155 & 0.94 & 0.82 & 0.004 & 0.099 & 0.95 & 0.84 \\
  & TWFE & -0.071 & 0.415 & 0.95 & 0.93 & -0.011 & 0.287 & 0.99 & 1.01 & -0.024 & 0.179 & 0.98 & 1.07 \\
  & Callaway--Sant'Anna & 0.049 & 0.358 & 0.95 & 0.89 & -0.000 & 0.226 & 0.97 & 1.10 & 0.000 & 0.171 & 0.97 & 1.10 \\
  & Gardner (did2s) & -0.071 & 0.415 & 0.95 & 0.92 & -0.011 & 0.287 & 0.99 & 1.01 & -0.022 & 0.185 & 0.99 & 1.03 \\
  & DoubleML DR-DiD & 0.014 & 0.540 & 0.96 & 1.04 & 0.005 & 0.288 & 0.99 & 1.18 & 0.003 & 0.201 & 0.99 & 1.12 \\
  & Synthetic DiD & -0.026 & 0.363 & 0.97 & 0.95 & 0.000 & 0.258 & 0.97 & 1.05 & -0.023 & 0.169 & 1.00 & 1.06 \\
  & grf-DiD (Wang) & -0.006 & 0.349 & 0.93 & 0.79 & -0.000 & 0.248 & 0.95 & 0.83 & -0.003 & 0.171 & 0.91 & 0.84 \\
\addlinespace[2pt]
\addlinespace
\multicolumn{14}{l}{\textit{AR(1) serial correlation}} \\
$\gatt(4,4)$ & DiD-BCF & -0.035 & 0.278 & 0.90 & 0.68 & 0.010 & 0.175 & 0.90 & 0.79 & -0.011 & 0.112 & 0.96 & 0.77 \\
  & TWFE & -0.062 & 0.384 & 0.96 & 0.90 & -0.005 & 0.240 & 0.98 & 1.04 & -0.034 & 0.163 & 0.99 & 1.01 \\
  & Callaway--Sant'Anna & 0.016 & 0.301 & 0.98 & 0.89 & 0.016 & 0.177 & 0.98 & 1.17 & -0.025 & 0.129 & 0.99 & 1.10 \\
  & Gardner (did2s) & -0.062 & 0.384 & 0.95 & 0.89 & -0.005 & 0.240 & 0.98 & 1.03 & -0.033 & 0.156 & 0.99 & 1.03 \\
  & DoubleML DR-DiD & 0.033 & 0.498 & 0.98$^{\dagger}$ & 0.99 & 0.033 & 0.270 & 0.99 & 1.15 & -0.018 & 0.168 & 0.97 & 1.08 \\
  & Synthetic DiD & -0.019 & 0.288 & 0.99 & 0.91 & 0.003 & 0.182 & 1.00 & 1.15 & -0.036 & 0.129 & 0.98 & 1.06 \\
  & grf-DiD (Wang) & -0.005 & 0.291 & 0.94 & 0.76 & 0.013 & 0.186 & 0.93 & 0.90 & -0.032 & 0.133 & 0.91 & 0.81 \\
\addlinespace[2pt]
$\att$ & DiD-BCF & -0.040 & 0.288 & 0.82 & 0.57 & 0.007 & 0.193 & 0.81 & 0.60 & 0.007 & 0.128 & 0.85 & 0.59 \\
  & TWFE & -0.092 & 0.446 & 0.98 & 0.95 & -0.017 & 0.310 & 0.97 & 0.99 & -0.022 & 0.198 & 0.98 & 1.04 \\
  & Callaway--Sant'Anna & 0.007 & 0.325 & 0.94 & 0.87 & 0.009 & 0.185 & 1.00 & 1.19 & -0.005 & 0.148 & 0.98 & 1.10 \\
  & Gardner (did2s) & -0.092 & 0.446 & 0.98 & 0.94 & -0.017 & 0.310 & 0.97 & 0.99 & -0.020 & 0.197 & 0.98 & 1.03 \\
  & DoubleML DR-DiD & -0.019 & 0.441 & 0.98 & 1.08 & 0.017 & 0.231 & 1.00 & 1.28 & -0.008 & 0.172 & 0.98 & 1.15 \\
  & Synthetic DiD & -0.040 & 0.357 & 0.97 & 0.95 & -0.003 & 0.250 & 0.96 & 1.07 & -0.024 & 0.166 & 0.99 & 1.08 \\
  & grf-DiD (Wang) & -0.036 & 0.320 & 0.93 & 0.79 & 0.005 & 0.221 & 0.93 & 0.86 & -0.009 & 0.150 & 0.92 & 0.84 \\
\addlinespace[2pt]
\addlinespace
\bottomrule
\end{tabular}
\par\smallskip\footnotesize \emph{Note.} 100 replications per cell. $^{\dagger}$computed on the replications in which the estimator returned an estimate. In these single-cohort designs ($G_i \in \{4,\infty\}$), the impact event-study coefficient ($k=0$) coincides with $\mathrm{GATT}(4,4)$ for TWFE and Gardner (did2s), while Synthetic DiD is evaluated on the subpanel $t \le 4$ for $\mathrm{GATT}(4,4)$. Pooled ATT for Callaway--Sant'Anna and DoubleML DR-DiD uses simple aggregation across post-treatment cells, Gardner (did2s) uses a static second stage, and grf-DiD aggregates doubly-robust cell scores with unit-clustered standard errors.
\end{table}

\begin{table}[htbp]
\centering
\scriptsize
\setlength{\tabcolsep}{4pt}
\caption{Individual-level CATT surface at the small-panel end of the sample-size sweeps, $d=1$. Within-replication RMSE and MAPE over the treated observations, and pointwise 95\% coverage, averaged across replications.}
\label{tab:smallnsurface}
\begin{tabular}{llrrrrrrrrr}
\toprule
  &   & \multicolumn{3}{c}{$n=50$} & \multicolumn{3}{c}{$n=100$} & \multicolumn{3}{c}{$n=200$} \\
\cmidrule(lr){3-5} \cmidrule(lr){6-8} \cmidrule(lr){9-11}
Design & Estimator & RMSE & MAPE & Cov. & RMSE & MAPE & Cov. & RMSE & MAPE & Cov. \\
\midrule
Independent & DiD-BCF & 0.465 & 0.117 & 0.92 & 0.345 & 0.086 & 0.95 & 0.249 & 0.060 & 0.97 \\
  & TWFE & 1.139 & 0.285 & 0.59 & 1.107 & 0.289 & 0.45 & 0.902 & 0.221 & 0.35 \\
  & Callaway--Sant'Anna & 1.126 & 0.291 & 0.57 & 1.094 & 0.288 & 0.41 & 1.069 & 0.285 & 0.29 \\
  & Gardner (did2s) & 1.139 & 0.285 & 0.58 & 1.107 & 0.289 & 0.45 & 1.071 & 0.284 & 0.32 \\
  & DoubleML DR-DiD & 1.305 & 0.324 & 0.75 & 1.137 & 0.295 & 0.56 & 1.082 & 0.288 & 0.38 \\
  & Synthetic DiD & 1.092 & 0.279 & 0.50 & 1.085 & 0.286 & 0.37 & 1.062 & 0.282 & 0.26 \\
  & grf-DiD (Wang) & 1.155 & 0.296 & 0.54 & 1.013 & 0.268 & 0.48 & 0.677 & 0.166 & 0.60 \\
\addlinespace
AR(1) serial & DiD-BCF & 0.504 & 0.126 & 0.85 & 0.387 & 0.097 & 0.88 & 0.299 & 0.073 & 0.89 \\
  & TWFE & 1.141 & 0.283 & 0.57 & 1.108 & 0.289 & 0.44 & 0.902 & 0.221 & 0.35 \\
  & Callaway--Sant'Anna & 1.102 & 0.284 & 0.49 & 1.079 & 0.286 & 0.35 & 1.063 & 0.284 & 0.25 \\
  & Gardner (did2s) & 1.141 & 0.283 & 0.57 & 1.108 & 0.289 & 0.44 & 1.071 & 0.284 & 0.32 \\
  & DoubleML DR-DiD & 1.197 & 0.301 & 0.70 & 1.120 & 0.295 & 0.49 & 1.073 & 0.287 & 0.32 \\
  & Synthetic DiD & 1.092 & 0.278 & 0.50 & 1.084 & 0.285 & 0.37 & 1.062 & 0.282 & 0.25 \\
  & grf-DiD (Wang) & 1.134 & 0.290 & 0.48 & 0.989 & 0.263 & 0.42 & 0.615 & 0.150 & 0.59 \\
\addlinespace
\bottomrule
\end{tabular}
\par\smallskip\footnotesize \emph{Note.} Every estimator other than DiD-BCF reports a single number per cohort-time cell, so its surface is that number broadcast to the individuals in the cell; the comparison measures what is lost by not modelling individual heterogeneity.
\end{table}

\begin{figure}[htbp]
\centering
\includegraphics[width=0.98\linewidth]{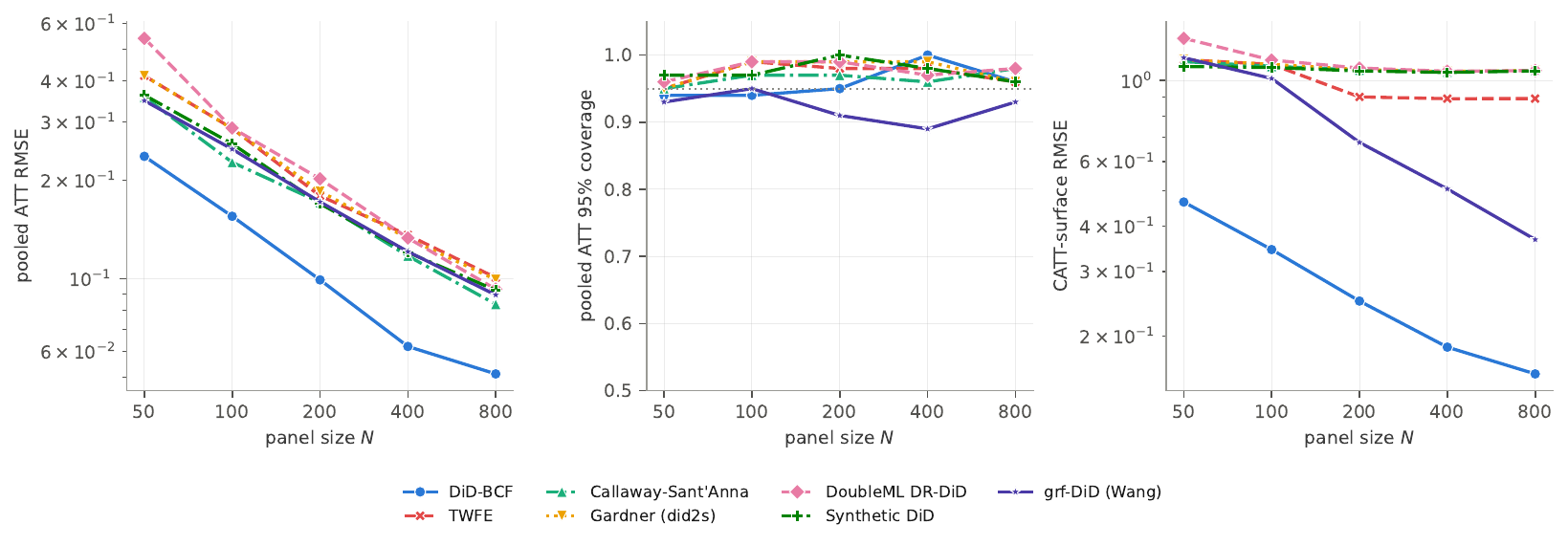}
\caption{Average-effect accuracy and coverage alongside conditional-surface accuracy in small panels.}\label{fig:smalln}
\end{figure}

\subsection{Pre-trend diagnostic}\label{app:pretrendres}
Pre-trend diagnostics use the auxiliary lead model and three rules specified in Appendix~\ref{supp:designs}. The diagnostic cutoff is $p_{\mathrm{tail}}<0.025$, equivalent to a two-sided posterior sign score below $0.05$, with Bonferroni correction for the any-$k$ rule. Tables~\ref{tab:pretrend} and \ref{tab:pretrendd3} report rejection percentages across all seven methods and all ten designs, together with available-replication counts where these differ from 200. These pre-trend experiments cover $d=1$ and $d=3$; no $d=2$ pre-trend simulations are available. Missing subgroup tests are labeled NA and are excluded from rejection-rate denominators.
\begin{landscape}
\begin{table}[p]
\centering\footnotesize
\caption{Pre-trend rejection percentages at $d=1$. Each cell reports slope / $X_1$ subgroup contrast / Bonferroni any-$k$ rejection. CS denotes Callaway--Sant'Anna; NA denotes an unavailable test. All available entries use 200 replications except the CS contrast under conditional PT (198). TWFE is evaluated on an independent set of 200 replications.}\label{tab:pretrend}
\resizebox{\linewidth}{!}{\setlength{\tabcolsep}{4pt}
\begin{tabular}{lrrrrrrr}
\toprule
Scenario & DiD-BCF & \shortstack{TWFE\\event study} & \shortstack{Synthetic\\DiD} & CS & did2s & DoubleML & grf-DiD \\
\midrule
Parallel trends hold & 4.0 / 0.5 / 2.5 & 4.5 / NA / 3.5 & 0.0 / 0.0 / 0.0 & 5.0 / 5.0 / 4.5 & 4.5 / 5.5 / 3.5 & 4.0 / 3.0 / 4.5 & 5.0 / 4.0 / 4.0 \\
Conditional PT only & 8.0 / 0.0 / 3.0 & 75.5 / NA / 71.0 & 5.5 / 0.0 / 3.5 & 7.0 / 8.1 / 7.5 & 75.0 / 8.5 / 71.0 & 5.5 / 2.0 / 6.5 & 10.0 / 3.0 / 9.5 \\
Group slope 0.05 & 20.5 / 0.5 / 9.0 & 10.0 / NA / 6.0 & 0.0 / 0.0 / 0.0 & 9.0 / 5.0 / 9.5 & 9.5 / 6.0 / 6.0 & 8.5 / 1.5 / 4.5 & 8.5 / 5.0 / 6.0 \\
Group slope 0.10 & 52.5 / 0.5 / 36.0 & 25.0 / NA / 20.0 & 0.5 / 0.0 / 0.0 & 32.0 / 5.0 / 27.5 & 24.0 / 5.5 / 19.0 & 18.0 / 2.0 / 15.0 & 28.5 / 4.0 / 23.0 \\
Group slope 0.20 & 93.0 / 0.5 / 86.5 & 70.0 / NA / 69.0 & 17.0 / 0.0 / 6.5 & 83.0 / 5.0 / 76.0 & 70.0 / 5.0 / 68.5 & 54.0 / 3.5 / 49.0 & 79.5 / 3.5 / 71.5 \\
Group slope 0.40 & 100.0 / 0.5 / 100.0 & 100.0 / NA / 100.0 & 85.0 / 0.0 / 76.5 & 100.0 / 5.0 / 100.0 & 100.0 / 5.0 / 100.0 & 98.5 / 2.5 / 99.0 & 100.0 / 4.0 / 100.0 \\
Unit-effect slope 0.20 & 91.5 / 0.5 / 84.5 & 70.5 / NA / 66.5 & 14.5 / 0.0 / 5.0 & 80.5 / 5.5 / 78.0 & 70.5 / 6.0 / 66.0 & 56.0 / 3.5 / 51.0 & 74.5 / 7.0 / 71.5 \\
Cancelling heterogeneity 0.10 & 4.0 / 3.0 / 2.0 & 4.5 / NA / 3.5 & 0.0 / 3.5 / 0.0 & 4.5 / 32.0 / 4.5 & 4.0 / 23.5 / 3.5 & 3.5 / 11.5 / 3.5 & 5.5 / 27.5 / 4.0 \\
Cancelling heterogeneity 0.20 & 5.0 / 52.5 / 4.0 & 5.5 / NA / 3.5 & 0.0 / 30.0 / 0.0 & 4.5 / 74.5 / 5.5 & 5.5 / 66.5 / 3.5 & 4.0 / 35.5 / 3.0 & 6.0 / 70.0 / 4.5 \\
Cancelling heterogeneity 0.40 & 11.0 / 100.0 / 7.0 & 6.5 / NA / 4.5 & 0.0 / 91.5 / 0.0 & 6.0 / 100.0 / 5.0 & 6.5 / 100.0 / 4.5 & 5.5 / 93.5 / 3.5 & 6.5 / 100.0 / 6.5 \\
\bottomrule
\end{tabular}
}
\end{table}
\clearpage
\begin{table}[p]
\centering\footnotesize
\caption{Pre-trend rejection percentages at $d=3$, in slope / $X_1$ subgroup contrast / Bonferroni any-$k$ order. CS denotes Callaway--Sant'Anna; NA denotes an unavailable test. Available counts are 200 except for CS under conditional PT (181 / 145 / 181), CS contrasts in all other designs (196), and TWFE under cancelling heterogeneity 0.20 (150 / NA / 150). Rates use available replications and are not adjusted for failed or missing fits.}\label{tab:pretrendd3}
\resizebox{\linewidth}{!}{\setlength{\tabcolsep}{4pt}
\begin{tabular}{lrrrrrrr}
\toprule
Scenario & DiD-BCF & \shortstack{TWFE\\event study} & \shortstack{Synthetic\\DiD} & CS & did2s & DoubleML & grf-DiD \\
\midrule
Parallel trends hold & 5.0 / 0.0 / 4.0 & 4.5 / NA / 3.5 & 0.0 / 0.0 / 0.0 & 5.5 / 7.7 / 4.5 & 4.5 / 5.5 / 3.5 & 2.0 / 2.0 / 2.5 & 5.0 / 6.0 / 4.0 \\
Conditional PT only & 8.5 / 0.0 / 4.5 & 75.5 / NA / 71.0 & 5.5 / 0.0 / 3.5 & 6.6 / 13.8 / 7.7 & 75.0 / 8.5 / 71.0 & 4.5 / 1.5 / 5.5 & 8.0 / 3.0 / 6.5 \\
Group slope 0.05 & 20.0 / 0.0 / 10.5 & 10.0 / NA / 6.0 & 0.0 / 0.0 / 0.0 & 10.5 / 7.7 / 8.0 & 9.5 / 6.0 / 6.0 & 6.5 / 3.0 / 4.0 & 11.5 / 6.0 / 8.0 \\
Group slope 0.10 & 48.5 / 0.0 / 34.5 & 25.0 / NA / 20.0 & 0.5 / 0.0 / 0.0 & 30.0 / 7.7 / 22.0 & 24.0 / 5.5 / 19.0 & 19.0 / 1.5 / 12.5 & 31.5 / 6.0 / 25.0 \\
Group slope 0.20 & 90.0 / 0.0 / 83.0 & 70.0 / NA / 69.0 & 17.0 / 0.0 / 6.5 & 76.5 / 7.7 / 72.0 & 70.0 / 5.0 / 68.5 & 53.0 / 0.5 / 44.5 & 75.0 / 6.5 / 70.5 \\
Group slope 0.40 & 100.0 / 0.0 / 100.0 & 100.0 / NA / 100.0 & 85.0 / 0.0 / 76.5 & 100.0 / 7.7 / 100.0 & 100.0 / 5.0 / 100.0 & 98.0 / 3.5 / 97.5 & 100.0 / 6.0 / 100.0 \\
Unit-effect slope 0.20 & 89.5 / 0.5 / 81.5 & 70.5 / NA / 66.5 & 14.5 / 0.0 / 5.0 & 76.0 / 8.7 / 70.5 & 70.5 / 6.0 / 66.0 & 50.5 / 4.0 / 44.0 & 74.5 / 7.5 / 69.0 \\
Cancelling heterogeneity 0.10 & 5.0 / 3.5 / 3.5 & 4.5 / NA / 3.5 & 0.0 / 3.5 / 0.0 & 5.0 / 32.7 / 5.0 & 4.0 / 23.5 / 3.5 & 3.5 / 11.0 / 3.5 & 5.0 / 29.5 / 4.0 \\
Cancelling heterogeneity 0.20 & 6.5 / 42.0 / 3.5 & 7.3 / NA / 4.0 & 0.0 / 30.0 / 0.0 & 4.0 / 69.9 / 4.5 & 5.5 / 66.5 / 3.5 & 2.5 / 33.0 / 3.5 & 5.5 / 70.5 / 5.0 \\
Cancelling heterogeneity 0.40 & 10.5 / 100.0 / 7.5 & 6.5 / NA / 4.5 & 0.0 / 91.5 / 0.0 & 4.5 / 100.0 / 5.5 & 6.5 / 100.0 / 4.5 & 5.0 / 87.5 / 4.5 & 6.5 / 100.0 / 7.0 \\
\bottomrule
\end{tabular}
}
\end{table}
\clearpage
\end{landscape}
\begin{figure}[htbp]
\centering
\includegraphics[width=0.98\linewidth]{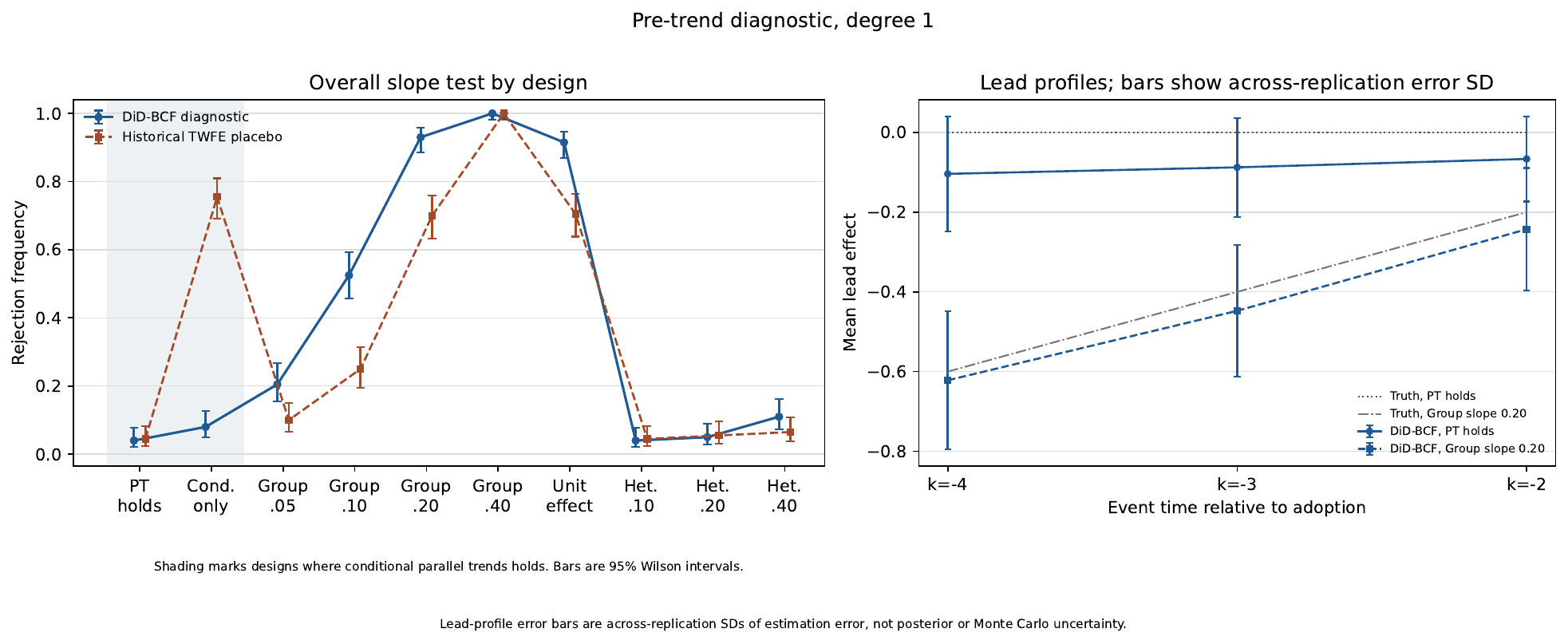}
\caption{Pre-trend diagnostic and TWFE rejection frequencies with 95\% Wilson Monte Carlo intervals (left). Estimated and true pre-treatment lead profiles across designs (right); error bars indicate across-replication standard deviations.}\label{fig:pretrend}
\end{figure}
At $d=1$, DiD-BCF slope rejection is $4.0\%$ under unconditional parallel trends and $8.0\%$ under conditional parallel trends only, compared with $4.5\%$ and $75.5\%$ for marginal TWFE and $4.5\%$ and $75.0\%$ for did2s. Under conditional PT, Callaway--Sant'Anna, DoubleML, Synthetic DiD, and grf-DiD reject in $7.0\%$, $5.5\%$, $5.5\%$, and $10.0\%$ of available runs. At $d=3$, DiD-BCF rejects in $5.0\%$ under unconditional PT and $8.5\%$ under conditional PT. Thus covariate adjustment greatly reduces false rejections, but does not guarantee exact nominal size in these finite samples.

Across linear differential slopes $0.05$, $0.10$, $0.20$, and $0.40$ at $d=1$, DiD-BCF power is $20.5\%$, $52.5\%$, $93.0\%$, and $100.0\%$, compared with $9.0\%$, $32.0\%$, $83.0\%$, and $100.0\%$ for Callaway--Sant'Anna and $8.5\%$, $28.5\%$, $79.5\%$, and $100.0\%$ for grf-DiD. Synthetic DiD is conservative, with zero rejection under unconditional PT and only $0.5\%$ power at slope $0.10$. The Bonferroni any-$k$ rule is less powerful than the slope rule against these linear alternatives, giving DiD-BCF rejection rates of $9.0\%$, $36.0\%$, $86.5\%$, and $100.0\%$. All power comparisons use nominal rather than size-adjusted thresholds.

For heterogeneous departures of $\pm0.20$ conditional on $X_1$ that cancel in the population aggregate, DiD-BCF's contrast rule rejects in $52.5\%$ of runs at $d=1$, exceeding DoubleML ($35.5\%$) but remaining below Callaway--Sant'Anna ($74.5\%$), grf-DiD ($70.0\%$), and did2s ($66.5\%$). DiD-BCF's null rejection rate is $0.5\%$, compared with $3.0\%$ for DoubleML, and its power at $\pm0.10$ is $3.0\%$, compared with $11.5\%$ for DoubleML. At $\pm0.40$, their power is $100.0\%$ and $93.5\%$, respectively. At $d=3$, DoubleML's null rejection rate is $2.0\%$, and its power at $\pm0.10$, $\pm0.20$, and $\pm0.40$ is $11.0\%$, $33.0\%$, and $87.5\%$; DiD-BCF reaches $42.0\%$ and $100.0\%$ at the latter two magnitudes. These results document a conservative DiD-BCF subgroup diagnostic, with rankings that depend on violation magnitude. Aggregate slope tests have little power against these cancelling alternatives. A small differential slope of $0.05$ induces Pooled ATT bias of $0.197$ and 95\% coverage of $57.0\%$ at $d=1$, despite detection in only $20.5\%$ of runs, motivating sensitivity analysis even after pre-test non-rejection.
\begin{figure}[htbp]
\centering
\includegraphics[width=0.98\linewidth]{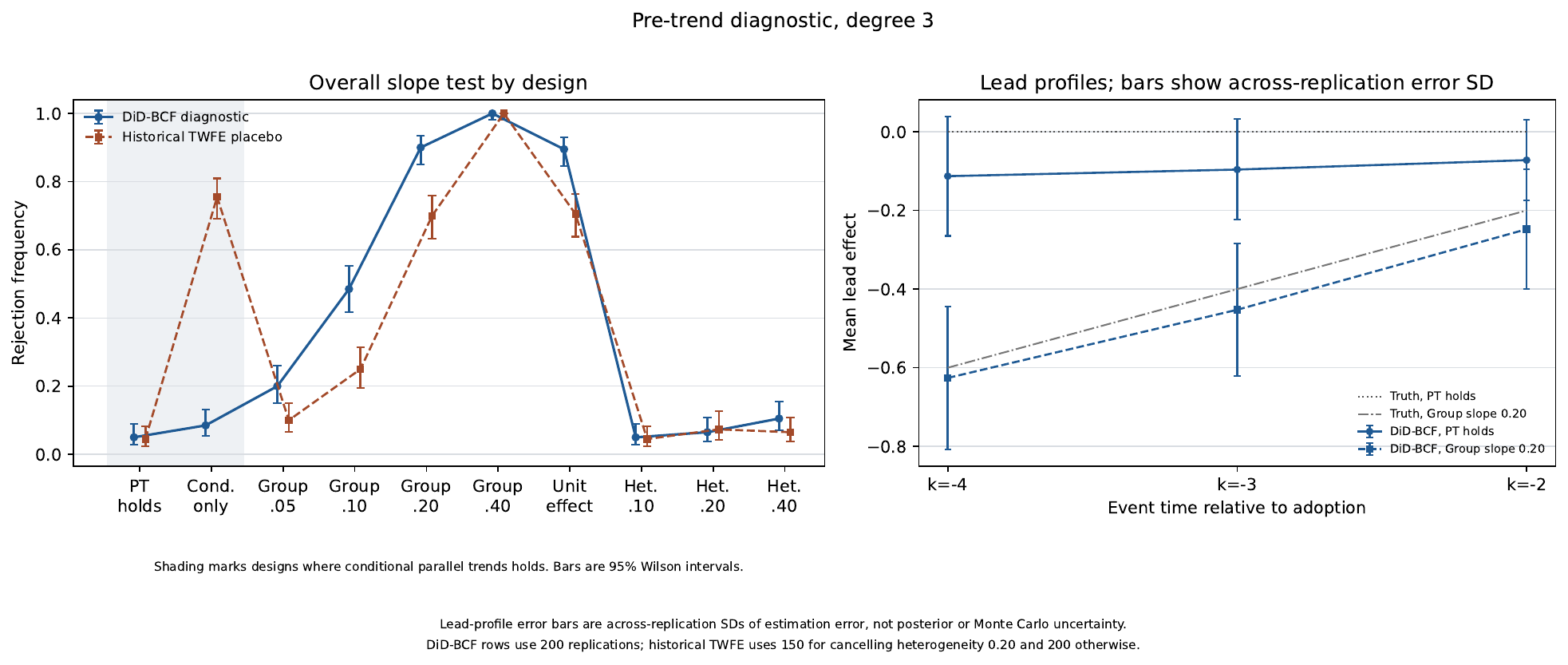}
\caption{Pre-trend diagnostic at $d=3$, with the same Monte Carlo uncertainty and lead-profile conventions as Figure~\ref{fig:pretrend}.}\label{fig:pretrendd3}
\end{figure}

\subsection{Working-error prior sensitivity}\label{ssec:prior_sensitivity}
\begin{figure}[htbp]
\centering
\includegraphics[width=0.93\linewidth]{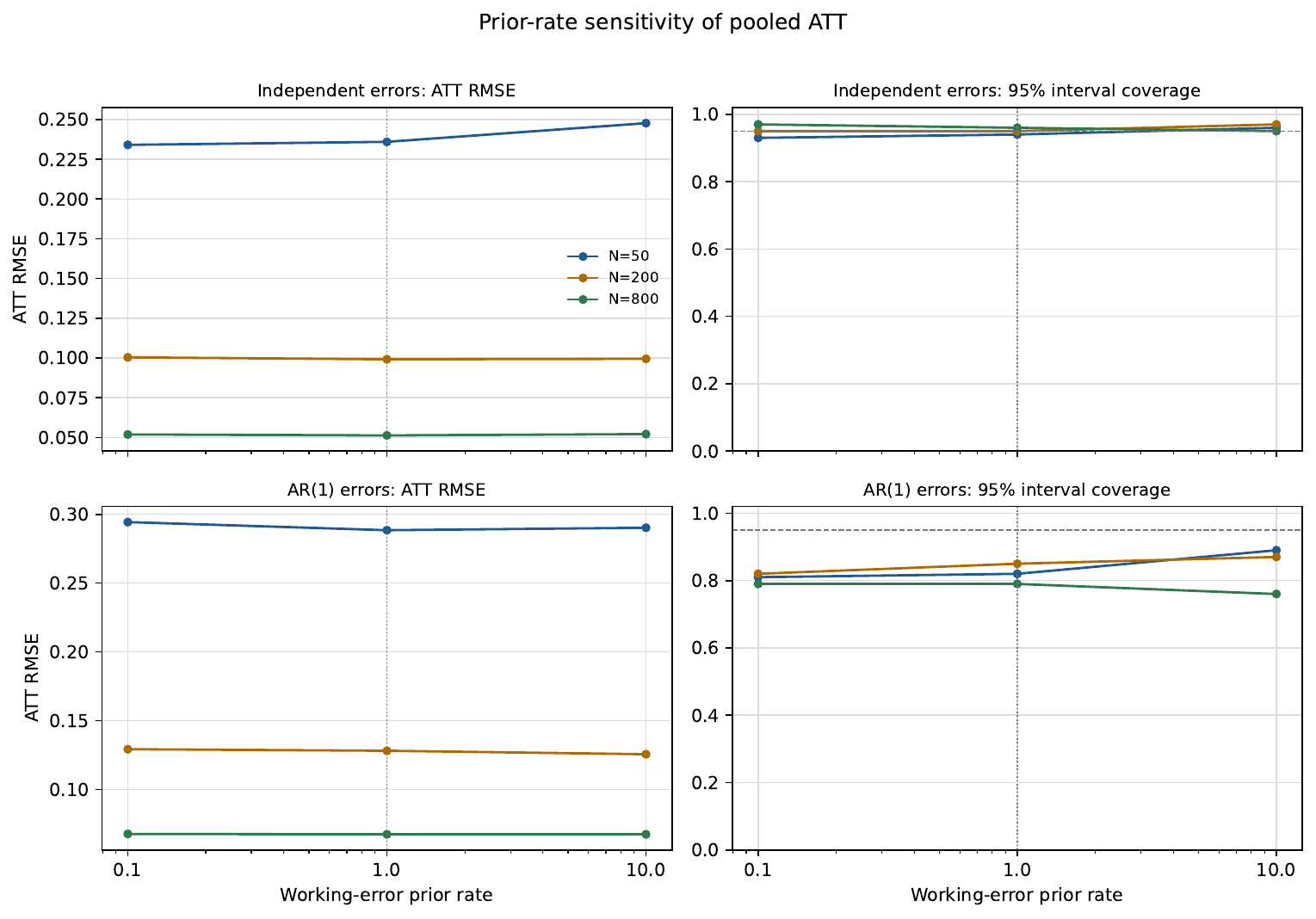}
\caption{Pooled-ATT RMSE and 95\% interval coverage across three working-error prior rates $\beta_\sigma \in \{0.1, 1, 10\}$, with $R = 100$ replications per cell. The horizontal reference line marks nominal 0.95 coverage.}\label{fig:prior_sensitivity}
\end{figure}

To assess sensitivity to the observation error variance prior, we vary the inverse-gamma rate parameter $\beta_\sigma \in \{0.1, 1, 10\}$ while holding the shape parameter fixed at $\alpha_\sigma = 2$. All tree priors and data-generating processes remain identical. Table~\ref{tab:prior_sensitivity} and Figure~\ref{fig:prior_sensitivity} present point accuracy and interval coverage across panel sizes $n \in \{50, 200, 800\}$ under independent and serial errors. While larger prior rate parameters widen credible intervals in small panels, the primary specification ($\beta_\sigma = 1$) provides balanced performance across sample sizes.
\begin{table}[htbp]
\centering
\scriptsize
\caption{Prior-rate sensitivity of pooled ATT estimates. Each cell uses 100 replications; the baseline rate-1 results match those from the canonical and sweep designs.}
\label{tab:prior_sensitivity}
\begin{tabular}{lrrr rrrr}
\toprule
Error process & $n$ & Prior rate & Bias & RMSE & MAE & 95\% coverage & Mean 95\% length \\
\midrule
AR(1) & 50 & 0.1 & -0.043 & 0.294 & 0.234 & 0.810 & 0.779 \\
AR(1) & 50 & 1.0 & -0.040 & 0.288 & 0.227 & 0.820 & 0.795 \\
\addlinespace[1pt]
AR(1) & 50 & 10.0 & -0.036 & 0.290 & 0.234 & 0.890 & 0.992 \\
AR(1) & 200 & 0.1 & 0.007 & 0.129 & 0.101 & 0.820 & 0.358 \\
AR(1) & 200 & 1.0 & 0.007 & 0.128 & 0.100 & 0.850 & 0.363 \\
\addlinespace[1pt]
AR(1) & 200 & 10.0 & 0.003 & 0.126 & 0.097 & 0.870 & 0.388 \\
AR(1) & 800 & 0.1 & -0.002 & 0.068 & 0.054 & 0.790 & 0.176 \\
AR(1) & 800 & 1.0 & -0.003 & 0.067 & 0.054 & 0.790 & 0.175 \\
\addlinespace[1pt]
AR(1) & 800 & 10.0 & -0.003 & 0.068 & 0.054 & 0.760 & 0.179 \\
Independent & 50 & 0.1 & -0.010 & 0.234 & 0.183 & 0.930 & 0.897 \\
Independent & 50 & 1.0 & -0.007 & 0.236 & 0.185 & 0.940 & 0.921 \\
\addlinespace[1pt]
Independent & 50 & 10.0 & 0.002 & 0.248 & 0.195 & 0.960 & 1.081 \\
Independent & 200 & 0.1 & 0.002 & 0.100 & 0.078 & 0.950 & 0.423 \\
Independent & 200 & 1.0 & 0.004 & 0.099 & 0.077 & 0.950 & 0.425 \\
\addlinespace[1pt]
Independent & 200 & 10.0 & 0.003 & 0.099 & 0.077 & 0.970 & 0.451 \\
Independent & 800 & 0.1 & -0.001 & 0.052 & 0.041 & 0.970 & 0.208 \\
Independent & 800 & 1.0 & -0.002 & 0.051 & 0.041 & 0.960 & 0.210 \\
\addlinespace[1pt]
Independent & 800 & 10.0 & -0.002 & 0.052 & 0.041 & 0.950 & 0.212 \\
\bottomrule
\end{tabular}
\end{table}

\subsection{Empirical application estimates, prior sensitivity, and pre-trend diagnostic}\label{ssec:application_sensitivity}
Table~\ref{tab:application_estimates} reports DiD-BCF treatment effect estimates for teen employment across the $N_{\mathrm{tr}} = 291$ treated county-year observations (191 unique treated counties) in the MPDTA dataset, contrasting the primary model equipped with county random intercepts against a specification without random intercepts. In the primary specification, DiD-BCF estimates an overall disemployment effect of $-4.72\%$ with an equal-tailed 95\% credible interval of $[-7.33\%, -2.03\%]$ and a two-sided Bayesian posterior sign score of $p_{\mathrm{Bayes}} = 0.001$. Omitting county random intercepts produces a nearly identical point estimate of $-4.66\%$ ($p_{\mathrm{Bayes}} = 0.009$), while widening the 95\% credible interval to $[-7.98\%, -1.38\%]$. Incorporating unit random intercepts absorbs persistent county-level unobservables and tightens posterior inference without shifting the location of the point estimate. Across baseline population terciles, point estimates indicate stronger disemployment in medium-population counties ($-5.75\%$, 95\% CI $[-9.07\%, -2.50\%]$, $p_{\mathrm{Bayes}} < 0.001$) than in small-population counties ($-4.42\%$, 95\% CI $[-8.42\%, -0.71\%]$, $p_{\mathrm{Bayes}} = 0.020$) or high-population counties ($-3.68\%$, 95\% CI $[-7.19\%, 0.58\%]$, $p_{\mathrm{Bayes}} = 0.071$). Although county random intercepts narrow credible intervals within each stratum relative to the unclustered model (where tercile estimates are $-4.30\%$, $-5.85\%$, and $-3.47\%$), pairwise posterior credible intervals for between-tercile contrasts consistently overlap zero, confirming that while aggregate disemployment is robustly detectable, population-size heterogeneity remains within sampling uncertainty.

\begin{table}[htbp]
\centering\spacingset{1}\small
\caption{DiD-BCF treatment effect estimates for treated county-year observations in the MPDTA dataset. The model with county random intercepts is the primary specification; the model without random intercepts serves as a robustness comparison. Point estimates (posterior means) and equal-tailed 95\% credible intervals are reported on the percentage scale ($100[\exp(\tau) - 1]$).}
\label{tab:application_estimates}
\resizebox{\linewidth}{!}{%
\begin{tabular}{llrrrr}
\toprule
Specification & Group & $N_{\mathrm{tr}}$ (county-years) & Effect & 95\% CI & $p_{\mathrm{Bayes}}$ \\
\midrule
County random intercept & All treated & 291 & $-4.72\%$ & $[-7.33\%,\ -2.03\%]$ & 0.001 \\
County random intercept & Small population & 72 & $-4.42\%$ & $[-8.42\%,\ -0.71\%]$ & 0.020 \\
County random intercept & Medium population & 119 & $-5.75\%$ & $[-9.07\%,\ -2.50\%]$ & \ensuremath{<0.001} \\
County random intercept & High population & 100 & $-3.68\%$ & $[-7.19\%,\ 0.58\%]$ & 0.071 \\
\addlinespace
No random intercept & All treated & 291 & $-4.66\%$ & $[-7.98\%,\ -1.38\%]$ & 0.009 \\
No random intercept & Small population & 72 & $-4.30\%$ & $[-9.01\%,\ 0.54\%]$ & 0.075 \\
No random intercept & Medium population & 119 & $-5.85\%$ & $[-9.69\%,\ -1.90\%]$ & 0.004 \\
No random intercept & High population & 100 & $-3.47\%$ & $[-7.81\%,\ 1.49\%]$ & 0.151 \\
\bottomrule
\end{tabular}}
\end{table}

Table~\ref{tab:application_prior_sensitivity} examines the sensitivity of all-treated ATT estimates to the observation error variance hyperprior rate $\beta_\sigma \in \{0.1, 1.0, 10.0\}$ (holding shape fixed at $\alpha_\sigma = 2$) across three independent random-seed initializations (Runs 1, 2, and 3) on the same panel. Across all nine configurations incorporating county random intercepts, point estimates remain tightly clustered between $-4.65\%$ and $-4.89\%$, and every 95\% credible interval strictly excludes zero ($p_{\mathrm{Bayes}} \le 0.023$). Even under the most diffuse prior rate ($\beta_\sigma = 10.0$), which places greater prior mass on larger observation noise, the county-random-intercept specification produces stable ATT estimates of $-4.66\%$, $-4.65\%$, and $-4.66\%$ across Runs 1, 2, and 3, with 95\% intervals $[-8.52\%, -0.81\%]$, $[-8.32\%, -0.63\%]$, and $[-8.59\%, -0.92\%]$. By contrast, omitting unit random intercepts at $\beta_\sigma = 10.0$ inflates posterior variance enough that the 95\% intervals cross zero ($[-9.48\%, 1.25\%]$ in Run 1, with $p_{\mathrm{Bayes}} \in [0.084, 0.100]$), illustrating how county random intercepts stabilize inference against prior diffuseness.

\begin{table}[htbp]
\centering\scriptsize
\caption{All-treated MPDTA treatment effect estimates across the residual error variance prior-rate sensitivity grid. Run 1 corresponds to the primary random-seed initialization, while Runs 2 and 3 represent independent random-seed initializations on the same panel. Posterior credible intervals are reported separately per run without pooling.}
\label{tab:application_prior_sensitivity}
\begin{tabular}{rllrrr}
\toprule
Prior rate & Specification & Run & ATT (\%) & 95\% interval (\%) & $p_{\mathrm{Bayes}}$ \\
\midrule
0.1 & No random intercept & 1 & $-4.66\%$ & $[-7.98\%,\ -1.25\%]$ & 0.012 \\
0.1 & No random intercept & 2 & $-4.72\%$ & $[-7.68\%,\ -1.42\%]$ & 0.007 \\
0.1 & No random intercept & 3 & $-4.53\%$ & $[-7.56\%,\ -1.35\%]$ & 0.004 \\
0.1 & County random intercept & 1 & $-4.89\%$ & $[-7.49\%,\ -2.33\%]$ & \ensuremath{<0.001} \\
0.1 & County random intercept & 2 & $-4.83\%$ & $[-7.41\%,\ -2.26\%]$ & \ensuremath{<0.001} \\
0.1 & County random intercept & 3 & $-4.85\%$ & $[-7.40\%,\ -2.11\%]$ & \ensuremath{<0.001} \\
1.0 & No random intercept & 1 & $-4.66\%$ & $[-7.98\%,\ -1.38\%]$ & 0.009 \\
1.0 & No random intercept & 2 & $-4.77\%$ & $[-8.31\%,\ -1.02\%]$ & 0.013 \\
1.0 & No random intercept & 3 & $-4.64\%$ & $[-8.32\%,\ -1.03\%]$ & 0.012 \\
1.0 & County random intercept & 1 & $-4.72\%$ & $[-7.33\%,\ -2.03\%]$ & 0.001 \\
1.0 & County random intercept & 2 & $-4.79\%$ & $[-7.54\%,\ -2.07\%]$ & 0.003 \\
1.0 & County random intercept & 3 & $-4.76\%$ & $[-7.48\%,\ -2.00\%]$ & \ensuremath{<0.001} \\
10.0 & No random intercept & 1 & $-4.46\%$ & $[-9.48\%,\ 1.25\%]$ & 0.100 \\
10.0 & No random intercept & 2 & $-4.57\%$ & $[-9.65\%,\ 0.92\%]$ & 0.091 \\
10.0 & No random intercept & 3 & $-4.66\%$ & $[-9.62\%,\ 0.62\%]$ & 0.084 \\
10.0 & County random intercept & 1 & $-4.66\%$ & $[-8.52\%,\ -0.81\%]$ & 0.023 \\
10.0 & County random intercept & 2 & $-4.65\%$ & $[-8.32\%,\ -0.63\%]$ & 0.023 \\
10.0 & County random intercept & 3 & $-4.66\%$ & $[-8.59\%,\ -0.92\%]$ & 0.012 \\
\bottomrule
\end{tabular}
\end{table}

Table~\ref{tab:application_pretrend} reports empirical pre-trend diagnostics across prior rates and independent random-seed initializations, evaluating both aggregate linear slopes and localized event-time lead departures via one-sided posterior tail probabilities $p_{\mathrm{tail}}$ against the prespecified $0.025$ threshold (equivalent to the two-sided criterion $p_{\mathrm{Bayes}} = 2\,p_{\mathrm{tail}} < 0.05$). In the primary specification ($\beta_\sigma = 1.0$, Run 1), the estimated pre-treatment linear slope is $-0.0057$ with a 95\% credible interval of $[-0.0202, 0.0071]$ and $p_{\mathrm{tail}} = 0.206$, indicating no secular pre-treatment divergence. Across all three initializations at $\beta_\sigma = 1.0$, estimated slopes remain close to zero ($-0.0057$ to $-0.0062$, with $p_{\mathrm{tail}} \in [0.161, 0.206]$). In the lead-specific diagnostic, unadjusted minimum lead tail probabilities dip below $0.025$ only under the tight error-variance prior $\beta_\sigma = 0.1$ ($p_{\mathrm{tail}} = 0.021$ in Run 1), whereas Bonferroni-adjusted lead tail probabilities across the $L = 3$ pre-treatment leads range from $0.064$ to $0.298$ across all nine runs ($0.098$ to $0.146$ at $\beta_\sigma = 1.0$), never rejecting conditional parallel trends at the $0.025$ cutoff.

\begin{table}[htbp]
\centering\scriptsize
\caption{Empirical MPDTA pre-trend slope and lead-specific diagnostic summaries across error variance prior rates and independent random-seed initializations. Run 1 corresponds to the primary random-seed initialization, while Runs 2 and 3 represent independent random-seed initializations on the same panel. Slope intervals are central 95\% posterior credible intervals. Columns 5--7 report one-sided posterior tail probabilities $p_{\mathrm{tail}}$ for the linear slope, the minimum unadjusted lead tail probability ($\min_{k \le -2} p_{\mathrm{tail},k}$), and its Bonferroni-adjusted counterpart ($\min\{1, L \min_{k \le -2} p_{\mathrm{tail},k}\}$ across $L = 3$ pre-treatment leads), each evaluated against the prespecified $0.025$ cutoff (equivalent to two-sided $p_{\mathrm{Bayes}} = 2\,p_{\mathrm{tail}} < 0.05$).}
\label{tab:application_pretrend}
\begin{tabular}{rrrrrrr}
\toprule
Prior rate & Run & Slope & 95\% interval & $p_{\mathrm{tail}}$ (slope) & $p_{\mathrm{tail}}$ (min lead) & $p_{\mathrm{tail}}$ (Bonf.\ lead) \\
\midrule
0.1 & 1 & $-0.0074$ & $[-0.0206,\ 0.0056]$ & 0.157 & 0.021 & 0.064 \\
0.1 & 2 & $-0.0067$ & $[-0.0207,\ 0.0060]$ & 0.171 & 0.036 & 0.108 \\
0.1 & 3 & $-0.0063$ & $[-0.0195,\ 0.0056]$ & 0.209 & 0.026 & 0.078 \\
1.0 & 1 & $-0.0057$ & $[-0.0202,\ 0.0071]$ & 0.206 & 0.046 & 0.138 \\
1.0 & 2 & $-0.0062$ & $[-0.0207,\ 0.0057]$ & 0.161 & 0.049 & 0.146 \\
1.0 & 3 & $-0.0062$ & $[-0.0194,\ 0.0061]$ & 0.181 & 0.033 & 0.098 \\
10.0 & 1 & $-0.0038$ & $[-0.0215,\ 0.0116]$ & 0.294 & 0.093 & 0.278 \\
10.0 & 2 & $-0.0030$ & $[-0.0204,\ 0.0114]$ & 0.311 & 0.090 & 0.270 \\
10.0 & 3 & $-0.0030$ & $[-0.0204,\ 0.0118]$ & 0.334 & 0.099 & 0.298 \\
\bottomrule
\end{tabular}
\end{table}

\subsection{Conditional surfaces across designs}\label{ssec:surfres}
DiD-BCF consistently achieves superior precision in estimating conditional response surfaces relative to scalar broadcast estimators across all canonical designs, with grf-DiD serving as the primary individual-level benchmark. At degree $d=3$, non-linear interactions increase the estimation difficulty for all methods, but DiD-BCF retains substantial efficiency gains.
{\footnotesize
\begin{longtable}{lrrrrrr}
\caption{CATT-surface accuracy: error of the estimated individual treatment effect over the treated observations, averaged across replications (Monte-Carlo SD of the within-replication RMSE in parentheses).}\label{tab:surface}\\
\toprule
Estimator & $d$ & RMSE (SD) & MAE & MAPE & Cov.95 & Len.95 \\
\midrule
\endfirsthead
\multicolumn{7}{l}{\emph{\footnotesize Table \ref{tab:surface} continued from the previous page.}}\\
\toprule
Estimator & $d$ & RMSE (SD) & MAE & MAPE & Cov.95 & Len.95 \\
\midrule
\endhead
\midrule
\multicolumn{7}{r}{\emph{\footnotesize continued on the next page}}\\
\endfoot
\bottomrule
\endlastfoot
\multicolumn{7}{l}{\textit{Canonical: baseline}} \\
\quad DiD-BCF & 1 & 0.249 (0.058) & 0.191 & 0.060 & 0.968 & 1.027 \\
\quad TWFE & 1 & 0.902 (0.041) & 0.762 & 0.221 & 0.352 & 0.974 \\
\quad Callaway--Sant'Anna & 1 & 1.069 (0.065) & 0.878 & 0.285 & 0.292 & 0.898 \\
\quad Gardner (did2s) & 1 & 1.071 (0.065) & 0.882 & 0.284 & 0.320 & 0.996 \\
\quad DoubleML DR-DiD & 1 & 0.914 (0.049) & 0.770 & 0.225 & 0.404 & 1.153 \\
\quad Synthetic DiD & 1 & 1.062 (0.063) & 0.874 & 0.282 & 0.259 & 0.797 \\
\quad grf-DiD (Wang) & 1 & 0.677 (0.093) & 0.572 & 0.166 & 0.600 & 1.325 \\
\quad DiD-BCF & 2 & 0.270 (0.066) & 0.210 & 0.066 & 0.960 & 1.053 \\
\quad TWFE & 2 & 0.902 (0.041) & 0.762 & 0.221 & 0.352 & 0.974 \\
\quad Callaway--Sant'Anna & 2 & 1.071 (0.066) & 0.880 & 0.285 & 0.302 & 0.934 \\
\quad Gardner (did2s) & 2 & 1.071 (0.065) & 0.882 & 0.284 & 0.320 & 0.996 \\
\quad DoubleML DR-DiD & 2 & 0.916 (0.052) & 0.772 & 0.226 & 0.399 & 1.138 \\
\quad Synthetic DiD & 2 & 1.062 (0.063) & 0.874 & 0.282 & 0.259 & 0.797 \\
\quad grf-DiD (Wang) & 2 & 0.677 (0.093) & 0.572 & 0.166 & 0.600 & 1.325 \\
\quad DiD-BCF & 3 & 0.471 (0.085) & 0.325 & 0.086 & 0.857 & 1.100 \\
\quad TWFE & 3 & 0.902 (0.041) & 0.762 & 0.221 & 0.352 & 0.974 \\
\quad Callaway--Sant'Anna & 3 & 1.082 (0.104) & 0.869 & 0.250 & 0.298 & 0.931 \\
\quad Gardner (did2s) & 3 & 1.082 (0.102) & 0.871 & 0.250 & 0.312 & 0.997 \\
\quad DoubleML DR-DiD & 3 & 0.915 (0.051) & 0.771 & 0.225 & 0.399 & 1.136 \\
\quad Synthetic DiD & 3 & 1.073 (0.101) & 0.865 & 0.248 & 0.250 & 0.797 \\
\quad grf-DiD (Wang) & 3 & 0.677 (0.093) & 0.572 & 0.166 & 0.600 & 1.325 \\
\addlinespace
\multicolumn{7}{l}{\textit{Canonical: strong confounder}} \\
\quad DiD-BCF & 1 & 0.244 (0.057) & 0.190 & 0.059 & 0.974 & 1.067 \\
\quad TWFE & 1 & 0.900 (0.044) & 0.760 & 0.220 & 0.352 & 0.974 \\
\quad Callaway--Sant'Anna & 1 & 1.064 (0.063) & 0.875 & 0.284 & 0.294 & 0.899 \\
\quad Gardner (did2s) & 1 & 1.070 (0.066) & 0.880 & 0.283 & 0.320 & 0.995 \\
\quad DoubleML DR-DiD & 1 & 0.915 (0.048) & 0.770 & 0.224 & 0.406 & 1.149 \\
\quad Synthetic DiD & 1 & 1.060 (0.063) & 0.873 & 0.281 & 0.257 & 0.795 \\
\quad grf-DiD (Wang) & 1 & 0.671 (0.090) & 0.567 & 0.165 & 0.608 & 1.324 \\
\quad DiD-BCF & 2 & 0.266 (0.065) & 0.211 & 0.065 & 0.965 & 1.091 \\
\quad TWFE & 2 & 0.900 (0.044) & 0.760 & 0.220 & 0.352 & 0.974 \\
\quad Callaway--Sant'Anna & 2 & 1.066 (0.065) & 0.876 & 0.284 & 0.304 & 0.932 \\
\quad Gardner (did2s) & 2 & 1.070 (0.066) & 0.880 & 0.283 & 0.320 & 0.995 \\
\quad DoubleML DR-DiD & 2 & 0.913 (0.051) & 0.769 & 0.225 & 0.404 & 1.146 \\
\quad Synthetic DiD & 2 & 1.060 (0.063) & 0.873 & 0.281 & 0.257 & 0.795 \\
\quad grf-DiD (Wang) & 2 & 0.671 (0.090) & 0.567 & 0.165 & 0.607 & 1.324 \\
\quad DiD-BCF & 3 & 0.435 (0.078) & 0.306 & 0.081 & 0.892 & 1.158 \\
\quad TWFE & 3 & 0.900 (0.044) & 0.760 & 0.220 & 0.352 & 0.974 \\
\quad Callaway--Sant'Anna & 3 & 1.077 (0.102) & 0.863 & 0.249 & 0.303 & 0.931 \\
\quad Gardner (did2s) & 3 & 1.080 (0.101) & 0.868 & 0.249 & 0.315 & 0.995 \\
\quad DoubleML DR-DiD & 3 & 0.912 (0.048) & 0.768 & 0.224 & 0.406 & 1.148 \\
\quad Synthetic DiD & 3 & 1.071 (0.100) & 0.862 & 0.247 & 0.255 & 0.796 \\
\quad grf-DiD (Wang) & 3 & 0.671 (0.090) & 0.567 & 0.164 & 0.607 & 1.324 \\
\addlinespace
\multicolumn{7}{l}{\textit{Canonical: serial correlation}} \\
\quad DiD-BCF & 1 & 0.299 (0.074) & 0.239 & 0.073 & 0.892 & 0.951 \\
\quad TWFE & 1 & 0.902 (0.044) & 0.762 & 0.221 & 0.348 & 0.967 \\
\quad Callaway--Sant'Anna & 1 & 1.063 (0.062) & 0.875 & 0.284 & 0.250 & 0.771 \\
\quad Gardner (did2s) & 1 & 1.071 (0.066) & 0.882 & 0.284 & 0.318 & 0.989 \\
\quad DoubleML DR-DiD & 1 & 0.903 (0.041) & 0.763 & 0.223 & 0.348 & 0.966 \\
\quad Synthetic DiD & 1 & 1.062 (0.061) & 0.875 & 0.282 & 0.254 & 0.792 \\
\quad grf-DiD (Wang) & 1 & 0.615 (0.098) & 0.518 & 0.150 & 0.589 & 1.180 \\
\quad DiD-BCF & 2 & 0.318 (0.088) & 0.252 & 0.078 & 0.877 & 0.976 \\
\quad TWFE & 2 & 0.902 (0.044) & 0.762 & 0.221 & 0.348 & 0.967 \\
\quad Callaway--Sant'Anna & 2 & 1.065 (0.064) & 0.877 & 0.284 & 0.261 & 0.809 \\
\quad Gardner (did2s) & 2 & 1.071 (0.066) & 0.882 & 0.284 & 0.318 & 0.989 \\
\quad DoubleML DR-DiD & 2 & 0.903 (0.044) & 0.764 & 0.224 & 0.345 & 0.962 \\
\quad Synthetic DiD & 2 & 1.062 (0.061) & 0.875 & 0.282 & 0.254 & 0.792 \\
\quad grf-DiD (Wang) & 2 & 0.614 (0.097) & 0.518 & 0.150 & 0.590 & 1.180 \\
\quad DiD-BCF & 3 & 0.465 (0.095) & 0.335 & 0.090 & 0.818 & 1.060 \\
\quad TWFE & 3 & 0.902 (0.044) & 0.762 & 0.221 & 0.348 & 0.967 \\
\quad Callaway--Sant'Anna & 3 & 1.076 (0.103) & 0.865 & 0.249 & 0.256 & 0.806 \\
\quad Gardner (did2s) & 3 & 1.082 (0.103) & 0.869 & 0.250 & 0.311 & 0.990 \\
\quad DoubleML DR-DiD & 3 & 0.901 (0.041) & 0.762 & 0.222 & 0.346 & 0.960 \\
\quad Synthetic DiD & 3 & 1.072 (0.101) & 0.864 & 0.248 & 0.251 & 0.793 \\
\quad grf-DiD (Wang) & 3 & 0.614 (0.098) & 0.518 & 0.150 & 0.590 & 1.181 \\
\addlinespace
\multicolumn{7}{l}{\textit{Canonical: selection on observables}} \\
\quad DiD-BCF & 1 & 0.266 (0.058) & 0.206 & 0.062 & 0.962 & 1.046 \\
\quad TWFE & 1 & 0.889 (0.061) & 0.743 & 0.211 & 0.366 & 0.971 \\
\quad Callaway--Sant'Anna & 1 & 1.062 (0.077) & 0.864 & 0.268 & 0.370 & 1.083 \\
\quad Gardner (did2s) & 1 & 1.204 (0.115) & 0.959 & 0.327 & 0.326 & 0.972 \\
\quad DoubleML DR-DiD & 1 & 0.912 (0.072) & 0.758 & 0.215 & 0.464 & 1.287 \\
\quad Synthetic DiD & 1 & 1.120 (0.091) & 0.893 & 0.302 & 0.283 & 0.778 \\
\quad grf-DiD (Wang) & 1 & 0.691 (0.089) & 0.583 & 0.161 & 0.607 & 1.370 \\
\quad DiD-BCF & 2 & 0.289 (0.065) & 0.226 & 0.067 & 0.949 & 1.070 \\
\quad TWFE & 2 & 0.889 (0.061) & 0.743 & 0.211 & 0.366 & 0.971 \\
\quad Callaway--Sant'Anna & 2 & 1.091 (0.129) & 0.887 & 0.279 & 0.399 & 1.225 \\
\quad Gardner (did2s) & 2 & 1.204 (0.115) & 0.959 & 0.327 & 0.326 & 0.972 \\
\quad DoubleML DR-DiD & 2 & 0.918 (0.072) & 0.763 & 0.217 & 0.469 & 1.311 \\
\quad Synthetic DiD & 2 & 1.120 (0.091) & 0.893 & 0.302 & 0.283 & 0.778 \\
\quad grf-DiD (Wang) & 2 & 0.691 (0.089) & 0.583 & 0.161 & 0.607 & 1.370 \\
\quad DiD-BCF & 3 & 0.489 (0.078) & 0.349 & 0.091 & 0.836 & 1.115 \\
\quad TWFE & 3 & 0.889 (0.061) & 0.743 & 0.211 & 0.366 & 0.971 \\
\quad Callaway--Sant'Anna & 3 & 1.106 (0.144) & 0.847 & 0.243 & 0.456 & 1.229 \\
\quad Gardner (did2s) & 3 & 1.218 (0.134) & 0.953 & 0.295 & 0.366 & 0.974 \\
\quad DoubleML DR-DiD & 3 & 0.914 (0.074) & 0.761 & 0.216 & 0.466 & 1.305 \\
\quad Synthetic DiD & 3 & 1.135 (0.116) & 0.863 & 0.265 & 0.370 & 0.780 \\
\quad grf-DiD (Wang) & 3 & 0.691 (0.089) & 0.583 & 0.161 & 0.607 & 1.370 \\
\addlinespace
\multicolumn{7}{l}{\textit{Staggered adoption (baseline dynamics)}} \\
\quad DiD-BCF & 1 & 0.674 (0.067) & 0.494 & 0.193 & 0.785 & 1.443 \\
\quad TWFE & 1 & 2.212 (0.091) & 1.784 & 0.414 & 0.191 & 1.145 \\
\quad Callaway--Sant'Anna & 1 & 2.113 (0.118) & 1.674 & 0.786 & 0.285 & 1.620 \\
\quad Gardner (did2s) & 1 & 2.466 (0.116) & 1.968 & 0.969 & 0.192 & 1.245 \\
\quad DoubleML DR-DiD & 1 & 1.799 (0.076) & 1.464 & 0.332 & 0.378 & 1.935 \\
\quad Synthetic DiD & 1 & 2.914 (0.138) & 2.223 & 0.814 & 0.145 & 0.922 \\
\quad grf-DiD (Wang) & 1 & 1.520 (0.068) & 1.242 & 0.283 & 0.384 & 1.714 \\
\quad DiD-BCF & 2 & 0.702 (0.074) & 0.513 & 0.198 & 0.777 & 1.470 \\
\quad TWFE & 2 & 2.212 (0.091) & 1.784 & 0.414 & 0.191 & 1.145 \\
\quad Callaway--Sant'Anna & 2 & 2.113 (0.119) & 1.674 & 0.789 & 0.290 & 1.651 \\
\quad Gardner (did2s) & 2 & 2.466 (0.116) & 1.968 & 0.969 & 0.192 & 1.245 \\
\quad DoubleML DR-DiD & 2 & 1.798 (0.071) & 1.464 & 0.331 & 0.378 & 1.947 \\
\quad Synthetic DiD & 2 & 2.914 (0.138) & 2.223 & 0.814 & 0.145 & 0.922 \\
\quad grf-DiD (Wang) & 2 & 1.520 (0.069) & 1.242 & 0.283 & 0.384 & 1.714 \\
\quad DiD-BCF & 3 & 0.890 (0.145) & 0.579 & 0.116 & 0.767 & 1.547 \\
\quad TWFE & 3 & 2.212 (0.091) & 1.784 & 0.414 & 0.191 & 1.145 \\
\quad Callaway--Sant'Anna & 3 & 2.120 (0.188) & 1.636 & 0.368 & 0.293 & 1.653 \\
\quad Gardner (did2s) & 3 & 2.464 (0.188) & 1.943 & 0.459 & 0.161 & 1.243 \\
\quad DoubleML DR-DiD & 3 & 1.804 (0.082) & 1.468 & 0.331 & 0.381 & 1.965 \\
\quad Synthetic DiD & 3 & 2.911 (0.200) & 2.120 & 0.401 & 0.158 & 0.924 \\
\quad grf-DiD (Wang) & 3 & 1.520 (0.069) & 1.242 & 0.283 & 0.384 & 1.713 \\
\addlinespace
\multicolumn{7}{l}{\textit{Staggered adoption (amplified dynamics)}} \\
\quad DiD-BCF & 1 & 0.904 (0.094) & 0.655 & 0.158 & 0.791 & 1.932 \\
\quad TWFE & 1 & 4.435 (0.167) & 3.521 & 0.580 & 0.130 & 1.674 \\
\quad Callaway--Sant'Anna & 1 & 3.519 (0.206) & 2.646 & 0.785 & 0.252 & 2.203 \\
\quad Gardner (did2s) & 1 & 4.748 (0.201) & 3.760 & 1.215 & 0.134 & 1.904 \\
\quad DoubleML DR-DiD & 1 & 2.970 (0.122) & 2.300 & 0.329 & 0.322 & 2.358 \\
\quad Synthetic DiD & 1 & 5.600 (0.219) & 4.131 & 1.074 & 0.098 & 1.133 \\
\quad grf-DiD (Wang) & 1 & 2.402 (0.105) & 1.851 & 0.269 & 0.357 & 2.218 \\
\quad DiD-BCF & 2 & 0.951 (0.100) & 0.686 & 0.171 & 0.781 & 1.969 \\
\quad TWFE & 2 & 4.435 (0.167) & 3.521 & 0.580 & 0.130 & 1.674 \\
\quad Callaway--Sant'Anna & 2 & 3.519 (0.213) & 2.647 & 0.787 & 0.254 & 2.221 \\
\quad Gardner (did2s) & 2 & 4.748 (0.201) & 3.760 & 1.215 & 0.134 & 1.904 \\
\quad DoubleML DR-DiD & 2 & 2.969 (0.125) & 2.298 & 0.329 & 0.319 & 2.346 \\
\quad Synthetic DiD & 2 & 5.600 (0.219) & 4.131 & 1.074 & 0.098 & 1.133 \\
\quad grf-DiD (Wang) & 2 & 2.402 (0.105) & 1.851 & 0.269 & 0.356 & 2.217 \\
\quad DiD-BCF & 3 & 1.344 (0.253) & 0.822 & 0.110 & 0.735 & 2.053 \\
\quad TWFE & 3 & 4.435 (0.167) & 3.521 & 0.580 & 0.130 & 1.674 \\
\quad Callaway--Sant'Anna & 3 & 3.527 (0.319) & 2.588 & 0.367 & 0.256 & 2.227 \\
\quad Gardner (did2s) & 3 & 4.737 (0.316) & 3.675 & 0.612 & 0.122 & 1.898 \\
\quad DoubleML DR-DiD & 3 & 2.972 (0.121) & 2.299 & 0.329 & 0.323 & 2.376 \\
\quad Synthetic DiD & 3 & 5.588 (0.323) & 3.991 & 0.569 & 0.091 & 1.136 \\
\quad grf-DiD (Wang) & 3 & 2.402 (0.105) & 1.851 & 0.269 & 0.357 & 2.218 \\
\addlinespace
\addlinespace
\end{longtable}
\noindent \emph{Note.} $n=200$, 100 replications. Estimators that produce only cell-level effects have their $\mathrm{GATT}(g,t)$ (or event-study, or overall ATT) estimate broadcast to every treated observation in the cell, which is the natural best case for a homogeneous-effect estimator. Cov.95 is \emph{pointwise} coverage of the individual effect.\par\vspace{6pt}
}

\begin{figure}[htbp]
\centering
\includegraphics[width=0.93\linewidth]{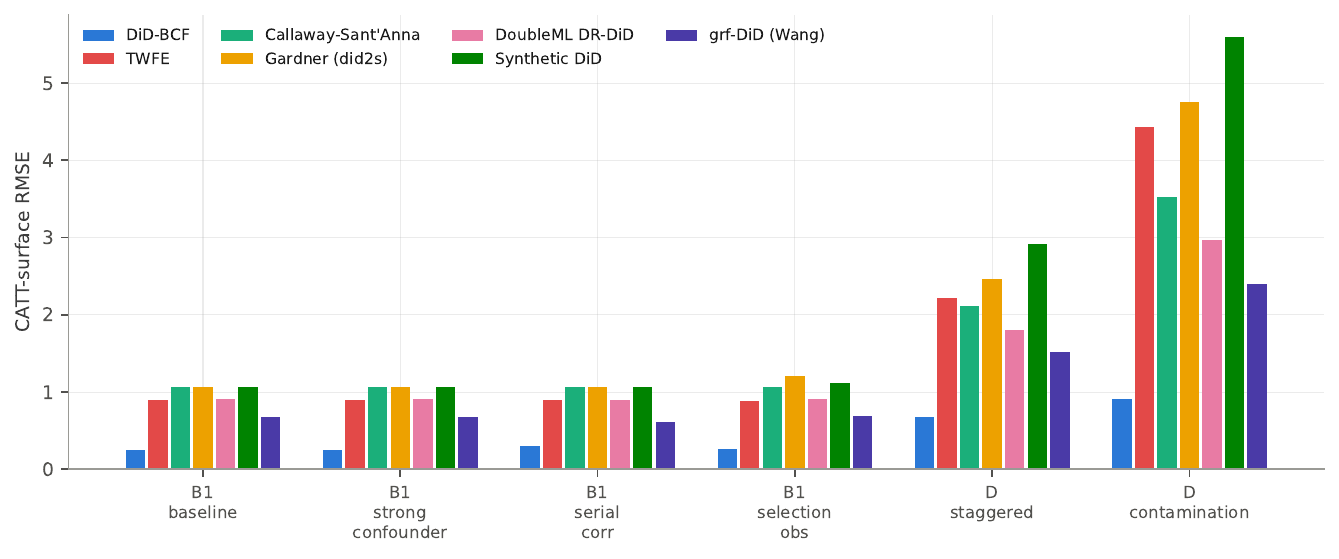}
\caption{Conditional-effect surface accuracy across canonical simulation designs.}\label{fig:surface}
\end{figure}
\begin{figure}[htbp]
\centering
\includegraphics[width=0.93\linewidth]{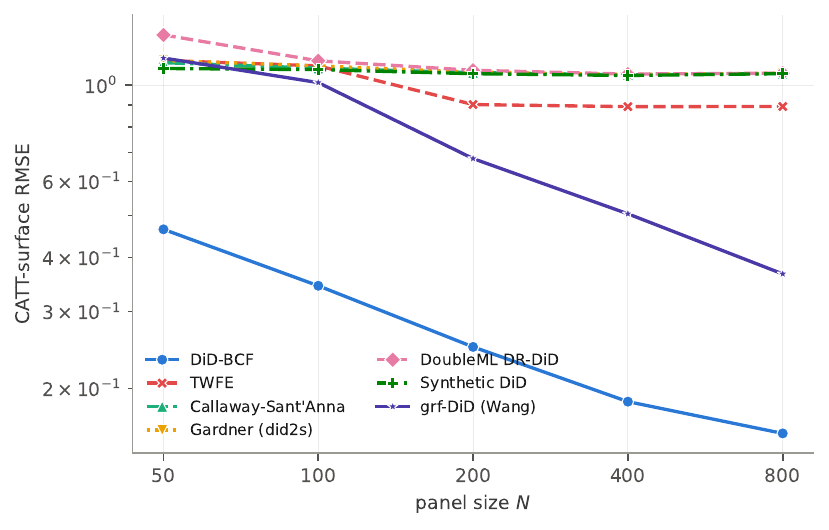}
\caption{Conditional-effect surface accuracy across sample sizes and observed covariate representations.}\label{fig:surfacesweep}
\end{figure}

\clearpage
\section{Theoretical Properties of DiD-BCF}\label{supp:theory}

\providecommand{\BL}{d_{\mathrm{BL}}}
\providecommand{\norm}[1]{\left\lVert#1\right\rVert}

\subsection{Scope and overview}\label{sec:raw_scope}

This section gives the formal results for the structured DiD-BCF posterior.
The first part establishes the causal contrast and the role of the two-way
baseline. The second part derives the efficient influence function when the
full panel is used as one observation. The third part gives a sufficient
Bernstein--von Mises result for the posterior mixture over partitions and
Gaussian scales. The fourth part proves a primitive result for adaptive,
depth-capped forests on finite covariate support, including the behavior at
simple truths. It then records covariance misspecification, empirical
standardization, the total-effect gauge, and realized versus population
averages. The final extensions include proper unknown scalar variance and
posterior means on finite support, and a continuous-covariate theorem for
growing saturated histogram mixtures with feasible inference at the raw center.
Sections~\ref{rfcons:inputs}, \ref{rfcons:scalar} and \ref{rfcons:fixed}
give the full proof closure for continuous-cut integrated effect
consistency at scalar smooth truth with growing depth caps and at joint finite-tree
truth with fixed capacities. Section~\ref{sec:imp_extension} records the matrix
representation of the three-forest hierarchy, conjugate inverse-gamma variance
updates, properness of the joint working posterior, quotient-space
identification of linear causal functionals, and Monte Carlo draw dispersion
under finite simulation budgets.

The positive posterior results are stated for the identifiable composite
treatment effect and for the exact mathematical posterior distribution, rather
than for unnormalized component parameters or finite-horizon Markov chain
approximations. Table~\ref{tab:twospeed} summarizes the scope and guarantees of
these theoretical results.

\begin{table}[htbp]
\centering
\caption{Scope of the DiD-BCF theory.}\label{tab:twospeed}
\begin{tabular}{@{}L{0.20\linewidth}L{0.54\linewidth}L{0.20\linewidth}@{}}
\toprule
Object & Guarantee & Result \\
\midrule
Identification & Conditional parallel trends is equivalent to an additive
cohort--time baseline; the total effect is the four-cell contrast &
Lemma~\ref{lem:twoway}, Theorem~\ref{thm:id} \\
Full-panel average & Efficient influence function under a Gaussian panel
experiment, including treated-composition uncertainty &
Theorem~\ref{thm:eif} \\
Adaptive posterior mixture & Sufficient bias, selection-complexity, and
variance conditions for a Gaussian limit & Theorem~\ref{thm:mixture} \\
Finite support & Generic-truth BvM with fixed depth caps; simple-truth
adaptation and its local nonuniformity &
Thm.~\ref{thm:finite}\newline Thm.~\ref{thm:simple} \\
Unknown scalar variance & Generic finite-support posterior mean and BvM,
empirical standardization, feasible cluster variance &
Thm.~\ref{uv:thm:unknown}\newline Cor.~\ref{uv:cor:cluster} \\
Continuous histograms & Growing balanced saturated partitions; functional
bias, posterior moments, cluster and population intervals &
Thm.~\ref{cc:thm:continuous}\newline Cor.~\ref{cc:cor:feasible} \\
Fixed-capacity continuous cuts & Integrated total-effect and fixed-average
consistency on a joint finite-tree truth class; conditional spherical errors &
Thm.~\ref{rfcons:f:thm:fixedeffects} \\
Scalar growing continuous cuts & Integrated effect consistency for bounded
H\"older truth of order above one-half; growing depth caps &
Thm.~\ref{rfcons:s:thm:scalar} \\
Hierarchical prior and quotient gauge & Proper working posterior, quotient-space functional identification, and finite-draw Monte Carlo dispersion &
Prop.~\ref{imp:prop:proper}\newline Prop.~\ref{imp:prop:quotient} \\
\bottomrule
\end{tabular}
\end{table}

\subsection{Sampling framework and estimands}\label{sec:sampling}

We observe independent units $O_i=(X_i,G_i,Y_i^\top)^\top$, where
$Y_i=(Y_{i1},\ldots,Y_{iT})^\top$, $T\ge2$ is fixed, and
$G_i\in\cG\cup\{\infty\}$ is an absorbing adoption cohort. Here
$\cG\subseteq\{2,\ldots,T\}$ and $\infty$ denotes never-treated units.
Treatment status is $D_{it}=\ind{G_i\le t}$ for finite $G_i$, and event time
is $k_{it}=t-G_i$. Covariates $X_i$ are measured before treatment and are
not affected by treatment.

Potential outcomes satisfy consistency and SUTVA. For $g\in\cG$, no
anticipation means
\[
\E[Y_{it}(g)\mid G_i=g,X_i=x]
=\E[Y_{it}(0)\mid G_i=g,X_i=x],\qquad t<g,
\]
and conditional parallel trends means
\[
\E[Y_{it}(0)-Y_{i,t-1}(0)\mid G_i=g,X_i=x]
=\E[Y_{it}(0)-Y_{i,t-1}(0)\mid G_i=\infty,X_i=x]
\]
for $t\ge2$ and almost every relevant $x$. We assume finite first absolute
moments within each cohort for the potential outcomes appearing in the
displayed conditional means and estimands, and overlap of every treated
cohort with the never-treated cohort. In particular, $P(G=\infty)>0$, $P(G=g)>0$, and
the conditional probability of never treatment in each two-group comparison
is bounded away from zero. Writing $P_g$ for the law of $X$ conditional on
$G=g$, overlap implies $P_g\ll P_\infty$: a covariate set with zero
never-treated probability also has zero probability in cohort $g$.
All support qualifications below are understood in this measure-theoretic
sense. Conditional means and their finite time differences are defined on
a common set of full $P_g$ measure for each treated cohort.

For $t\ge g$, define
\begin{align*}
\CATT(x;g,t)&=\E[Y_{it}(g)-Y_{it}(0)\mid G_i=g,X_i=x],\\
\GATT(g,t)&=\E[\CATT(X_i;g,t)\mid G_i=g].
\end{align*}
When effects are pooled by event time, we assume that there is a function
$\vartheta_0(x,k)$ such that
$\CATT(x;g,g+k)=\vartheta_0(x,k)$ on the relevant treated supports. This is
the cross-cohort restriction represented by a treatment forest indexed by
$(x,k)$ rather than by $(g,x,k)$.

\subsection{Identification and the structured mean}\label{sec:id}

Let
\[
\mu_0(g,t,x)=\E[Y_{it}(0)\mid G_i=g,X_i=x].
\]
The following lemma is the exact population content of the baseline
architecture. It permits cohort-specific levels and covariate-dependent
levels, while excluding cohort-specific untreated trends.

\begin{lemma}[Two-way characterization of conditional parallel trends]
\label{lem:twoway}
Under the sampling and overlap conditions above, conditional parallel trends
holds if and only if there exist functions $a_0(g,x)$ and $b_0(t,x)$ such that
\[
\mu_0(g,t,x)=a_0(g,x)+b_0(t,x),\qquad b_0(1,x)=0,
\]
on the relevant supports. This representation is unique under the displayed
normalization, up to the corresponding covariate-null sets.
\end{lemma}

\begin{proof}
For the forward implication, set
$a_0(g,x)=\mu_0(g,1,x)$ and
$b_0(t,x)=\mu_0(\infty,t,x)-\mu_0(\infty,1,x)$. Summing the conditional
parallel-trend increments from period one to $t$ gives
$\mu_0(g,t,x)-\mu_0(g,1,x)=b_0(t,x)$, which proves the representation.
For the reverse implication, the increment from $t-1$ to $t$ is
$b_0(t,x)-b_0(t-1,x)$ for every cohort, so conditional parallel trends
follows. For uniqueness, any normalized representation must satisfy
$a_0(g,x)=\mu_0(g,1,x)$ by evaluating it at period one. Evaluating it
for the never-treated cohort then gives
$b_0(t,x)=\mu_0(\infty,t,x)-\mu_0(\infty,1,x)$. Thus both functions
are uniquely determined on their relevant supports. All equalities are
understood up to covariate-null sets.
\end{proof}

\begin{proposition}[Treatment-status reparameterization]
\label{prop:reparam}
On the observed panel, the conditional-mean class
\[
f(g,t,x)=a(g,x)+b(t,x)+D(g,t)\vartheta(x,t-g)
\]
is unchanged if the effect surface is assigned arbitrary values at rows for
which $D(g,t)=0$. Consequently imposing zero values at inactive rows changes
no observable conditional mean. Under the two-way baseline, the four-cell
contrast is exactly the total effect:
\[
f(g,t,x)-f(g,g-1,x)-f(\infty,t,x)+f(\infty,g-1,x)
=\vartheta(x,t-g).
\]
This is an equality of identified mean functions, not a statement that
posterior distributions of extrapolated inactive-row surface values are
invariant.
\end{proposition}

\begin{proof}
The multiplier $D(g,t)$ is zero at all control and pre-treatment rows, so
the value of $\vartheta$ there cannot change $f$ on the observed design. For
$t\ge g$, the two treated cells contribute
$a(g,x)+b(t,x)+\vartheta(x,t-g)$ and $a(g,x)+b(g-1,x)$, while the two
never-treated cells contribute $a(\infty,x)+b(t,x)$ and
$a(\infty,x)+b(g-1,x)$. Subtracting cancels both $a$ and $b$ and leaves the
effect term.
\end{proof}

\begin{theorem}[Nonparametric DiD identification]
\label{thm:id}
Under consistency, no anticipation, conditional parallel trends, and overlap,
for every $g\in\cG$, $t\ge g$, and almost every relevant $x$,
\begin{equation}\label{eq:id}
\CATT(x;g,t)=
\E[Y_{it}-Y_{i,g-1}\mid G_i=g,X_i=x]
-\E[Y_{it}-Y_{i,g-1}\mid G_i=\infty,X_i=x].
\end{equation}
Hence $\GATT(g,t)$ and every fixed-weight aggregate are point identified.
Under pooled event-time effects, the structured DiD-BCF total-effect surface
equals the contrast in Proposition~\ref{prop:reparam}.
\end{theorem}

\begin{proof}
Fix $g\in\cG$ and $t\in\{g,\ldots,T\}$, and write $b=g-1$.
Since $g\ge2$, the baseline $b$ is observed and precedes treatment in
cohort $g$. Work on a set of full $P_g$ measure on which the relevant
conditional means and the parallel-trend equalities hold. Such a common
set exists because the cohort and period sets are finite, and overlap
transfers never-treated covariate-null sets to $P_g$-null sets.

First, telescope the untreated increments from $b$ to $t$. Conditional
parallel trends at each period $s=g,\ldots,t$ and linearity of
conditional expectation give
\begin{align}
&\E[Y_{it}(0)-Y_{ib}(0)\mid G_i=g,X_i=x] \notag\\
&\quad=\sum_{s=g}^{t}
\E[Y_{is}(0)-Y_{i,s-1}(0)\mid G_i=g,X_i=x] \notag\\
&\quad=\sum_{s=g}^{t}
\E[Y_{is}(0)-Y_{i,s-1}(0)\mid G_i=\infty,X_i=x] \notag\\
&\quad=\E[Y_{it}(0)-Y_{ib}(0)\mid G_i=\infty,X_i=x].
\label{eq:id_untreated_drift}
\end{align}
All sums are finite, and the moment conditions justify every conditional
expectation. For never-treated units, consistency gives observed untreated
outcomes at both dates. Hence the last line is the observable conditional
mean $\E[Y_{it}-Y_{ib}\mid G_i=\infty,X_i=x]$.

For the treated cohort, consistency at $t$ and no anticipation at $b$
imply, in conditional mean,
\begin{align*}
\E[Y_{it}-Y_{ib}\mid G_i=g,X_i=x]
&=\E[Y_{it}(g)-Y_{ib}(0)\mid G_i=g,X_i=x]\\
&=\E[Y_{it}(g)-Y_{it}(0)\mid G_i=g,X_i=x]\\
&\quad+\E[Y_{it}(0)-Y_{ib}(0)\mid G_i=g,X_i=x]\\
&=\CATT(x;g,t)
+\E[Y_{it}-Y_{ib}\mid G_i=\infty,X_i=x],
\end{align*}
where the last equality uses \eqref{eq:id_untreated_drift}. Rearranging
proves \eqref{eq:id}. This argument needs only the conditional-mean
no-anticipation restriction stated above; almost-sure no anticipation is
also sufficient.

To establish point identification, let
$m_{h,t,b}(x)=\E[Y_{it}-Y_{ib}\mid G_i=h,X_i=x]$ for
$h\in\{g,\infty\}$. The observed-data law determines $m_{g,t,b}$
$P_g$-almost everywhere and $m_{\infty,t,b}$ $P_\infty$-almost
everywhere. Absolute continuity $P_g\ll P_\infty$ therefore makes
their difference uniquely determined $P_g$-almost everywhere, regardless
of the versions chosen for these conditional expectations. Thus any two
admissible potential-outcome laws inducing the same observed-data law
have the same $\CATT(x;g,t)$ on this support. The observed-data law
also determines $P_g$, so
\[
\GATT(g,t)=\int
\{m_{g,t,b}(x)-m_{\infty,t,b}(x)\}\,dP_g(x)
\]
is identified. The assumed moments make this integral finite: its
integrand equals the integrable conditional treatment effect. Every
finite linear combination of these identified cells with fixed known
weights is consequently identified. The same conclusion applies when
the weights are themselves identified features of the observed-data law,
such as cohort shares.

Finally, write $f(h,s,x)=\E[Y_{is}\mid G_i=h,X_i=x]$.
Conditional-expectation linearity rewrites \eqref{eq:id} as
\[
\CATT(x;g,t)=f(g,t,x)-f(g,b,x)-f(\infty,t,x)+f(\infty,b,x).
\]
Under event-time pooling this left-hand side is
$\vartheta_0(x,t-g)$. In the structured mean, the four right-hand
terms are respectively $a_0(g,x)+b_0(t,x)+\vartheta_0(x,t-g)$,
$a_0(g,x)+b_0(b,x)$, $a_0(\infty,x)+b_0(t,x)$, and
$a_0(\infty,x)+b_0(b,x)$. Their contrast cancels both nuisance
functions and leaves the total effect. Without pooling the same
calculation leaves the cohort-specific effect $\CATT(x;g,t)$.
\end{proof}

\subsection{Full-panel canonical gradient}\label{sec:full_panel}

The preceding result is nonparametric identification. Efficiency depends on
which observations are retained. The following calculation treats the full
panel as one observation and therefore permits all periods and cohorts to
contribute to an average effect.

Let $\beta(x)$ collect the distinct cohort levels, common time effects, and
cell effects after the baseline normalization, and let fixed incidence
matrices $B_g$ satisfy
$\E[Y\mid X=x,G=g]=B_g\beta(x)$. Write
$e_g(x)=P(G=g\mid X=x)$ and assume
\[
Y\mid X=x,G=g\sim N(B_g\beta(x),\Omega_g(x)),
\]
where each $\Omega_g(x)$ is positive definite with eigenvalues uniformly
bounded above and below. Define
\[
Q(x)=\sum_g e_g(x)B_g^\top\Omega_g(x)^{-1}B_g.
\]
For a fixed target cell $(g_\star,t_\star)$, let $c$ select the corresponding
effect coordinate, $p_\star=P(G=g_\star)$, and
\[
\theta=\E[c^\top\beta(X)\mid G=g_\star],\qquad
l(x)=e_{g_\star}(x)c/p_\star.
\]
Assume that $\E[l(X)^\top Q(X)^{-1}l(X)]$ and
$\Var\{c^\top\beta(X)\mid G=g_\star\}$ are finite. These are
the square-integrability conditions for the efficient influence function.

\begin{lemma}[Full-panel information]
\label{lem:rank}
If every treated cohort has a pre-treatment period, the never-treated cohort
is observed throughout, and every relevant $e_g(x)>0$, then $Q(x)$ is positive
definite. If the cohort probabilities are uniformly bounded below, its
smallest eigenvalue is uniformly bounded away from zero under the covariance
eigenvalue bounds.
\end{lemma}

\begin{proof}
If $v^\top Q(x)v=0$, positive definiteness of each covariance matrix and
positivity of every $e_g(x)$ imply $B_gv=0$ for every cohort. The
never-treated period-one equation sets the never-treated level component to
zero, and its remaining periods set all common time components to zero.
Period one for each treated cohort then sets its level component to zero, and
its post-treatment equations set all of its effect components to zero. Thus
$v=0$. The matrices have finite dimension, and uniform overlap and covariance
bounds turn this pointwise rank statement into a uniform eigenvalue bound.
\end{proof}

\begin{theorem}[Full-panel efficient influence function]
\label{thm:eif}
In the Gaussian conditional-mean model above, with unrestricted distribution
of $(X,G)$, the efficient influence function of $\theta$ is
\begin{equation}\label{eq:full-eif}
\phi(O)=l(X)^\top Q(X)^{-1}B_G^\top\Omega_G(X)^{-1}
\{Y-B_G\beta(X)\}
+\frac{\ind{G=g_\star}}{p_\star}
\{c^\top\beta(X)-\theta\}.
\end{equation}
Its variance is
\begin{equation}\label{eq:full-var}
V_{\mathrm{full}}=\E[l(X)^\top Q(X)^{-1}l(X)]
+\frac{1}{p_\star}\Var\{c^\top\beta(X)\mid G=g_\star\}.
\end{equation}
The gradient is unchanged when finite-dimensional Gaussian covariance tangent
directions are added independently of the mean tangent directions.
\end{theorem}

\begin{proof}
For a mean path $\beta_s=\beta+sh$, the score is
$S_h=h(X)^\top B_G^\top\Omega_G^{-1}(Y-B_G\beta)$ and the target derivative
is $\E[l(X)^\top h(X)]$. Conditional covariance of two mean scores is
$h(X)^\top Q(X)v(X)$. Taking $v=Q^{-1}l$ therefore represents the derivative
on the mean tangent space. For a path that perturbs the distribution of
$(X,G)$ with mean-zero score $s(X,G)$, differentiating the ratio defining
$\theta$ gives
\[
\E\left[\frac{\ind{G=g_\star}}{p_\star}
\{c^\top\beta(X)-\theta\}s(X,G)\right].
\]
The residual term in \eqref{eq:full-eif} is orthogonal to this distribution
tangent space because the residual has conditional mean zero. Thus the sum
represents every derivative. Its conditional variance gives the first term
of \eqref{eq:full-var}, and the distribution term gives the second; their
cross covariance is zero. Finally, Gaussian mean scores are linear in the
residual and covariance scores are centered quadratic forms. Their covariance
is a third centered Gaussian moment and is therefore zero, so adding the
covariance nuisance does not change the gradient.
\end{proof}

\subsection{Adaptive Gaussian-mixture BvM}\label{sec:bvm}

Condition on a fixed design. Let $Y=f_0+u$, where
$u\sim N(0,\Omega_n)$. Let $H$ index forest partitions, leaf scales, and any
conditioned covariance parameters. Conditional on $H$, suppose the working
likelihood and proper Gaussian coefficient prior are
\[
Y\mid\beta,H\sim N(F_H\beta,W_H^{-1}),\qquad
\beta\mid H\sim N(0,\Lambda_H^{-1}),
\]
with $\Lambda_H$ positive definite. For a linear total-effect functional
$T_H=l_H^\top\beta$, set
\begin{equation}\label{eq:matrices}
V_H=(F_H^\top W_HF_H+\Lambda_H)^{-1},\qquad
a_H=W_HF_HV_Hl_H,\qquad v_H=l_H^\top V_Hl_H.
\end{equation}
Completing the square gives
\[
T_H\mid H,Y\sim N(a_H^\top Y,v_H).
\]
The posterior weights on $H$ are allowed to depend on $Y$.

Let $\theta_{0,n}$ denote the realized-design target, let $w_n$ be a
design-measurable oracle vector, and write
$\theta_n^*=\theta_{0,n}+w_n^\top u$. For deterministic design-measurable
sets $\cH_n$, define
\[
\mathfrak g_n=\E_u\sup_{H\in\cH_n}
\left|\sqrt n(a_H-w_n)^\top u\right|.
\]

\begin{theorem}[Adaptive DiD-BCF Gaussian-mixture BvM]
\label{thm:mixture}
Suppose $\Pi(\cH_n\mid Y)\to1$ in probability and, for some
$v,s^2\in(0,\infty)$,
\begin{align}
\sup_{H\in\cH_n}\sqrt n|a_H^\top f_0-\theta_{0,n}|&\to0,
\label{eq:bias}\\
\mathfrak g_n&\to0,\label{eq:complexity}\\
\sup_{H\in\cH_n}|nv_H-v|&\to0,\qquad
nw_n^\top\Omega_nw_n\to s^2.\label{eq:variance}
\end{align}
Then
\[
\BL\left(\Pi\left(\sqrt n(T-\theta_n^*)\in\cdot\mid Y\right),
N(0,v)\right)\to0
\]
in probability. The posterior median is
$\theta_n^*+o_P(n^{-1/2})$, and its central $(1-\alpha)$ intervals have
limiting coverage
\[
2\Phi\left(z_{1-\alpha/2}\sqrt{v/s^2}\right)-1
\]
for $\theta_{0,n}$. In particular, nominal coverage requires $v=s^2$.
\end{theorem}

\begin{proof}
For every $H$ the conditional center has the exact decomposition
\[
\sqrt n(a_H^\top Y-\theta_n^*)
=\sqrt n(a_H^\top f_0-\theta_{0,n})
+\sqrt n(a_H-w_n)^\top u.
\]
The supremum of the absolute value of the second term is $o_P(1)$ by
Markov's inequality and \eqref{eq:complexity}; the first term is uniformly
$o(1)$ by \eqref{eq:bias}. Couple each conditional Gaussian draw to
$\sqrt v Z$, with $Z\sim N(0,1)$. For every bounded one-Lipschitz function,
the difference of expectations is bounded by the center discrepancy plus
$\E|Z|\,|\sqrt{nv_H}-\sqrt v|$, uniformly on $\cH_n$. Integrating this
bound against arbitrary data-dependent posterior weights is valid because
the bound is uniform; the complement contributes at most twice its posterior
mass. This proves the bounded-Lipschitz limit without treating an
outcome-selected partition as fixed. Nondegeneracy gives convergence of
fixed posterior quantiles. Finally,
$\sqrt n(\theta_n^*-\theta_{0,n})$ is Gaussian with variance tending to
$s^2$, which yields the coverage formula. Neither posterior-mean nor
posterior-variance convergence is implied by this weak limit without
the corresponding uniform integrability.
\end{proof}

\begin{lemma}[Finite-class selection bound]
\label{lem:width}
If $\cH_n$ has $M_n$ elements and
\[
\delta_n^2=\sup_{H\in\cH_n}n(a_H-w_n)^\top\Omega_n(a_H-w_n),
\]
then
\[
\mathfrak g_n\le\delta_n\sqrt{2\log(2M_n)}.
\]
Thus $\delta_n\sqrt{\log(2M_n)}\to0$ implies
\eqref{eq:complexity}.
\end{lemma}

\begin{proof}
The $2M_n$ signed Gaussian coordinates have exponential moments bounded by
$\exp(t^2\delta_n^2/2)$. Summing these bounds, applying Jensen's inequality
to the maximum, and minimizing over $t>0$ gives the display. No independence
among the coordinates is used.
\end{proof}

\begin{proposition}[Projection and functional shrinkage]
\label{prop:projection}
Hold the distribution of $(X,G)$ fixed and equip structured coefficient
functions with the information inner product
\[
\langle h,k\rangle_Q=\E[h(X)^\top Q(X)k(X)].
\]
For the linear target $\mathcal T(h)=\E[l(X)^\top h(X)]$, let
$r=Q^{-1}l$. If $S$ is a closed linear subspace and $P_S$ is its orthogonal
projection, then
\[
\mathcal T(P_S\beta_0)-\mathcal T(\beta_0)
=-\langle r-P_Sr,\beta_0-P_S\beta_0\rangle_Q,
\]
and hence the absolute value is bounded by
$\|r-P_Sr\|_Q\|\beta_0-P_S\beta_0\|_Q$. In a fixed identifiable design,
with $G=F^\top WF>0$, $b^\circ=G^{-1}F^\top Wf_0$, the Gaussian posterior
center satisfies
\[
a^\top f_0-\theta_0
=l^\top b^\circ-\theta_0
-l^\top(G+\Lambda)^{-1}\Lambda b^\circ.
\]
\end{proposition}

\begin{proof}
The functional is the $Q$ inner product with $r$. Write
$(I-P_S)\beta_0$ for the approximation error and use orthogonality of
$r-P_Sr$ to $P_S\beta_0$ to obtain the first identity. Cauchy--Schwarz
gives the bound. For the matrix identity,
$F^\top Wf_0=Gb^\circ$ and
$(G+\Lambda)^{-1}G=I-(G+\Lambda)^{-1}\Lambda$. The final term is the
explicit Gaussian shrinkage contribution. Redundant parameter columns are
first reduced to their identifiable mean space.
\end{proof}

\subsection{Primitive finite-support BvM}\label{sec:finite}

Assume now that $X$ has fixed finite support
$\{x_1,\ldots,x_J\}$, every group--covariate stratum has a positive limiting
frequency, and $T$ and the cohort set are fixed. Push the forest posterior
forward to $\gamma\in\R^q$, the identifiable structured mean coordinates,
including the total effect and with the baseline gauge removed. Let $A_n$ be
the incidence matrix in the observed panel, let $W_n$ be a known block
precision matrix, and suppose
\[
Y=A_n\gamma_0+u,\qquad u\sim N(0,W_n^{-1}),\qquad
n^{-1}A_n^\top W_nA_n\to Q_*>0.
\]
Fixed depth caps and finite support induce finitely many distinct linear mean
subspaces $L_1,\ldots,L_M\subseteq\R^q$. Group configurations inducing the
same subspace together. Assume that at least one supported configuration spans
$\R^q$ and has a fixed positive prior weight. Conditional on every full-span
configuration, the induced prior density on $\gamma$ is proper, continuous,
and positive near $\gamma_0$; the remaining component priors are proper. The
finite collection and its positive weights are fixed after the finite support
is observed and do not depend on outcomes or on $n$.

\begin{theorem}[Primitive DiD-BCF BvM at generic truths]
\label{thm:finite}
Under the preceding finite-support conditions, suppose
\[
\gamma_0\notin\bigcup_{j:L_j\ne\R^q}L_j.
\]
For any bounded design-measurable vector $c_n\to c\ne0$, let
\[
\widehat\gamma_n=(A_n^\top W_nA_n)^{-1}A_n^\top W_nY.
\]
Then
\[
\BL\left(\Pi\left(\sqrt n(c_n^\top\gamma-c_n^\top\widehat\gamma_n)\in\cdot
\mid Y\right),N(0,c^\top Q_*^{-1}c)\right)\to0
\]
in probability. The posterior median is efficient for the conditional-design
target $c_n^\top\gamma_0$, and central intervals have asymptotically nominal
coverage. All supported forest configurations are integrated out.
\end{theorem}

\begin{proof}
The finite union of proper subspaces is closed. The genericity condition and
$Q_*>0$ therefore give a strictly positive limiting information-distance
between $\gamma_0$ and that union. Completing the square in the Gaussian
likelihood shows that the maximized likelihood on every proper subspace is
smaller than the unrestricted maximum by $Cn+O_P(\sqrt n)$ for some $C>0$.
Integration against a proper prior cannot increase this likelihood bound.
In a full-span component, integration over a ball of radius $n^{-1/2}$ around
$\widehat\gamma_n$ gives a lower bound of order $n^{-q/2}$ times a positive
tight factor, because the prior density is continuous and positive near
$\gamma_0$. Thus the posterior mass of all proper subspaces tends to zero.

On the full space, the likelihood ratio relative to the unrestricted GLS
estimate is exactly
\[
\frac{L_n(\gamma)}{L_n(\widehat\gamma_n)}
=\exp\left\{-\frac12(\gamma-\widehat\gamma_n)^\top
A_n^\top W_nA_n(\gamma-\widehat\gamma_n)\right\}.
\]
With $h=\sqrt n(\gamma-\widehat\gamma_n)$, the prior density converges
uniformly on compact $h$-sets to a positive constant, while the quadratic
kernel converges to $\exp(-h^\top Q_*h/2)$. Uniform eigenvalue bounds control
the Gaussian tails inside a fixed neighborhood of $\gamma_0$; outside that
neighborhood the likelihood is exponentially small and prior propriety
controls the integral. Consequently the coefficient posterior, after
centering at $\widehat\gamma_n$ and scaling by $\sqrt n$, converges in total
variation to $N(0,Q_*^{-1})$. Applying the bounded linear map $c_n^\top$ gives
the display. The GLS center is exactly unbiased with matching conditional
Gaussian covariance, so its sampling variance has the same limit and the
median and coverage claims follow.
\end{proof}

\begin{remark}[A concrete nonempty configuration]
\label{rem:finite_instance}
Take $T=8$, one treated cohort at $g=4$, and a binary baseline covariate with
positive probability in both groups. The identifiable structured mean has
$4$ group--covariate levels, $14$ time--covariate coefficients after setting
$b(1,\cdot)=0$, and $10$ post-period--covariate effects, for $q=28$. A level
tree of depth two, a trend tree of depth four, and an effect tree of depth
four can span these tables, while the default budgets and depth caps allow
such configurations. Proper Gaussian leaf priors then give a positive density
on the full mean space. This is a finite-support witness, not a claim that a
continuous-covariate data set is effectively discrete.
\end{remark}

\begin{theorem}[Simple truths and local alternatives]
\label{thm:simple}
Suppose every supported subspace $L_j$ has a proper density relative to its
own volume, continuous and positive near $\gamma_0$ whenever
$\gamma_0\in L_j$. Suppose there is a unique smallest-dimensional supported
subspace $L_*$ containing $\gamma_0$, with dimension $k>0$. Let $U$ have
orthonormal columns spanning $L_*$ and assume $c^\top U\ne0$. Then the
posterior probability of $L_*$ tends to one and the target has a BvM limit
centered at
\[
\widehat\gamma_{*,n}=U(U^\top A_n^\top W_nA_nU)^{-1}
U^\top A_n^\top W_nY,
\]
with limiting posterior Gaussian variance on the $\sqrt n$ scale
\[
v_*=c^\top U(U^\top Q_*U)^{-1}U^\top c.
\]
The central intervals are pointwise nominal at $\gamma_0$. For each fixed
$h$ and $\gamma_n=\gamma_0+h/\sqrt n$, the same model selection conclusion
holds, but
\[
\sqrt n\{c_n^\top\widehat\gamma_{*,n}-c_n^\top\gamma_n\}
\rightsquigarrow N(b_h,v_*),
\quad b_h=c^\top(P_*-I)h,
\quad P_*=U(U^\top Q_*U)^{-1}U^\top Q_*.
\]
The corresponding limiting coverage is
\[
\Phi\left(z_{1-\alpha/2}-\frac{b_h}{\sqrt{v_*}}\right)
-\Phi\left(-z_{1-\alpha/2}-\frac{b_h}{\sqrt{v_*}}\right),
\]
which need not equal $1-\alpha$.
\end{theorem}

\begin{proof}
Subspaces not containing $\gamma_0$ are separated from it and are excluded
exponentially. For a subspace containing $\gamma_0$, completing the square
in its $d_j$ coordinates gives an integrated likelihood of order
$n^{-d_j/2}\exp\{O_P(1)\}$. The unique smallest dimension therefore selects
$L_*$. Conditional on that space, the likelihood is exactly quadratic and
gives the restricted GLS BvM and variance $v_*$. At a local alternative, all
subspaces containing $\gamma_0$ still have only an $O_P(1)$ likelihood
comparison, so the same polynomial dimension comparison selects $L_*$. The
restricted center has expectation equal to the empirical information
projection of $\gamma_n$, giving limiting bias
$c^\top(P_*-I)h$ and variance $v_*$. Substitution into the Gaussian interval
gives the coverage formula.
\end{proof}

If several smallest-dimensional subspaces contain the truth, a nontrivial
mixture can remain. The simple-truth result is pointwise and does not provide
uniform coverage over local alternatives in the larger mean space.

\subsection{Covariance, standardization, and gauges}\label{sec:transfer}

\begin{proposition}[Working covariance and sandwich variance]
\label{prop:sandwich}
Retain the finite-support design, fixed finite prior mixture, full-span
prior regularity and genericity conditions of Theorem~\ref{thm:finite}.
Let the true law be $Y=A_n\gamma_0+u$, $u\sim N(0,\Omega_n)$, while
the working Gaussian likelihood uses precision $W_n$. Suppose
\[
n^{-1}A_n^\top W_n\Omega_nW_nA_n\to J_*>0,
\]
and write $Q_*=\lim n^{-1}A_n^\top W_nA_n>0$. For fixed $c\ne0$,
the posterior mean and median of $c^\top\gamma$ equal
$c^\top\widehat\gamma_n+o_P(n^{-1/2})$, where $\widehat\gamma_n$ is
the working GLS estimator. Moreover,
\[
n\Var_\Pi(c^\top\gamma\mid Y)\longrightarrow
V_{\mathrm{post}}=c^\top Q_*^{-1}c
\]
in probability, whereas
\[
\sqrt n(c^\top\widehat\gamma_n-c^\top\gamma_0)
\rightsquigarrow N(0,s^2),\qquad
s^2=c^\top Q_*^{-1}J_*Q_*^{-1}c.
\]
In particular, $n\Var(c^\top\widehat\gamma_n)\to s^2$.
Raw central $(1-\alpha)$ posterior intervals have limiting coverage
\[
2\Phi\left(z_{1-\alpha/2}\sqrt{V_{\mathrm{post}}/s^2}\right)-1.
\]
A consistent estimate of $s^2/n$ applied to the GLS, posterior-mean or
posterior-median center gives asymptotically nominal Gaussian intervals.
\end{proposition}

\begin{proof}
Set $G_n=A_n^\top W_nA_n$. The working GLS estimator is exactly Gaussian,
unbiased, and has covariance
\[
G_n^{-1}A_n^\top W_n\Omega_nW_nA_nG_n^{-1}.
\]
Multiplication by $n$ gives the limit $Q_*^{-1}J_*Q_*^{-1}$, so
$\widehat\gamma_n-\gamma_0=O_P(n^{-1/2})$ and the stated sampling
limit follows. The working likelihood ratio remains exactly
\[
\frac{L_n(\gamma)}{L_n(\widehat\gamma_n)}
=\exp\{-\tfrac12(\gamma-\widehat\gamma_n)^\top
G_n(\gamma-\widehat\gamma_n)\}.
\]
The separation and local-prior argument of Theorem~\ref{thm:finite}
therefore applies unchanged: proper subspaces are exponentially excluded,
and the full-space posterior for
$h=\sqrt n(\gamma-\widehat\gamma_n)$ converges in total variation
to $N(0,Q_*^{-1})$.

To justify posterior moments, fix a neighborhood of $\gamma_0$ disjoint
from every proper supported subspace, on which all full-span component
densities are bounded above and one is bounded below. Its local evidence
is at least a positive tight multiple of
$n^{-q/2}L_n(\widehat\gamma_n)$. On this neighborhood, the eigenvalue
bound for $G_n/n$ gives a Gaussian envelope in $h$, also after multiplication
by $\|h\|^4$. Outside it, $\|h\|\ge c\sqrt n$ with probability tending
to one, and
\[
\sup_{\|h\|\ge c\sqrt n}\|h\|^4e^{-c_1\|h\|^2}
\le C e^{-c_2n}
\]
for fixed positive constants and all sufficiently large $n$. Integrating
against the proper component priors and dividing by the polynomial evidence
bound makes this tail negligible, including proper subspace components.
Thus the first two posterior moments of $h$ converge to the Gaussian
moments. This proves the posterior-mean equivalence and the normalized
posterior variance limit without requiring prior moments. Quantile
convergence gives the median equivalence. Finally the sampling Gaussian
limit gives raw coverage, and Slutsky's theorem gives the intervals with
the consistently estimated sandwich variance. No sampling second-moment
claim for the posterior mean is inferred merely from its $o_P(n^{-1/2})$
equivalence to GLS.
\end{proof}

\begin{lemma}[Affine transfer under empirical standardization]
\label{lem:standardization}
Let $S_n>0$ be data-measurable, $S_n\to S\in(0,\infty)$ in probability,
and suppose that, for some $v_z>0$,
\[
\BL\left(\Pi\{\sqrt n(T^z-\widehat\theta_n^z)\in\cdot\mid\cD_n\},
N(0,v_z)\right)\longrightarrow0
\]
in probability. Define $T_n=S_nT^z$ and
$\widehat\theta_n=S_n\widehat\theta_n^z$. Then the corresponding
conditional posterior for $\sqrt n(T_n-\widehat\theta_n)$ converges
to $N(0,S^2v_z)$ in bounded-Lipschitz distance. This is a distributional
statement; convergence of posterior variances additionally requires the
corresponding scaled posterior second moments to converge.

For the sampling expansion, suppose additionally that $\theta^z$ is fixed,
$\theta=S\theta^z$, $\widehat\theta_n^z-\theta^z=o_P(1)$, and
\[
\sqrt n(S_n-S)=\Delta_n+o_P(1),\qquad \Delta_n=O_P(1).
\]
Then
\[
\sqrt n(\widehat\theta_n-\theta)
=S\sqrt n(\widehat\theta^z_n-\theta^z)+\theta^z\Delta_n+o_P(1).
\]
If $(\sqrt n(\widehat\theta_n^z-\theta^z),\Delta_n)$ has a joint
Gaussian limit $(Z,\Delta)$, the limiting sampling variance is
$S^2\Var(Z)+(\theta^z)^2\Var(\Delta)+2S\theta^z\operatorname{Cov}(Z,\Delta)$.
\end{lemma}

\begin{proof}
Conditionally on the data,
$\sqrt n(T_n-\widehat\theta_n)=S_n\sqrt n(T^z-\widehat\theta_n^z)$.
Rescaling a bounded one-Lipschitz test function changes its bounded-Lipschitz
norm by at most $\max\{1,S_n\}$, which is bounded in probability.
The assumed posterior distance therefore tends to zero after this rescaling.
The comparison Gaussian has variance $S_n^2v_z\to S^2v_z$, proving
the conditional distributional conclusion. This argument alone gives no
posterior moment convergence.

For the center, the exact identity is
\[
S_n\widehat\theta^z_n-S\theta^z
=S(\widehat\theta^z_n-\theta^z)
+\theta^z(S_n-S)
+(S_n-S)(\widehat\theta^z_n-\theta^z).
\]
The scale expansion and tightness of $\Delta_n$ imply
$S_n-S=O_P(n^{-1/2})$. Consistency of the standardized center makes
the last product $o_P(n^{-1/2})$. Multiplication by $\sqrt n$ proves
the displayed expansion. Under the stated joint limit, continuous mapping
gives the limiting variable $SZ+\theta^z\Delta$, whose variance is
the stated expression.
\end{proof}

This sampling calculation concerns a fixed standardized population target.
For empirical outcome standardization, the original-scale likelihood and
induced prior can instead be analyzed through the exact affine representation
in Lemma~\ref{uv:lem:affine}. The later original-scale theorems already
account for that transformation and require no additional scale-variance
term appended to their stated limits.

\begin{proposition}[The total effect is identified, not its decomposition]
\label{prop:gauge}
In a linear mean model $F\beta$, a linear functional $l^\top\beta$ is
identified if and only if $l^\top h=0$ for every $h\in\ker F$. If a global
effect intercept $\alpha$ and a constant effect-forest contribution $k$ enter
only through $\alpha+k$, neither component is identified, while their sum is.
For $Y_i=\alpha+k+\epsilon_i$ with independent unit-variance errors and
independent Gaussian priors of precisions $\lambda_\alpha,\lambda_k>0$,
\[
\Var(\alpha\mid Y)=
\frac{n+\lambda_k}{n(\lambda_\alpha+\lambda_k)+\lambda_\alpha\lambda_k}
\longrightarrow\frac1{\lambda_\alpha+\lambda_k},
\]
whereas
\[
\Var(\alpha+k\mid Y)=
\frac{\lambda_\alpha+\lambda_k}
{n(\lambda_\alpha+\lambda_k)+\lambda_\alpha\lambda_k}
\sim n^{-1}.
\]
\end{proposition}

\begin{proof}
Two coefficient vectors generate the same conditional mean exactly when
their difference lies in $\ker F$. A functional is therefore identified if
and only if it annihilates that kernel. For the two-component example, the
likelihood depends only on $\alpha+k$, giving the stated kernel and identified
sum. Inverting the posterior precision matrix with diagonal entries
$n+\lambda_\alpha$ and $n+\lambda_k$ and off-diagonal entries $n$ gives the
two displayed variances.
\end{proof}

The implementation therefore reports the total effect
$\taug+\tau_{\mathrm{forest}}$. A proper prior can regularize the likelihood
gauge, but it does not turn either component into a data-identified object.

\subsection{Realized and population averages}\label{sec:population}

For a target cell $(g_\star,t_\star)$, let
\[
\theta_{0,n}=\frac{1}{n_\star}\sum_{i:G_i=g_\star}
\vartheta_0(X_i,t_\star-g_\star),\qquad
n_\star=\sum_i\ind{G_i=g_\star},
\]
and let
$\theta=\E[\vartheta_0(X,t_\star-g_\star)\mid G=g_\star]$. If the effect
has a finite conditional second moment, a ratio expansion gives
\[
\theta_{0,n}-\theta
=\frac1n\sum_{i=1}^n\frac{\ind{G_i=g_\star}}{p_\star}
\{\vartheta_0(X_i,t_\star-g_\star)-\theta\}+o_P(n^{-1/2}).
\]
The composition term is orthogonal to the full-panel residual term in
\eqref{eq:full-eif}. A conditional-design posterior interval quantifies
outcome noise around $\theta_{0,n}$; a population interval must add the
composition variance in \eqref{eq:full-var}, analytically or through a
jointly coupled design bootstrap. This distinction is about the estimand and
does not require changing the DiD-BCF regression center.

\subsection{Implementation scope and candidate extensions}\label{sec:prior}

The implementation standardizes the outcome by its sample mean and
standard deviation. On the standardized outcome scale, its default prior uses
125 level trees and 125 trend trees, each with split probability hyperparameters
$(\alpha,\beta)=(0.95,2)$, maximum depth 10, and sampled leaf-variance scales
equipped with a conjugate inverse-gamma prior $\operatorname{IG}(3/2, 1/250)$
(Proposition~\ref{imp:prop:scale}). It uses 100 effect trees with
$(\alpha,\beta)=(0.25,3)$, maximum depth 5, and a fixed effect leaf variance
$0.5/100$, which is not sampled and provides stronger shrinkage on treatment
effect heterogeneity. The default observation error variance prior has shape 2
and rate 1 on the standardized outcome scale. The global intercept $\taug$ is
assigned a zero-mean, unit-variance Gaussian prior on the standardized outcome
scale ($\taug \sim \mathcal{N}(0, 1)$). Optional unit or cohort random
intercepts are time-invariant additions. The default effect forest may also
split on cohort, which enlarges the effect class from $\vartheta(x,k)$ to
$\vartheta(g,x,k)$.

The preceding primitive results use known working covariance and fixed proper
induced priors. The following extensions prove the generic finite-support
case with unknown scalar variance, empirical standardization and posterior
means, and a continuous-covariate result for growing saturated histogram
mixtures. Their feasible cluster intervals preserve the raw point estimator.
The continuous-cut results below additionally prove exact-posterior integrated
effect consistency on their fixed-capacity joint finite-tree and scalar
growing-capacity truth classes, while Section~\ref{sec:imp_extension}
establishes the matrix and quotient-space representation of the hierarchical
working posterior and characterizes Monte Carlo approximation error across
finite posterior draw budgets. For an intended posterior $\pi$ and tree-space
proposal kernel $q$, exact Metropolis--Hastings invariance evaluates the full
ratio $\pi(H')q(H',H)/\{\pi(H)q(H,H')\}$, accounting for tree topology priors,
boundary constraints, and forward and reverse proposal probabilities.
Random-effect variance components, smooth multivariate truth, Bernstein--von
Mises conditions for adaptive continuous-cut forests, and Markov chain mixing
rates remain avenues for further theoretical analysis.

\providecommand{\Pp}{\mathbb P}
\providecommand{\one}{\mathbf 1}
\subsection{Unknown scalar variance and empirical standardization}\label{sec:uv_extension}

The following results extend the generic finite-support raw posterior to unknown scalar variance, posterior means, and feasible cluster uncertainty. The assumptions are restrictions on the identifiable induced prior and design, rather than universal implementation guarantees.

\subsubsection{The exact experiment and its implementation boundary}

Condition on a balanced design of $n$ units and $T$ periods, with fixed $T$.
Write $m=nT$ and stack the observations as
\[
Y=A_n\gamma_0+u,\qquad u\sim N_m(0,v_0I_m),\qquad v_0>0,
\qquad Q_n=A_n^\top A_n/n\longrightarrow Q>0.
\]
The dimension $q$ of $\gamma$ is fixed. Its coordinates are the identifiable
structured conditional means on fixed finite covariate support, including the
total treatment effect, after removing the baseline and intercept/forest
gauges. The target is $c_n^\top\gamma$, with bounded design-measurable
$c_n\to c\ne0$. For an effect contrast, $c_n^\top r=0$, where
$A_nr=\mathbf1_m$ is the constant-outcome direction.

The forest prior induces a finite union of distinct linear mean subspaces
$L_j\subseteq\R^q$. Every $L_j$ contains $r$, and at least one equals
$\R^q$. Forest configurations with the same mean subspace can be grouped.
The truth is generic relative to this collection:
\[
\gamma_0\notin\bigcup_{j:L_j\ne\R^q}L_j.
\]
The conditional priors on $(\gamma,v)$ are proper and supported on
$L_j\times(0,\infty)$. Their weights are bounded below for at least one
full-span component, in probability. Within every full-span component, the
joint prior density $\rho_{j,n}$ converges locally uniformly in probability,
on a neighborhood of $(\gamma_0,v_0)$, to a finite continuous density
$\rho_j$ strictly positive there. There are finitely many such components.
These assumptions permit an empirical-Bayes prior, but exclude an increasingly
concentrated or exponentially downweighted full-span component.

This specification is nonempty. For binary $X$, $T=8$, and one cohort adopting
at period four, the normalized structured mean has dimension $28$. Trees of
depth two, four, and four span the level, trend, and cell-effect tables.
Specifically, at each of the two covariate values there are two cohort levels,
seven normalized common time effects, and five post-treatment cell effects,
so $q=2(2+7+5)=28$. A level tree first splits on $X$ and then on the binary
cohort label. A trend tree first splits on $X$ and then uses three balanced
splitting levels to separate the eight periods. An effect tree first splits
on $X$ and then uses at most three splitting levels to separate event times
$0,1,2,3,4$. The remaining trees can be stumps. The implemented level/trend
depth caps are 10 and the effect depth cap is 5, with respective tree counts
125, 125, and 100, so these configurations are supported when the discrete
separating cutpoints are available. Positive limiting group--$X$ frequencies
eventually satisfy the minimum leaf size of five. The raw table has
$4+16+10=30$ coefficients and two baseline gauges; its observation map has
rank 28. The separate global treatment intercept is redundant with constant
effect leaves and adds no identifiable dimension.
Direct rational elimination on this
observation map confirms its full identifiable rank.
Generic coefficients exclude
only finitely many proper subspaces. A positive Gaussian density on the
identifiable coordinates and a proper inverse-gamma variance satisfy the
prior assumptions. This witness is not a continuous-covariate argument.

\subsubsection{Standardization as a prior transformation}

Let $M_n=\overline Y$, $S_n>0$ be the empirical outcome standard deviation,
and $Z=(Y-M_n\mathbf1_m)/S_n$. Suppose $M_n\to M$ and
$S_n\to S\in(0,\infty)$ in probability. Conditional on the observed data,
the original and standardized coordinates are related by
\[
\gamma=M_nr+S_n\gamma^z,\qquad v=S_n^2v^z.
\]
The working standardized likelihood is Gaussian, and its noise prior is
$v^z\sim\operatorname{IG}(a,b)$ with $a,b>0$, in the convention
$p(v^z)\propto(v^z)^{-a-1}\exp(-b/v^z)$. The implemented default prior specifies $(a,b)=(2,1)$.

\begin{lemma}[Exact original-scale representation]\label{uv:lem:affine}
For each realized data set the transformed standardized posterior equals the
posterior obtained from the original-scale Gaussian likelihood and the
transformed, data-dependent prior. In a component of dimension $d_j$, its
density with respect to subspace volume and $dv$ is
\[
\rho_{j,n}(\gamma,v)=S_n^{-(d_j+2)}
\rho_j^z\left((\gamma-M_nr)/S_n,v/S_n^2\right).
\]
In particular the original-scale variance prior is
$\operatorname{IG}(a,bS_n^2)$. If $\rho_j^z$ is continuous and strictly
positive near the limiting transformed truth, the transformed densities
satisfy the local uniform condition above.
\end{lemma}
\begin{proof}
The standardized likelihood evaluated at the transformed parameters is
$S_n^m$ times the original likelihood. For a fixed data set this factor is
constant in the parameters and component index and cancels on normalization.
Translation by $M_nr$ preserves $L_j$, while scaling its $d_j$ coordinates
has Jacobian $S_n^{d_j}$; the variance change has Jacobian $S_n^2$.
This proves the density formula and the inverse-gamma rate transformation.
On a compact neighborhood of the truth and an event where $M_n,S_n$ lie
near their limits, the transformed arguments lie in a compact subset of the
continuity region. Uniform continuity there proves local uniform convergence.
\end{proof}

Gaussian leaf and intercept priors push forward to positive Gaussian densities
on any full-span identifiable space. Integrating proper inverse-gamma leaf
scales preserves local continuity when the inverse covariance moments needed
to dominate these densities are finite. For a fixed full-rank leaf map and
fixed-shape inverse-gamma scales this follows by bounding its induced minimum
eigenvalue below by a positive constant times the smallest leaf scale; all
negative moments of these scales are finite. This verifies the density
condition for a stabilized finite library of the stated Gaussian forest
priors. The claim does not verify changing cutpoint-library probabilities or
the stationary distribution of an implementation's transition kernel.

\subsubsection{Joint mean and variance limit}

Define
\[
\widehat\gamma_n=(A_n^\top A_n)^{-1}A_n^\top Y,
\quad R_n=\|Y-A_n\widehat\gamma_n\|^2,
\quad \widehat v_n=R_n/m.
\]
\begin{theorem}[Unknown-variance raw BvM]\label{uv:thm:unknown}
Under the stated finite-support design, genericity, and prior assumptions,
the posterior probability of the full mean space tends to one. In total
variation, in sampling probability,
\[
\Pi\left(\sqrt n
\begin{pmatrix}\gamma-\widehat\gamma_n\\v-\widehat v_n\end{pmatrix}
\in\cdot\mid Y\right)
\longrightarrow
N_{q+1}\left(0,
\begin{pmatrix}v_0Q^{-1}&0\\0&2v_0^2/T\end{pmatrix}\right).
\]
The first two posterior moments of $\sqrt n(\gamma-\widehat\gamma_n)$
also converge to those of the displayed Gaussian. Consequently both the
posterior mean and median of the realized-design total effect equal
$c_n^\top\widehat\gamma_n+o_P(n^{-1/2})$, its posterior variance multiplied
by $n$ converges to $v_0c^\top Q^{-1}c$, and its central intervals have
asymptotically nominal coverage. These are exact-posterior results, not
claims about fixed-length MCMC.
\end{theorem}

\begin{proof}
\emph{Separation.} For any $\gamma$, orthogonal least-squares decomposition
gives the exact identity
\[
\|Y-A_n\gamma\|^2=R_n+
(\gamma-\widehat\gamma_n)^\top A_n^\top A_n
(\gamma-\widehat\gamma_n).
\]
For each proper supported subspace the minimum second term divided by $n$
converges in probability to the positive squared $Q$-distance $\delta_j$
from $\gamma_0$ to that subspace. Also $R_n/n\to Tv_0$.
Maximizing over $v>0$, the likelihood ratio of its best fit to the
unrestricted maximum is
\[
\left(1+\frac{n\delta_j+o_P(n)}{R_n}\right)^{-m/2}
=\exp\left\{-\frac{nT}{2}
\log\left(1+\frac{\delta_j}{Tv_0}\right)+o_P(n)\right\}.
\]
Every proper component integral is at most this maximum because its prior
is proper. In a full component, local positivity gives an integrated
likelihood at least the unrestricted maximum times
$n^{-(q+1)/2}$ times a positive factor bounded away from zero in probability.
To see this, integrate on a bounded $(q+1)$-dimensional $n^{-1/2}$ ball
about $(\widehat\gamma_n,\widehat v_n)$ and use the local expansion below.
Finite many components and a positive full-component weight imply posterior
exclusion of every proper subspace.

\emph{Local calculation.} Put $h=\sqrt n(\gamma-\widehat\gamma_n)$,
$k=\sqrt n(v-\widehat v_n)$. The exact log likelihood ratio is
\[
-\frac{m}{2}\left\{\log(v/\widehat v_n)
+\widehat v_n/v-1\right\}-\frac{h^\top Q_nh}{2v}.
\]
Taylor expansion uniformly on bounded $(h,k)$ sets gives
\[
-\frac{h^\top Qh}{2v_0}-\frac{Tk^2}{4v_0^2}+o_P(1).
\]
There is no first-order cross term. The prior density is uniformly its
positive value at the truth plus $o_P(1)$ on these sets, even though that
prior is computed from the data.

\emph{Tails and normalization.} On a fixed small neighborhood of the truth,
the exact ratio is bounded above by $\exp\{-C(\|h\|^2+k^2)\}$ on
events whose probability tends to one, because $Q_n$ has eigenvalues bounded
away from zero and $v$ stays in a compact subset of $(0,\infty)$.
The prior densities are uniformly bounded there. Outside that neighborhood,
the supremum likelihood ratio is exponentially small in $n$: the scalar
function $\log x+x^{-1}-1$ is positive off $x=1$ and diverges at zero
and infinity; with $v$ bounded near $v_0$, a fixed displacement of $\gamma$
instead gives a positive quadratic loss. Prior propriety bounds the
remaining integral by that supremum. Dividing by the polynomial lower
normalizer, then increasing a fixed radius in $(h,k)$, establishes tight
Gaussian tails. The local expansion therefore yields total-variation
convergence in each full component and hence their finite mixture.

\emph{Moment control.} Total variation alone is insufficient for the mean
claim, so we check uniform integrability separately. On the local
neighborhood, multiply the Gaussian tail bound by $\|h\|^4$; its integral
still tends to zero as the radius increases. Outside that neighborhood,
the supremum of $\|h\|^4$ times the exact likelihood ratio is exponentially
small in $n$. Indeed, with $v$ in a compact interval, a displaced mean gives
$\|h\|^4e^{-C\|h\|^2}$ with $\|h\|^2$ of order at least $n$.
When $v$ leaves that interval, maximization over $h$ bounds the mean factor
by a constant times $v^2$, while the variance factor is
$\exp[-m\{\log(v/\widehat v_n)+\widehat v_n/v-1\}/2]$.
The latter dominates $v^2$ uniformly on both variance tails for sufficiently
large $m$. Prior propriety and the polynomial normalizer therefore control
the fourth moment outside the local neighborhood, including proper mean
subspaces. This gives uniform integrability of the first two scaled mean
moments and proves their convergence.

\emph{Coverage.} Conditional on the design,
$\widehat\gamma_n-\gamma_0$ is exactly Gaussian with covariance
$v_0(A_n^\top A_n)^{-1}$. It is independent of $R_n/v_0$, which has
$\chi^2_{m-q}$ distribution. These facts give the stated matching sampling
variance for the contrast. Convergence of posterior quantiles to those of
the nondegenerate Gaussian gives the median and interval claims.
\end{proof}

\subsubsection{Correlated errors and inference without replacing the center}

Partition $A_n$ into $T\times q$ unit blocks $A_i$. Retain the finite
mean spaces, genericity, and prior conditions, but now let the true errors
$u_i$ be independent across units, mean zero conditional on the design,
with covariance $\Omega_i$. Assume uniformly bounded $A_i$, uniformly
bounded fourth error moments, and
\[
v_* = \lim_n\frac1{nT}\sum_i\operatorname{tr}(\Omega_i)>0,
\qquad J=\lim_n\frac1n\sum_i A_i^\top\Omega_i A_i>0.
\]
Use the same Gaussian independence working likelihood with unknown $v$.
The prior local condition is now imposed at $(\gamma_0,v_*)$.

\begin{corollary}[Working posterior and feasible cluster variance]
\label{uv:cor:cluster}
The joint posterior limit in Theorem~\ref{uv:thm:unknown} holds with $v_*$
in place of $v_0$, centered at the same OLS and residual-variance estimators.
Its mean and median conclusions also hold. For the realized-design contrast,
the limiting posterior and sampling variances, in units of $1/n$, are
\[
V_{\mathrm{post}}=v_*c^\top Q^{-1}c,
\qquad V_{\mathrm{samp}}=c^\top Q^{-1}JQ^{-1}c.
\]
The unmodified raw central $(1-\alpha)$ intervals therefore have limiting
coverage
\[
2\Phi\left(z_{1-\alpha/2}
\sqrt{V_{\mathrm{post}}/V_{\mathrm{samp}}}\right)-1.
\]
Let $\overline\gamma_n$ be the exact posterior mean and set
\[
\widehat u_i=Y_i-A_i\overline\gamma_n,
\quad \widehat J_n=\frac1n\sum_i A_i^\top\widehat u_i\widehat u_i^\top A_i,
\quad \widehat V_n=c_n^\top Q_n^{-1}\widehat J_nQ_n^{-1}c_n.
\]
Then $\widehat V_n\to V_{\mathrm{samp}}$ in probability and
$c_n^\top\overline\gamma_n\pm z_{1-\alpha/2}\sqrt{\widehat V_n/n}$
has asymptotically nominal coverage. This last interval changes uncertainty
quantification, not the point estimator; it is not the raw posterior interval.
\end{corollary}
\begin{proof}
Bounded design blocks and the error moment conditions imply the independent
unit laws of large numbers and Lindeberg central limit theorem. Consequently
$\widehat\gamma_n-\gamma_0=O_P(n^{-1/2})$,
$R_n/(nT)\to v_*$, and
\[
\sqrt n(\widehat\gamma_n-\gamma_0)
\ \rightsquigarrow\ N(0,Q^{-1}JQ^{-1}).
\]
Indeed, the unprojected squared residual average converges to $v_*$;
subtracting the fixed-dimensional fitted-noise projection changes $R_n$ by
$O_P(1)$. The working likelihood decomposition, separation, local expansion,
and posterior moment bounds in the theorem depend on these consistency
facts, not on a Gaussian true law. They thus prove the same posterior limit
with $v_*$, while the displayed unit central limit theorem gives its actual
sampling covariance. Quantile convergence yields the coverage expression.

For feasibility, write $d_n=\overline\gamma_n-\gamma_0=O_P(n^{-1/2})$.
Expanding $\widehat u_i=u_i-A_id_n$, bounded $A_i$ and
$n^{-1}\sum_i\|u_i\|=O_P(1)$ bound the cross terms in $\widehat J_n$
by $O_P(\|d_n\|)$ and the quadratic terms by $O_P(\|d_n\|^2)$.
The uniform fourth moment bound implies
$n^{-1}\sum_i A_i^\top(u_iu_i^\top-\Omega_i)A_i=o_P(1)$ entrywise.
Thus $\widehat J_n\to J$. The posterior-mean equivalence to OLS and
Slutsky's theorem prove the interval statement.
\end{proof}

This corollary supplies a feasible variance estimator in a fixed-dimensional
identifiable regime. It does not license using estimated forest leaf residuals
and a selected partition as though they were a fixed-dimensional design when
continuous-covariate partitions remain adaptive. That extension is separate.

\providecommand{\Pp}{\mathbb P}
\providecommand{\one}{\mathbf 1}
\subsection{Continuous covariates and growing histogram mixtures}\label{sec:cc_extension}

This extension concerns a growing, saturated histogram subclass of structured forests. It preserves the raw total-effect average. The default fixed tree budgets, adaptive partition localization, and induced-prior precision bounds require separate verification.

\subsubsection{Experiment, target, and a nonempty regime}

Observe independent identically distributed units $(X_i,G_i,Y_i)$, with
$X_i\in[0,1]^p$, a fixed finite cohort set including a never-treated cohort,
and $Y_i\in\R^T$, with fixed $T$. Each cohort has a pre-treatment period.
After setting the first common time effect to zero, let $\beta_0(x)\in\R^d$
collect all cohort levels, common time effects, and post-treatment cell
effects. Fixed incidence matrices $B_g$ give the model
\[
Y_i=B_{G_i}\beta_0(X_i)+u_i,\qquad
u_i\mid X_i,G_i\sim N_T(0,v_0I_T),\qquad v_0>0.
\]
Explicitly, write the cohort set as
$\mathcal G\subseteq\{2,\ldots,T,\infty\}$, with $\infty\in\mathcal G$,
and order the coordinates as
\[
\beta(x)=\big((a_g(x))_{g\in\mathcal G},
(b_t(x))_{t=2}^T,
(\tau_{g,t}(x))_{g\in\mathcal G\setminus\{\infty\},\ t\ge g}\big).
\]
Row $t$ of $B_g$ has a one in coordinate $a_g$, a one in $b_t$
when $t\ge2$, and a one in $\tau_{g,t}$ when $g<\infty$ and
$t\ge g$, with zeros elsewhere. For a vector in the kernel of the
stacked incidence matrices, the never-treated period-1 row first gives
$a_\infty=0$; its remaining rows give all $b_t=0$. Each treated
cohort's period-1 row then gives $a_g=0$, and its post-treatment rows
give every $\tau_{g,t}=0$. This proves full column rank. The vector
having ones in all level coordinates and zeros elsewhere yields a
constant outcome in every cohort and period. The selector $c_{g,t}$
is the coordinate vector for $\tau_{g,t}$.

The coefficient $c\in\R^d$ selects one total effect, or a fixed linear
combination of effects averaged over a specified target cohort $g_*$.
Write $D_i=\one\{G_i=g_*\}$, $p_*=P(G=g_*)$, $p_n=n^{-1}\sum_iD_i$,
and
\[
\theta_{0,n}=\frac{1}{np_n}\sum_iD_i c^\top\beta_0(X_i),
\qquad \theta_0=E[c^\top\beta_0(X)\mid G=g_*].
\]
Statements are made on $p_n>0$, an event of probability tending to one.
The first target is conditional on the realized treated covariates.
Throughout, $P_nf=n^{-1}\sum_i f(X_i,G_i,Y_i)$ denotes empirical averaging.

\noindent\textbf{Assumption ledger.}
The density of $X$ is bounded above and below by positive constants.
The probabilities $e_g(x)=P(G=g\mid X=x)$ are bounded below by a positive
constant. The function $\beta_0$ is bounded and $\alpha$-H\"older, with
$0<\alpha\le1$. Define
\[
C_g=B_g^\top B_g,\quad Q(x)=\sum_g e_g(x)C_g,
\quad l(x)=e_{g_*}(x)c/p_*,\quad r(x)=Q(x)^{-1}l(x).
\]
Assume $r$ is bounded and $\zeta$-H\"older, $0<\zeta\le1$.
For example, $\zeta$-H\"older cohort probabilities imply this property.
The never-treated observations and the pre-treatment period make the stacked
$B_g$ full column rank; overlap then bounds the eigenvalues of $Q(x)$
away from zero. Every constant in this note may depend on the fixed
incidence matrices, overlap, density, smoothness, and prior bounds.

For each $n$, use a deterministic finite family $\mathcal P_n$ of
$M_n$ measurable partitions of $[0,1]^p$, specified without the outcomes
or the sampled covariates. Each partition has $K_n$ cells, cell diameter
at most $h_n\to0$, and cell probabilities between $a/K_n$ and $A/K_n$
for fixed $0<a<A<\infty$. On every cell the model includes all $d$
structured coefficients. Thus its identifiable dimension is $dK_n$.
Set $t_n=\log(2nM_n)$ and impose
\begin{equation}\label{cc:eq:rates}
\sqrt n h_n^{\alpha+\zeta}\to0,\qquad
h_n^{\min(\alpha,\zeta)}\sqrt{t_n}\to0,\qquad
K_nt_n/\sqrt n\to0.
\end{equation}
The prior assigns arbitrary nonnegative weights summing to one to these
partitions. Conditional on a partition, its identifiable coefficient vector
$b\in\R^{dK_n}$ has prior $N(\mu_H,\Lambda_H^{-1})$, independently of
$v\sim\operatorname{IG}(a_n,b_n)$, in the density convention
$v^{-a_n-1}\exp(-b_n/v)$. With probability tending to one, all coordinate
means are bounded, all eigenvalues of $\Lambda_H$ lie in a common compact
subset of $(0,\infty)$, and $a_n,b_n$ lie in a common compact subset of
$(0,\infty)$. These bounds hold uniformly over $H\in\mathcal P_n$.
The parameters and weights may be data dependent, provided these bounds
hold. All posterior statements and conditional Gaussian formulas refer
to the conditional posterior kernel computed with that realized prior.

This model is nonempty. Take $p=1$, a positive bounded covariate density,
Lipschitz cohort probabilities bounded away from zero, and Lipschitz
structured means. Equal-width partitions with $K_n\asymp n^{1/3}$,
or polynomially many comparable regular partitions, satisfy
\eqref{cc:eq:rates}. Fixed positive Gaussian precisions and
$\operatorname{IG}(2,1)$ satisfy the prior conditions.
More generally, when $K_n\asymp h_n^{-p}$ and $M_n$ is polynomial,
one may take $K_n\asymp n^\kappa$ with
\[
\frac{p}{2(\alpha+\zeta)}<\kappa<\frac12,
\]
provided $\alpha+\zeta>p$.

\noindent\textbf{Dimensional feasibility of the full rate ledger.}
The restriction $\alpha+\zeta>p$ is necessary for \eqref{cc:eq:rates},
including for non-grid measurable partitions. Let $\bar f$ bound the
covariate density above and let $\omega_p$ be the volume of the unit ball
in $\R^p$. A cell of diameter at most $h_n$ lies in a ball of radius
$h_n$, so $P(X\in I)\le\bar f\omega_p h_n^p$. Summing over the cells
gives $K_nh_n^p\ge1/(\bar f\omega_p)$. If $\alpha+\zeta\le p$, then
for all sufficiently large $n$,
\[
\sqrt n h_n^{\alpha+\zeta}
\ge\sqrt n h_n^p
\ge\frac{\sqrt n}{\bar f\omega_p K_n}
\longrightarrow\infty,
\]
because the third rate implies $K_n/\sqrt n\to0$.
This contradicts the first rate. Since $\alpha,\zeta\le1$, the full
ledger is therefore empty for $p\ge2$. This restriction concerns the
stated theorem conditions; it does not assert impossibility under other
conditions or for other methods. Corollary~\ref{cc:cor:randomized} removes
the first rate and admits every fixed $p$.

\subsubsection{Functional bias before the posterior calculation}

For a cell $I$ of $H$, define
\begin{align*}
Q_I&=E[C_G\mid X\in I],\qquad e_I=P(G=g_*\mid X\in I),\\
b_I^\circ&=Q_I^{-1}E[C_G\beta_0(X)\mid X\in I],\\
r_I&=Q_I^{-1}e_Ic/p_*.
\end{align*}
These are population projections in the information inner product
\[
\langle f,g\rangle_Q=E[f(X)^\top Q(X)g(X)].
\]
Let $b_H^\circ(x)=b_I^\circ$ and $r_H(x)=r_I$ on $I$.

\begin{lemma}[Population projection bias]\label{cc:lem:population}
The raw population average of $b_H^\circ$ satisfies
\[
E[l(X)^\top\{b_H^\circ(X)-\beta_0(X)\}]
=-\langle r-r_H,\beta_0-b_H^\circ\rangle_Q.
\]
Uniformly over the family, its absolute value is bounded by
$C h_n^{\alpha+\zeta}$, while
$\sup_x\|\beta_0(x)-b_H^\circ(x)\|\le C h_n^\alpha$
and $\sup_x\|r(x)-r_H(x)\|\le C h_n^\zeta$.
\end{lemma}
\begin{proof}
The cell normal equations give
$E[Q(X)(\beta_0-b_H^\circ)\mid X\in I]=0$ and
$Q_Ir_I=E[Q(X)r(X)\mid X\in I]$. Thus $b_H^\circ$ and
$r_H$ are the corresponding orthogonal projections. Subtracting the
projection of $r$ from the inner product with the mean residual proves
the identity. For any $x\in I$,
\[
b_I^\circ-\beta_0(x)
=Q_I^{-1}E[Q(X)\{\beta_0(X)-\beta_0(x)\}\mid X\in I].
\]
Uniform eigenvalue bounds and the cell diameter give the first sup-norm
bound. The identical argument with $r$ gives the other. Cauchy--Schwarz
in the information norm proves the bias bound.
\end{proof}

For the empirical design, let $n_I=\sum_i\one\{X_i\in I\}$ and
\[
\widehat Q_I=\frac1{n_I}\sum_{i:X_i\in I}C_{G_i},\quad
\widehat e_I=\frac1{n_I}\sum_{i:X_i\in I}D_i,\quad
\widehat r_I=\widehat Q_I^{-1}\widehat e_Ic/p_n.
\]
Define $F_H$ by its unit blocks, so the rows for a unit in $I$ multiply
the cell coefficient by $B_{G_i}$. Let $G_H=F_H^\top F_H$, and let
$L_H$ have cell blocks $(n_I\widehat e_I/(np_n))c$.
The posterior target is exactly $T_H=L_H^\top b$.
On the joint posterior over $(H,b,v)$, denote this scalar target by
$\mathcal T:=T_H$; the symbol $T$ continues to denote the panel length.
Denote the unregularized mean fit to the true conditional means by
$\widetilde b_H=G_H^{-1}F_H^\top f_0$, where
$f_{0,i}=B_{G_i}\beta_0(X_i)$.

\begin{lemma}[Empirical functional bias and design weights]\label{cc:lem:empirical}
With probability tending to one, all empirical cell matrices are invertible,
$n_I\asymp n/K_n$, and the eigenvalues of $G_H$ are between
$c n/K_n$ and $C n/K_n$, uniformly over $H$ and its cells. Furthermore,
\begin{align}
\sup_H|L_H^\top\widetilde b_H-\theta_{0,n}|
&=O_P\left(h_n^{\alpha+\zeta}
+h_n^\alpha\sqrt{t_n/n}+h_n^\alpha K_nt_n/n\right),\label{cc:eq:empbias}\\
\sup_{H,i}\|\widehat r_{H,I(i)}-r(X_i)\|
&=O_P\left(h_n^\zeta+\sqrt{K_nt_n/n}+n^{-1/2}\right).\label{cc:eq:weights}
\end{align}
\end{lemma}
\begin{proof}
Bernstein bounds for bounded variables, applied to the finitely many cell
counts, entries of $C_G$, and cohort indicators, and a union bound over at
most $M_nK_n$ cells give deviations of order $\sqrt{K_nt_n/n}$ for
conditional cell averages. In particular $n/K_n\gg t_n$, so counts are
uniformly comparable to their expectations and the inverses are bounded.
This proves the eigenvalue claim. Matrix inversion on the resulting compact
positive-definite set, $p_n-p_*=O_P(n^{-1/2})$, and
Lemma~\ref{cc:lem:population} prove \eqref{cc:eq:weights}.

For the bias, set $R_{H,i}=\beta_0(X_i)-b_{H,I(i)}^\circ$ and
$\bar r_I=(p_*/p_n)r_I$. Exact normal equations yield
\[
L_H^\top\widetilde b_H-\theta_{0,n}
=\frac1n\sum_i
\left(\widehat r_{I(i)}^\top C_{G_i}-D_i c^\top/p_n\right)R_{H,i}.
\]
Define $\xi_H=(r_{I(X)}^\top C_G-Dc^\top/p_*)
\{\beta_0(X)-b_{H,I(X)}^\circ\}$. The last display equals
\[
\frac{p_*}{p_n}\{(P_n-P)\xi_H+P\xi_H\}
+\frac1n\sum_{I\in H}(\widehat r_I-\bar r_I)^\top
\sum_{i:X_i\in I}C_{G_i}R_{H,i}.
\]
The population term is the bias in Lemma~\ref{cc:lem:population}, and
$|\xi_H|\le C h_n^\alpha$. Bernstein and a union bound give
$\sup_H|(P_n-P)\xi_H|=O_P(h_n^\alpha\sqrt{t_n/n})$.
Conditional cell normal equations give $E[C_GR_H\mid X\in I]=0$.
The same concentration calculation therefore bounds each inner sum by
$O_P(h_n^\alpha\sqrt{(n/K_n)t_n})$, uniformly over cells, and
$\widehat r_I-\bar r_I=O_P(\sqrt{K_nt_n/n})$ uniformly.
Their product is $O_P(h_n^\alpha t_n)$ per cell. Summing $K_n$
cells and dividing by $n$ proves \eqref{cc:eq:empbias}. All bounds are
uniform before the outcomes choose a partition.
\end{proof}

\subsubsection{Unknown variance and the adaptive raw posterior}

Let $\widehat b_H=G_H^{-1}F_H^\top Y$. Given $H,v,Y$, direct
Gaussian conditioning gives
\begin{align*}
E[b\mid H,v,Y]&=\widehat b_H-
v(G_H+v\Lambda_H)^{-1}\Lambda_H(\widehat b_H-\mu_H),\\
\operatorname{Var}(b\mid H,v,Y)&=v(G_H+v\Lambda_H)^{-1}.
\end{align*}
The empirical target thus has a univariate Gaussian conditional posterior,
mixed over both $H$ and $v$. The raw center is its posterior mean.

\begin{lemma}[Uniform variance concentration and moments]\label{cc:lem:variance}
Under the ledger and \eqref{cc:eq:rates}, for every $\varepsilon>0$
and every fixed positive integer $s$,
\[
\sup_H\Pi(|v-v_0|>\varepsilon\mid H,Y)\to0,
\qquad
\sup_H\left|E[v^s\mid H,Y]-v_0^s\right|\to0
\]
in probability. Also
$\sup_H\|\widehat b_H\|=O_P(\sqrt{K_n})$.
\end{lemma}
\begin{proof}
Let $P_H=F_HG_H^{-1}F_H^\top$ and $R_H=\|(I-P_H)Y\|^2$.
Projection cannot increase the squared error of the population histogram
approximation, so $\sup_H\|(I-P_H)f_0\|^2\le Cnh_n^{2\alpha}$.
Conditional on the design, $u^\top P_Hu/v_0$ is chi-squared with
$dK_n$ degrees of freedom. Its exponential-moment bound and a union
bound give $\sup_Hu^\top P_Hu=O_P(K_n+t_n)$.
The cross term is bounded by
$2\|(I-P_H)f_0\|\|u\|=O_P(nh_n^\alpha)$.
Since $\|u\|^2/(nT)\to v_0$, it follows that
$\sup_H|R_H/(nT)-v_0|\to0$.
The fitted noise coefficient vector has norm squared at most
$C(K_n/n)\sup_Hu^\top P_Hu=o_P(K_n)$;
the true fitted coefficients are bounded in norm by $C\sqrt{K_n}$
using the cell eigenvalue bounds. This proves the last assertion.

For clarity, concentration is obtained from the likelihood rather than
assumed. For a fixed $H$, its likelihood maximum is attained at
$(\widehat b_H,\widehat v_H)$, $\widehat v_H=R_H/(nT)$.
The exact likelihood ratio at $(b,v)$ is
\[
\exp\left[-\frac{nT}{2}
\left\{\log(v/\widehat v_H)+\widehat v_H/v-1\right\}
-\frac{(b-\widehat b_H)^\top G_H(b-\widehat b_H)}{2v}\right].
\]
On the event just established, integration over the coordinate box
$|b_j-\widehat b_{H,j}|\le n^{-1/2}$ and a variance interval
of radius $n^{-1/2}$ around $\widehat v_H$ bounds the evidence below
by its likelihood maximum times $\exp\{-O_P(K_n\log n+\log n)\}$,
uniformly in $H$. The Gaussian density in this box is at least
$\exp(-C K_n)$, its volume is $(2/\sqrt n)^{dK_n}$, and its
quadratic likelihood loss is bounded. The inverse-gamma density is
uniformly bounded below on the relevant compact variance interval.
Here $K_n\log n=o(n)$ follows from \eqref{cc:eq:rates}.

For $|v-v_0|>\varepsilon$, the supremum ratio is $\exp(-cn)$
uniformly in $H$ on this event. Prior propriety bounds the integral above
by that supremum. Division by the preceding subexponential lower bound
proves concentration. For $v$ on either noncompact tail, multiplication
by $v^s$ leaves an exponentially small supremum for all sufficiently
large $n$: the scalar loss diverges as $v\downarrow0$ and grows like
$\log v$ as $v\uparrow\infty$. On a compact interval, concentration
and boundedness imply convergence of $v^s$ moments. These two arguments
prove the moment assertion without requiring finite variance of the
\emph{prior} $\operatorname{IG}(2,1)$.
\end{proof}

Set
\[
w_{n,i}=\frac1n B_{G_i}r(X_i),\qquad
\theta_n^*=\theta_{0,n}+\sum_iw_{n,i}^\top u_i,
\qquad V_0=v_0 E[r(X)^\top Q(X)r(X)]>0.
\]
The positivity restriction excludes a zero target $c$.

\begin{theorem}[Continuous-covariate raw histogram-mixture BvM]
\label{cc:thm:continuous}
Under the ledger and \eqref{cc:eq:rates}, the exact posterior, with its
outcome-dependent weights over $H$, satisfies
\[
\left\|\Pi\{\sqrt n(\mathcal T-\theta_n^*)\in\cdot\mid Y,X,G\}
-N(0,V_0)\right\|_{\rm TV}\to0
\]
in sampling probability. Its first two scaled moments converge to those
of this Gaussian. In particular, the raw posterior mean and median equal
$\theta_n^*+o_P(n^{-1/2})$, $n\operatorname{Var}(\mathcal T\mid Y,X,G)\to V_0$,
and raw central intervals have asymptotically nominal coverage for
$\theta_{0,n}$. No correction of the point estimator is used.
\end{theorem}
\begin{proof}
First consider the unregularized center $L_H^\top\widehat b_H$.
Its bias relative to $\theta_{0,n}$ is \eqref{cc:eq:empbias}.
All three terms are $o_P(n^{-1/2})$ by \eqref{cc:eq:rates}.
Its noise weights are $w_{H,i}=B_{G_i}\widehat r_{H,I(i)}/n$.
Conditional on the design, the differences
$\sqrt n\sum_i(w_{H,i}-w_{n,i})^\top u_i$ are centered Gaussians
with standard deviations bounded in probability by
\[
C\{h_n^\zeta+\sqrt{K_nt_n/n}+n^{-1/2}\}.
\]
The Gaussian maximum bound, valid without independence across partitions,
multiplies this expression by at most $\sqrt{2\log(2M_n)}$.
The result tends to zero: $h_n^\zeta\sqrt{t_n}\to0$ and
$\sqrt{K_nt_n^2/n}\to0$ by \eqref{cc:eq:rates}.
Thus the unregularized centers all equal $\theta_n^*+o_P(n^{-1/2})$
uniformly, before taking any outcome-dependent mixture.

On every fixed compact variance interval, the regularization shift is
bounded by
\[
\|L_H\|\,v\|(G_H+v\Lambda_H)^{-1}\|\,\|\Lambda_H\|
\|\widehat b_H-\mu_H\|
=O_P(K_n/n),
\]
since $\|L_H\|\le C/\sqrt{K_n}$ and
$\|\widehat b_H-\mu_H\|=O_P(\sqrt{K_n})$ uniformly.
Its root-$n$ multiple tends to zero. The same bound for arbitrary $v$
is $O_P(vK_n/n)$, which also controls its posterior expectation by
Lemma~\ref{cc:lem:variance}.

For the variance, the unregularized identity is
\[
nL_H^\top G_H^{-1}L_H
=\frac1n\sum_i\widehat r_{H,I(i)}^\top C_{G_i}
\widehat r_{H,I(i)}.
\]
Uniform convergence in \eqref{cc:eq:weights} and the ordinary law of large
numbers make this tend uniformly to $E[r^\top Qr]$.
The relative operator difference between $(G_H+v\Lambda_H)^{-1}$
and $G_H^{-1}$ is $O(K_n/n)$ for compact $v$, so regularization
has no first-order variance effect. Lemma~\ref{cc:lem:variance} restricts
the variance mixture to $v\to v_0$. Univariate Gaussian densities are
continuous in total variation in their center and positive variance;
the uniform center and variance bounds therefore establish the displayed
mixture limit.

Moment convergence needs more than total variation. On a compact variance
interval the scaled conditional means tend uniformly to zero and their
scaled variances stay uniformly bounded, so conditional Gaussian fourth
moments are uniformly bounded and the first two moments converge. Away
from that interval, the operator
$(G_H+v\Lambda_H)^{-1}G_H$ is uniformly bounded for $v>0$
by the uniform condition numbers of $G_H$ and $\Lambda_H$.
Hence the conditional target center is $O_P(1)$ uniformly in $v,H$,
and its scaled fourth power is $O_P(n^2)$.
Also $n\operatorname{Var}(T_H\mid H,v,Y)\le Cv$.
The exponentially small variance-tail integrals, including their $v^2$
moments proved above, dominate these polynomial factors. This gives
uniform integrability and the mean and variance claims.

Finally, conditional on the design,
$\sqrt n(\theta_n^*-\theta_{0,n})$ is exactly Gaussian with variance
\[
\frac{v_0}{n}\sum_i r(X_i)^\top C_{G_i}r(X_i)\longrightarrow V_0.
\]
Posterior quantile convergence proves the median and coverage statements.
\end{proof}

\begin{corollary}[Constant cohort probabilities]\label{cc:cor:randomized}
If all $e_g(x)$ are constant, $r$ is constant and belongs to every saturated
histogram space. The population bias is exactly zero. All conclusions of
Theorem~\ref{cc:thm:continuous} hold after deleting the first condition in
\eqref{cc:eq:rates} and retaining the remaining ledger. For polynomial $M_n$,
$K_n\asymp h_n^{-p}\asymp n^\kappa$ with any $0<\kappa<1/2$
works for every fixed $p$ and every $0<\alpha\le1$.
\end{corollary}
\begin{proof}
Lemma~\ref{cc:lem:population} has $r-r_H=0$. In the empirical and
Gaussian-weight bounds the representer approximation term is also zero.
The remaining powers of $h_n$ only require $h_n^\alpha\sqrt{t_n}\to0$;
the empirical product and shrinkage terms are controlled by
$K_nt_n/\sqrt n\to0$. These hold for the stated polynomial choice.
\end{proof}

\subsubsection{Correlated panels and population averages at the same center}

Now replace the true conditional covariance by $\Omega_g(x)$, still
Gaussian and independent across units, with eigenvalues uniformly bounded
above and below. Retain the independence working likelihood and all mean,
partition, and prior conditions. Put
\[
v_*=E[\operatorname{tr}\Omega_G(X)]/T,\quad
J(x)=\sum_g e_g(x)B_g^\top\Omega_g(x)B_g,
\quad V_{\rm samp}=E[r(X)^\top J(X)r(X)],
\]
and $V_{\rm post}=v_*E[r^\top Qr]$.

\begin{corollary}[Feasible inference at the raw posterior mean]
\label{cc:cor:feasible}
The theorem's posterior limit and moments hold with $V_{\rm post}$,
while the sampling variance of its center is $V_{\rm samp}$.
Consequently raw central intervals have limiting realized-target coverage
\[
2\Phi\left(z_{1-\eta/2}\sqrt{V_{\rm post}/V_{\rm samp}}\right)-1.
\]
Let $\bar\theta_n$ and $\bar\beta_n(x)$ denote the exact raw posterior
mean of the target and coefficient function. Choose any prespecified
reference partition $H_0\in\mathcal P_n$, compute its design-only
$\widehat r_{H_0,I(i)}$, and set
\[
\widehat u_i=Y_i-B_{G_i}\bar\beta_n(X_i),\qquad
\widehat V_n=\frac1n\sum_i
\left\{\widehat r_{H_0,I(i)}^\top B_{G_i}^\top\widehat u_i\right\}^2.
\]
Then $\widehat V_n\to V_{\rm samp}$ and
$\bar\theta_n\pm z_{1-\eta/2}\sqrt{\widehat V_n/n}$ has
asymptotically nominal coverage for $\theta_{0,n}$.

For the population target define
\[
\widehat C_n=\frac1n\sum_i\frac{D_i}{p_n^2}
\{c^\top\bar\beta_n(X_i)-\bar\theta_n\}^2,
\qquad C_0=\frac{\operatorname{Var}(c^\top\beta_0(X)\mid G=g_*)}{p_*}.
\]
Then $\widehat C_n\to C_0$, and the same center with radius
$z_{1-\eta/2}\sqrt{(\widehat V_n+\widehat C_n)/n}$ has nominal
coverage for $\theta_0$. These are frequentist intervals using the raw
point estimator, rather than unmodified posterior credible intervals.
\end{corollary}
\begin{proof}
Uniformly bounded conditional Gaussian covariance supplies the same Gaussian
maximum bound for weight differences. The projection of the error has
quadratic-form tails bounded by those of a constant multiple of a
$dK_n$-dimensional chi-squared variable, so the variance proof now has
$R_H/(nT)\to v_*$. All posterior calculations use the unchanged
working likelihood. Conditional Gaussian noise has sampling variance
$n^{-1}\sum_i r(X_i)^\top B_{G_i}^\top\Omega_{G_i}(X_i)
B_{G_i}r(X_i)\to V_{\rm samp}$, which proves the coverage expression.

For feasibility, the cell normal equations and Gaussian maximum bound
give unregularized fitted coefficient error in empirical $L^2$ norm of
order $O_P(h_n^\alpha+\sqrt{K_nt_n/n})$ uniformly in $H$.
The posterior-mean regularization contribution in that norm is at most
$O_P(K_n/n)E[v\mid H,Y]$, uniformly. Variance tails are negligible
by the preceding proof. Convexity thus gives
$n^{-1}\sum_i\|\bar\beta_n(X_i)-\beta_0(X_i)\|^2=o_P(1)$.
The reference representer is uniformly bounded and converges uniformly
at the sampled covariates to $r$ by \eqref{cc:eq:weights}. Expanding
the squared residual score and applying Cauchy--Schwarz, empirical
coefficient consistency, and $n^{-1}\sum_i\|u_i\|^2=O_P(1)$ shows
that replacing the fitted residual and representer by $u_i$ and $r(X_i)$
changes $\widehat V_n$ by $o_P(1)$.
Uniform fourth moments give the law of large numbers for these true
squared scores. This proves $\widehat V_n\to V_{\rm samp}$.
The same empirical $L^2$ argument and bounded true coefficients prove
$\widehat C_n\to C_0$.

For the population center, the treated ratio expansion gives
\[
\sqrt n(\bar\theta_n-\theta_0)
=\frac1{\sqrt n}\sum_i\left[
r(X_i)^\top B_{G_i}^\top u_i+
\frac{D_i}{p_*}\{c^\top\beta_0(X_i)-\theta_0\}\right]+o_P(1).
\]
The two terms are uncorrelated by the conditional zero mean of $u_i$.
Their iid central limit theorem has variance $V_{\rm samp}+C_0$.
Slutsky's theorem proves both interval assertions.
\end{proof}

\begin{corollary}[The implemented person-period ATT]\label{cc:cor:personperiod}
For every treated cohort let $z_g=T-g+1$ and
$s_g=\sum_{t\ge g}c_{g,t}$, where $c_{g,t}$ selects its total effect.
For the never-treated cohort set $z_g=0$ and $s_g=0$.
Let $d_*=E[z_G]>0$, $d_n=P_nz_G$, and
\[
\theta_{0,n}^{\rm pp}=\frac{P_n[s_G^\top\beta_0(X)]}{d_n},
\qquad \theta_0^{\rm pp}=\frac{E[s_G^\top\beta_0(X)]}{d_*},
\qquad r^{\rm pp}(x)=Q(x)^{-1}\frac{\sum_g e_g(x)s_g}{d_*}.
\]
Assume the representer ledger for $r^{\rm pp}$ in place of $r$.
The preceding posterior and sampling results then hold for the raw
person-period average, with this representer and target. Its population
composition score and feasible variance estimator are
\[
\frac{s_G^\top\beta_0(X)-\theta_0^{\rm pp}z_G}{d_*},
\qquad
\widehat C_n^{\rm pp}=\frac1n\sum_i
\left\{\frac{s_{G_i}^\top\bar\beta_n(X_i)-
\bar\theta_n^{\rm pp}z_{G_i}}{d_n}\right\}^2.
\]
The within-panel noise score uses $r^{\rm pp}$; the feasible noise
variance uses its empirical reference-partition version. This target
weights each treated unit by its number of post-treatment periods.
\end{corollary}
\begin{proof}
Replace the bounded vector $D_ic$ in every population and empirical
normal equation by $s_{G_i}$, and $p_*,p_n$ by $d_*,d_n$.
The target cell block becomes
$L_{H,I}=\sum_{i:X_i\in I}s_{G_i}/(nd_n)$; the representer cell block
becomes $Q_I^{-1}E[s_G\mid X\in I]/d_*$.
These marks have finite fixed bounds and $d_n-d_*=O_P(n^{-1/2})$,
so all empirical, shrinkage, Gaussian maximum, and posterior variance
calculations are unchanged. Differentiating the population ratio, or
expanding its empirical numerator and denominator, gives the displayed
composition score. The empirical $L^2$ argument in
Corollary~\ref{cc:cor:feasible} proves consistency of its squared-score
estimator. Conditional zero error means make it orthogonal to the
noise score, so the two variances add at the unchanged raw center.
\end{proof}

\subsubsection{Transfer boundaries and fixed capacity}

Let $m_n$ denote the empirical standardization location and $S_n$ its
positive scale. Outcome standardization fits the prior ledger when $m_n$
has a finite probability limit and $S_n$ converges to a strictly positive
finite limit. In every saturated histogram model the constant-outcome
direction is available. The exact affine transformation
$b=m_nr_H^{\rm const}+S_nb^z$, $v=S_n^2v^z$ gives original-scale
Gaussian mean $m_nr_H^{\rm const}+S_n\mu_H^z$ and precision
$\Lambda_H^z/S_n^2$, which satisfy the same uniform bounds,
and turns $\operatorname{IG}(a,b)$ into
$\operatorname{IG}(a,bS_n^2)$. Here $r_H^{\rm const}$ denotes the
coefficient vector yielding an outcome constant, not the functional
representer $r$. At corresponding outcomes and parameters, the standardized
likelihood satisfies $\mathcal L_z=S_n^{nT}\mathcal L_Y$, where
$\mathcal L_Y$ is the original-scale likelihood. This factor is
parameter-constant. No separate scale variance penalty should be appended
after the original-scale calculation.

When the relevant forest components can split on the required finite
cohort and time labels, saturated histograms on tree-representable
covariate partitions can be represented with enough leaves to separate
the cells and those labels. Finite axis-aligned rectangular grids provide
examples of such covariate partitions.
The abstract histogram theorem also permits measurable cells without such
a finite tree representation; no forest support assertion is made for them.
The tree case is a function-class representability statement, conditional
on available splits and sufficient depth and leaves. Positive prior mass
would additionally require positive probabilities for the required tree
structures and neighborhoods of leaf values; it is not asserted here.
Gaussian priors induced by the actual redundant leaves
need not have the uniform identifiable precision bounds in the ledger.
The default prior also mixes coarse, unbalanced, and incomplete mean spaces;
this note does not prove it concentrates on the stated histogram class.
Conditioning on a single fitted forest would not verify that requirement.

\begin{proposition}[A quantitative fixed-capacity obstruction]
\label{cc:prop:capacity}
For $X\sim\operatorname{Unif}[0,1]$, every sum of $M$ univariate
piecewise constant trees of depth at most $D$, plus an intercept, has at
most $L=M(2^D-1)$ distinct jumps. Its squared $L^2$ error in
approximating $f(x)=x$ is at least $1/[12(L+1)^2]$.
\end{proposition}
\begin{proof}
There are at most $2^D-1$ splitting nodes per tree; taking their union
produces at most $L+1$ intervals. On an interval of length $a_j$, the
best constant is its midpoint and the integrated squared error is
$a_j^3/12$. Since $\sum_j a_j=1$, Jensen gives
$\sum_j a_j^3\ge1/(L+1)^2$. Splits in additional variables fixed at
a particular label do not increase the number of jumps in $x$.
\end{proof}

This establishes a surface approximation floor, including for a linear
effect function when all other predictors are constant. It does not prove
that every average-effect functional is biased: when its representer
belongs to the restricted space, population projection bias can be zero
even with positive surface error. Lemmas~\ref{cc:lem:population} and
\ref{cc:lem:empirical} locate the separate average-effect question. The
remaining default-sampler work is to prove appropriate posterior localization,
induced-prior energy bounds, and functional calibration for its actual
adaptive partitions, or to delimit a smaller truth class without claiming
generic continuous-surface consistency.

\begin{proposition}[A fixed-histogram average-effect counterexample]
\label{cc:prop:functionalcapacity}
Let $T=2$, $X\sim\operatorname{Unif}[0,1]$, and let $D$ indicate adoption
in period 2, with $P(D=1\mid X=x)=1/4+x/2$. Set both cohort levels to
zero, the common untreated change to $x$, and the true treatment effect
to zero. Add independent $N_2(0,v_0I_2)$ errors. For any fixed partition
$H$ into $J$ intervals of positive length, fit the saturated structured histogram
with both cohort levels, the period-2 common trend, and the period-2
effect on every interval. Its population least-squares raw treated-effect
average $\theta_H^\circ$ satisfies
\[
\theta_H^\circ
=\sum_{I\in H}\frac{|I|\operatorname{Cov}(e(X),X\mid X\in I)}
{p_*(1-e_I)}
\ge \frac{1}{12J^2}>0,
\qquad p_*=\frac12.
\]
For a fixed finite family of these partitions with $J\le J_{\max}$,
fixed proper Gaussian coefficient priors and a proper inverse-gamma
variance prior, the exact raw posterior mean has probability tending to
one of being at least $1/(24J_{\max}^2)$. Raw central credible intervals
have coverage tending to zero for the true zero target.
\end{proposition}
\begin{proof}
Each interval has positive probability, and overlap gives positive
conditional probability to both cohorts. Within a cell, the four
structured coefficients parametrize the four
cohort-period mean observations without a constraint. The least-squares
fit therefore matches their conditional means. Its effect is the cell
treated change minus the cell untreated change, namely
\[
\tau_I^\circ=E[X\mid D=1,X\in I]-E[X\mid D=0,X\in I]
=\frac{\operatorname{Cov}(e(X),X\mid X\in I)}{e_I(1-e_I)}.
\]
The target cell weight is $|I|e_I/p_*$. For an interval of length $a_I$,
$\operatorname{Cov}(e(X),X\mid I)=a_I^2/24$, so summing these
weights gives the identity. Since $1-e_I\le1$,
$\theta_H^\circ\ge\sum_I a_I^3/12\ge1/(12J^2)$ by Jensen.

For a fixed finite family, cell counts diverge, the empirical Gram
matrices divided by $n$ converge to positive-definite limits, and their
OLS coefficients converge to these population projections, uniformly
over the family. Residual variance converges to a positive finite value
in each component. The likelihood separation argument in
Lemma~\ref{cc:lem:variance} applies with fixed dimension and these possibly
different variance limits; the proper variance posterior has bounded
moments and each coefficient posterior concentrates at its OLS limit.
Conditional Gaussian formulas make its target mean differ from the
OLS target by $O_P(n^{-1})$. These statements are uniform over the
finite family, so arbitrary posterior mixture weights preserve the
strict positive lower bound. Each component puts probability tending
to one above $1/(24J_{\max}^2)$; hence the mixture's lower central
quantile is positive with probability tending to one. This proves the
coverage assertion without assuming that the mixture selects one
partition or has vanishing between-partition variance.
\end{proof}

This is an average-effect counterexample within fixed saturated interval
histograms, not a theorem about every restricted forest subspace. Its
truth satisfies the continuous mean, overlap, and Gaussian-error conditions,
but a fixed $H$ violates the shrinking-diameter rate of
Theorem~\ref{cc:thm:continuous}. A center-preserving variance adjustment
alone cannot remove this positive limiting bias.

\providecommand{\PiN}{\Pi_n}
\providecommand{\normN}[1]{\left\|#1\right\|_N}

\subsection{Continuous-cut forest consistency: likelihood and prior inputs}\label{rfcons:inputs}

The following results concern the exact full-panel posterior under the specified
proper statistical prior. They concern total effects, including the global
intercept when present. Initialized finite sampling runs, interval coverage
and random-effect covariance models require separate justification.

\subsubsection{A likelihood argument that does not require bounded leaf scales}

Condition on a design and let $N=nT$, where $T$ is fixed. The true experiment is
\[
 Y=f_0+u,\qquad u\sim N_N(0,v_0 I_N),\quad v_0>0,
 \qquad \normN{h}^2=N^{-1}\|h\|^2.
\]
The working likelihood is $L_N(f,v)=(2\pi v)^{-N/2}
\exp\{-\|Y-f\|^2/(2v)\}$, for $v>0$.
The candidate means lie in a design-measurable union of $M_n$ linear spaces
$S_H$, each of dimension at most $r_n$. Redundant leaf coordinates are allowed.
Write $k_n=r_n+1+\log M_n$. A proper prior $\PiN$ may depend on the design,
and may even depend on $Y$ if the local mass condition below holds with
probability tending to one. This last case defines an empirical-Bayes posterior
by the same normalized likelihood-times-prior formula; no claim about a
fixed-prior joint Bayesian experiment is needed.

\begin{lemma}[Uniform projection control]\label{rfcons:c:lem:projection}
For every $x>0$, with probability at least $1-e^{-x}$,
\[
 \max_H \|P_{\operatorname{span}(S_H,f_0)}u\|^2
 \le 4v_0\{r_n+1+\log M_n+x\}.
\]
Consequently, uniformly over candidate means, if $z=\normN{f-f_0}$,
\[
 |N^{-1}u^\top(f-f_0)|\le \Delta_n z,
 \qquad \Delta_n=O_P(\sqrt{k_n/N}).
\]
\end{lemma}
\begin{proof}
For a projection of rank $d\le r_n+1$, $\|Pu\|^2/v_0$ has a chi-square
distribution of rank $d$. Its exponential moment at $1/4$ is $2^{d/2}$.
Markov's inequality at $4(d+\log M_n+x)$ gives probability at most
$\exp[-d-\log M_n-x+(d/2)\log2]\le M_n^{-1}e^{-x}$.
A union bound proves the first assertion, including rank zero.
Since $f-f_0$ belongs to the projected space, Cauchy--Schwarz proves the
second assertion. Conditioning on the design preserves the Gaussian law;
the constants do not depend on the design or the choice of space.
\end{proof}

\begin{theorem}[Empirical contraction and posterior moments]\label{rfcons:c:thm:union}
Suppose $a_n\downarrow0$ satisfies
\begin{equation}
 k_n=o(Na_n^2),\qquad \sqrt N a_n^2\longrightarrow\infty.
 \label{rfcons:c:eq:rates}
\end{equation}
For fixed positive constants $C_b,C_\pi$, assume, with probability tending to
one, that
\begin{equation}
 \PiN\{\normN{f-f_0}\le C_ba_n,\ |v-v_0|\le a_n\}
 \ge \exp(-C_\pi Na_n^2). \label{rfcons:c:eq:mass}
\end{equation}
Then a fixed sufficiently large $L$ exists such that
\begin{equation}
 \PiN\{\normN{f-f_0}>La_n\text{ or }|v-v_0|>La_n\mid Y\}
 \longrightarrow0
 \quad\text{in }P_0\text{-probability}.
 \label{rfcons:c:eq:contraction}
\end{equation}
Also
\begin{equation}
 \int\normN{f-f_0}^2\,d\PiN(f,v\mid Y)=O_P(a_n^2).\label{rfcons:c:eq:moments}
\end{equation}
In particular the posterior mean of the prediction vector exists with
probability tending to one and has empirical squared error $O_P(a_n^2)$.
These conclusions require no exponential prior tail bound or prior second
moment for the coefficient hierarchy.
\end{theorem}
\begin{proof}
Put $q=N^{-1}\|u\|^2$, $c=N^{-1}u^\top(f-f_0)$, and $z=\normN{f-f_0}$.
Lemma~\ref{rfcons:c:lem:projection} and \eqref{rfcons:c:eq:rates} give $|c|\le\Delta_n z$ with
$\Delta_n=o_P(a_n)$. The chi-square mean and variance give
$q-v_0=O_P(N^{-1/2})=o_P(a_n^2)$. All ensuing bounds hold on events whose
probabilities tend to one. The exact identity is
\begin{equation}
 {1\over N}\log{L_N(f,v)\over L_N(f_0,v_0)}
 =-{1\over2}\log(v/v_0)-{q+z^2-2c\over2v}+{q\over2v_0}.\label{rfcons:c:eq:lr}
\end{equation}
For $z\le C_ba_n$ and $|v-v_0|\le a_n$, Taylor expansion of
$\log(v/v_0)+v_0/v-1$ on a fixed compact neighborhood of $v_0$, together
with $|c|\le\Delta_n z$, bounds \eqref{rfcons:c:eq:lr} below by $-C_1a_n^2$.
The additional term from $q-v_0$ is
$(q-v_0)(1/v_0-1/v)/2=o_P(a_n^2)$ uniformly on this set.
Thus the likelihood-ratio denominator is at least
\begin{equation}
 D_N:=\int{L_N(f,v)\over L_N(f_0,v_0)}d\PiN
 \ge\exp[-C_DNa_n^2],\qquad C_D=C_\pi+C_1+1.\label{rfcons:c:eq:denom}
\end{equation}

For $z\ge La_n$, eventually $\Delta_n\le z/4$, so the residual average
$R=q+z^2-2c$ is at least $q+z^2/2$. Profiling \eqref{rfcons:c:eq:lr} over all $v>0$ yields
\begin{equation}
 \sup_{v>0}{L_N(f,v)\over L_N(f_0,v_0)}
 \le \exp\!\left[{N\over2}\{q/v_0-1-\log(q/v_0)\}\right]
       (1+z^2/(2q))^{-N/2}.
 \label{rfcons:c:eq:profile}
\end{equation}
The exponential prefactor is $\exp[o_P(Na_n^2)]$, and eventually
$v_0/2\le q\le2v_0$. For fixed $L$, \eqref{rfcons:c:eq:profile} is therefore at most
$\exp[-c_1NL^2a_n^2+o_P(Na_n^2)]$, uniformly over $z\ge La_n$,
with $c_1>0$ depending only on $v_0$.
No quadratic decay in $z$ is asserted for arbitrarily large $z$;
the exact logarithmic bound \eqref{rfcons:c:eq:profile} controls that range.

To control $v$, complete the square:
$z^2-2c\ge-\Delta_n^2$. Hence \eqref{rfcons:c:eq:lr} is at most
\begin{equation}
 -\tfrac12\log(v/v_0)-(q-\Delta_n^2)/(2v)+q/(2v_0).
 \label{rfcons:c:eq:varprofile}
\end{equation}
On $v\in[v_0/2,2v_0]$ this equals
$-\{\log w+1/w-1\}/2+o_P(a_n^2)$, uniformly, where $w=v/v_0$.
On this compact interval $\log w+1/w-1\ge c_2(w-1)^2$.
Thus $|v-v_0|\ge La_n$ gives the same type of exponential bound.
For the omitted tails, put $b=q-\Delta_n^2$.
The derivative of $-\log(v/v_0)/2-b/(2v)$ is $(b-v)/(2v^2)$.
Since $b\to v_0$, the upper bound \eqref{rfcons:c:eq:varprofile} increases up to $b$ and decreases
afterward. Its supremum on either tail is consequently at the corresponding
compact endpoint, where it is a fixed negative number eventually.
Integrating these uniform bounds against a probability prior and using \eqref{rfcons:c:eq:denom}
proves \eqref{rfcons:c:eq:contraction} by choosing a fixed $L$ sufficiently large.

For \eqref{rfcons:c:eq:moments}, set $b_0=1/(4v_0)$ and use \eqref{rfcons:c:eq:profile} to bound the weighted tail numerator
by the supremum of
\[
 \exp[o_P(Na_n^2)]\,z^2(1+b_0z^2)^{-N/2},\qquad z\ge La_n.
\]
The function $x(1+b_0x)^{-N/2}$ decreases for
$x\ge[b_0(N/2-1)]^{-1}$. Since $Na_n^2\to\infty$, this holds for
$x\ge L^2a_n^2$ eventually. The weighted tail numerator is at most
$L^2a_n^2\exp[-c_3NL^2a_n^2+o_P(Na_n^2)]$ for a fixed $c_3>0$.
After division by \eqref{rfcons:c:eq:denom}, it is $o_P(a_n^2)$ for sufficiently large fixed $L$.
The interior contributes at most $L^2a_n^2$. This also proves integrability
on the stated events. Jensen's inequality gives the posterior-mean claim.
\end{proof}

\subsubsection{Primitive conditions for the full-panel product forest}

Observe independent, identically distributed units $(X_i,G_i,Y_i)$,
with fixed $T$, fixed positive
integer $p$, baseline
$X_i\in[0,1]^p$, fixed finite cohorts including a never-treated cohort.
Conditional on all observed $(X_i,G_i)$, the stacked errors have law
$N_{nT}(0,v_0I_{nT})$, with $v_0>0$; thus the errors are independent of
the design as well as independent across units and periods.
Assume the marginal density of $X$ is
bounded above and below by positive constants and every cohort probability
given $X$ is at least $e_*>0$. The first observed period is pre-treatment for every treated cohort. The raw conditional mean is
\begin{equation}
 f_{gt}(x)=a_g(x)+b_t(x)+D_{gt}\tau_{gt}(x),\qquad
 D_{gt}=1\{g\ne\infty,\ t\ge g\}. \label{rfcons:c:eq:structured}
\end{equation}
The truth is bounded and each component is H\"older-$\alpha$, $0<\alpha\le1$,
in its permitted continuous features. Effect modifiers are a fixed subset of
the nuisance features; cohort and event-time dependence must match the effect
design. The initial branch permits cohort-specific effects. There is no
propensity-score estimate, random-effect block, missing period, or time-varying
covariate in this theorem.

Use three independent forests with fixed positive tree counts, an optional
Gaussian effect intercept, and the feature restrictions in \eqref{rfcons:c:eq:structured}. Conditional
on a scale, leaf coefficients are independent centered Gaussians. The two
nuisance scales have the conjugate proper law
$\operatorname{IG}(3/2,1/(2m))$ for their respective fixed tree counts $m$.
The effect scale is fixed positive, and the noise variance has a proper
$\operatorname{IG}(2,1)$ prior, on the chosen fixed outcome scale.
More generally, all these hyperparameters may stay in fixed compact subsets
of their positive domains. The leaf hierarchy is independent of the noise
variance before observing data.

At depth $d$, the unnormalized topology factor is
$\alpha_F/(1+d)^{\beta_F}$ for a split and its complement for a terminal node,
where $0<\alpha_F<1$ and $0\le\beta_F<\infty$ are fixed for each forest.
Split-feature probabilities are fixed and bounded below on permitted features.
For a continuous feature a threshold is uniform between the node's observed
extrema. Discrete time/cohort features use their numeric separating gaps.
Restrict this law to trees satisfying the depth cap and a fixed positive
minimum leaf-row count, and normalize it. Terminal nodes at the cap retain
their raw terminal factor. This is the specified intended statistical prior,
not an inference about the invariant distribution of a compiled sampler.

Set $J_n=2^{\lfloor\kappa\log_2 n\rfloor}$ and
\begin{equation}
 0<\kappa<\min\{(p+2\alpha)^{-1},(4\alpha)^{-1}\},\qquad
 D_n=p\log_2J_n+c_0,\quad a_n=J_n^{-\alpha}.\label{rfcons:c:eq:schedule}
\end{equation}
The fixed integer $c_0$ is large enough to separate the finitely many discrete
cohort, time and event-time labels before a balanced covariate partition.
For example $\kappa=1/[2(p+2)]$ satisfies the strict inequalities
simultaneously for every $0<\alpha\le1$, without using unknown smoothness.
The caps in all forests grow according to \eqref{rfcons:c:eq:schedule}. Fixed caps 10 and 5 do not
implement this sequence.

\begin{lemma}[Pattern count and local prior mass]\label{rfcons:c:lem:forestmass}
Under these conditions the observed prediction vectors lie in at most
$M_n$ linear spaces with
\[
 r_n\le C J_n^p,\qquad \log M_n\le C J_n^p\log n.
\]
With probability tending to one the joint prior assigns mass at least
$\exp[-C J_n^p\log n]$ to
$\{\normN{f-f_0}\le C_ba_n,\ |v-v_0|\le a_n\}$.
\end{lemma}
\begin{proof}
Across the fixed number of trees the depth caps bound the total number of
nodes and coefficient coordinates by $C J_n^p$. A binary tree with $s$ nodes
has at most $4^s$ possible ordered topologies. Every internal node has a
fixed finite number of feature choices and at most $N+1$ observed routing
patterns per feature. Summing over $s\le C J_n^p$ gives the stated bounds.
Continuous threshold values within one observed gap induce identical
prediction columns and belong to the same linear prediction space.

For approximation, partition each permitted continuous feature into $J_n$
equal intervals. Separate observed discrete labels first. For the effect
block, separate cohort and then the event-time labels possible for that cohort;
impossible cohort--event-time pairs need no leaf. A balanced binary tree then
represents the component's grid values with depth at most
$p\log_2J_n+c_0$ and at most $C J_n^p$ leaves. Put this approximating tree
in one tree of each block, and use zero-valued stumps for the remaining trees
and a zero effect intercept. H\"older regularity gives maximum mean error
$C J_n^{-\alpha}$ at all observations. All required label--grid cells have
probability at least $cJ_n^{-p}$. A binomial Chernoff bound and a union bound
show that their counts exceed any fixed minimum eventually, since
$nJ_n^{-p}/\log n\to\infty$.

To give the continuous-cut witness positive prior mass, perturb every
deterministic covariate threshold by less than $n^{-3}$. The probability
that any observed continuous feature lies within $n^{-3}$ of any of the
$O(J_n)$ global grid boundaries is at most $C nJ_n n^{-3}=o(1)$.
Thus, with probability tending to one, these intervals preserve all observed
routes. Every threshold interval has conditional prior mass at least
$c n^{-3}$ because the feature range is at most one. Discrete separating
intervals have fixed positive length after the corresponding labels appear.
The raw depth factors cost at most $C J_n^p\log(D_n+1)$ in negative log
mass, and feature/threshold choices cost at most $C J_n^p\log n$.
Restricting and normalizing an admissible subset of the underlying probability
law can only increase the mass of the admissible witness event.

Restrict each nuisance scale to a fixed compact interval with positive prior
mass. The conditional Gaussian variances then have fixed positive upper and
lower bounds. Perturb every bounded witness coefficient by at most
$a_n/C_m$, where $C_m$ is a fixed multiple of the total number of trees.
A bounded-center Gaussian small interval has probability at least $ca_n$.
Multiplication over $O(J_n^p)$ coordinates costs at most
$C J_n^p\log(1/a_n)+C J_n^p$. This controls the prediction error by
$C_ba_n$. The noise prior has positive continuous density near $v_0$, so its
interval contributes at least $ca_n$. Combining the bounds proves the claim.
\end{proof}

\begin{theorem}[Growing continuous-cut forest prediction]\label{rfcons:c:thm:continuous}
The exact full-panel posterior specified above satisfies \eqref{rfcons:c:eq:contraction} and \eqref{rfcons:c:eq:moments}, with
$a_n=J_n^{-\alpha}$, for the observed panel mean and noise variance.
The posterior-mean panel prediction has empirical squared error $O_P(a_n^2)$.
\end{theorem}
\begin{proof}
Lemma~\ref{rfcons:c:lem:forestmass} supplies the model count and prior mass.
The strict inequalities in \eqref{rfcons:c:eq:schedule} give
$J_n^p\log n=o(na_n^2)$ and $\sqrt n a_n^2\to\infty$.
Apply Theorem~\ref{rfcons:c:thm:union}, conditionally on the design and then
unconditionally. The design events have probability tending to one.
\end{proof}

\paragraph{Coding rank and total effects.}\label{rfcons:c:coding}
Normalize $b_1(x)=0$ and write the structured mean as $f_g(x)=B_g\beta(x)$.
The coordinate vector $\beta$ contains cohort levels, the remaining common
time effects, and all post-adoption total effects. The first period is assumed
pre-treatment for all treated cohorts. With a never-treated cohort observed
in every period, the stacked matrix with rows $B_g$ has full column rank:
zero never-treated means give its level and all common time effects zero;
each treated cohort's first-period mean then gives its level zero, and its
post-period means give all its total effects zero. Consequently
\[
 Q(x)=\sum_g e_g(x)B_g^\top B_g\succeq
 e_*\sum_g B_g^\top B_g\succeq\lambda_*I,\quad \lambda_*>0.
\]
The normalization removes a function-valued baseline gauge; it does not
remove or change the total treatment effects or their intercept/forest gauge.

\paragraph{Causal interpretation.}\label{rfcons:c:causal}
Let $Y_t(g)$ be the potential outcome under adoption at $g$ and $Y_t(\infty)$
under never adopting. Outcome consistency means $Y_t=Y_t(G)$.
For a pre-period $s<g$, no anticipation requires equality of the conditional
means of $Y_s(g)$ and $Y_s(\infty)$ given $G=g,X$.
The required conditional parallel-trend restriction is, for each $t\ge g$,
\[
 E[Y_t(\infty)-Y_s(\infty)\mid G=g,X=x]
 =E[Y_t(\infty)-Y_s(\infty)\mid G=\infty,X=x].
\]
Thus the identified contrast is
\[
 \tau_{gt,0}(x)=f_{gt,0}(x)-f_{gs,0}(x)-f_{\infty t,0}(x)+f_{\infty s,0}(x)
 =E[Y_t(g)-Y_t(\infty)\mid G=g,X=x].
\]
Its cohort-weighted and subgroup-treated averages are ATT targets.
To interpret $\int\tau_{gt,0}dP_X$ as an ATE, or its population subgroup
average as a GATE, additionally require
\[
 E[Y_t(g)-Y_t(\infty)\mid G=h,X=x]=\tau_{gt,0}(x)
\]
for every cohort $h$ in the population being averaged. Under this explicit
mean-effect transportability assumption, Theorems~\ref{rfcons:f:thm:fixedeffects} and~\ref{rfcons:s:thm:scalar} give
consistency of those targets too. Integrated CATE consistency describes a
surface loss; it is not consistency at every specified covariate point.
Shrinking-probability groups are outside the theorem.

\paragraph{Empirical standardization.}\label{rfcons:c:scaling}
Let the empirical outcome mean and standard deviation be $\bar Y_n,s_n$,
converging to finite constants with $s_n\to s_\infty>0$.
If starred variables denote standardized parameters, push the prior back by
\[
 f=\bar Y_n\mathbf1_N+s_nf^*,\quad
 \tau=s_n\tau^*,\quad v=s_n^2v^*,\quad\lambda_F=s_n^2\lambda_F^*.
\]
The effect intercept and every leaf height are multiplied by $s_n$; the
location offset is added to the nuisance sum only. Every candidate prediction
space contains $\mathbf1_N$, through the nuisance leaf indicators, so this
shift preserves the union of spaces. The standardized likelihood's Jacobian
depends on the data and $s_n$ only and cancels in posterior normalization.
An $\operatorname{IG}(a,b)$ variance becomes $\operatorname{IG}(a,s_n^2b)$;
a Gaussian coefficient variance becomes $s_n^2$ times its starred variance.
On compact events the transformed prior has bounded centers and its positive
hyperparameters remain in fixed compact intervals. The small-ball proof is
uniform on these events, including noise variance and nuisance-scale priors.
The likelihood envelope allows this outcome-dependent proper prior. This
transfers the stated results to empirical standardization without any claim
of posterior equality with a differenced-outcome model.

\subsection{A scalar smooth-truth extension}\label{rfcons:scalar}

\subsubsection{Setup and local dependencies}

Use the primitive continuous-cut panel model in Section~\ref{rfcons:inputs}, with $p=1$.
Units are independent and identically distributed, $T$ and the finite cohort
set are fixed, $X\in[0,1]$
has density bounded above and below, every cohort probability given $X$ is
at least $e_*>0$, and a never-treated cohort and an initial common
pre-treatment period are observed. The conditional mean is
\[
 f_{gt,0}(x)=a_{g,0}(x)+b_{t,0}(x)+D_{gt}\tau_{gt,0}(x).
\]
The components are bounded H\"older-$\alpha$ functions in their permitted
features, with $1/2<\alpha\le1$. Conditional on all observed $(X_i,G_i)$,
the stacked errors have law $N_{nT}(0,v_0I_{nT})$, with $v_0>0$; the
errors are independent of the design, units and periods.
Forest counts are fixed positive integers. Leaf
coefficients are Gaussian, the nuisance scales have the proper hierarchical
inverse-gamma priors specified in Section~\ref{rfcons:inputs}, the effect variance is
fixed positive, the optional effect intercept has fixed Gaussian variance,
and the noise prior is proper. All topology, feature, threshold, admissibility
and normalization rules are those of its continuous-cut model.

Set $J=2^{\lfloor\kappa\log_2 n\rfloor}$, $a=J^{-\alpha}$, and use caps
$\log_2J+c_0$, with
\[
 0<\kappa<\min\{(1+2\alpha)^{-1},(4\alpha)^{-1}\}.
\]
For example, $\kappa=1/6$ works for every $1/2<\alpha\le1$.
Write $N=nT$ and $\|z\|_N^2=N^{-1}\|z\|^2$.
Lemma~\ref{rfcons:c:lem:forestmass} and Theorem~\ref{rfcons:c:thm:continuous} give, on events tending to one,
\begin{equation}
 \begin{aligned}
 &k_n=O(J\log n)=o(Na^2),\\
 &D_N:=\int \frac{L_N(f,v)}{L_N(f_0,v_0)}d\PiN
 \ge e^{-C_DNa^2},\\
 &\int\|f-f_0\|_N^2d\PiN(\cdot\mid\mathcal D_n)=O_P(a^2).
 \end{aligned}
 \label{rfcons:s:eq:inputs}
\end{equation}
Its approximation construction supplies a deterministic regular-grid
structured function $\beta^*$ within $Ca$ in maximum coordinate norm of
the normalized truth $\beta_0$. This grid is used only to approximate the
truth in the proof; it restricts no posterior draw.
Normalize $b_1=0$, write $f_g=B_g\beta$, and retain the coding bound in Section~\ref{rfcons:c:coding} $\sum_g e_g(x)B_g^\top B_g\succeq\lambda_*I$.

\subsubsection{Uniform occupancy without a fixed posterior partition}

\begin{lemma}[All-interval count comparison]\label{rfcons:s:lem:intervals}
For a sufficiently large fixed $A$, put $h_n=A\log n/n$.
With probability tending to one, simultaneously for every interval
$I\subset[0,1]$ and cohort $g$, the following hold.
If $P_X(I)\ge h_n$, then
\begin{equation}
 \tfrac12 P(X\in I,G=g)\le n^{-1}n_{Ig}
 \le\tfrac32 P(X\in I,G=g).
 \label{rfcons:s:eq:dense}
\end{equation}
If $P_X(I)<h_n$, then $n^{-1}n_I\le C h_n$, where $C$ is fixed.
The interval and its endpoints may depend on the data or on a posterior draw.
\end{lemma}
\begin{proof}
Take a deterministic mesh with $M=\lceil n^3\rceil$ intervals.
The density upper bound gives maximum cell probability $C/M$.
The probability that any cell contains two observed covariates is at most
\[
 \binom{n}{2}\sum_{j=1}^M P_X(C_j)^2\le C n^2/M=o(1).
\]
There are at most $(M+1)^2=O(n^6)$ mesh-endpoint intervals.
For one such interval/cohort event with probability $p$, centered Bernoulli
exponential moments obey
\[
 \log E e^{s(B-p)}\le p(e^s-1-s)\le p s^2,\qquad |s|\le1.
\]
The first inequality is $\log(1+y)\le y$; the second follows by the
exponential series, since $\sum_{j\ge2}1/j!<1$.
For the sum of $n$ indicators, Chernoff's inequality with
$|s|=\min\{1,\sqrt{x/(np)}\}$ gives
\[
 P\{|n^{-1}n_{Ig}-p|>2\sqrt{px/n}+2x/n\}\le2e^{-x}.
\]
When $p=0$ the count is zero almost surely.
Use $x=10\log n$ and a union bound over mesh intervals, the fixed cohort
set and the marginal-$X$ events. All these count bounds hold with probability
tending to one.

Enclose and inscribe an arbitrary interval by mesh-endpoint intervals.
At most two mesh cells differ. On the no-collision event their empirical
mass is at most $2/n$ and their population mass is at most $2C/M$.
Thus uniformly over all intervals and cohorts,
\begin{equation}
 |n^{-1}n_{Ig}-p_{Ig}|
 \le C_1\{\sqrt{p_{Ig}\log n/n}+\log n/n\}.
 \label{rfcons:s:eq:relativeerror}
\end{equation}
The same statement holds for marginal interval counts.
For $P_X(I)\ge h_n$, overlap gives $p_{Ig}\ge e_*h_n$.
Choose $A$ so that the right side of \eqref{rfcons:s:eq:relativeerror} is at most $p_{Ig}/2$;
this is possible since its ratio to $p_{Ig}$ is at most
$C_1\{(e_*A)^{-1/2}+(e_*A)^{-1}\}$.
For $P_X(I)<h_n$, the marginal bound gives $n_I/n\le C h_n$.
These events quantify over every interval before any partition is chosen.
\end{proof}

\subsubsection{A posterior bound on Gaussian effect coefficients}

Let $\gamma_\tau$ collect every effect-leaf coefficient and the optional
effect intercept, and put $R_\tau=\|\gamma_\tau\|^2$. Its dimension depends
on the trees but is at most $C J$. Every total effect at every covariate
value is a sum of a fixed number of its coordinates, so
\begin{equation}
 \sup_{g,t,x}|\tau_{gt}(x)|^2\le C R_\tau.\label{rfcons:s:eq:height}
\end{equation}

\begin{lemma}[Effect coefficient moment]\label{rfcons:s:lem:coefficients}
Under the continuous-cut posterior above,
\begin{equation}
 \int R_\tau\,d\PiN(\cdot\mid\mathcal D_n)=O_P(Na^2).
 \label{rfcons:s:eq:coeffmoment}
\end{equation}
No bound on nuisance leaf heights or exponential nuisance-scale tail is used.
\end{lemma}
\begin{proof}
Conditional on any admissible tree collection, the effect coefficients are
independent centered Gaussians with variances bounded by a fixed positive
constant. The optional Gaussian intercept is included in this bound.
For a fixed sufficiently small $t_0>0$, their Gaussian exponential moments
therefore give
\begin{equation}
 E_{\PiN}\exp(t_0R_\tau)\le\exp(C_0J),\qquad
 \PiN(R_\tau>r)\le\exp(C_0J-t_0r),\quad r\ge0.
 \label{rfcons:s:eq:gaussianprior}
\end{equation}
This bound is uniform in the trees and hence survives their prior mixture.

Lemma~\ref{rfcons:c:lem:projection} gives $|u^\top(f-f_0)|/N\le\Delta_n z$,
where $\Delta_n^2=O_P(k_n/N)$, $q=\|u\|^2/N=v_0+O_P(N^{-1/2})$ and
$z=\|f-f_0\|_N$. Every residual average is at least $q-\Delta_n^2>0$
on events tending to one. Profile the likelihood over $v>0$ to obtain
\begin{equation}
 \sup_{f,v}\log\frac{L_N(f,v)}{L_N(f_0,v_0)}
 \le \frac N2\left\{q/v_0-1-\log(q/v_0)
                  +\log\frac q{q-\Delta_n^2}\right\}
 =O_P(1+k_n)=o_P(Na^2).
 \label{rfcons:s:eq:globalsup}
\end{equation}
Combine \eqref{rfcons:s:eq:inputs}, \eqref{rfcons:s:eq:gaussianprior} and \eqref{rfcons:s:eq:globalsup}. For every $r\ge0$, the posterior tail is bounded
by the minimum of one and
\[
 \exp\{C_DNa^2+o_P(Na^2)+C_0J-t_0r\}.
\]
Since $J=o(Na^2)$ and $Na^2\to\infty$, integrate this bound after a fixed
large multiple of $Na^2$. The interior integral is at most that multiple,
and the remaining exponential tail has integral at most $1/t_0$ eventually.
The identity $E R=\int_0^\infty P(R>r)dr$ proves \eqref{rfcons:s:eq:coeffmoment}, including finiteness
on the stated high-probability events.
\end{proof}

\subsubsection{Integrated and average-effect consistency}

\begin{theorem}[Scalar continuous-cut total effects]\label{rfcons:s:thm:scalar}
For every fixed post-adoption total-effect coordinate,
\begin{equation}
 \int\|\tau_{gt}-\tau_{gt,0}\|_{L_2(P_X)}^2
       d\PiN(\cdot\mid\mathcal D_n)
 =O_P\{J^{1-2\alpha}\log n\}=o_P(1).
 \label{rfcons:s:eq:scalarresult}
\end{equation}
The same bound holds for its all-unit empirical covariate squared error.
The exact-posterior mean total effect is consistent in integrated loss.
Each fixed averaging distribution $Q$ with $dQ/dP_X\in L_2(P_X)$ has a
consistent posterior-mean average, and its posterior squared average error
tends to zero. Realized equal-row ATT and finitely many fixed positive-
probability subgroup-treated averages satisfy the same conclusions centered
on their realized true averages. Causal interpretation uses the explicit no-anticipation, parallel-trend and transportability conditions in Section~\ref{rfcons:c:causal}.
\end{theorem}
\begin{proof}
For each posterior tree collection, take the common partition generated by
every continuous cut in every block and discrete-label branch, and by the
$J$ regular-grid boundaries used for $\beta^*$. All forest coordinates,
$\beta$ and $\beta^*$ are constant on its intervals. The total number of
cuts is at most $C J$, so the number of refined intervals is at most $C J$.
This count is specific to a scalar covariate. Call a refined interval dense
if $P_X(I)\ge h_n$, and sparse otherwise. Lemma~\ref{rfcons:s:lem:intervals} applies
to every such partition simultaneously.

On a dense interval, \eqref{rfcons:s:eq:dense} compares the empirical and population weights of
the nonnegative energies $\|B_g(\beta_I-\beta_I^*)\|^2$.
Together with coding coercivity, this gives
\begin{equation}
 \int_{\mathrm{dense}}\|\beta-\beta^*\|^2dP_X
 \le C\|f-f^*\|_N^2
 \le C\{\|f-f_0\|_N^2+a^2\}.\label{rfcons:s:eq:denseenergy}
\end{equation}
The analogous empirical coordinate bound follows from \eqref{rfcons:s:eq:dense} and the
population bound. Adding the bounded truth-approximation error preserves
the right side. This argument requires no nuisance-height bound.

The union of sparse refined intervals has population mass at most
$C J h_n$ and empirical covariate mass at most $C J h_n$.
Bounded true total effects and \eqref{rfcons:s:eq:height} imply, for either measure,
\begin{equation}
 \int_{\mathrm{sparse}}|\tau_{gt}-\tau_{gt,0}|^2
 \le C J h_n(R_\tau+1).\label{rfcons:s:eq:sparseenergy}
\end{equation}
Integrate \eqref{rfcons:s:eq:denseenergy}-\eqref{rfcons:s:eq:sparseenergy} under the posterior and use \eqref{rfcons:s:eq:inputs} and \eqref{rfcons:s:eq:coeffmoment}. The resulting
bound is
\[
 O_P(a^2+J h_nNa^2+J h_n)
 =O_P(J^{1-2\alpha}\log n)=o_P(1),
\]
since $T$ is fixed, $Na^2\to\infty$, $J$ is a positive power of $n$ up to
a factor of two, and $\alpha>1/2$.
Markov's inequality gives posterior integrated consistency, and Jensen's
inequality gives posterior-mean integrated consistency. Cauchy--Schwarz
with the fixed $L_2$ density ratio gives fixed average-effect consistency.
For realized treated or subgroup averages, Cauchy--Schwarz bounds squared
average error by the all-unit empirical squared effect error divided by the
row set's positive limiting fraction. Fixed $T$, fixed finite cohorts and
the law of large numbers ensure these denominators are bounded away from
zero with probability tending to one. Jensen gives the mean claims.
\end{proof}

The construction does not require an admissible constant multiple of the
empirical norm to control the population norm everywhere. Sparse intervals
are instead charged to a posterior Gaussian coefficient moment. The argument instead controls gap widths and posterior heights jointly.
For multiple covariates, the number and geometry of common-refinement cells
must be controlled more sharply; this proof does not establish that extension.
The rate is a sufficient consistency bound, not an asserted optimal rate.

\paragraph{Empirical standardization.}
The original-scale pushforward in Section~\ref{rfcons:c:scaling} also applies here.
On events where the empirical outcome scale lies in a positive compact
interval, the Gaussian effect variances remain uniformly bounded, so \eqref{rfcons:s:eq:gaussianprior}
is uniform. The location shift stays in the nuisance block. The interval
count event and approximation arguments are unchanged. Thus all conclusions
also apply to that specified outcome-dependent proper prior.

\subsection{Fixed-capacity continuous-cut effects}\label{rfcons:fixed}

\subsubsection{Statistical configuration and truth class}

Use the raw structured mean $f_{gt}=a_g+b_t+D_{gt}\tau_{gt}$ and the proper
continuous-cut prior specified in Section~\ref{rfcons:inputs}.
Units are independent and identically distributed, with fixed $T$, finite
cohorts including never treated and an initial common pre-treatment period,
baseline $X\in[0,1]^p$ for a fixed positive integer $p$, density bounded above
and below, and all conditional cohort probabilities at least $e_*>0$.
Conditional on all observed $(X_i,G_i)$, the stacked errors have law
$N_{nT}(0,v_0I_{nT})$, with $v_0>0$; the errors are independent of
the design, units and periods.
Forest counts, positive Gaussian effect/intercept variances, proper noise and
nuisance inverse-gamma hyperparameters, permitted features and minimum leaf
row count are fixed. Unlike the growing-capacity prediction theorem, all
depth caps here are fixed integers. The prior uses continuous numeric cuts
between observed node extrema and the same normalized admissibility rule.
The constants may include the counts 125, 125 and 100 and caps 10, 10 and 5;
those particular settings are not required.

Assume the bounded true structured mean and total effects admit one joint
fixed representation in the specified level, time and effect forest blocks,
using those counts, caps, feature restrictions and optional intercept.
Every leaf in that witness has positive probability
under the observed rows on which that forest's minimum leaf-row rule is
enforced, and every split uses a permitted
feature and separates positive-probability children. Split locations and
coefficients of this truth witness are fixed. Impossible cohort/event-time
cells need no leaf. This is a structural truth assumption, not a posterior
localization assumption. It includes covariate-constant models and
heterogeneous effects on finitely many covariate rectangles whenever their
trees fit within the declared caps. It need not include linear or arbitrary
smooth surfaces exactly.

Let $K_*$ bound the total nodes and coefficients of any candidate forest
collection; it is finite and independent of $n$. Normalize the nuisance
gauge by $b_1=0$, write $f_g=B_g\beta$, and use the coding bound in Section~\ref{rfcons:c:coding}
\[
 \sum_g e_g(x)B_g^\top B_g\succeq\lambda_*I,\qquad\lambda_*>0.
\]
The normalization preserves total effects. Set $N=nT$ and
$a_n=n^{-\eta}$ for any fixed $0<\eta<1/4$.

\subsubsection{Empirical prediction and uniform rectangular information}

\begin{lemma}[Fixed-capacity prediction inputs]\label{rfcons:f:lem:fixedinputs}
The observed prediction spaces have rank at most $C_{K_*}$ and
$\log M_n\le C_{K_*}\log n$. The proper prior assigns mass at least
$\exp(-C\log n)$ to
\[
 \{\|f-f_0\|_N\le C_ba_n,\ |v-v_0|\le a_n\}
\]
with probability tending to one. Consequently Theorem~\ref{rfcons:c:thm:union} gives
\begin{equation}
 \int\|f-f_0\|_N^2d\PiN(\cdot\mid\mathcal D_n)=O_P(a_n^2),
 \qquad D_N\ge\exp(-C_DNa_n^2)
 \label{rfcons:f:eq:fixedprediction}
\end{equation}
on events tending to one, where $D_N$ is its likelihood-ratio normalizer.
\end{lemma}
\begin{proof}
There are finitely many ordered topologies and feature choices with at most
$K_*$ nodes. Each split feature has at most $N+1$ observed routing patterns,
so counting routing patterns gives the stated rank and model count.

Use the fixed truth witness. Positive observed-row leaf probabilities and
independent units imply that every witness leaf count diverges in probability.
All child observed ranges eventually contain their fixed cut locations.
The probability that any observed continuous feature is within $n^{-3}$
of any of its finitely many true cuts is at most $C n n^{-3}=o(1)$.
On the complementary event, cut neighborhoods of width $n^{-3}$ preserve
all sample routes and are admissible. Their uniform threshold probabilities
are at least $cn^{-3}$. Required discrete-label gaps have fixed positive
mass after the corresponding labels appear. The fixed topology and feature
factors have fixed positive mass. Restricting and normalizing admissible trees
can only increase the mass of this admissible witness event.

On a compact nuisance-scale event of positive prior mass, each bounded truth
coefficient has a Gaussian interval of radius $a_n/C$ with probability at
least $ca_n$. There are a fixed number of coefficients, and each prediction
sums a fixed number of tree values. The proper noise density gives an
additional interval mass at least $ca_n$. These factors cost at most
$C\log n$ in negative log mass. No H\"older approximation is needed.
Finally $\log n=o(Na_n^2)$ and $\sqrt N a_n^2\to\infty$ because
$0<\eta<1/4$. Apply Theorem~\ref{rfcons:c:thm:union}, rather than the growing-forest prediction theorem. Its denominator proof supplies \eqref{rfcons:f:eq:fixedprediction}.
\end{proof}

\begin{lemma}[All-rectangle information]\label{rfcons:f:lem:rectangles}
For a sufficiently large fixed $A$, let $h_n=A\log n/n$.
With probability tending to one, uniformly over every axis-aligned rectangle
$C\subset[0,1]^p$ and every cohort $g$, if $P_X(C)\ge h_n$ then
\begin{equation}
 \tfrac12 P(X\in C,G=g)\le n_{Cg}/n
 \le\tfrac32 P(X\in C,G=g).
 \label{rfcons:f:eq:rectdense}
\end{equation}
If $P_X(C)<h_n$, then its empirical covariate mass is at most $C_1h_n$.
Rectangles may be chosen after observing data or a posterior forest draw.
\end{lemma}
\begin{proof}
Apply the mesh proof in Lemma~\ref{rfcons:s:lem:intervals} in every
coordinate. A coordinate slab of width $1/\lceil n^3\rceil$ has marginal
probability at most $C/n^3$, by the joint density upper bound. The probability
that any marginal mesh slab contains two sample points is $O(p/n)=o(1)$.
Inscribing or enclosing an arbitrary rectangle by mesh-endpoint rectangles
changes at most $2p$ boundary slabs, with empirical mass at most $2p/n$ and
population mass at most $C p/n^3$ on that event.

There are $O(n^{6p})$ mesh-endpoint rectangles. The elementary Bernoulli
exponential-moment calculation in that lemma applies to each rectangle/cohort
indicator and each marginal rectangle indicator. Use
$x=(6p+4)\log n$ in its Chernoff bound. A union bound and the boundary-slab
comparison give uniformly
\[
 |n_{Cg}/n-p_{Cg}|
 \le C_2\{\sqrt{p_{Cg}\log n/n}+\log n/n\}.
\]
Overlap yields $p_{Cg}\ge e_*h_n$ for dense rectangles, and a sufficiently
large fixed $A$ makes the relative error at most one-half. The marginal
bound gives the sparse-rectangle conclusion. Fixed $p$ is essential to the
constant and mesh exponent, but no coordinate independence is assumed.
\end{proof}

\subsubsection{Total effects and averages}

\begin{theorem}[Fixed-capacity effect consistency]\label{rfcons:f:thm:fixedeffects}
For each fixed post-adoption total-effect coordinate,
\begin{equation}
 \int\|\tau_{gt}-\tau_{gt,0}\|_{L_2(P_X)}^2
 d\PiN(\cdot\mid\mathcal D_n)=O_P(a_n^2\log n)=o_P(1).
 \label{rfcons:f:eq:fixedresult}
\end{equation}
The same bound holds for its all-unit empirical covariate squared error.
Posterior-mean total effects are consistent in integrated loss, and fixed
averages with $L_2(P_X)$ density ratios and realized equal-row ATT/subgroup-
treated averages with positive limiting fractions are consistent.
Population ATE/GATE and integrated CATE interpretation uses the explicit
causal assumptions in Section~\ref{rfcons:c:causal}, including transportability when
averaging beyond the treated cohort. This statement does not assert
pointwise CATE or coverage.
\end{theorem}
\begin{proof}
Let $R_\tau$ be the squared norm of all effect-leaf coefficients and the
optional Gaussian intercept. The number of these coefficients is at most
$C_{K_*}$, and their fixed Gaussian variances have a fixed upper bound.
For some fixed $t_0>0$, uniformly over all candidate trees,
$E_{\PiN}e^{t_0R_\tau}\le e^{C_{K_*}}$.
Lemma~\ref{rfcons:c:lem:projection} gives $\Delta_n^2=O_P(\log n/N)$.
The global variance-profile calculation in Lemma~\ref{rfcons:s:lem:coefficients} therefore
gives a likelihood-ratio supremum with log $O_P(1+\log n)=o_P(Na_n^2)$.
Combining this with \eqref{rfcons:f:eq:fixedprediction}, the Gaussian prior tail and its tail integral yields
\begin{equation}
 \int R_\tau\,d\PiN(\cdot\mid\mathcal D_n)=O_P(Na_n^2).
 \label{rfcons:f:eq:fixedcoeff}
\end{equation}
This is the argument of Lemma~\ref{rfcons:s:lem:coefficients} with a fixed coordinate bound;
it does not require a nuisance-height or nuisance-scale tail bound.
Each total effect is a fixed sum of coefficients, so its squared supremum
is at most $C R_\tau$.

For each candidate draw, refine all its continuous cut hyperplanes and all
cut hyperplanes of the fixed truth witness. There are at most $K_*+K_0$
cut locations, with $K_0$ fixed. The resulting rectangular partition has
at most $L_*=(K_*+K_0+1)^p$ cells, a fixed finite number. Both normalized
$\beta$ and $\beta_0$ are constant on each cell. On cells of population
mass at least $h_n$, \eqref{rfcons:f:eq:rectdense} compares population and empirical panel energies.
The coding bound gives both population and empirical dense-region coordinate
error at most $C\|f-f_0\|_N^2$.

The remaining cells have total population mass at most $L_*h_n$ and
empirical mass at most $C L_*h_n$. Bounded true effects and the Gaussian
coefficient supremum bound give sparse-region total-effect error at most
$C L_*h_n(R_\tau+1)$ for either measure. Integrate both region bounds
and use \eqref{rfcons:f:eq:fixedprediction} and \eqref{rfcons:f:eq:fixedcoeff}:
\[
 O_P(a_n^2+h_nNa_n^2+h_n)=O_P(a_n^2\log n)=o_P(1).
\]
Fixed $T$ absorbs $N/n$; $Na_n^2\to\infty$ absorbs the last term.
Markov and Jensen give posterior and posterior-mean integrated consistency.
Cauchy--Schwarz gives fixed-average consistency. For realized treated or
fixed positive-probability subgroup row sets, divide the all-unit empirical
error bound by their fraction, bounded away from zero by the iid law of
large numbers. Fixed finite combinations of period/cohort averages preserve
the conclusion.
\end{proof}

The theorem establishes a nonempty continuous-covariate truth class for
fixed capacities. It does not remove the known non-density of fixed-capacity
forests over unrestricted smooth functions, characterize their pseudo-target
outside this class, or certify a finite run of the source sampler.
Empirical outcome standardization transfers through the compact original-scale pushforward in Section~\ref{rfcons:c:scaling}; the Gaussian effect exponential-moment constants
remain uniform. Constants in this proof can be very large for production
forest counts, so the displayed bound is an asymptotic consistency statement
rather than a practical sample-size or sampling-budget recommendation.

\begingroup
\small
\providecommand{\Pp}{\mathbb{P}}
\providecommand{\one}{\mathbf{1}}

\subsection{Algebraic Representation of the Forest Hierarchy and Posterior Functionals}\label{sec:imp_extension}

This section records the matrix representation of the three-forest hierarchy on the standardized outcome scale, establishes the conjugate variance updates and properness of the working posterior, characterizes the quotient space of likelihood-identified linear functionals together with chart-level shrinkage bounds, and quantifies Monte Carlo approximation dispersion under a finite posterior draw budget.

\subsubsection{Matrix Representation and Target Functionals}

Let $N_{\mathrm{obs}} = nT$ denote the total number of stacked panel observations across $n$ cross-sectional units and $T \ge 2$ fixed panel periods, with baseline covariates $X_i$ measured prior to treatment adoption and a nonempty set of treated post-treatment observations ($N_{\mathrm{tr}} = \sum_{i=1}^n \sum_{t=1}^T D_{it} \ge 1$). For an admissible forest chart $H$ indexing the tree topologies and split thresholds across the three ensembles, let $U_a \in \{0,1\}^{N_{\mathrm{obs}} \times K_a}$, $U_b \in \{0,1\}^{N_{\mathrm{obs}} \times K_b}$, and $U_\tau \in \{0,1\}^{N_{\mathrm{obs}} \times K_\tau}$ denote the concatenated terminal-leaf indicator matrices of the cohort-level forest $a(G_i, X_i)$, the calendar-time trend forest $b(t, X_i)$, and the treatment-effect forest $\tau_{\mathrm{forest}}(G_i, k_{it}, X_i)$, where $K_a$, $K_b$, and $K_\tau$ denote the respective total leaf counts across trees in each ensemble. Writing $D \in \{0,1\}^{N_{\mathrm{obs}}}$ for the stacked post-treatment indicator vector with row entries $D_r = \ind{G_{i(r)} \le t(r),\, G_{i(r)} < \infty}$, define the composite design matrix and stacked coefficient vector
\[
 F_H = \big[\, U_a,\; U_b,\; D,\; \operatorname{diag}(D)U_\tau \,\big], \qquad
 \gamma = (\gamma_a^\top,\, \gamma_b^\top,\, \taug,\, \gamma_\tau^\top)^\top \in \R^{p_H},
\]
where $p_H = K_a + K_b + 1 + K_\tau$. Let $\bar{Y} = N_{\mathrm{obs}}^{-1}\sum_{r=1}^{N_{\mathrm{obs}}} Y_r$ and $s_Y = \{N_{\mathrm{obs}}^{-1}\sum_{r=1}^{N_{\mathrm{obs}}} (Y_r - \bar{Y})^2\}^{1/2} > 0$ denote the empirical mean and standard deviation of all panel outcomes, and define the standardized outcome vector $Y^z = (Y - \bar{Y}\one_{N_{\mathrm{obs}}})/s_Y$. Conditional on $(\gamma, v, H)$, the Gaussian working likelihood is
\[
 Y^z \mid \gamma, v, H \sim \mathcal{N}(F_H\gamma,\, v I_{N_{\mathrm{obs}}}).
\]
In this architecture, the level forest partitions $(G_i, X_i)$, the trend forest partitions $(t, X_i)$, and the treatment-effect forest partitions $(G_i, k_{it}, X_i)$ (or a designated subset of baseline effect modifiers together with event time and cohort).

For any nonempty realized target row set $\cS \subseteq \{r \in \{1, \dots, N_{\mathrm{obs}}\} : D_r = 1\}$ and nonnegative weights $(\pi_r)_{r \in \cS}$ satisfying $\sum_{r \in \cS} \pi_r = 1$, define the chart-specific linear contrast vector
\[
 q_H = \big(0_{K_a}^\top,\, 0_{K_b}^\top,\, 1,\, (w_{j\ell})_{j,\ell}^\top\big)^\top, \qquad
 w_{j\ell} = \sum_{r \in \cS} \pi_r \ind{\text{row } r \text{ belongs to leaf } (j, \ell) \text{ of } \tau_{\mathrm{forest}}}.
\]
Because the sample mean $\bar{Y}$ is restored solely to the baseline prognostic prediction, the corresponding causal contrast on the original outcome scale is $s_Y q_H^\top \gamma$. Taking $\cS = \{r : D_r = 1\}$ with uniform weights $\pi_r = 1/N_{\mathrm{tr}}$ yields the realized overall $\ATT$, weighting treated units in proportion to their observed post-treatment follow-up periods; restricting $\cS$ to a single cohort-period cell $(g, t)$ or event-time cell $k = t - g$ yields the realized $\GATT(g, t)$ or event-study $\ATT(k)$, respectively.

On the standardized outcome scale, conditional on the forest chart $H$ and the shared baseline leaf-variance parameters $\eta = (\lambda_a, \lambda_b)$, the hierarchical Gaussian prior specifies
\[
 \gamma \mid H, \eta \sim \mathcal{N}(0, D_H), \qquad
 D_H = \operatorname{diag}\!\left(\lambda_a I_{K_a},\; \lambda_b I_{K_b},\; 1,\; \tfrac{1}{200} I_{K_\tau}\right),
\]
paired with independent inverse-gamma variance priors
\[
 \lambda_a, \lambda_b \stackrel{\mathrm{ind}}{\sim} \operatorname{IG}(3/2, 1/250), \qquad
 v \sim \operatorname{IG}(2, 1),
\]
where $\operatorname{IG}(a, b)$ has density $b^a x^{-a-1}\exp(-b/x)/\Gamma(a)$ for $x > 0$. In particular, the global intercept $\taug$ is assigned a zero-mean, unit-variance Gaussian prior on the standardized outcome scale ($\taug \sim \mathcal{N}(0, 1)$), corresponding to prior variance $s_Y^2$ on the original outcome scale, and the prior covariance $D_H$ is specified independently of the standardized residual variance $v$. In the default specification, the treatment-effect leaf variance is fixed at $0.5/100 = 1/200$, the baseline leaf-variance scales $(\lambda_a, \lambda_b)$ are initialized at $2/125$ prior to Gibbs updates, the three ensembles comprise $(125, 125, 100)$ trees with maximum depth caps $(10, 10, 5)$, and each terminal leaf is constrained to contain at least five panel observations.

\subsubsection{Conjugate Scale Updates, Posterior Properness, and Quotient Identification}

\begin{proposition}[Conjugate leaf-variance and residual-variance updates]\label{imp:prop:scale}
Consider a forest ensemble with $K \ge 1$ terminal leaves, conditional Gaussian leaf prior $\gamma_{\mathrm{leaf}} \mid \lambda \sim \mathcal{N}(0, \lambda I_K)$, and squared leaf norm $Q_\lambda = \norm{\gamma_{\mathrm{leaf}}}^2$. Under the half-parameterized inverse-gamma hyperprior $\lambda \sim \operatorname{IG}(a_0/2, b_0/2)$ with $a_0, b_0 > 0$, the conditional posterior distribution of $\lambda$ given $\gamma_{\mathrm{leaf}}$ is $\operatorname{IG}((a_0 + K)/2, (b_0 + Q_\lambda)/2)$. For the baseline hyperparameters $(a_0, b_0) = (3, 1/125)$, the induced hyperprior $\lambda \sim \operatorname{IG}(3/2, 1/250)$ has prior mean $1/125$ and infinite variance. Separately, under the standardized Gaussian likelihood $Y^z \mid \gamma, v, H \sim \mathcal{N}(F_H\gamma, v I_{N_{\mathrm{obs}}})$ and residual variance prior $v \sim \operatorname{IG}(a_v, b_v)$ with $a_v, b_v > 0$, the conditional posterior of $v$ given $(Y^z, \gamma, H)$ is $\operatorname{IG}(a_v + N_{\mathrm{obs}}/2, b_v + \mathrm{RSS}/2)$, where $\mathrm{RSS} = \norm{Y^z - F_H\gamma}^2$.
\end{proposition}
\begin{proof}
Conditional on $\lambda$, the independent Gaussian leaf parameters contribute the scale likelihood factor $\lambda^{-K/2}\exp\{-Q_\lambda/(2\lambda)\}$. Multiplying by the inverse-gamma prior kernel $\lambda^{-a_0/2 - 1}\exp\{-b_0/(2\lambda)\}$ yields the conditional posterior kernel $\lambda^{-(a_0 + K)/2 - 1}\exp\{-(b_0 + Q_\lambda)/(2\lambda)\}$, which is the density kernel of $\operatorname{IG}((a_0 + K)/2, (b_0 + Q_\lambda)/2)$. When $(a_0, b_0) = (3, 1/125)$, the shape parameter is $3/2 > 1$ and the rate parameter is $1/250$, giving prior mean $(1/250)/(3/2 - 1) = 1/125$, whereas the second moment diverges because the shape parameter does not exceed $2$. Analogously, multiplying the Gaussian likelihood kernel $v^{-N_{\mathrm{obs}}/2}\exp\{-\mathrm{RSS}/(2v)\}$ by $v^{-a_v - 1}\exp(-b_v/v)$ yields $\operatorname{IG}(a_v + N_{\mathrm{obs}}/2, b_v + \mathrm{RSS}/2)$.
\end{proof}

\begin{proposition}[Properness of the hierarchical working posterior]\label{imp:prop:proper}
For any normalized prior probability measure over the depth-capped admissible forest charts $H$, the hierarchical prior specification above induces a proper joint working posterior for every finite standardized outcome vector $Y^z \in \R^{N_{\mathrm{obs}}}$ with $N_{\mathrm{obs}} \ge 1$.
\end{proposition}
\begin{proof}
Conditional on $(\gamma, v, H)$, the Gaussian likelihood density is strictly positive and bounded above by $(2\pi v)^{-N_{\mathrm{obs}}/2}$. Applying Tonelli's theorem to the nonnegative joint integrand, integrating first over the normalized conditional Gaussian law of $\gamma \mid H, \eta$, next over the normalized prior laws of $\eta = (\lambda_a, \lambda_b)$ and $H$, and finally over $v \sim \operatorname{IG}(a_v, b_v)$ yields the marginal evidence bound
\[
 0 < Z(Y^z) \le \int_0^\infty (2\pi v)^{-N_{\mathrm{obs}}/2} \frac{b_v^{a_v}}{\Gamma(a_v)} v^{-a_v - 1} e^{-b_v/v} \, dv
 = (2\pi b_v)^{-N_{\mathrm{obs}}/2} \frac{\Gamma(a_v + N_{\mathrm{obs}}/2)}{\Gamma(a_v)} < \infty.
\]
Strict positivity of $Z(Y^z)$ follows because $Y^z$ is finite and $D_H$ and $v$ are positive definite and strictly positive almost everywhere under the prior; under the default $(a_v, b_v) = (2, 1)$, the upper bound simplifies to $(2\pi)^{-N_{\mathrm{obs}}/2}\Gamma(2 + N_{\mathrm{obs}}/2) < \infty$.
\end{proof}

\begin{proposition}[Quotient parameterization and unidentified target remainder]\label{imp:prop:quotient}
Fix an admissible forest chart $H$ and positive variance parameters $(\eta, v)$. Because $D \ne 0$ is a column of $F_H$, $r_H = \operatorname{rank}(F_H) \ge 1$. Factor $F_H = A R$, where $A \in \R^{N_{\mathrm{obs}} \times r_H}$ has full column rank $r_H$ and $R \in \R^{r_H \times p_H}$ has full row rank $r_H$, and set $\Sigma = R D_H R^\top$. Then $\Sigma$ is symmetric positive definite, $b = R\gamma \sim \mathcal{N}(0, \Sigma)$ under the conditional prior, and $\gamma$ decomposes orthogonally as
\[
 \gamma = D_H R^\top \Sigma^{-1} b + \xi, \qquad
 \xi \perp b, \qquad
 \xi \sim \mathcal{N}(0, C_H), \qquad
 C_H = D_H - D_H R^\top \Sigma^{-1} R D_H.
\]
For any linear functional vector $q \in \R^{p_H}$, the conditional posterior given $(Y^z, H, \eta, v)$ decomposes as
\[
 q^\top \gamma = t^\top b + q^\top \xi, \qquad
 t = \Sigma^{-1} R D_H q,
\]
where $\xi \mid Y^z, H, \eta, v \sim \mathcal{N}(0, C_H)$ retains its conditional prior distribution independently of $b$. Consequently, $\Var(q^\top \gamma \mid Y^z, H, \eta, v) \ge q^\top C_H q$. Moreover, $q^\top \gamma$ is a likelihood-identified linear functional of $b$ if and only if $q \perp \ker F_H$, equivalently $q = R^\top L$ for a unique $L \in \R^{r_H}$, in which case $q^\top C_H q = 0$ and $t = L$.
\end{proposition}
\begin{proof}
Because $R$ has full row rank, $R^\top z \ne 0$ for every nonzero $z \in \R^{r_H}$, and positive definiteness of $D_H$ implies $z^\top \Sigma z = (R^\top z)^\top D_H (R^\top z) > 0$. Setting $\xi = \gamma - D_H R^\top \Sigma^{-1} R\gamma$, direct calculation gives $\operatorname{Cov}(\xi, b) = D_H R^\top - D_H R^\top \Sigma^{-1}\Sigma = 0$ and $\Var(\xi) = C_H$; joint Gaussianity therefore yields $\xi \perp b$ under the conditional prior. Because the Gaussian likelihood depends on $\gamma$ solely through $F_H\gamma = Ab$, conditioning on $Y^z$ updates only $b$ and leaves $\xi \sim \mathcal{N}(0, C_H)$ independent of $b$, establishing the variance lower bound. Factoring
\[
 C_H = D_H^{1/2}(I_{p_H} - P_H)D_H^{1/2}, \qquad
 P_H = D_H^{1/2} R^\top (R D_H R^\top)^{-1} R D_H^{1/2},
\]
notes that $P_H$ is the orthogonal projector onto $\operatorname{col}(D_H^{1/2} R^\top)$. Hence $q^\top C_H q = 0$ if and only if $D_H^{1/2} q \in \operatorname{col}(D_H^{1/2} R^\top)$, which holds if and only if $q = R^\top L$ for some $L \in \R^{r_H}$. Injectivity of $A$ implies $\ker F_H = \ker R$, so $q \in \operatorname{col}(R^\top)$ is equivalent to $q \perp \ker F_H$; uniqueness of $L$ follows from injectivity of $R^\top$, and substituting $q = R^\top L$ gives $t = \Sigma^{-1}\Sigma L = L$.
\end{proof}

Additive constants (or covariate functions supported on the intersection of the level and trend partitions) shifted between $U_a\gamma_a$ and $U_b\gamma_b$, as well as additive constants transferred between the global intercept $\taug$ and constant directions of the effect forest $U_\tau\gamma_\tau$, belong to $\ker F_H$ and annihilate $q_H$. By contrast, if a trend leaf indicator coincides with $D$ on an empirically cohort-pure covariate cell in a finite sample, shifting mass between that baseline leaf and a treatment leaf alters $q_H^\top \gamma$ while preserving $F_H\gamma$; chart-level identification of $q_H^\top \gamma$ therefore requires the overlap condition $q_H \perp \ker F_H$.

\begin{lemma}[Pathwise drift and target-energy shrinkage]\label{imp:lem:energy}
On a chart $H$ where the target functional is identified ($q_H = R^\top L$), put $G = A^\top A$, $\Lambda = \Sigma^{-1}$, $K = G + v\Lambda$, $\widehat{b} = G^{-1}A^\top Y$, and $b^\circ = G^{-1}A^\top f_0$. Under a Gaussian prior $b \sim \mathcal{N}(\mu, \Lambda^{-1})$ and working likelihood $Y \mid b \sim \mathcal{N}(Ab, v I_{N_{\mathrm{obs}}})$, the conditional posterior mean and variance of $L^\top b$ are
\[
 m = L^\top \widehat{b} - v L^\top K^{-1}\Lambda(\widehat{b} - \mu), \qquad
 s^2 = v L^\top K^{-1} L,
\]
and for $Y = f_0 + u$ with $a = A K^{-1} L$, the error decomposes pathwise as
\[
 m - \theta_{0,n} = (L^\top b^\circ - \theta_{0,n}) - v L^\top K^{-1}\Lambda(b^\circ - \mu) + a^\top u.
\]
If $v\Lambda \preceq \epsilon G$ in Loewner order for some $\epsilon > 0$, then for every $z \in \R^{r_H}$,
\[
 |v L^\top K^{-1}\Lambda z|^2 \le \epsilon s^2 (z^\top \Lambda z).
\]
Consequently, $n s^2 = O(1)$ and $\epsilon (b^\circ - \mu)^\top \Lambda (b^\circ - \mu) = o(1)$ suffice for negligible $\sqrt{n}$-scale prior-shrinkage drift on that chart.
\end{lemma}
\begin{proof}
Conjugate Gaussian conditioning yields the posterior precision $v^{-1}(G + v\Lambda) = v^{-1}K$, variance $s^2 = v L^\top K^{-1} L$, and mean $m$. Substituting $\widehat{b} = b^\circ + G^{-1}A^\top u$ and applying the matrix identity $I_{r_H} - v K^{-1}\Lambda = K^{-1}G$ gives the noise term $L^\top K^{-1} G (G^{-1}A^\top u) = a^\top u$ and the pathwise decomposition. When $v\Lambda \preceq \epsilon G$, Cauchy--Schwarz in the $\Lambda$-inner product together with $G \preceq K$ implies
\[
 |v L^\top K^{-1}\Lambda z|^2
 \le \big(v^2 L^\top K^{-1}\Lambda K^{-1} L\big)(z^\top \Lambda z)
 \le \big(\epsilon v L^\top K^{-1} G K^{-1} L\big)(z^\top \Lambda z)
 \le \epsilon s^2 (z^\top \Lambda z),
\]
from which the $\sqrt{n}$-scale bound follows upon setting $z = b^\circ - \mu$.
\end{proof}

\subsubsection{Monte Carlo Approximation Under Finite Draw Budgets}

\begin{proposition}[Monte Carlo approximation from finite posterior draws]\label{imp:prop:finite}
Suppose an exact scalar target posterior conditional on the sample is $\mathcal{N}(c_n, V_p/n)$ with $V_p > 0$, and let $\theta^{(1)}, \dots, \theta^{(S)}$ denote $S \ge 1$ independent posterior draws generated conditionally on the data using auxiliary simulation randomness independent of the sample. Let $\widehat{\theta}_{S,n} = S^{-1}\sum_{s=1}^S \theta^{(s)}$ denote the Monte Carlo posterior mean estimator. Then, conditionally on the sample for every $n \ge 1$ (and hence unconditionally),
\[
 \sqrt{n}(\widehat{\theta}_{S,n} - c_n) \sim \mathcal{N}(0, V_p/S).
\]
If in addition $\sqrt{n}(c_n - \theta_{0,n}) \rightsquigarrow \mathcal{N}(0, V_s)$ as $n \to \infty$, then unconditionally
\[
 \sqrt{n}(\widehat{\theta}_{S,n} - \theta_{0,n}) \rightsquigarrow \mathcal{N}(0, V_s + V_p/S).
\]
For any fixed draw budget $S$, empirical-quantile credible interval endpoints likewise retain nondegenerate simulation randomness on the $n^{-1/2}$ scale, whereas both sources of Monte Carlo error vanish as $S \to \infty$.
\end{proposition}
\begin{proof}
Represent the conditional draws as $\theta^{(s)} = c_n + \sqrt{V_p/n}\, Z_s$ for $s = 1, \dots, S$, where $Z_1, \dots, Z_S \stackrel{\mathrm{i.i.d.}}{\sim} \mathcal{N}(0, 1)$ independently of the sample data. Averaging over $s = 1, \dots, S$ gives $\sqrt{n}(\widehat{\theta}_{S,n} - c_n) = \sqrt{V_p}\,\bar{Z}_S \sim \mathcal{N}(0, V_p/S)$ for every $n \ge 1$. Writing $\sqrt{n}(\widehat{\theta}_{S,n} - \theta_{0,n}) = \sqrt{n}(c_n - \theta_{0,n}) + \sqrt{V_p}\,\bar{Z}_S$, independence between the sample data and $(Z_1, \dots, Z_S)$ implies via characteristic functions that the sum converges weakly to $\mathcal{N}(0, V_s + V_p/S)$. Finally, for any empirical quantile functional $Q_S$ equivariant under strictly increasing affine transformations, $Q_S(\theta^{(1)}, \dots, \theta^{(S)}) = c_n + \sqrt{V_p/n}\, Q_S(Z_1, \dots, Z_S)$, where $Q_S(Z_1, \dots, Z_S)$ is nondegenerate at fixed $S$ and converges in probability to the corresponding standard normal quantile as $S \to \infty$.
\end{proof}

Proposition~\ref{imp:prop:finite} isolates pure Monte Carlo draw dispersion under independent sampling from the Gaussian limit posterior; under Markov chain Monte Carlo updates, serial correlation among retained draws replaces $S$ with an effective sample size governed by the mixing rate of the transition kernel.

\endgroup

\end{document}